\documentclass[12pt,oneside,onecolumn,amsmath,amssymb,prc]{report}
\usepackage{amsmath}
\usepackage{hyperref}
\usepackage{caption}
\usepackage{csquotes}
\usepackage{indentfirst}
 
\usepackage{graphicx,bm,floatflt,amssymb,epsf,psfig,rotate,dcolumn,color}
\newsavebox{\fmbox}                       

\def\e{\epsilon}

\def\bc{\begin{center}}
\def\ec{\end{center}}
\def\be{\begin{equation}}
\def\ee{\end{equation}}
\def\bi{\begin{itemize}}
\def\ei{\end{itemize}}

\newcommand{\br}{\begin{array}}
\newcommand{\er}{\end{array}}
\newcommand{\ba}{\begin{eqnarray}}
\newcommand{\ea}{\end{eqnarray}}

\def\bpmat{\begin{pmatrix}}
\def\epmat{\end{pmatrix}}

\def\lim{\raisebox{-.9ex}{\rlap{\tiny $\e \rightarrow {\tiny 0}$}} \raisebox{.9ex}{lim}}

\begin{document}
\setcounter{page}{0}
\thispagestyle{empty}

\begin{center}
%
%
\vspace*{25pt}
{
\Large\textsc\scshape\bf\ STATISTICAL AND DYNAMICAL MODEL\\
\vspace{.4cm}
STUDIES OF NUCLEAR MULTIFRAGMENTATION\\
\vspace{.6cm}
REACTIONS AT INTERMEDIATE ENERGIES}\\
\vspace{1.2cm}

{\large \emph{By}}\\

\vspace{0.4cm}

{\bf {\Large SWAGATA MALLIK}}\\
 \vspace{0.5cm}
{\bf {\large Enrolment No : PHYS04201204002}}\\
\vspace{0.5cm}
{\large \bf {Variable Energy Cyclotron Centre, Kolkata}}
\vspace{1.0cm}

\emph{\text {A thesis submitted to the}} \\
\vspace{3pt}
\emph{\text{ Board of Studies in Physical Sciences}}\\
\vspace{3pt}
\emph{\text {In partial fulfillment of requirements}} \\
\vspace{3pt}
\emph{\text{for the Degree of}}\\
\vspace{1.0cm}
\textbf{DOCTOR OF PHILOSOPHY}

\vspace{0.3cm}

            \emph{of}\\

\vspace{0.3cm}

            \large{HOMI BHABHA NATIONAL INSTITUTE}

\vspace{1cm}
\begin{figure}[hbt]
\begin{center}
\includegraphics[width=3cm,keepaspectratio=true]{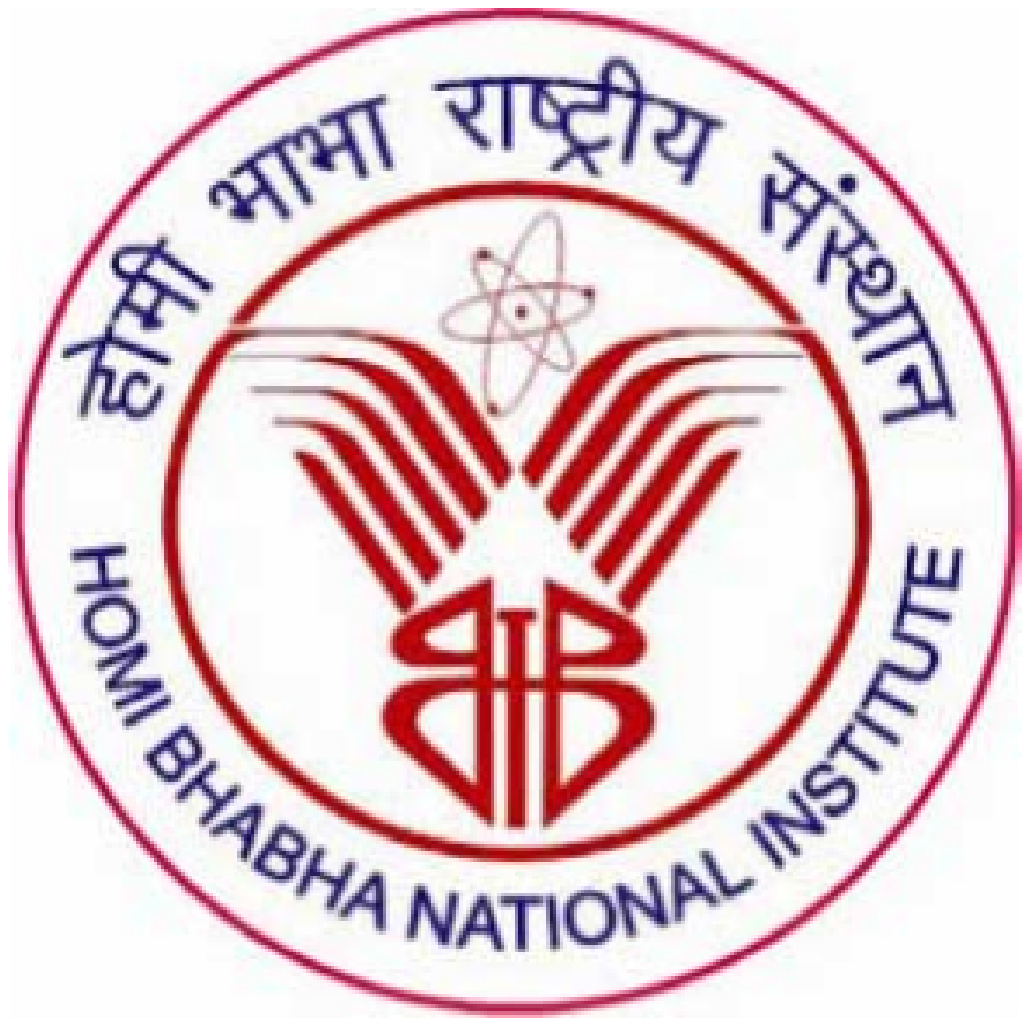}
\end{center}
\end{figure}

\vspace{1cm}

{\large February, 2016}

\end{center}

\setcounter{page}{0}
\pagenumbering{roman}
\newpage
\linespread{1.4}
\thispagestyle{empty}

\vskip5cm
\begin{center}
{\Large \bf{STATEMENT BY AUTHOR\\}}

\vskip 2\baselineskip

\end{center}
\begin{normalsize}
This dissertation has been submitted in partial fulfillment of requirements for an advanced degree at Homi Bhabha National Institute (HBNI) and is deposited in the Library to be made available to borrowers under rules of the HBNI.\\
Brief quotation from this dissertation are allowable without special permission, provided that accurate acknowledgement of source is made. Requests for permission for extended quotation from or reproduction of this manuscript in whole or in part may be granted by the Competent Authority of HBNI when in his or her judgement the proposed use of the material is in the interests of scholarship. In all other instances, however, permission must be obtained from the author.\\

\vskip3cm
\begin{flushright}
Swagata Mallik
\end{flushright}
\end{normalsize} 
\newpage
\thispagestyle{empty}

\vskip5cm
\begin{center}
{\Large \bf{DECLARATION\\}}

\vskip 2\baselineskip

\end{center}
\begin{normalsize}
I, hereby declare that the investigation presented in the thesis has been carried out by me. The work is original and has not been submitted earlier as a whole or in part for a degree/diploma at this or any other Institution/University.
\vskip3cm
\begin{flushright}
Swagata Mallik
\end{flushright}
\end{normalsize} 
\newpage
\begin{center}
{\Large \bf LIST OF PUBLICATIONS}\\
\vskip0.9cm
{\large \bf \underline{(A) Relevant to the present Thesis}}
\vskip0.4cm
{\bf \underline{In refereed journals}}
\end{center}

\begin{normalsize}

\begin{enumerate}
\item \textbf{Model for projectile fragmentation: Case study for Ni on Ta and Be, and Xe on Al,}\\
\emph{S. Mallik}, G. Chaudhuri and S. Das Gupta,\\
Physical Review C \textbf{83}, 044612 (2011).
\item \textbf{Improvements to a model of projectile fragmentation,}\\
\emph{S. Mallik}, G. Chaudhuri and S. Das Gupta,\\
Physical Review C \textbf{84}, 054612 (2011).
\item \textbf{“Conditions for equivalence of statistical ensembles in nuclear multifragmentation,}\\
\emph{S. Mallik}, and G. Chaudhuri,\\
Physics Letters B \textbf{718}, 189 (2012).
\item \textbf{Symmetry energy from nuclear multifragmentation,}\\
\emph{S. Mallik}, and G. Chaudhuri,\\
Physical Review C \textbf{87}, 011602 (2013) (Rapid Communication).
\item \textbf{“Transformation between statistical ensembles in the modeling of nuclear fragmentation,}\\
G. Chaudhuri, F. Gulminelli and \emph{S. Mallik},\\
Physics Letters B \textbf{724}, 115 (2013).
\item \textbf{Temperature of projectile like fragments in heavy ion collisions,}\\
 S. Das Gupta, \emph{S. Mallik} and G. Chaudhuri,\\
Physics Letters B \textbf{726}, 427 (2013).
\item \textbf{Effect of particle fluctuation on isoscaling and isobaric yield ratio of nuclear multifragmentation,}\\
\emph{S. Mallik} and G. Chaudhuri,\\
Physics Letters B \textbf{727}, 282 (2013).
\item \textbf{Estimates for temperature in projectile-like fragments in geometrical and transport models,}\\
\emph{S. Mallik}, S. Das Gupta and G. Chaudhuri,\\
Physical Review C \textbf{89}, 044614 (2014).
\item \textbf{Event simulations in a transport model for intermediate energy heavy ion collisions: Applications to multiplicity distributions,}\\
\emph{S. Mallik}, S. Das Gupta and G. Chaudhuri,\\
Physical Review C \textbf{91}, 034616 (2015).
\item \textbf{Hybrid model for studying nuclear multifragmentation around the Fermi energy domain: The case of central collisions of Xe on Sn,}\\
\emph{S. Mallik}, G. Chaudhuri and S. Das Gupta,\\
Physical Review C \textbf{91}, 044614 (2015).
\end{enumerate}
\vskip0.9cm
\newpage
\begin{center}
{\bf \underline{In conferences}}
\end{center}
\begin{enumerate}
\item \textbf{Variation of multiplicity of intermediate mass fragments and differential charge distributions with Zbound in projectile fragmentation reactions,}\\
\emph{S. Mallik}, G. Chaudhuri and S. Das Gupta\\
Proceedings of the DAE-BRNS Symposium on Nuclear Physics \textbf{56}, 862 (2011)
\item \textbf{Study of the charge, mass and isotopic distribution in projectile fragmentation reactions}\\
G. Chaudhuri, \emph{S. Mallik} and S. Das Gupta\\
Proceedings of the DAE-BRNS Symposium on Nuclear Physics \textbf{56}, 760 (2011)
\item \textbf{Study of symmetry energy to temperature ratio in in projectile fragmentation reaction}\\
\emph{S. Mallik}, G. Chaudhuri and S. Das Gupta\\
Proceedings of the DAE-BRNS Symposium on Nuclear Physics \textbf{57}, 744 (2012)
\item \textbf{Universality of projectile fragmentation model}\\
G. Chaudhuri, \emph{S. Mallik} and S. Das Gupta\\
Proceedings of the DAE-BRNS Symposium on Nuclear Physics \textbf{57}, 746 (2012)
\item \textbf{Isoscaling parameter in nuclear multifragmentation}\\
\emph{S. Mallik} and G. Chaudhuri\\
Proceedings of the DAE-BRNS Symposium on Nuclear Physics \textbf{57}, 748 (2012)
\item \textbf{Nuclear multifragmentation: Basic Research and Application}\\
\emph{S. Mallik} and G. Chaudhuri\\
Proceedings of the Conference on \enquote{Challenges of Basic Research in Innovating Technologies-2013}, Page No-31, Narosa Publishing House (2015)
\item \textbf{A model for projectile fragmentation}\\
G. Chaudhuri, \emph{S. Mallik} and S. Das Gupta\\
NN 2012 conference proceeding, Journal of Physics: Conference Series \textbf{420}, 012098 (2013)
\item \textbf{Searching the conditions of convergence of statistical ensembles for fragmentation of finite nuclei}\\
G. Chaudhuri and \emph{S. Mallik}\\
Proceedings of the International Symposium on Nuclear Physics \textbf{58}, 540 (2013)
\item \textbf{Determination of temperature profile in projectile fragmentation: A microscopic static model approach}\\
\emph{S. Mallik}, G. Chaudhuri and S. Das Gupta\\
Proceedings of the International Symposium on Nuclear Physics \textbf{58}, 542 (2013)
\item \textbf{Nuclear multifragmentation: Basic concepts}\\
G. Chaudhuri and \emph{S. Mallik} and S. Das Gupta\\
Proceedings of the National Conference on Nuclear Physics-2013, Pramana-Journal of physics \textbf{82}, 907 (2014)
\item \textbf{Determination of initial conditions of projectile fragmentation from transport model calculations}\\
\emph{S. Mallik}, G. Chaudhuri and S. Das Gupta\\
Proceedings of the DAE-BRNS Symposium on Nuclear Physics \textbf{59}, 522 (2014)
\item \textbf{A hybrid model for studying nuclear multifragmentation around the Fermi energy domain}\\
\emph{S. Mallik}, G. Chaudhuri and S. Das Gupta\\
Proceedings of the DAE-BRNS Symposium on Nuclear Physics \textbf{59}, 348 (2014)
\item \textbf{Signatures of nuclear liquid gas phase transition from transport model calculations for
intermediate energy heavy ion collisions}\\
\emph{S. Mallik}, S. Das Gupta and G. Chaudhuri\\
Proceedings of the DAE-BRNS Symposium on Nuclear Physics \textbf{60}, 510 (2015)
\end{enumerate}
\vskip0.9cm
\newpage
\begin{center}
{\large \bf \underline{(B) Other publications}}
\vskip0.4cm
{\bf \underline{In refereed journals}}
\end{center}
\begin{enumerate}
\item \textbf{Effect of secondary decay on isoscaling : Result from canonical thermodynamical model,}\\
G. Chaudhuri and \emph{S. Mallik},\\
Nucl. Phys. \textbf{A} \textbf{849}, 190 (2011).
\item \textbf{Liquid Gas Phase transition in hypernuclei,}\\
\emph{S. Mallik}, and G. Chaudhuri,\\
Physical Review C \textbf{91}, 054603 (2015).
\item \textbf{Finite size effects on the phase diagram of the thermodynamical cluster model,}\\
\emph{S. Mallik}, F. Gulminelli and G. Chaudhuri,\\
Physical Review C \textbf{92}, 064605 (2015).
\end{enumerate}
\begin{center}
{\bf \underline{In conferences}}
\end{center}
\begin{enumerate}
\item \textbf{Effect of Secondary Decay on Isoscaling from Canonical Thermodynamical Model,}\\
G. Chaudhuri and \emph{S. Mallik},\\
Proceedings of the DAE-BRNS Symposium on Nuclear Physics \textbf{55}, 482 (2010)
\item \textbf{Hypernuclear liquid gas phase transition,}\\
P. Das, G. Chaudhuri and \emph{S. Mallik},\\
Proceedings of the DAE-BRNS Symposium on Nuclear Physics \textbf{60}, 362 (2015)
\end{enumerate}
\end{normalsize} 
\newpage
%
%
%
%
%
%
%
%
\thispagestyle{empty}
\vspace*{225pt}
\begin{center}
{\Huge\textsc\scshape{\bf\ {\it Dedicated to my family}}} \\
\end{center}
\vskip3cm 
\newpage
\thispagestyle{empty}

\vskip5cm
\begin{center}
{\Large \bf{ACKNOWLEDGEMENTS\\}}

\vskip 2\baselineskip

\end{center}
\begin{normalsize}
 \vskip 1cm
I am highly delighted for having this opportunity to thank some persons whose blessings, love and guidance help me in making my research work fruitful. First and foremost, I would like to acknowledge invaluable contribution and guidance of Prof. Gargi Chaudhuri, my thesis supervisor. Her useful suggestions and encouragement greatly inspire me in concentrating on my research work. It would have been impossible for me to develop my thesis work without her help.\\
\indent
I have no words to thank Prof. Subal Das Gupta, McGill University, Canada for his kind support, great encouragement and counsel. He continuously helps me in pursuing my research work with great motivation. I am fortunate enough for being a collaborator of him in this field of research.\\
\indent
I would like to express my gratitude to Prof. Francesca Gulminelli, University of Caen, France for the pleasurable experience in working together. I am thankful to M. Mocko, M. B. Tsang, W. Trautmann, Kelic-Heil Aleksandra and Karl-Heinz Schmidt for giving me the access to experimental data.\\
\indent
I am very much indebted to Prof. Santanu Pal, former head, Theoretical Physics Division, Variable Energy Cyclotron Centre (VECC) and Prof. Dinesh Kumar Srivastava, Director, VECC for their kind initiative in creating the proper atmosphere and providing enough facilities during my research work. They constantly rendered their careful support and guidance in the progress of my work.\\
\indent
I am sincerely thankful to Prof. Asish Kumar Chaudhuri, Head, Theoretical Nuclear Physics Group, VECC and Prof. Jan-e Alam, Dean, Physical Science, Homi Bhabha National Institute for encouraging and suggesting me to make my work more interesting.\\
\indent
I am very much grateful to Prof. Devasish Narayan Basu and Prof. Asish Dhara, VECC and Prof. Debades Bandyopadhyay, Saha Institute of Nuclear Physics for their useful advices and comments.\\
\indent
I would love to acknowledge the active co-operation of my immediate senior colleague Dr. Jhilam Sadhukhan. I have learnt a lot from the countless discussion with him. I am thankful for getting the company of Dr. Partha Pratim Bhaduri, Dr. Parnika Das, Dr. Tilak Kumar Ghosh, Dr. Shashi Srivastava, Dr. Rupa Chatterjee and Sushant Kumar Singh.\\
\indent
I would like to mention the names of Sanjib Muhuri, Dipta Pratim Dutta, Sabir Ali, Tapan Kumar Mandi and Debashis Banerjee for spending with me some joyful moments during my research work.\\
\indent
With great pleasure I express my heartiest thanks to my first Physics teachers Mrs. Manasi Pal (Karmakar) and Mr. Basudev Karmakar for making me devoted to this subject. I pay my due respect to Mr. Dilip Kumar Hore and Mr. Jagabandhu Das of Balagarh High School, Prof. Bikash De of Hooghly Mohsin College and Prof. Bishnu Charan Sarkar, Prof. Souranshu Mukhopadhyay and Prof. Tanmoy Banerjee of Burdwan University for awakening in me the inspiration and motivating me towards the future research on this subject.\\
\indent
I recall the inspiration I got from the ex-colleagues of my first working place Samudragarh High School. Special thanks to my intimate friend Brindaban Modak who helped me in various ways from my college life. I am thankful to my friends and seniors Biplab Debnath, Subrata Nandy, Amit Bhattacharyya, Tanmoy Majumder, Safiul Islam, Asif Islam, Hirakendu Basu, Sanat Kumar Pandit, Sukanta Maity and Biswaranjan Nayak whose love and good wishes encouraged me to proceed in my life spontaneously.\\
\indent
Last but not the least, I express my deep respect to my parents Mr. Gobinda Lal Mallik and Mrs. Lalima Mallik who have continuously energized me not only to pursue my research work but to go forward into each step of life. I would like to appreciate enthusiastic company and cheerful motivation of my wife Salini who has supported me heartily for the completion of my research work. I acknowledge affectionate encouragement of my maternal aunt Mrs. Tanima Bandyopadhyay and maternal uncle Mr. Amar Nath Bhattacharyya who stands beside my family for my betterment. I am also grateful to my all other teachers, relatives, friends, neighbours and colleagues whose good willing helps me to proceed in this journey.

\vskip3cm
\begin{flushright}
Swagata Mallik
\end{flushright}
\end{normalsize}

\newpage
\tableofcontents
\newpage
\addcontentsline{toc}{chapter}{Synopsis}
\thispagestyle{empty}

\vskip5cm
\begin{center}
{\Large \bf{SYNOPSIS\\}}

\vskip 2\baselineskip

\end{center}
\begin{normalsize}
{\it {\bf Introduction:-}}
Nuclear Multifragmentation is an important area of research where an excited system is formed in the collision between two nuclei and when its excitation energy is greater than a few MeV per nucleon, it breaks into many nuclear fragments of different masses. For throwing light on the nuclear multifragmentation reaction and for explaining the relevant experimental data different theoretical models have been developed, which can be classified into two main categories: (i) dynamical models and (ii) statistical models. In this thesis, using statistical and dynamical model calculations, we have concentrated mainly on the following three aspects of multifragmentation reactions namely (i) Production of exotic nuclei which are normally not available in the laboratory (ii) Nuclear liquid-gas phase transition and (iii) Nuclear symmetry energy from heavy ion collisions at intermediate energies. In addition to these equivalence of statistical ensembles under different conditions in the framework of multifragmentation has also been studied.\\
{\it {\bf Development of a model for projectile fragmentation:-}}
Projectile fragmentation reaction is very useful for producing radioactive ion beams. A model for projectile fragmentation has been developed which involves the traditional concepts of heavy-ion reaction (abrasion) plus the well known statistical model of multifragmentation (Canonical thermodynamical Model) and evaporation model based on Weisskopf theory. This model is in general applicable and implementable in the limiting fragmentation region (for beam energies of 100 MeV/nucleon or higher). A very simple impact parameter dependence of freeze-out temperature has been incorporated in the model which helps to analyze the more peripheral collisions. The projectile fragmentation model has been applied successfully to calculate the production cross-sections for a wide range of exotic as well as stable nuclei of different projectile fragmentation reactions at different energies. Different important observables of projectile fragmentation like intermediate mass fragments, largest cluster size, differential charge distribution etc have also been calculated from this model. The calculations have been repeated for reactions with different projectile-target combinations with widely varying projectile energy and have been compared with the experimental data.\\
The initial conjecture for the mass and excitation of projectile like fragment can be approximated by a microscopic static model. The microscopic static model has been further expanded in order to include dynamic effects using a transport model based on Boltzmann-Uehling-Uhlenbeck (BUU) equation. Then the Canonical thermodynamical model has been used to deduce the freeze-out temperature from the calculated excitation. It has been observed that the PLF masses at different impact parameters calculated from the microscopic static model and the BUU transport model are comparable to that obtained from the geometric abrasion model calculation. Nice agreement between the deduced temperature profiles and the previously used parameterized temperature profile has been obtained for different projectile fragmentation reactions at varying energies.\\
{\it {\bf Development of a hybrid model for multifragmentation around Fermi energy domain:-}}
In addition to the projectile fragmentation model, a hybrid model has been also developed separately for explaining the multifragmentation reaction around Fermi energy domain. In the hybrid model, initially the excitation of the colliding system has been calculated by using the dynamical BUU approach with proper consideration of pre-equilibrium emission. Then the fragmentation of this excited system has been calculated by the Canonical thermodynamical model and finally the decay of the excited fragments, which are produced in multifragmentation stage, has been calculated by the evaporation model. This model has been used calculate the freeze-out temperature of the central collision multifragmentation reactions. In order to check the accuracy of the model, different observables of nuclear multifragmentation like charge distribution, largest cluster probability distribution, average size of largest cluster has been calculated theoretically for for $^{129}$Xe on $^{119}$Sn reaction at beam energies of 32, 39, 45 and  50 MeV/nucleon and compared with the experimental data.\\
{\it {\bf Equivalence of statistical ensembles:-}}
Another important aspect studied in this thesis is the equivalence of statistical ensembles under different conditions. The underlying physical assumption behind the canonical and the grand canonical ensembles is fundamentally different, and in principle they agree only in the thermodynamical limit when the number of particles become infinite. In any statistical physics problem it is easier to compute any observable using grand canonical ensemble where total number of particles can fluctuate. For finite nuclei in intermediate energy heavy ion reactions there is no fluctuation in the total number of particles, therefore canonical or microcanonical ensembles are better suited. For the nuclear multifragmentation of finite nuclei the total charge distribution has been calculated in the framework of both canonical and grand canonical ensembles. It is observed that when the fragmentation is more, i.e. the production of larger fragment is less, the particle fluctuation in grand canonical model is less and the results from canonical and grand canonical model have been found to converge. This condition can be achieved by increasing the temperature or freeze-out volume or the source size or by decreasing the asymmetry of the source. When the results calculated from the two models based on canonical and grand canonical ensemble formalisms are different, an analytical formula has been derived which enables one to extract canonical results from a grand canonical calculation and vice versa. The conditions under which the equivalence holds are amenable to present day experiments.\\
{\it {\bf Nuclear symmetry energy from heavy ion collisions:-}}
Study of the nuclear symmetry energy in intermediate energy heavy ion reactions is an important area of research for determining the nuclear equation of state. In this thesis, the symmetry energy coefficient has been determined by different ways (isoscaling source method, isoscaling fragment method, fluctuation method and isobaric yield ratio method) in the framework of the canonical and grand canonical models. Source dependence of isoscaling parameters and source and isospin dependence of isobaric yield ratio parameters have been examined from the canonical and the grand canonical model calculation. Since the formulae that have been used for the deduction of symmetry energy coefficient have all been derived in the framework of grand canonical ensemble, therefore it is better to use the model based on this ensemble rather than canonical one (but canonical models are physically more acceptable for explaining intermediate energy heavy ion reactions). The ratio of the symmetry energy coefficient to temperature ($C_{sym}/T$) has been extracted using the different prescriptions in the framework of the projectile fragmentation model for (i) $^{58}$Ni and $^{64}$Ni on $^{9}$Be at 140 MeV/nucleon and (ii) $^{124}$Xe and $^{136}$Xe on $^{208}$Pb at 1GeV/nucleon and the results have been compared with the experimental data. It has been observed that, the extracted $C_{sym}/T$ values from the primary  fragments are close to each other for all the four prescriptions mentioned above. The values of $C_{sym}/T$ obtained from the secondary fragments are close to those obtained from experimental yields but they differ from those obtained from the primary fragments and the input value used of Csym/T in the model. The main message of this part of the thesis is that the experimental yields which are from the 'cold' fragments should not be used to deduce the value of the symmetry energy coefficient since the formulae used for the deduction are all valid at the break-up stage of the reaction and secondary decay disturbs the equilibrium scenario of the break-up stage.\\
{\it {\bf Nuclear liquid-gas phase transition from dynamical model calculation:-}}
An enormous amount of experimental and theoretical work exists on phase coexistence or liquid-gas phase transition in heavy ion collisions at intermediate energy. The standard methods of theoretical studies on liquid-gas phase transition at intermediate energy collisions assume that because of two body collisions nucleons equilibrate in a given volume and then dissociate into composites of different sizes according to the availability of phase space. This work of the thesis focuses on whether the results of the transport model calculations (BUU) at intermediate energy can reveal signatures of phase transition. This has never been attempted before by using any transport model. In order to study that, a simplified yet accurate method of BUU transport model has been developed which allows calculation of fluctuations in systems much larger than what was considered feasible in a well-known and already existing model. The distribution of clusters obtained from this model has been found to be remarkably similar to that obtained in the equilibrium statistical model and provides evidence of first-order phase transition.\\
\vskip3cm
\end{normalsize} 
\newpage
\listoffigures
\addcontentsline{toc}{chapter}{List of Figures}
\newpage
\listoftables
\addcontentsline{toc}{chapter}{List of Tables}
\newpage
\setcounter{page}{0}
\pagenumbering{arabic}
\chapter{Overview}
\begin{normalsize}
\section{Introduction}
\indent
The journey of understanding the fundamental nature of matter started in the $6^{th}$ century B. C. when philosopher \enquote{Democritus} opined that each kind of material could be subdivided in to the \enquote{smallest indivisible elements invisible to the naked eye} called the \enquote{atom}. The philosophical theory of atom was first scientifically elucidated in 1908 by chemist John Dalton. The journey was boosted with the discovery of radioactivity by \enquote{Becquerel} in $1896$ \cite{Becquerel} and electron by \enquote{J. J. Thomson} in $1897$ \cite{Thomson}. The \enquote{existence of the nucleus as the tiny central part of an atom} was first proposed by \enquote{Rutherford} in $1911$ \cite{Rutherford} which marked the beginning of nuclear physics. In order to understand the stability of the atom and to explain its emission spectra, in $1913$ \enquote{Niels Bohr} prescribed the quantum mechanical analogue of the Rutherford's model. With the discovery of the neutron, the neutral particle by \enquote{Chadwick} in 1932 \cite{Chadwick}, it was established that the nucleus is made up of neutrons and protons and the electrons are moving around the the nucleus. In $1960$'s it was discovered that even the neutrons and protons are not fundamental particles they are made up of quarks. Though, information on the actual nature of nuclear force and the different nuclear properties is still limited and not well established, however, much progress has been made in last seven decades towards its understanding.\\
\indent
The study of nuclear reactions is a diverse field, allowing to address a wide range of nuclear properties and other areas of science and technology. In nuclear reaction, in usual cases, there is a nucleus at rest in the laboratory frame (the target) and another nucleus (the projectile) is accelerated towards the target and hits it. Then, due to collision of the projectile and target nuclei, an excited nuclear system is formed. In $1919$ \enquote{Rutherford} performed the first artificial nuclear reaction by bombarding ordinary nitrogen ($^{14}$N) with $7.68$ MeV $\alpha$ particle emitted from a $^{214}$Po, resulting in emission of proton and production of unstable $^{17}$O nucleus. With time, the accelerators have been built which could produce light as well as heavy ion beams with energy varying from few MeV/nucleon to several TeV/nucleon. The heavy-ion physics is a relatively new domain. Heavy ions are generally defined as the nuclei having mass number greater than $4$. In heavy ion reactions, at each domain of beam energies ranging from the Coulomb barrier of colliding nuclei to ultra-relativistic regime, reaction mechanisms can widely vary.\\
\indent
At low bombarding energy (below $10$ MeV/nucleon), complete fusion or compound nucleus reaction results from most central collision, and binary dissipative (also known as \enquote{deep-inelastic} reactions, following multi-nucleon transfers between the colliding nuclei) and quasi-elastic reactions for increasingly peripheral collisions. When the energy is more than $10$ MeV/nucleon, incomplete fusion process occurs. From $20$ MeV/nucleon to $2$ GeV/nucleon the most dominant reaction channel is nuclear multifragmentation. High energy nuclear reaction (few GeV/nucleon to few TeV/nucleon) focuses on the creation and observation of a new state of matter namely quark-gluon-plasma.\\
\indent
Hence, depending upon the bombarding energies a widely varying phenomena are exhibited by heavy-ion collisions. In this broad scenario, only multifragmentation reaction will be concentrated upon in this thesis. This will be introduced in the next section.
\section{Nuclear Multifragmentation}
The study of nuclear multifragmentation is important for understanding the reaction mechanism at intermediate and high energies. In this case, due to collision of projectile and target nuclei, an excited nuclear system is formed. If its excitation energy is greater than a few MeV/nucleon ($\approx$ 3 to 10 MeV/nucleon), then it breaks into many nuclear fragments of different masses. This is known as nuclear multifragmentation. The name 'multifragmentation' was introduced by \enquote{J. P. Bondorf}. Here 'multi' indicates 'more than two'. Generally at low excitation ($\approx$ 1 to 2 MeV/nucleon) the compound nucleus decays by evaporation of light particles, or if the system is heavy, it breaks  into two fragments by binary fission process. Therefore multifragmentation can be considered as the higher energy version of fission and particle evaporation. It is important to mention that disintegration into more than two fragments also happens in lower energy nuclear reactions (i.e. low excitation) but there the process proceeds sequentially i.e. after one decay the residual system gets time to relax(relaxation time $\tau \approx \frac{2R}{c_s}$, where $R$ is the radius of the compound nuclear system and $c_s$ is the velocity of sound) in a new equilibrium state before the next decay occurs. But in intermediate energy nuclear reactions, the excitation of the system is greater than a few MeV/nucleon ($\approx$ 3 to 10 MeV/nucleon), therefore the time interval between successive emissions is comparable or sometimes lesser to the relaxation time ($\tau$) and the existence of long-lived compound nuclear system is unlikely which leads to the scenario of explosion like process of the whole excited nuclear system. This leads to multifragmentation \cite{Bondorf1,Moretto,Fuchs,Borderie1,JNDe,Cole,Das,Chomaz_report,Durand,Lynch_intro}. The pictorial view of the fission, evaporation and multifragmentation is given in Fig. \ref{Intro}. Multifragmentation is mainly observed in three kinds of reaction (i) light ion induced reactions at large incident energies (in the GeV region) (ii) central heavy ion collisions between 25 MeV/nucleon to 200 MeV/nucleon and (iii) peripheral heavy ion collision from 25 MeV/nucleon to 2GeV/nucleon or above.\\
\begin{figure}[!h]
\begin{center}
\includegraphics[width=13.0cm,keepaspectratio=true,clip]{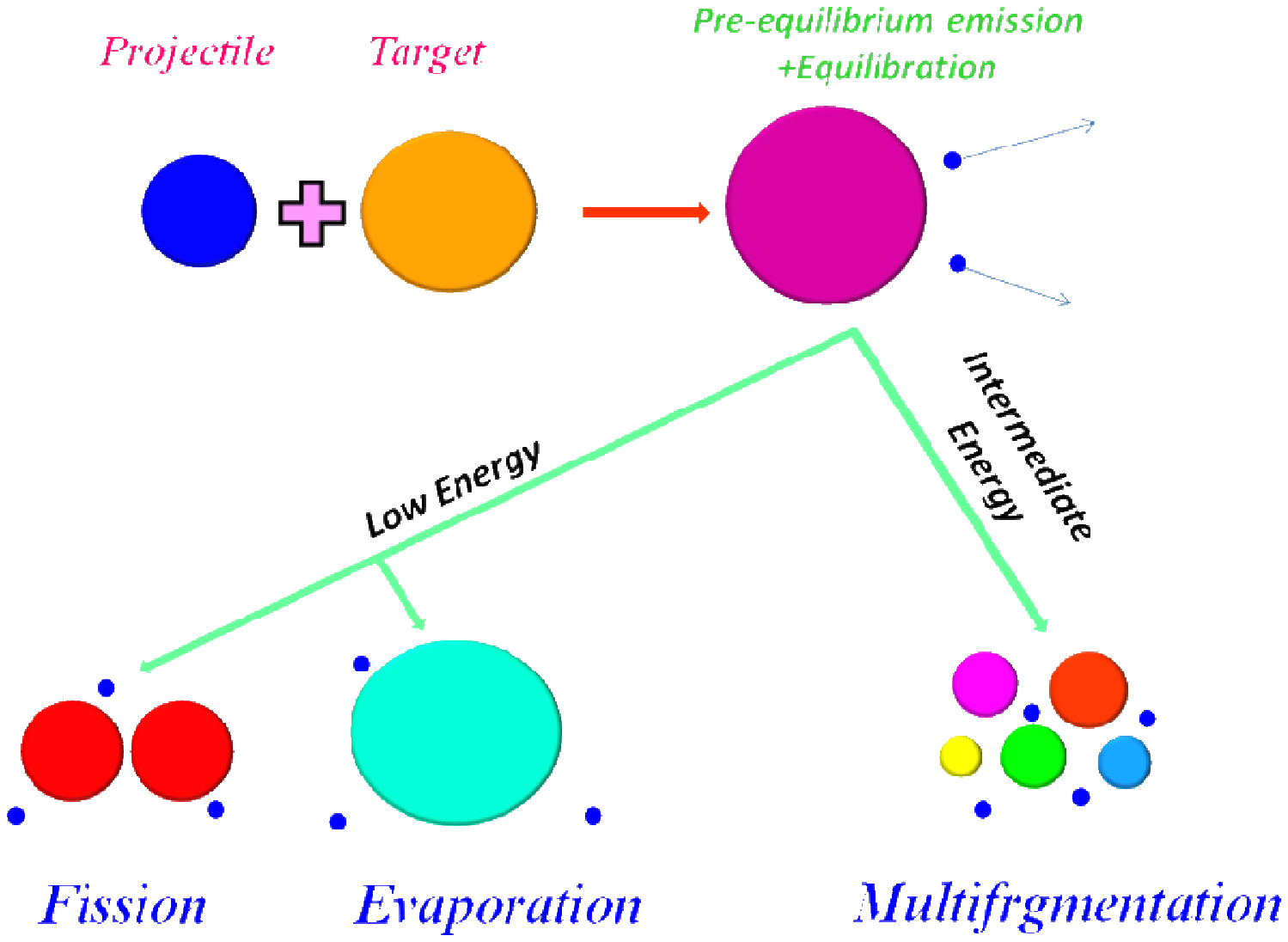}
\caption[Schematic diagram of fission, evaporation and multifragmentation]{Schematic diagram of fission, evaporation and multifragmentation.}
\label{Intro}
\end{center}
\end{figure}
\indent
The evolution of fission (or evaporation) to multifragmentation at higher excitation \cite{Randrup_intro,Engmann} can be easily understood by the fragment mass distribution. For example, mass distribution of different fragments produced from the system of mass $A_0=168$ and charge $Z_0=75$ (it represents $^{112}Sn + ^{112}Sn$ central collisions after pre-equilibrium particle emission) obtained from canonical thermodynamical model \cite{Das} calculation , is shown in Fig. \ref{Intro_mass_distribution}.(a). This model uses the concept of temperature which is quite familiar in heavy ion physics. In this thesis canonical thermodynamical model will be discussed later. At temperature $T=3$ MeV (lower excitation of compound nuclear system) fission is the dominating channel i.e. the multiplicity (total number of fragments) is about 2. But at $T=5$ MeV (moderate excitation), fission channel disappears and multi-fragmentation is the dominant process with a large number of intermediate mass fragments being formed. With further increase of temperature from $5$ MeV to $7$ MeV (very high excitation) the system mainly breaks into a larger number of smaller mass fragments. The variation of total fragment multiplicity with temperature is shown in the right panel of Fig. \ref{Intro_mass_distribution}.(b).\\
\begin{figure}[!h]
\begin{center}
\includegraphics[width=6.1in,height=2.9in,clip]{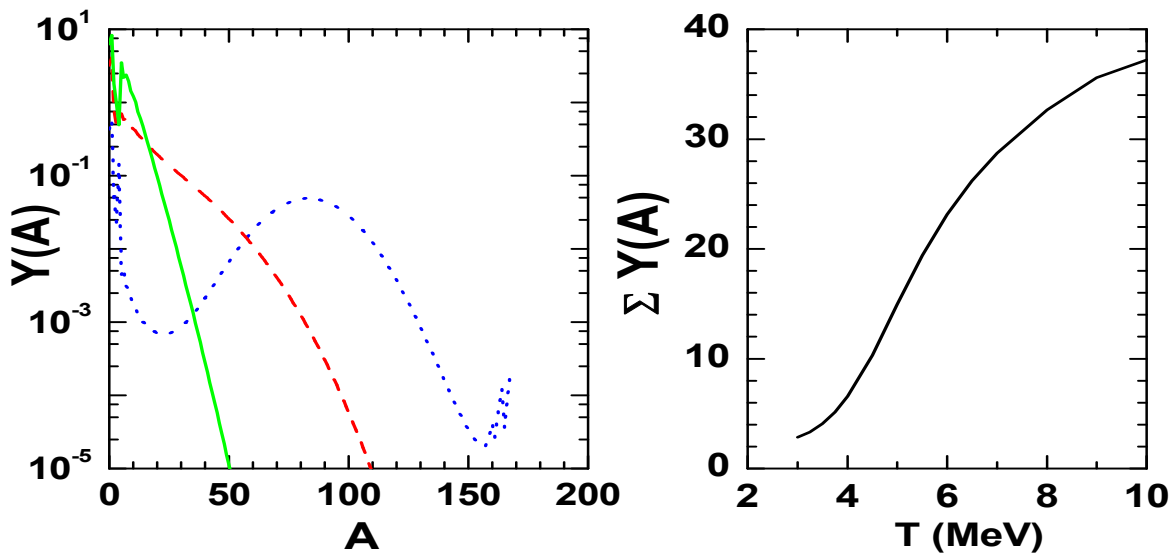}
\caption[Mass distribution: From fission to multifragmentation]{Left panel: Mass distribution from $A_0=168$ and $Z_0=75$ system studied at T=3 MeV (blue dotted line), 5 MeV (red dashed line) and 7 MeV (green solid line). Right panel: Variation of total multiplicity with temperature.}
\label{Intro_mass_distribution}
\end{center}
\end{figure}
\section{Experimental Overview}
The glorious discovery of nuclear multifragmentation happened about 75 years ago from the study of cosmic rays \cite{Gurevich,Schopper}. But cosmic radiation samples a wild mixture of projectiles having different energies, masses and charges. So one had to wait for particle accelerators which can provide sufficiently higher energy projectile beams. Nuclear multifragmentation reactions was studied in the accelerator experiments \cite{Lozhkin,Perfilov,Jacobsson} in 1950's. But at that time the reaction mechanism was completely unclear and this field of research progressed slowly up to the end of 1970's. The situation turns into dramatic progression in 1982 through the observation of multiple intermediate mass fragment (fragments having charge between 3 to 20) emission in Bavalac experiment (at Lawrence Berkeley Laboratory, USA)  of 250 MeV/nucleon Carbon beam on emulsion target \cite{Jacobsson}. After that, over the last thirty years experimental methodology for intermediate energy heavy ion reactions has been developed at National Superconducting Cyclotron Laboratory (NSCL) at Michigan State University (MSU, USA), Superconducting Cyclotron at Texas A$\&$M university (USA), Grand Accelerateur National D'ions Lourds (GANIL, France), Heavy-ion Synchrotron SIS accelerator at  Gesellschaft fur Schwerionenforschung mbH (GSI, Germany), Superconducting Cyclotron at Laboratori Nazionali del Sud in INFN, Catania (Italy), Riken (Japan) etc. In India, beam from superconducting cyclotron for performing experiments of nuclear multifragmentation  will be available soon at Variable Energy Cyclotron Centre, Kolkata.\\
\indent
In order to study the multifragmentation phenomena in the Fermi energy regime, it is desirable to detect the reaction products in a wide angular range, ideally in the $4\pi$ geometry. But comparatively higher energies lead to a strong kinematical focussing of the projectile like reaction products in the forward direction only. Usually the energy, charge and mass of the produced fragments are measured by radiochemical and electronic methods. From this information , total, isotopic and isobaric fragment mass yield, isotope specific fragment angular distribution, kinetic energy spectra etc can be constructed \cite{Veselsky,RDeSouza,Trautmann}. To identify the collision products over the entire mass range two methods are commonly attempted-(i) to develop the detector telescopes with time of flight(ToF) or pulse shape analysis and (ii) to develop high resolution magnetic spectrometers.\\
\indent
Substantial experimental progress on nuclear multifragmentation also sparked theoretical activities. The behaviour of nuclear system at intermediate energy collisions is a fascinating and multi-faceted story and the dynamics of the breaking up of nuclei is not as simple as it appears. There is no unique theory for explaining the proper mechanism of nuclear multifragmentation. Over the years, many theoretical models have been proposed to understand the complete reaction scenario and to explain the experimental observations. A brief survey of different theoretical models will be presented in the next section.
\section{Theoretical models of multifragmentation}
Different theoretical models have been developed for throwing light on the nuclear multifragmentation reaction and for explaining the relevant experimental data. These models differ from one another by the respective physical pictures and mathematical foundations adopted by the authors. The theoretical models can be classified into two main categories: (i) Dynamical models and (ii) statistical models. In next sections the theoretical models will be introduced briefly.
\subsection{Statistical models}
Nuclear multifragmentation reactions at intermediate energies are successfully described by statistical models based on equilibrium scenario of different excited fragments at freeze-out condition \cite{Bondorf1,Das,Gross1,Gross_freeze-out}. Statistical models are computationally much less intensive and clusterizations are done from direct phase space calculation. These models can nicely handle different kinds of experimental data like fragment production cross-section, largest cluster probability, isoscaling etc. In statistical models, one assumes that depending upon the original beam energy, the disintegrating system may undergo an initial compression and then begins to decompress. As the density of the system decreases, each nucleon is no longer able to interact with all its neighbours by means of attractive nuclear forces because, the nuclear forces are of short range. So higher density regions will develop into composites. As this collection of nucleons begins to move outward, rearrangements, mass transfers, nuclear coalescence and most physics will happen until the density decreases so much that the mean free paths for such processes become larger than the dimension of the system. This condition is termed as freeze-out \cite{Bondorf1}.\\
\indent
The disintegration of excited nuclei can be studied by implementation of different statistical ensembles at freeze-out condition. The finite system suggests that calculations by microcanonical and canonical ensembles should be more realistic. Initially Randrup and Koonin developed a microcanonical model based on Metropolis Monte Carlo methods \cite{Randrup,Randrup1}. Gross and his collaborators further developed microcanonical statistical multifragmentation model \cite{Gross1,Gross2} for explaining the nuclear multifragmentation process. Bondorf and his collaborators proposed an alternative statistical treatment known as statistical multifragmentation model (SMM) \cite{Bondorf1}. A large number of comparisons to experimental observables have been done with this model. In SMM, all possible partitions of the system into fragments are considered without invoking a Monte Carlo method but division of energy into between kinetic and internal part of the fragments needs Monte Carlo procedures. Therefore this model also needs very high computation. In addition to that, the internal excitation energy is divided up amongst the fragments in proportion to the fragment mass, fluctuation of excitation energy is not treated as one would expect in a true microcanonial treatment. Furthermore, explicit $N$-body correlations and interactions are ignored.\\
\indent
Canonical Thermodynamical Model (CTM)\cite{Das} which was introduced later can be easily implemented analytically by calculating statistical partition functions using recursion relations without involving the Monte Carlo sampling. The disadvantage of this model is that explicit $N$-body interactions (beyond mean field and Wigner-Seitz treatments) are ignored. It will be discussed in details in this thesis.\\
\indent
Different versions of grandcanonical models can be easily solved and they are more commonly used. In grand canonical models total mass or total charge fluctuation is allowed which may not be present in actual experiments.
\subsection{Dynamical models}
Though the statistical model calculations are very successful for explaining some observables of nuclear multifragmentation, it is applicable at the time of equilibrium (at freeze-out condition) only. But dynamical calculations are needed to explain real nuclear reaction completely i.e. how the system evolves with time. Freeze-out conditions, which are necessary for statistical models can only be obtained from the study of dynamical models. In addition to that, dynamical models have explained some important observables of multifragmentation like collective flow \cite{Gustafsson}, nuclear stopping \cite{Bauer_stopping}, balance energy \cite{Ogilvie_transverse_momenta} etc. which are unobtainable by statistical models.\\
\indent
For low energy nuclear reactions, due to unavailability of free states, almost all nucleon-nucleon collisions are blocked. Therefore the whole dynamics is governed by the nuclear mean field i.e. time dependent Hartee-Fock approach is appropriate to describe it. At very high energies, the nuclear mean field becomes unimportant and the Pauli blocking is negligible. i.e. the reaction dynamics is dominated by collisions (as well as particle production and annihilation), hence internuclear cascade calculations are suitable for explaining the dynamical behaviour. But in between these two energy regimes i.e. at intermediate energies the reaction mechanism is governed by nuclear mean field as well as collisions. Different dynamical models have been developed to explain the intermediate energy heavy ion reactions with proper consideration of nuclear mean field, Fermi momenta, nucleon-nucleon collision and Pauli blocking. These models are mainly classified into two main categories (i) Boltzmann-Uehling-Uhlenbeck (BUU) models and (ii) Quantum Molecular dynamics (QMD) models.\\
\indent
The BUU model for intermediate energy heavy ion collisions was first proposed by G. F. Bertsch and S. Das Gupta \cite{Dasgupta_BUU1}. In BUU approach, in order to approximate the continuous phase-space density, each nucleon is represented by many point-like test particles and the time evolution of the test particles are studied. Further isospin effect has been included \cite{Li-ibuu} with in the original BUU formalism. In this thesis BUU model will be discussed in details.\\
\indent
Quantum molecular dynamics approach \cite{Aichelin,Aichelin1,Beauvais} gives a prescription for quantum extension of the classical molecular approach \cite{Wilets,Vicentini}. In QMD models, individual nucleons are expressed as Gaussian wave packets with a finite, usually fixed, width and the time evolution of the wave packets is studied. These different formalisms affect the calculated dynamics. In molecular dynamics approach due to overlapping of the wave packets many body correlation is obtained. There are different improved versions of quantum molecular dynamics model like Antysymmetrized molecular dynamics model (AMD) \cite{Ono}, isospin dependent molecular dynamics model (IQMD) \cite{Hartnack} etc. These are different in spirit to the model used (BUU) in the thesis.  Closer in spirit yet quite distinct are some studies based on a Langevin model \cite{Ayik,Randrup2,Chomaz,Rizzo,Napolitani1}.\\
In addition to the statistical and dynamical models mentioned above, percolation model \cite{Campi,Bauer_percolation} and lattice gas model \cite{Pan} are also widely used for explaining the multifragmentation data. The percolation model is based on the bond percolation concept of condensed matter physics and successfully applied in nuclear physics to obtain clusters. But in percolation model there is no equation of state in the usual sense. The lattice gas model was developed later, has an equation of state as in Hartee-Fock theory as well as the capability of predicting clusters as in the percolation model. At this point it should be mentioned that only some of the important models have been touched upon in this thesis out of the vast literature available and several important works have been left out.\\
\indent
Nuclear multifragmentation is an important tool for basic research as well as for brevity for a wide variety of other applications. In the next sections of this chapter, some of the basic research related important applications of nuclear multifragmentation will be discussed.
\section{Production of stable and exotic nuclei}
Nuclei are composed of neutrons and protons. Due to fundamental laws of nature, which are still being investigated, all combinations of neutrons and protons are not allowed in the formation of nuclei. Fig. \ref{Nuclear_landscape} represents the nuclear landscape where several thousands of nuclei are expected to be found by the strong force. But only fewer than $300$ isotopes exist in nature (indicated by black squares). There are about $3000$ short lived nuclei which have been produced in the laboratories (shown by yellow region). But many thousands of radioactive nuclei having very small or very large neutron to proton ratio are yet to be explored (terra incognitica marked by green region). To understand the basic nuclear properties, nuclear physics experiments have been performed initially by using stable nuclei. But in order to understand the nuclear matter, one need to study the properties of stable as well as well as exotic nuclei. Detailed investigation of the properties of different exotic nuclei are also very important from astrophysical point of view. It will be very interesting to study about the modification of existing theoretical models in order to describe the properties of exotic nuclei and to know the actual positions of neutron and proton driplines. For studying many new phenomena like neutron and proton skins \cite{Tanihata2}, neutron halo \cite{Tanihata1}, large deformations of neutron rich isotopes \cite{Motobayashi} etc, exotic beam is essential. Nuclear multifragmentation is an efficient method for the production of stable as well as exotic nuclei. In general it has been found that in fragmentation reactions, the isotopic distribution (the variation of cross-section with respect to mass number at a fixed value of proton number) is approximately Gaussian in shape and the $N/Z$ of the centroid of isotopic distribution is close to the $N/Z$ of the fragmenting system. Therefore for lighter elements (whose stable isotopes have lower $N/Z$ compared to those of the source) the production cross-section of exotic neutron rich isotopes will be sufficiently high. For producing a wide variety of stable and exotic nuclei, projectile fragmentation reactions are commonly used \cite{Geissel_PF}.\\
\begin{figure}[h!]
\begin{center}
\includegraphics[width=12cm,keepaspectratio=true]{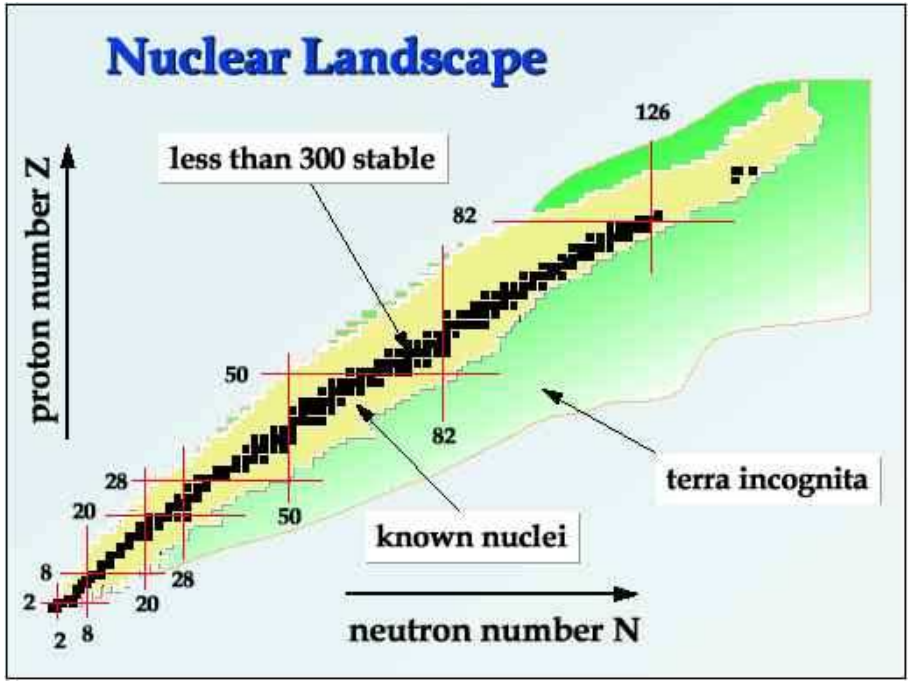}
\caption[Nuclear landscape]{Nuclear landscape — nuclei shown in proton versus neutron number representation \cite{Landscape}}
\label{Nuclear_landscape}
\end{center}
\end{figure}
\indent
In heavy ion collisions, if the beam energy is high enough, the participant-spectator scenario can be envisaged. For a general impact parameter, part of the
projectile will overlap with part of the target.  This is the participant region where violent collisions occur. In addition there are two mildly
excited remnants: projectile like fragment (PLF) or projectile spectator, with rapidity close to that of the projectile rapidity and target like fragment (TLF) or target spectator with rapidity near zero. These three parts break into fragments separately depending on their excitation energies. Pictorially this is shown in Fig. \ref{Projectile_fragmentation}.\\
\begin{figure}[h!]
\begin{center}
\includegraphics[width=14cm,keepaspectratio=true]{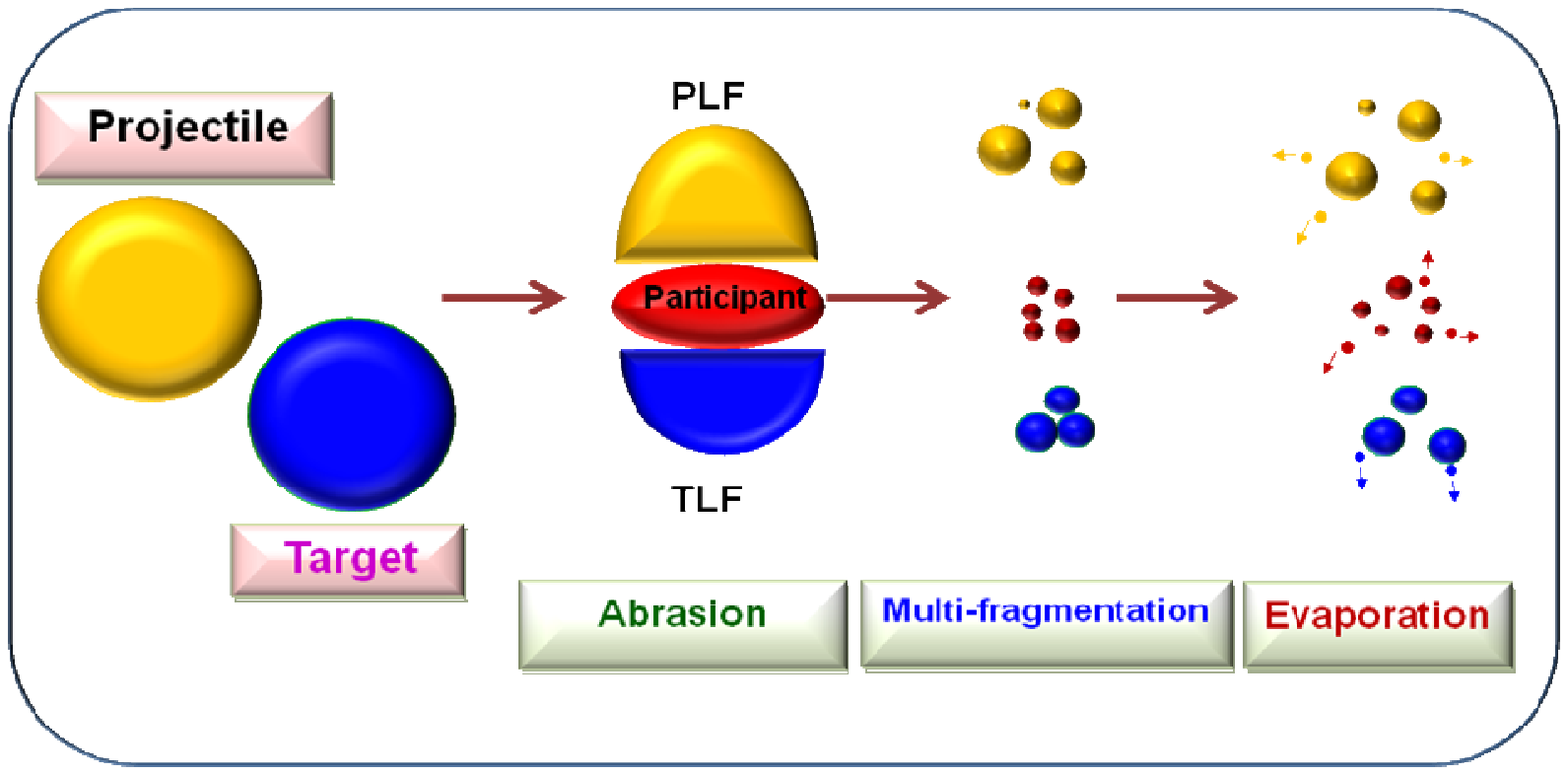}
\caption[Pictorial view of PLF, TLF and participant fragmentation]{Pictorial view of projectile spectator (PLF), target spectator (TLF) and participant fragmentation.}
\label{Projectile_fragmentation}
\end{center}
\end{figure}
\indent
Experimentally the fragments produced from projectile spectator move in the forward direction with almost projectile velocities. Therefore these fragments are easier to detect and analyse. In some of radioactive beam facilities around the world, the desired reaction products are subsequently transported for further experiments after mass, charge and momentum selection in a fragment separator. The high energy that the fragments automatically carry from the primary beam in this production method, eliminates the need for post-acceleration.\\
\indent
In addition to projectile fragmentation, central collision fragmentation reactions are also used in different laboratories for producing rare isotopes. In these cases the fragments move in all directions. Another prominent technique for producing exotic isotopes is Isotope Separation On-Line (ISOL) \cite{Kofoed} which is a separate topic and will not be discussed in this thesis.
\section{Nuclear Phase Transition}
One of the most exciting challenges in modern nuclear physics is to understand the behaviour of nuclear matter under extreme conditions of density and temperature. Symmetric nuclear matter is an idealized extrapolation of the atomic nucleus to infinite size (i.e. without surface or other finite-size effects) at the known saturation density of the nucleus, without Coulomb interaction and with equal proton and neutron densities (i.e. zero isospin). In the ground state, nuclear matter can be described as a many-body system with constant saturation density, constituted of nucleons at zero temperature and pressure and strongly interacting nuclear force is responsible for binding of the nucleons. The quantitative description of such a many-body strongly interacting system when it is far away from the saturation state relies on the knowledge of the nuclear equation of state (EoS) i.e. the dependence of the pressure or , alternatively, of the energy per nucleon on the temperature and the density.\\
\indent
Phase transition \cite{Stanley} is a process in which a thermodynamic system changes from one phase or state to another by transfer of energy. The study of phase transition is an interesting topic of research both theoretically and experimentally in different areas of physics like statistical mechanics, atomic and molecular physics, magnetism, superconductivity etc. Presently one of the most important motivation of experimental and theoretical nuclear physics studies is probing the liquid-gas coexistence region in the phase diagram of nuclear matter \cite{Siemens,Dasgupta_Phase_transition,Borderie2,Gross_phase_transition,Gross_book,Chomaz_phase_transition,Nayak,Curtin}.  The nuclear liquid gas  phase transition plays an important role in estimating the nuclear equation of state at finite temperatures. The EoS of nuclear matter is a fundamental ingredient to describe the dynamics of stellar collapse and supernovae explosion \cite{Lattimer}, as well as for the formation and structure of neutron stars \cite{Glendenning} or more complex systems such as “strange stars” \cite{Li} and “binary mergers” (neutron stars and black holes) \cite{Lattimer}.\\
\indent
The most common example of phase transition is water to vapour transition \cite{Huang_stat_mech}. The Lenard-Jones potential for water molecules is repulsive at very short range due to overlapping of the electron cloud and then at comparatively higher intermolecular separation it becomes attractive. Now, if one takes certain amount of water and starts to heat it, then initially the supplied heat energy is converted into kinetic energy and the temperature increases. But when the temperature ($T$) becomes $100^0 C$, then the supplied energy (latent heat) is wholly used to overcome the attractive potential, therefore the temperature remains constant and the water is converted into vapour. After completion of the conversion from water to vapour, again the temperature of vapour starts to increase. Turning to nuclear physics, the nuclear EoS provides a way to describe the bulk properties of a nuclear many body system in thermodynamical equilibrium, governed at the microscopic level by the two-body nucleon-nucleon (NN) interaction. If one studies the nucleon-nucleon interaction potential it is observed that its variation with separating distance is similar to the Lenard-Jones potential (though the scales are completely different as shown in Fig. \ref{Phase_transition_introductory}). Therefore one can expect similar kind of phase transition in nuclear physics. This phase transition occurs at subnormal densities and at a temperature of few MeV ($1$ MeV$ =1.2\times 10^{10}$ Kelvin).\\
\begin{figure} [ht]
\begin{center}
\includegraphics[width=2.25in,height=2.25in]{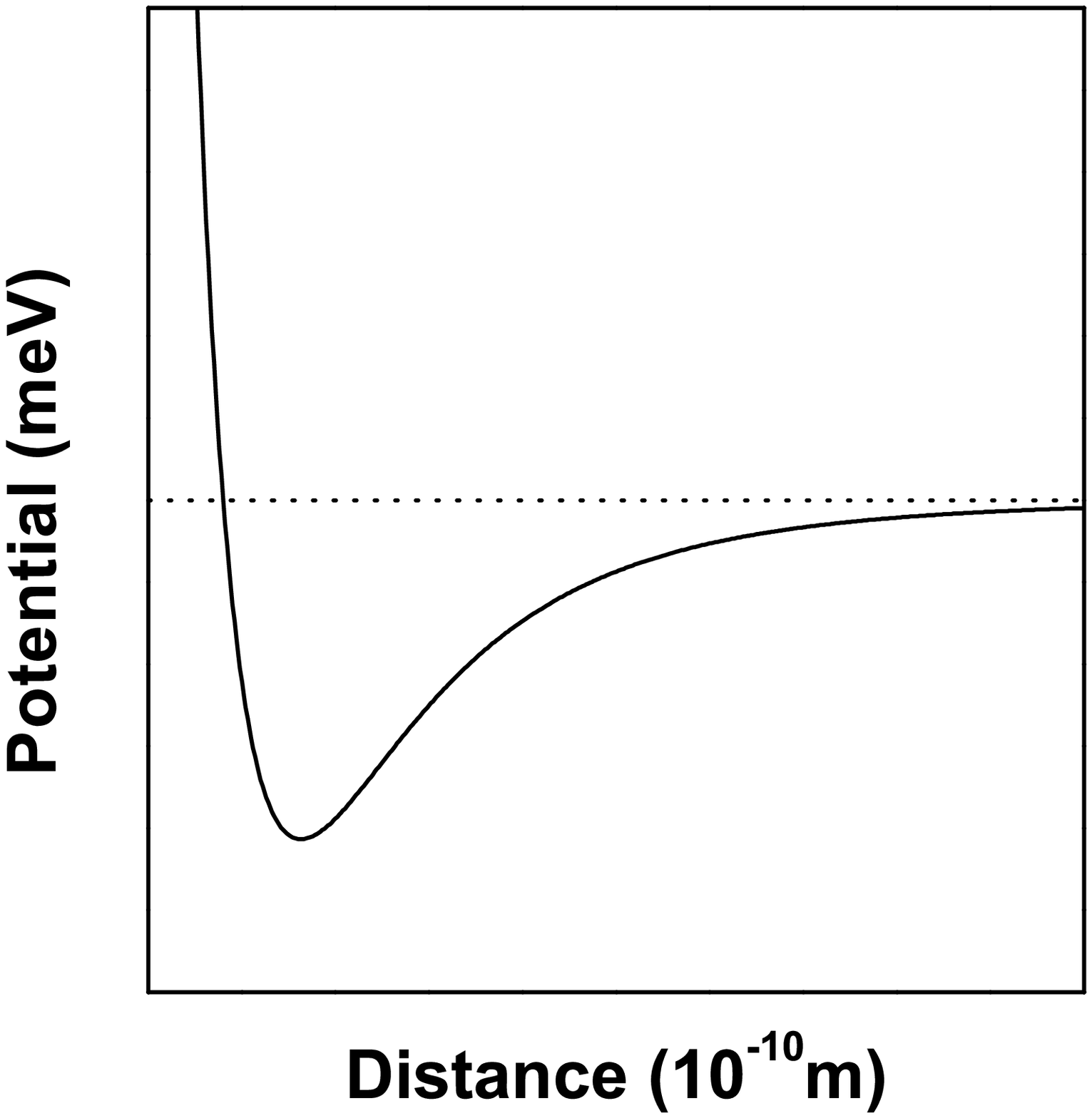}
\includegraphics[width=2.25in,height=2.25in,clip]{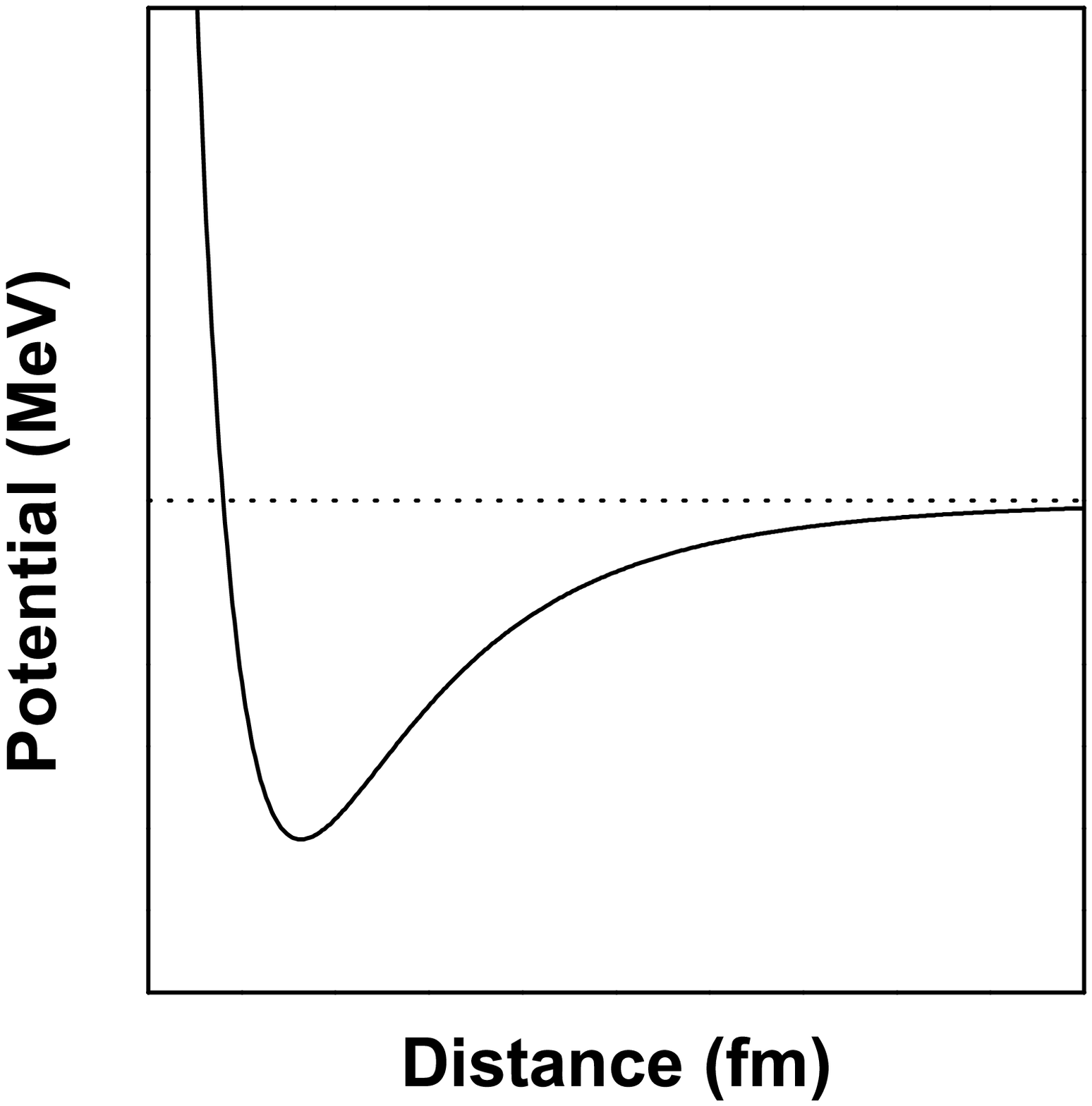}\\
\includegraphics[width=2.5in,height=2.3in,clip]{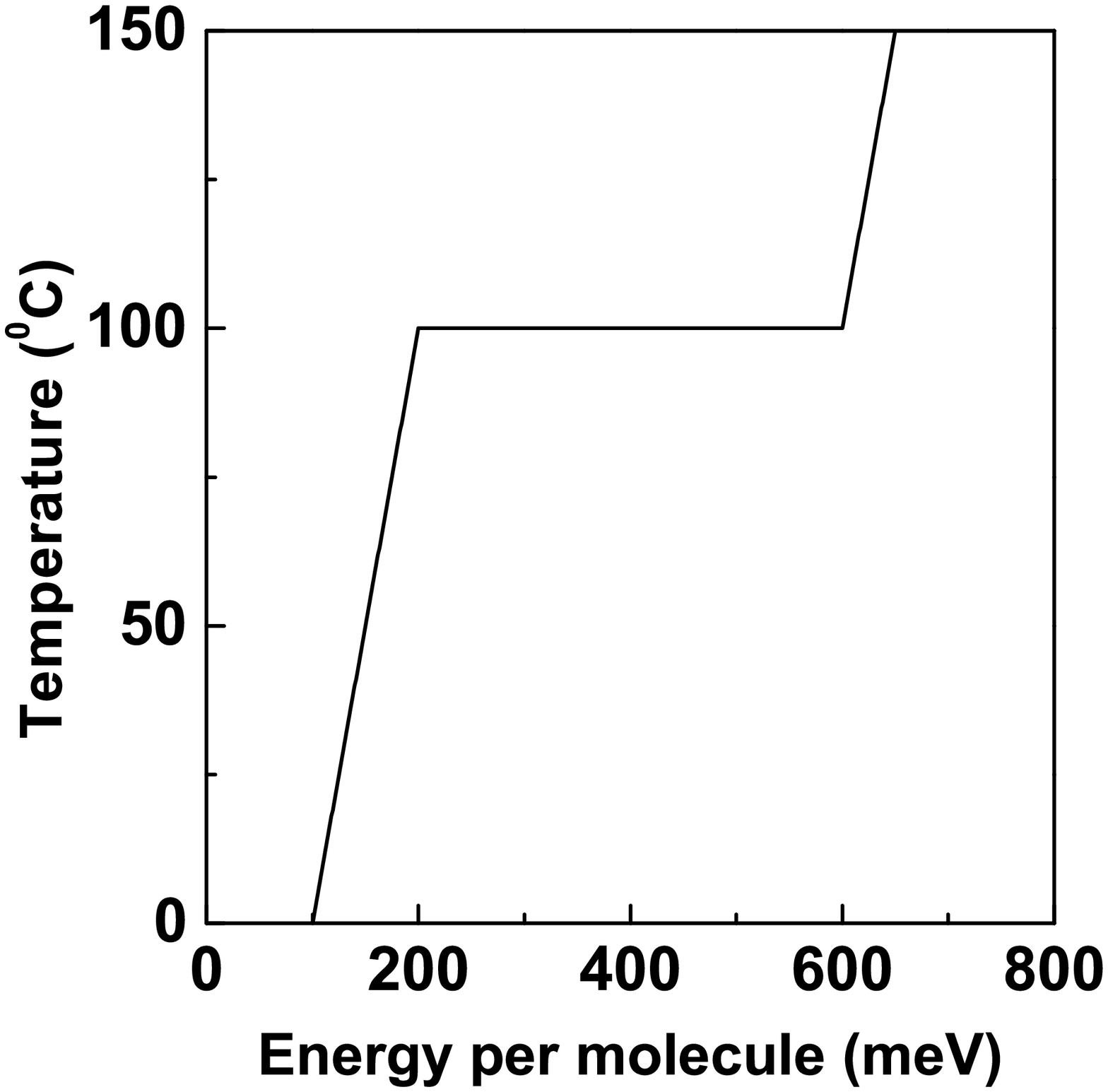}
\includegraphics[width=2.6in,height=2.3in,clip]{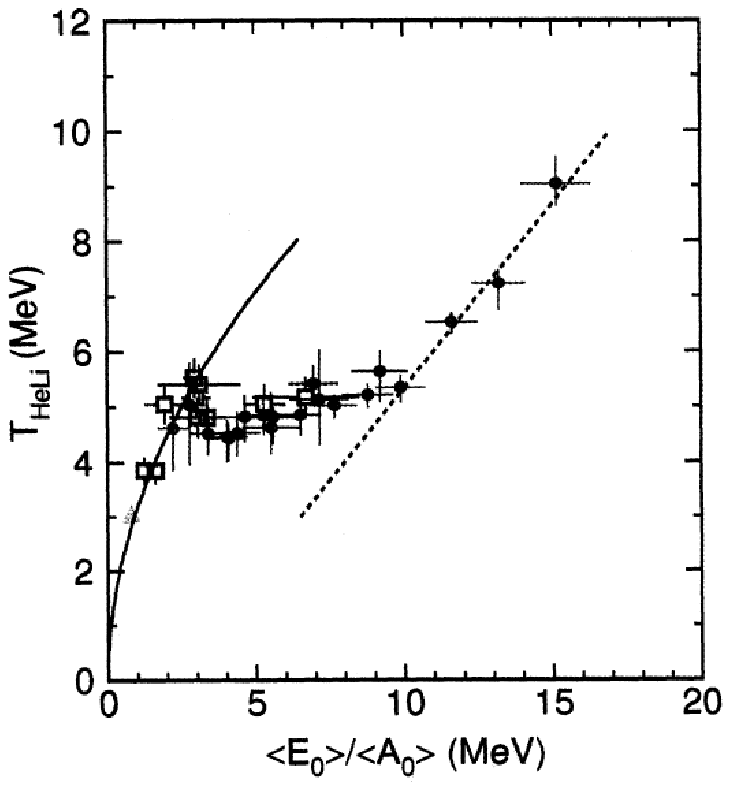}
\caption[Molecular and nuclear liquid-gas phase transition]{Upper Panels: Schematic view of radial dependence of molecular (left) and nucleon-nucleon interaction (right) potential.
Lower Panels: Caloric curves for water to vapour phase transition (left) and nuclear liquid-gas phase transition (right). The diagram of lower right panel is taken from Ref. \cite{Pochodzalla_phase_transition}}
\label{Phase_transition_introductory}
\end{center}
\end{figure}
\indent
Unfortunately, it is not possible to prepare infinite nuclear matter and to heat it to such a high temperature (of the order of MeV), In the laboratory the only possible way to achieve such high temperatures is through collisions between atomic nuclei (which can be considered as the “chunks” of nuclear matter) at intermediate energies. Nuclei at normal density and zero temperature behave like Fermi liquid therefore this transition is a liquid to gas phase transition. Also there are no direct probes to measure this high temperature in experiments. Indirect methods based on models are used to measure it. The collisions between the nuclei are over in $10^{-22}$ seconds, therefore one can not keep the matter in an exotic state long enough to study its properties. The detectors measures only the products of these collisions where all the final products are in normal states. Hence one need to extrapolate from the end products to what happened during disassembly. Traditionally phase transition is studied in the thermodynamic limit and for normal liquid or normal gas the number of particles is very high ($~10^{23}$). But, in laboratories, one can get a system containing at most a few hundreds nucleons which is far away from the thermodynamic limit. Also the signals of phase transition are affected due to presence of the Coulomb force between the protons. Hence both theoretical and experimental research on nuclear liquid gas phase transition is very interesting and highly challenging. Before going to the details of nuclear liquid gas phase transition, one has to know what is nuclear liquid and what is nuclear gas. Due to the different limitations mentioned above, the conventional definition of normal liquid and gas is not applicable here directly. Generally a large nucleus (size almost same as that of the fragmenting system) is termed as nuclear liquid, in addition to it there may be few nucleons. On the other side, a large number of free nucleons and few very light fragments is referred to as  nuclear gas.\\
\indent
There is an enormous amount of theoretical and experimental work done on nuclear liquid-gas phase transition in the last three decades. Calculation based on the lattice gas model first concluded that nuclear phase transition is first order in nature \cite{Pan_phase_transition}. Similar results are also obtained from statistical multifragmentation model (SMM), canonical thermodynamical model (CTM) etc. Flattening of the nuclear caloric curve \cite{Dasgupta_Phase_transition,Ma_Phase_transition}, bimodal distribution of the specific order parameter \cite{Chomaz_bimodality}, negative micro-canonical heat capacity \cite{Chomaz3,Agostino,Chomaz4}, spinodal decomposition \cite{Borderie} etc are useful signatures for supporting the first order behaviour of nuclear phase transition which are obtained both theoretically and experimentally. However the phase transition observed in percolation model calculations is second order in nature. Several experimental signals of second order phase transition are also reported, such as critical behavior like power laws in the charge distribution, $\Delta$ scaling, maximal fluctuation etc \cite{Hauger,Elliott1,Elliott2}.\\
\indent
In addition to the liquid-gas phase transition, at very high energy and high baryon density the nucleons themselves undergo phase transition and produces quark-gluon plasma (QGP) i.e. the transition between hadronic phase and QGP phase \cite{Shuryak}. A detailed knowledge of the quark-hadron phase transition is important for the study of the dynamics of the early universe (deconfined nuclear matter). This is a separate detailed topic and the discussions in the thesis will be restricted to nuclear liquid gas phase transition only.
\section{Probing Nuclear Symmetry energy by nuclear multifragmentation}
Isospin-dependent phenomena in nuclear physics has been an active area of research \cite{Bao-an-li1, Bao-an-li2, Bao-an-li5} in recent years with the aim of enriching the knowledge about the symmetry term of the nuclear equation of state. In addition to the symmetric nuclear matter, the study of asymmetric nuclear matter properties at different regimes of density and temperature is also a topic of great interest \cite{Tsang4, Lehaut, Samaddar} in the nuclear physics community.
\indent
The Equation of State (EoS) of hot neutron-rich matter at temperature $T$ and isospin asymmetry $\delta=(\rho_n-\rho_p)/(\rho_n+\rho_p)$ ($\rho_n$, $\rho_p$ and $\rho$ are the neutron, proton and nucleon densities respectively) can be written as \cite{Zuo,Chen1}
\begin{equation}
E(\rho,T,\delta)=E_0(\rho,T,\delta=0)+E_{sym}(\rho,T)\delta^2+O(\delta^4)
\label{Equation_of_state}
\end{equation}
\begin{figure}[b!]
\begin{center}
\includegraphics[width=10cm,height=12cm,clip]{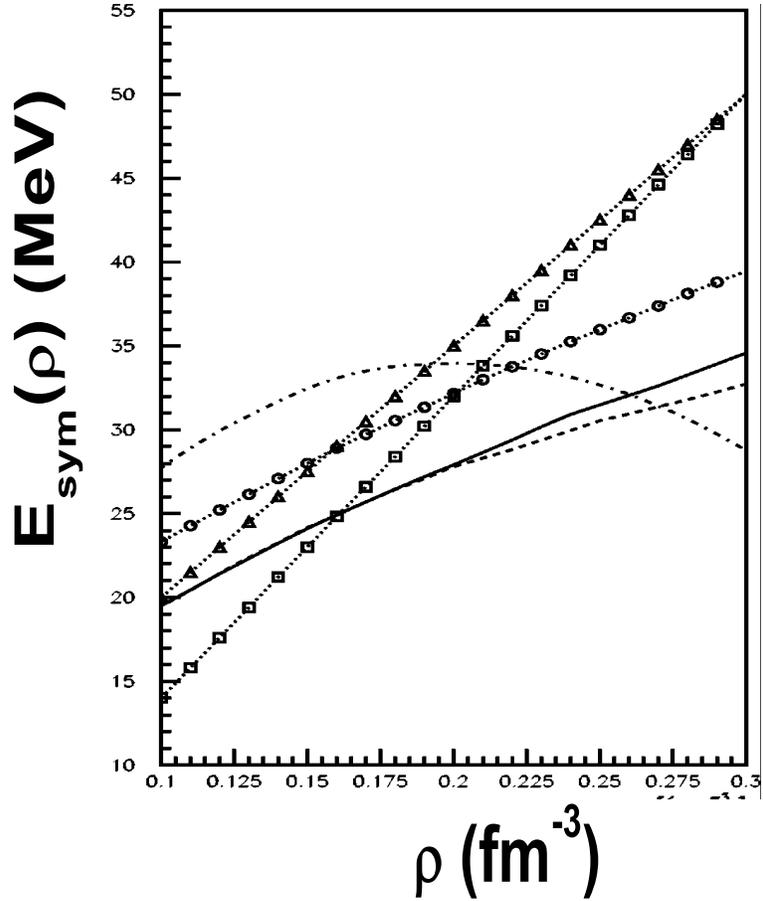}
\caption[Density dependence of symmetry energy]{Theoretical prediction of density dependence of symmetry energy from the continuous choice of Brueckner-Hartree-Fock (continuous choice) with Reid93 potential (circles), self-consistent Green function theory with Reid93 potential (full line), variational calculation from Ref. \cite{Wiringa} with Argonne Av14 potential (dashed line), Dirac-Brueckner-Hartree-Fock calculation from Ref. \cite{CHLee} (triangles), relativistic mean-field model from Ref. \cite{BLiu} (squares), effective field theory from Ref. \cite{Kaiser}. The diagram is taken from Ref. \cite{Diepernick}.}
\label{Symmetry_energy_density_dependence}
\end{center}
\end{figure}
where the first term represents the energy per nucleon of symmetric matter with equal fractions of neutrons and protons and $E_{sym}(\rho,T)$ in second term is the symmetry energy i.e. the energy cost to convert all protons of symmetric matter to neutrons at the fixed temperature $T$ and density $\rho$. In Eq. \ref{Equation_of_state} odd-order terms ($\delta$, $\delta^3$,...) are absent because of the charge invariance of the nuclear interaction (Coulomb is treated apart) and higher order-even terms as $\delta^4$, $\delta^6$,...are neglected since $\delta$ is rather small ($<<1$) (obviously this is not true fore pure neutron matter). Traditionally, Bethe-Weizacker binding energy formula \cite{Bethe,Weizsacker} from the liquid drop model can provide useful information about the nuclear symmetry energy of stable nuclei. But, stable nuclei are found at zero temperature and at saturation density ($\rho_0=0.16 fm^-3$). Therefore, it can not predict how the symmetry energy changes with temperature and density (i.e. away from the normal nuclear conditions). The density and/or temperature dependence of nuclear symmetry energy plays an important role in areas of astrophysical interest such as the study of supernova explosions and the properties of neutron stars \cite{Steiner,Lattimer2,Danielewicz_science}. This also has significant influence in deciding the structure of neutron-rich and neutron-deficient nuclei \cite{Brown}. Unfortunately the density and temperature dependence of symmetry energy is poorly known from microscopic many body theories. For example, Fig. \ref{Symmetry_energy_density_dependence} shows the density dependence of symmetry energy (at $T=0$ MeV) obtained from most widely used many body techniques. Different theoretical predictions are widely divergent at both low and high densities. The temperature dependence will make the scenario more complicated. The theoretical uncertainties are large due to lack of knowledge about the isospin dependence of nuclear effective interactions and the short-comings of existing many body techniques. On the other hand, during the nuclear multifragmentation process at intermediate energies, the nuclear system is compressed (and then expanded) and heated. Therefore, the study of nuclear multifragmentation provides a unique opportunity to extract the information about the symmetry energy at various densities and temperatures, and this has created much interest in the nuclear physics community in recent years.\\
\indent
In nuclear multifragmentation reactions, the neutron-proton composition of the break-up fragments is dictated by the symmetry term of the equation of state and hence the study of the multifragmentation process allows one to obtain information about the symmetry term. Isoscaling \cite{Tsang1}, isobaric yield ratio \cite{Huang} measurements etc. are standard methods which can connect the measurable fragment yields of multifragmentation reactions to the symmetry energy of excited nuclei and these have been applied to the analysis of heavy-ion collision data. The density dependence of nuclear symmetry energy can be extracted from different observables like free neutron to proton ratio of pre-equilibrium nucleons \cite{Bao-an-li3}, isospin fractionation $\pi^-/\pi^+$ ratio \cite{Benenson,Bao-an-li4} etc.\\
\indent
In addition to the three main applications, nuclear multifragmentation can also be used for spallation reaction (nuclear power production) \cite{Napolitani}, nuclear waste management (environment protection) \cite{Brandt}, proton and ion therapy (medical applications) \cite{Agodi,Pshenichnov}, radiation protection of space missions (space research) \cite{Townsend} etc.
\section{Motivation and Organization of the thesis}
In this thesis, the following three aspects of multifragmentation reactions namely (i) production of exotic nuclei which are normally not available in the laboratory (ii) nuclear symmetry energy from heavy ion collisions at intermediate energies and (iii) Nuclear liquid-gas phase transition will be discussed in details using statistical and dynamical models. In addition to these equivalence of statistical ensembles under different conditions in the framework of multifragmentation will also be studied.\\
\indent
The thesis is organized as follows. Chapter 2 describes the development of the model for projectile fragmentation and its application for calculating different important observables and the comparison with experimental data. A very simple impact parameter dependence of freeze-out temperature profile is introduced for understanding the reaction mechanism in the limiting fragmentation region. Chapter 3 contains the microscopic static model and dynamical Boltzmann-Uehling-Uhlenbeck calculations for determining the initial conditions (mass and excitation) of projectile fragmentation reactions. Chapter 4 is dedicated to formulate a hybrid model for studying the central collision multifragmentation reactions around the Fermi energy regime. The conditions for convergence of the statistical ensembles for the fragmentation of finite nuclei is described in chapter 5. In chapter 6, the symmetry energy coefficient is determined by different ways (isoscaling source method, isoscaling fragment method, fluctuation method and isobaric yield ratio method) in the framework of canonical and grand canonical model. The ratio of the symmetry energy coefficient to temperature ($C_{sym}/T$) has been extracted using the different prescriptions in the framework of the projectile fragmentation model and the results have been compared with the available experimental data. Signatures of nuclear liquid gas phase transition obtained from the dynamical model calculation and its comparison with already existing statistical model results are discussed in chapter 7. The thesis work is summarized in chapter 8, which also contains the possible future outlook of the work. The references are given at the end of the thesis.
\vskip3cm
\end{normalsize} 
\chapter{A model for projectile fragmentation}
\begin{normalsize}
\section{Introduction}
Projectile fragmentation is an important phenomenon, the study of which can reveal reaction mechanism in heavy ion collisions at intermediate and high energies. It is an efficient method for the production of different exotic nuclei and is used by many radioactive beam facilities around the world. Recently it is also widely used for studying liquid-gas phase transition and nuclear equation of state.\\
\indent
The aim of the chapter is to develop a model for projectile fragmentation in the limiting fragmentation region \cite{LeBrun}. This model \cite{Mallik2, Mallik3,Mallik101} involves concepts of heavy ion reaction plus the well known statistical model of multifragmentation (Canonical Thermodynamical Model) and evaporation.  Our model is computationally much less intensive  than  heavy ion phase-space exploration (HIPSE) model \cite{Lacroix} and antisymmetrized molecular dynamics (AMD) \cite{Ono} which are based on transport calculation. Our model is less phenomenological than EPAX \cite{Summerer2,Summerer} which is based on the empirical parametrization of fragmentation cross sections. An impact parameter dependent temperature profile has been developed in order to better account for the results at different $Z_{bound}$ ranges and also to confront with data from different projectile fragmentation reactions at different energies. Here $Z_{bound}$  \cite{Kreutz} is the number of charges measured in the extreme forward direction minus the sum of all $Z=1$ particles. Since PLF moves with a velocity close to that of the projectile, $Z_{bound}$ is a measure of the charges (hence indirectly of the size) of the PLF. For peripheral collisions, $Z_{bound}$ is large, but as the impact parameter decreases, $Z_{bound}$ falls reflecting a smaller size of PLF. The model is in general applicable and implementable above 100 MeV/nucleon.\\
\indent
The organization of this chapter is as follows. In Section 2.2 we describe the theoretical formulation of the model where as the impact parameter dependence of temperature is explained in Section 2.3. Section 2.4 contains the results obtained from theoretical calculation and comparison with experimental data of different projectile fragmentation reactions with different projectile-target combinations and varying projectile energies. Finally the results are summarised in Section 2.5.
\section{Formulation of Model}
The model for projectile fragmentation reaction consists of three stages: (i) abrasion, (ii) multifragmentation and (iii) evaporation. In heavy ion collision, if the beam energy is high enough, then in the abrasion stage at a particular impact parameter, three different regions are assumed to be formed: (i) projectile spectator or projectile like fragment (PLF), (ii) participant and (iii) target spectator or target like fragment (TLF). In this work, focus is in the fragmentation of the PLF. The number of neutrons and protons in the projectile spectator at different impact parameters are determined from abrasion stage. Then the break up of each abraded projectile spectator is separately calculated by using canonical thermodynamical model (CTM) \cite{Das}. Finally, the decay of excited fragments are calculated by evaporation model \cite{Mallik1} based on Weisskopf's formalism. The details of the three different stages are described below.
\subsection{Abrasion}
In abrasion stage, the projectile and targets are assumed as two hard spheres of radius $R_P=1.2A_P^{1/3}$ and $R_T=1.2A_T^{1/3}$ respectively ($A_P$ is projectile mass and $A_T$ is target mass). The projectile beam energy is assumed to be high enough so that straight-line geometry can be used for classifying projectile spectator, target spectator and participant region. The volume of the projectile that goes into the participant region $V_{Pc}(b)$ is calculated  at different impact parameters (b) ranging from central collision to peripheral collision. For calculation of $V_{Pc}(b)$, refer to Appendix A. Therefore the projectile spectator volume is $V_s(b)=V_P-V_{Pc}(b)$, where $V_P$ is the original volume of the projectile. If the original projectile contains $N_P$ neutrons and $Z_P$ protons, then the average number of neutrons in the PLF is $\langle N_s(b)\rangle=(V_s(b)/V_P)N_P$ and the average number of protons is $\langle Z_s(b)\rangle=(V(_sb)/V_P)Z_P$. These will usually be non-integers. Since in any event only integral numbers for neutrons and protons can materialise in the projectile spectator, one has to guess about the distribution of $N_s,Z_s$ which produces these average values.\\
\indent
Two distributions which can be applied are as follows. One is a minimal distribution model. Let $N_s^{min}(b)$ and $N_s^{max}(b)$ be the two nearest integers of $\langle N_s(b)\rangle$ and the probability of getting these from $\langle N_s(b)\rangle$ are $P_N^{max}(b)$ and $P_N^{min}(b)$ respectively so, $P_N^{max}(b)+P_N^{min}(b)=1$. Therefore, $P_N^{max}(b)=\langle N_s(b)\rangle-N_s^{min}(b)$ and $P_N^{min}(b)=N_s^{max}(b)-\langle N_s(b)\rangle$. From $\langle Z_s\rangle$, $P_{Z_s}(b)$ can be defined in the similar way. Therefore the probability of getting a PLF with $N_s$ neutrons and $Z_s$ protons at impact parameter $b$ is $P_{N_{s},Z_{s}}(b)=P_{N_{s}}(b)P_{Z_{s}}(b)$. Hence, in general, for each impact parameter, four possibilities of PLF's \{($N^{max}$,$Z^{max}$), ($N^{max}$,$Z^{min}$), ($N^{min}$,$Z^{max}$) and ($N^{min}$,$Z^{min}$)\}are calculated in the abrasion stage, each with different probability.\\
\indent
The alternative is a binomial distribution which has a long tail. There $P_{N_s}(b)$ is defined by $P_{N_s}(b)=(^{N_0}_{N_s})(occ(b))^{N_s}(1-occ(b))^{N_0-N_s}$ (see also \cite{Gaimard}). Here $occ(b)=V_s(b)/V_0$. Similarly one can define $P_{Z_s}(b)$ for binomial distribution and finally $P_{{N_s},{Z_s}}(b)=P_{N_s}(b)P_{Z_s}(b)$. The binomial distribution would be appropriate if the projectile is viewed as a collection of non-interacting neutron and proton gas with constant density throughout its volume.  This is oversimplification and it turns out that the binomial distribution is too long tailed. For extreme peripheral collision (with only 1 or 2 nucleons lost to the participant) the temperature of the PLF should be very low and the cross-sections can be directly confronted with data. The calculation gives a far too wide distribution. Hence the minimal distribution has been used for subsequent calculation, which is also easier to work with. The abrasion cross-section when there are $N_s$ neutrons and $Z_s$ protons in the PLF is labeled by $\sigma_{a,N_{s},Z_{s}}$:
\begin{equation}
\sigma_{a,N_{s},Z_{s}}=\int 2\pi bdbP_{N_{s},Z_{s}}(b)
\label{Abrasion_Eq1}
\end{equation}
where the suffix $a$ denotes abrasion. The limits of integration in Eq. \ref{Abrasion_Eq1} are $b_{min}$ and $b_{max}=R_T+ R_P$. $b_{min}$ is either $0$ (if the projectile is larger than the target) or $R_T-R_P$ (if the target is larger than the projectile, in this case at lower value of $b$ there is no PLF left).\\
\indent
Fig. \ref{Abrasion_Fig1} shows the variation of PLF mass with impact parameter calculated from abrasion model for two different reactions  $^{58}$Ni on $^{9}$Be and $^{58}$Ni on $^{181}$Ta. The projectile-target combinations are chosen such that it can highlight different aspects. In $^{58}$Ni on $^{9}$Be, for example the projectile is significantly larger than the target. In such a case, the abraded projectile has a lower limit on $A_s=34$ (as $^{9}$Be can drive out only some nucleons, not all).  But in experiments, significant cross-sections exist for composites with $A$=10 to 25 which can not come from either abrasion or evaporation. These therefore must arise from multifragmentation (stage 2) of an abraded system (stage 1). On the other hand for $^{58}$Ni on $^{181}$Ta (projectile smaller than target) the abraded system itself covers most of the range of composites seen in the experiment. The role of the multifragmentation and evaporation stage is to modify the cross-sections.\\
\begin{figure}[h!]
\begin{center}
\includegraphics[width=14cm,keepaspectratio=true,clip]{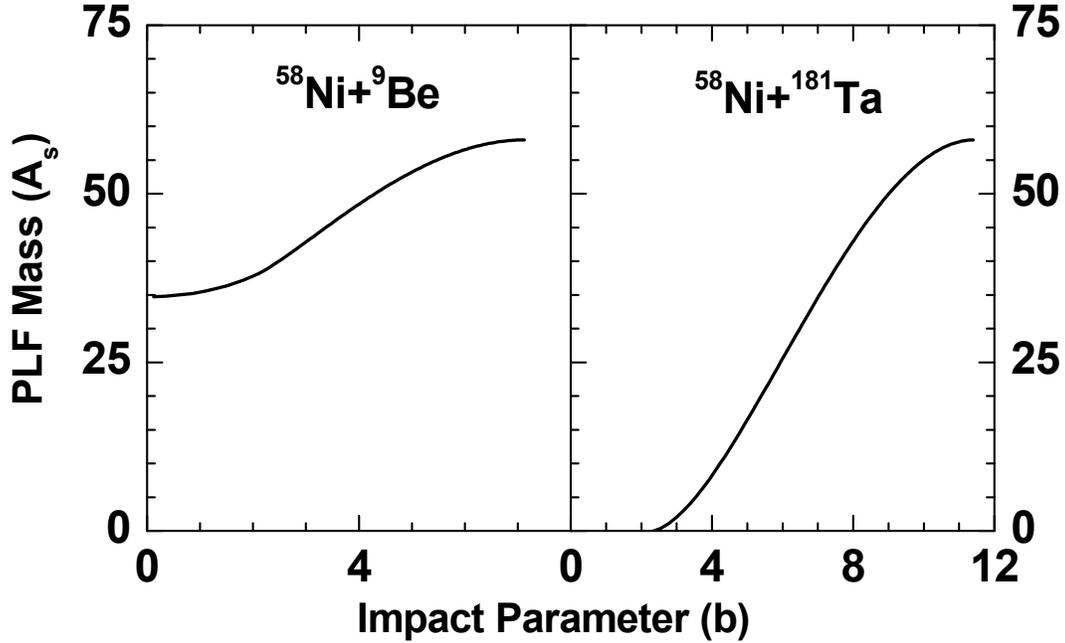}
\caption[Variation of PLF mass with impact parameter after abrasion stage]{Variation of PLF mass ($A_s$) with impact parameter (b) obtained from abrasion model for $^{58}$Ni on $^{9}$Be (left panel) and $^{58}$Ni on $^{181}$Ta (right panel) reaction.}
\label{Abrasion_Fig1}
\end{center}
\end{figure}
\indent
Actually there is an extra parameter that needs to be specified for abrasion cross-section.  The complete labeling is $\sigma_{a,N_s,Z_s,T}$ where $T$ is the temperature of the PLF, which is also a function of impact parameter.  Here this has been extended to the more general case where the temperature is dependent on the impact parameter $b$.  In evaluating Eq. \ref{Abrasion_Eq1} the integration is replaced by a sum.  The interval $b_{min}$ to $b_{max}$ is divided into small segments of length $\Delta b$. If the mid-point of the $i$-th bin is $<b_i>$ and the temperature for collision at $<b_i>$ is $T_i$ then
\begin{equation}
\sigma_{a,N_s,Z_s}=\sum_i\sigma_{a,N_s,Z_s,T_i}
\label{Abrasion_Eq2}
\end{equation}
where
\begin{equation}
\sigma_{a,N_s,Z_s,T_i}=2\pi<b_i>\Delta bP_{N_S,Z_s}(<b_i>)
\label{Abrasion_Eq3}
\end{equation}
PLF's with the same $N_s,Z_s$ but different $T_i$'s are treated independently for further calculations.
\subsection{Multifragmentation}
The abraded system of $N_s$ neutrons and $Z_s$ protons created at impact parameter $b$ will have an excitation which we characterize by a temperature $T$. The impact parameter dependent temperature profile can be obtained from (i) microscopic static model calculation (ii)  transport model based on BUU calculation (both described in Chapter 3) and (iii) parametrization using experimental data of different projectile fragmentation reactions (discussed in the next section). The abraded system with size $N_s,Z_s$ and temperature $T$ will break up into different composites and nucleons depending on the temperature or excitation. The canonical thermodynamic model (CTM)  \cite{Das} is used for calculating this break up.\\
\indent
It is assumed that the system with $N_s$ neutrons and $Z_s$ protons at temperature $T$, has expanded to a higher than normal volume and the partitioning into different composites can be calculated according to the rules of equilibrium statistical mechanics.  In a canonical model, the partitioning is done such that all partitions obey the total neutron and proton conservation.\\
The canonical partition function is given by
\begin{equation}
Q_{N_s,Z_s}=\sum\prod \frac{\omega_{N,Z}^{n_{N,Z}}}{n_{N,Z}!}
\label{Multifragmentation_Eq1}
\end{equation}
Here $\omega_{N,Z}$ is the partition function of one composite with neutron number $N$ and proton number $Z$ respectively and $n_{N,Z}$ is the number of this composite in the given channel. The product ($\prod$) is for one break-up channel whereas the sum ($\sum$) is over all possible channels of break-up (the number of such channels is enormous) which satisfy $N_s=\sum N\times n_{N,Z}$ and $Z_s=\sum Z\times n_{N,Z}$. Therefore, the actual expression of partition function is
\begin{equation}
Q_{N_s,Z_s}=\sum\prod \frac{\omega_{N,Z}^{n_{N,Z}}}{n_{N,Z}!}\delta(N_s-\Sigma Nn_{N,Z})\delta(Z_s-\Sigma Zn_{N,Z})
\label{Multifragmentation_Eq2}
\end{equation}
Computationally it is very difficult to solve Eq. \ref{Multifragmentation_Eq2} directly.\\
The probability of a given channel $P(\vec n_{N,Z})\equiv P(n_{0,1},n_{1,0},n_{1,1}......n_{I,J}.......)$ is given by
\begin{equation}
P(\vec n_{N,Z})=\frac{1}{Q_{N_s,Z_s}}\prod\frac{\omega_{N,Z}^{n_{N,Z}}}{n_{N,Z}!}
\label{Multifragmentation_Eq3}
\end{equation}
Therefore the average number of composites having $N$ neutrons and $Z$ protons is easily seen from the above equation to be
\begin{eqnarray}
\langle n_{N,Z}\rangle&=&\sum n_{N,Z}P(\vec n_{N,Z})\nonumber\\
&=&\frac{1}{Q_{N_s,Z_s}}\sum n_{N,Z}\prod_{i,j}\frac{\omega_{i,j}^{n_{i,j}}}{n_{i,j}!}\nonumber\\
&=&\frac{1}{Q_{N_s,Z_s}}\sum n_{N,Z}\frac{\omega_{N,Z}^{n_{N,Z}}}{n_{N,Z}!}\prod_{i\neq N,j\neq Z}\frac{\omega_{i,j}^{n_{i,j}}}{n_{i,j}!}\nonumber\\
&=&\frac{\omega_{N,Z}}{Q_{N_s,Z_s}}\sum\frac{\omega_{N,Z}^{(n_{N,Z}-1)}}{(n_{N,Z}-1)!}\prod_{i\neq N,j\neq Z}\frac{\omega_{i,j}^{n_{i,j}}}{n_{i,j}!}\nonumber\\
&=&\frac{\omega_{N,Z}Q_{N_s-N,Z_s-Z}}{Q_{N_s,Z_s}}
\label{Multifragmentation_Eq4}
\end{eqnarray}
Taking average on the both sides of neutron conservation relation
\begin{eqnarray}
\langle N_s\rangle=\langle \sum N\times n_{N,Z}\rangle\nonumber\\
N_s=\sum N\times \langle n_{N,Z}\rangle
\label{Multifragmentation_Eq4a}
\end{eqnarray}
Substituting Eq. \ref{Multifragmentation_Eq4} in Eq. \ref{Multifragmentation_Eq4a} one can obtain the recursion relation \cite{Chase}
\begin{equation}
Q_{N_s,Z_s}=\frac{1}{N_s}\sum_{N,Z}N\omega_{N,Z}Q_{N_s-N,Z_s-Z}
\label{Multifragmentation_Eq5}
\end{equation}
Similarly, by averaging on the both sides of proton conservation relation and substituting Eq. \ref{Multifragmentation_Eq4} in it, one can construct another recursion relation,
\begin{equation}
Q_{N_s,Z_s}=\frac{1}{Z_s}\sum_{N,Z}Z\omega_{N,Z}Q_{N_s-N,Z_s-Z}
\label{Multifragmentation_Eq5Z}
\end{equation}
Therefore, within very short time, partition functions of different nuclei can be calculated very easily by using any one of the above two recursion relations. Finally, by knowing the partition functions, the average multiplicity $\langle n_{N,Z}\rangle$ of different fragments can be calculated from Eq. \ref{Multifragmentation_Eq4}.\\
\indent
The one-body partition function $\omega_{N,Z}$ is a product of two parts: one arising from the translational motion of the center of mass of the composite and another from the intrinsic partition function of the composite:
\begin{equation}
\omega_{N,Z}=\frac{V(b)}{h^3}(2\pi mT)^{3/2}A^{3/2}\times z_{N,Z}(int)
\label{Multifragmentation_Eq6}
\end{equation}
Here $A=N+Z$ is the mass number of the composite and $V(b)$ is the volume available for translational motion (or excluded volume). During freeze-out condition \cite{Gross_freeze-out}, different particles and composites are not supposed to overlap with each other, hence the available volume within which the particles and composites move freely should be less compared to the freeze-out volume (the volume to which the system has expanded at break up stage) $V_f(b)$. Typically \enquote{Hanbury Brown Twiss pion interferometry method} gives a measure of the freeze-out volume. In the projectile fragmentation model a fairly typical value $V_f(b)=3V_0(b)$ have been used where $V_0(b)$ is the normal volume of projectile spectator created at impact parameter $b$ with $Z_s$ protons and $N_s$ neutrons. The \enquote{available volume} which is considered in the present calculation is $V(b)=V_f(b)-V_0(b)$. The detailed study on excluded volume in statistical models of nuclear multifragmentation can be found can be found in Ref. \cite{Majumder}.\\
\indent
The properties of the composites used in this work is being described here.  The proton and the neutron are fundamental building blocks thus $z_{1,0}(int)=z_{0,1}(int)=2$ where 2 takes care of the spin degeneracy.  For deuteron, triton, $^3$He and $^4$He $z_{N,Z}(int)=(2s_{N,Z}+1)\exp(-\beta E_{N,Z}(gr))$ where $\beta=1/T, E_{N,Z}(gr)$ is the experimental ground state energy of the composite and $(2s_{N,Z}+1)$ is the experimental spin degeneracy of the ground state. Excited states for these very low mass nuclei are not included. For mass number $A=5$ and greater, the liquid-drop formula has been used for the ground state. For nuclei in isolation, this reads ($A=N+Z$)
\begin{equation}
z_{N,Z}(int)=\exp\frac{1}{T}[W_0A-\sigma(T)A^{2/3}-a^{*}_c\frac{Z^2}{A^{1/3}}-C_{sym}\frac{(N-Z)^2}{A}+\frac{T^2A}{\epsilon_0}]
\label{Multifragmentation_Eq7}
\end{equation}
This follows from well-known thermodynamic identity $Z=\exp^{-F/T}$, where $F$ is the Helmholtz Free energy (as equilibrium is considered at constant volume). $F$ is,
\begin{eqnarray}
F&=&E(T)-TS(T)\nonumber\\
&=&E_0+E_{ex}(T)-TS(T)
\label{Multifragmentation_free_energy}
\end{eqnarray}
and liquid-drop model is used for calculating the binding energy $E_0$ and Fermi-gas model is applied for studying excitation energy $E_{ex}(T)$ and entropy $S(T)$. Therefore the expression of \ref{Multifragmentation_Eq7} includes the volume energy [$W_0=15.8$ MeV], the temperature dependent surface energy [$\sigma(T)=\sigma_{0}\{\frac{(T_{c}^2-T^2)}{(T_{c}^2+T^2)}\}^{5/4}$ with $\sigma_{0}=18.0$ MeV and $T_{c}=18.0$ MeV, $T_{c}$ is the critical temperature where surface tension vanishes \cite{Bondorf1}], the Coulomb energy, the symmetry energy ($C_{sym}=23.5$ MeV) and the term due to excitation. Since nuclear force is short range, it is considered that the nucleons of a given fragment are interacting through nuclear force, but it is assumed that during freeze-out the fragments are well separated such that there is no nuclear interaction among the nucleons of different fragments. Since the Coulomb interaction is long range, one have to consider some approximations in the Coulomb term in order to account for the Coulomb interaction between the fragments. This is done through Wigner-Seitz approximation \cite{Bondorf1}. It is assumed that during the break up process a uniform dilute charge distribution within the freeze-out radius $R_f(b)$ (greater than normal radius $R_s(b)$ of the projectile spectator formed at impact parameter $b$) contracts successively due to density fluctuation into denser fragments (at normal nuclear density) of radius $R_{N,Z}$ and hence one can write the Coulomb energy as
\begin{equation}
E_c=\frac{3}{5}\frac{Z_s^2 e^2}{R_f(b)}+\sum_{N,Z}\frac{3}{5}\frac{Z^2 e^2}{R_{N,Z}}\bigg{(}1-\frac{R_s(b)}{R_f(b)}\bigg{)}
\end{equation}
\indent
Since, the canonical model calculation is done at a fixed freeze-out volume $V_f(b)$, the constant term $\frac{3}{5}\frac{Z_s^2 e^2}{R_f(b)}$ is not significant and one can write $a^{*}_c=a_{c}\{{1-(V_0(b)/V_f(b)})^{1/3}\}$ with $a_{c}=0.72$ MeV. The term $\frac{T^2A}{\epsilon_0}$ ($\epsilon_{0}=16.0$ MeV) represents contribution from excited states since the composites are at a non-zero temperature.\\
In order to specify which nuclei are considered in computing$ Q_{N_s,Z_s}$ [Eq.(\ref{Multifragmentation_Eq5})] a ridge has been included along the line of stability.  The liquid-drop formula above also gives neutron and proton drip lines and the results shown here include all nuclei within the boundaries.\\
The entire break up calculation is repeated for each projectile spectator created after abrasion stage with different temperatures at different impact parameters. Let, $\langle n_{N.Z}^{N_s,Z_s,T_i}\rangle$ be the average number of fragments having $N$ neutrons and $Z$ protons created after the multifragmentation of a projectile spectator ($N_s,Z_s$) at temperature $T_i$, then cross-section after multifragmentation stage can be expressed as
\begin{equation}
\sigma_{m,N,Z,T_i}=\sum_{N_s,Z_s}\langle n_{N,Z}^{N_s,Z_s,T_i}\rangle \sigma_{a,N_s,Z_s,T_i}
\label{Multifragmentation_Eq8}
\end{equation}
\indent
This is the most important stage of the model and this stage can be replaced by another statistical multifragmentation model(SMM) \cite{Bondorf1} but the results are expected to be very similar \cite{Chaudhuri_plb}. The main advantage of CTM is that, in CTM there is no Monte-Carlo simulation (like SMM), the multiplicities are calculated by using simple recursion relation (Eq. \ref{Multifragmentation_Eq4}), therefore isotopes having very small multiplicities can be  produced very accurately using CTM.
\subsection{Evaporation}
The canonical thermodynamical model described above calculates the properties of the collision averaged system that can be approximated by an equilibrium ensemble. Ideally, one would like to measure the properties of excited primary fragments after emission in order to extract information about the collisions and compare directly with the equilibrium predictions of the model. However, the time scale of a nuclear reaction($10^{-20}s$) is much shorter than the time scale for particle detection ($10^{-9}s$). Before reaching the detectors, most fragments decay to stable isotopes in their ground states. Thus before any model simulations can be compared to experimental data, it is indispensable to have a model that simulates sequential decays \cite{Mallik1,Chaudhuri}. The fragments can $\gamma$-decay to shed their energy but may also decay by light particle emission to lower mass nuclei.  The emissions of $n,p,d,t,^3$He and $^4$He are included in addition to $\gamma$. Particle decay widths are obtained using the Weisskopf's evaporation theory \cite{Weisskopf}. Fission is also included as a de-excitation channel though for the nuclei of mass $<100$, its role will be quite insignificant.\\
\indent
The  CTM calculation gives average multiplicities (or cross-sections) for different $N$ and $Z$ but the evaporation model involves Monte-Carlo simulations, hence event by event description becomes mandatory. In order to do that, the average cross-sections are multiplied by a large number and each is considered as a separate event. The decay scheme of each of this event is studied and finally an averaging is done over all the events.\\
\indent
According to Weisskopf's conventional evaporation theory, the partial decay width for emission of a light particle of type $\nu$ is given by
\begin{equation}
\Gamma_{\nu} =\frac {sm\sigma_0}{\pi^{2}\hbar^{2}} \frac {(E^*-E_0-V_\nu)}{a_R} \exp({2\sqrt{a_R(E^*-E_0-V_\nu)}-2\sqrt{a_PE^*}})
\label{Evaporation_Eq1}
\end{equation}
Here  $m$ is the mass of the emitted particle, $s$ is its spin degeneracy. $E_0$ is the particle separation energy which is calculated from the binding energies of the parent nucleus, daughter nucleus and the binding energy of the emitted particle and the liquid drop model is used to calculate the binding energies. The subscript $\nu$ refers to the emitted particle, $P$ refers to the parent nuclei and $R$ refers to the residual(daughter) nuclei. $a_P$  $\&$ $a_R$ are the level density parameters of the parent and residual nuclei respectively. The level density  parameter is given by $a = A/16  MeV^{-1}$ and it connects the excitation energy $E^*$ and temperature $T$ through the following relations.
\begin{eqnarray}
E^* = a_PT_P^2 \nonumber\\
(E^*-E_0-V_\nu) = a_RT_R^2.
\label{Evaporation_Eq2}
\end{eqnarray}
where $T_P$ $\&$ $T_R$ are the temperatures of the emitting(parent) and the final(residual) nucleus respectively. $V_\nu$ is the Coulomb barrier which is zero for neutral particles and non-zero for charged particles. In order to calculate the Coulomb barrier for charged particles of mass $A \ge 2$ a touching sphere approximation is used \cite{Friedman},
\begin{eqnarray}
V_\nu &=&
\frac{Z_\nu(Z_P-Z_\nu)e^2}{r_i\{A_\nu^{1/3}+{(A_P-A_\nu)}^{1/3}\}}\hspace{1cm}
\mbox{for} \hspace{1cm} A_\nu \ge 2\nonumber\\
&=&  \frac{(Z_P-1)e^2}{r_iA_P^{1/3}}  \hspace{2.9cm} \mbox{for}
\hspace{1cm} protons
\label{Evaporation_Eq3}
\end{eqnarray}
where $r_i$ is taken as 1.44m.
$\sigma_0$ is the geometrical cross-section (inverse cross section) associated with the formation of the compound nucleus (parent) from the emitted particle and the daughter nucleus and is given by $\sigma_0=\pi R^2$ where,
\begin{eqnarray}
R &=& r_0\{{(A_P-A_\nu)}^{1/3} + {A_\nu}^{1/3}\}\hspace{1.2cm}\mbox{for}\hspace{0.3cm} A_\nu \ge 2\nonumber\\
&=& r_0({A_P-1})^{1/3} \hspace{2.8cm}\mbox{for}\hspace{0.3cm} A_\nu = 1.
\label{Evaporation_Eq4}
\end{eqnarray}
where $r_0$ = 1.2 fm.
For the emission of giant dipole $\gamma$-quanta, the formula is taken from Ref. \cite{Lynn}
\begin{equation}
\Gamma_{\gamma}=\frac{3}{\rho_{P}(E^{*})}\int_{0}^{E^*}d\varepsilon\rho_{R}(E^*-\varepsilon)f(\varepsilon)
\label{Evaporation_Eq5}
\end{equation}
with
\begin{equation}
f(\varepsilon)= \frac{4}{3\pi}\frac{1+\kappa}{m_{n}c^{2}}\frac{e^{2}}
{\hbar c}\frac{N_PZ_P}{A_P}\frac{\Gamma_{G}\varepsilon^{4}}
{(\Gamma_{G}\varepsilon)^{2}+(\varepsilon^{2}-E_{G}^{2})^{2}}
\label{Evaporation_Eq6}
\end{equation}
with $\kappa=0.75$, and $E_{G}$ and $\Gamma_{G}$ are the position and width of the giant dipole resonance.\\
For the fission width, the simplified Bohr-Wheeler formula is used which is given by
\begin{equation}
\Gamma_f = \frac{T_P}{2\pi}\exp{(-B_f/T_P)}
\label{Evaporation_Eq7}
\end{equation}
where $B_f$ is the fission barrier of the compound nucleus given
by\cite{Guaraldo}
\begin{equation}
B_f (MeV)= -1.40Z_P + 0.22(A_P-Z_P) +101.5.
\label{Evaporation_Eq8}
\end{equation}
\indent
Once the emission widths ($\Gamma$'s) are known, it is required to establish the emission algorithm which decides whether a particle is being emitted from the compound nucleus or not. This is done \cite{Chaudhuri} by first calculating the ratio $x=\tau / \tau_{tot}$ where $\tau_{tot}= \hbar / \Gamma_{tot}$, $\Gamma_{tot}=\sum_{\nu}\Gamma_{\nu}$ and $\nu =n,p,d,t,He^3,\alpha,\gamma$ or fission and then performing Monte-Carlo sampling from a uniformly distributed set of random numbers. In the case that a particle is emitted, the type of the emitted particle is next decided by a Monte Carlo selection with the weights $\Gamma_{\nu}/\Gamma_{tot}$ (partial widths). The kinetic energy of the emitted particle is subsequently determined by a third Monte-Carlo sampling of its energy spectrum. The energy, mass and charge of the nucleus is adjusted after each emission and the entire procedure is repeated until the resulting products are unable to undergo further decay. This procedure is followed for each of the primary fragment produced at a fixed temperature and then repeated over a large ensemble and the observables are calculated from the ensemble averages. The number and type of particles emitted and the final decay product in each event is registered and are taken into account properly keeping in mind the overall charge and baryon number conservation. This is the third and final stage of the calculation. The same calculation is repeated for each set of fragments produced after multifragmentation at different temperatures.
\section{Parameterized Temperature profile}
Initially with the increase of the projectile beam energy, the temperature of the projectile spectator also increases. But above a certain energy of the projectile beam the temperature of the projectile spectator will not increase. This is known as limiting fragmentation \cite{LeBrun}. This projectile fragmentation model is valid in the limiting fragmentation region. The main reasons behind the excitation of projectile spectator are its highly non-spherical shape and migration of some nucleons from participant to the projectile spectator.\\
\begin{figure}[h!]
\begin{center}
\includegraphics[width=3.25in,height=3.05in]{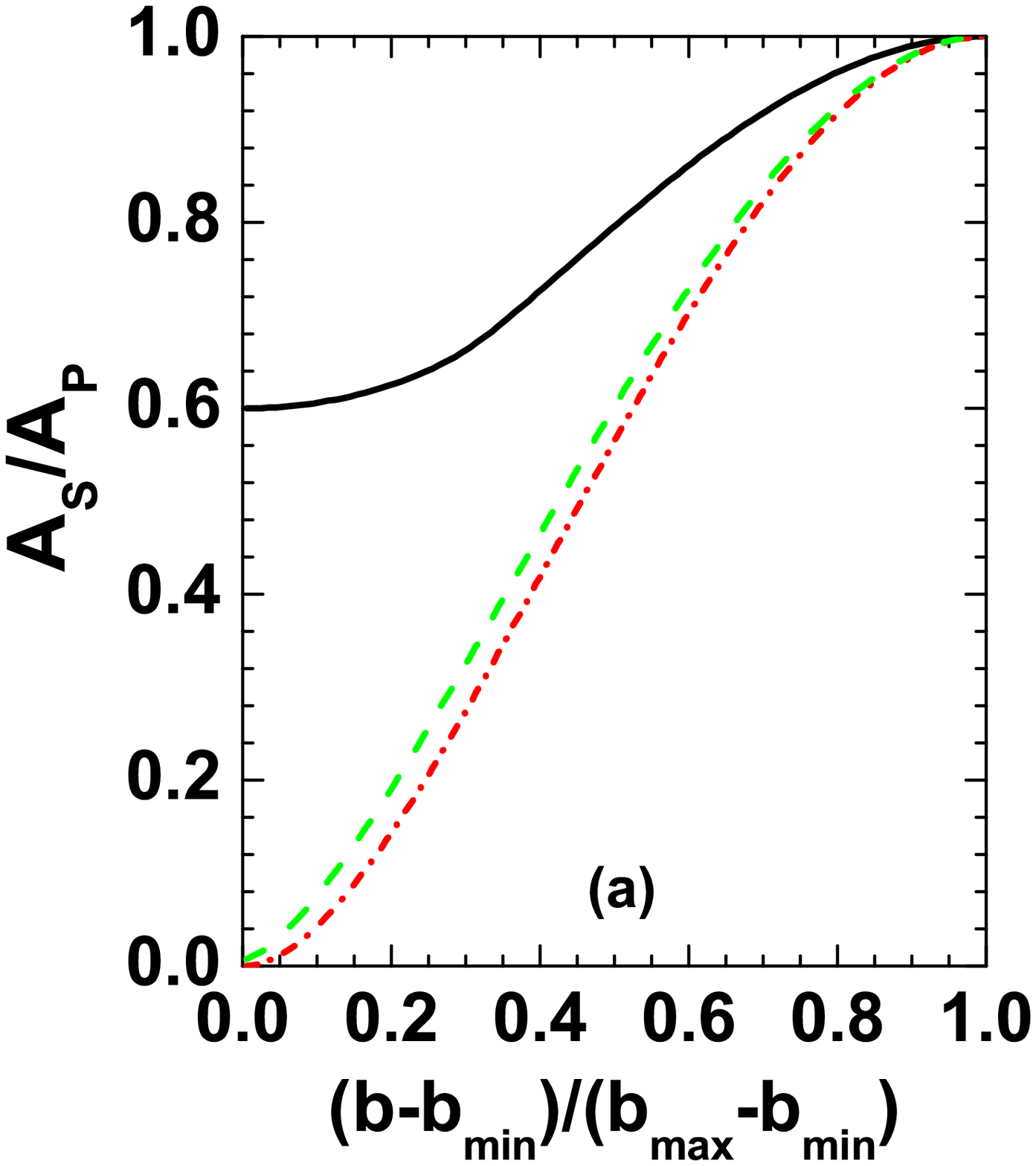}
\includegraphics[width=3.05in,height=3.036in]{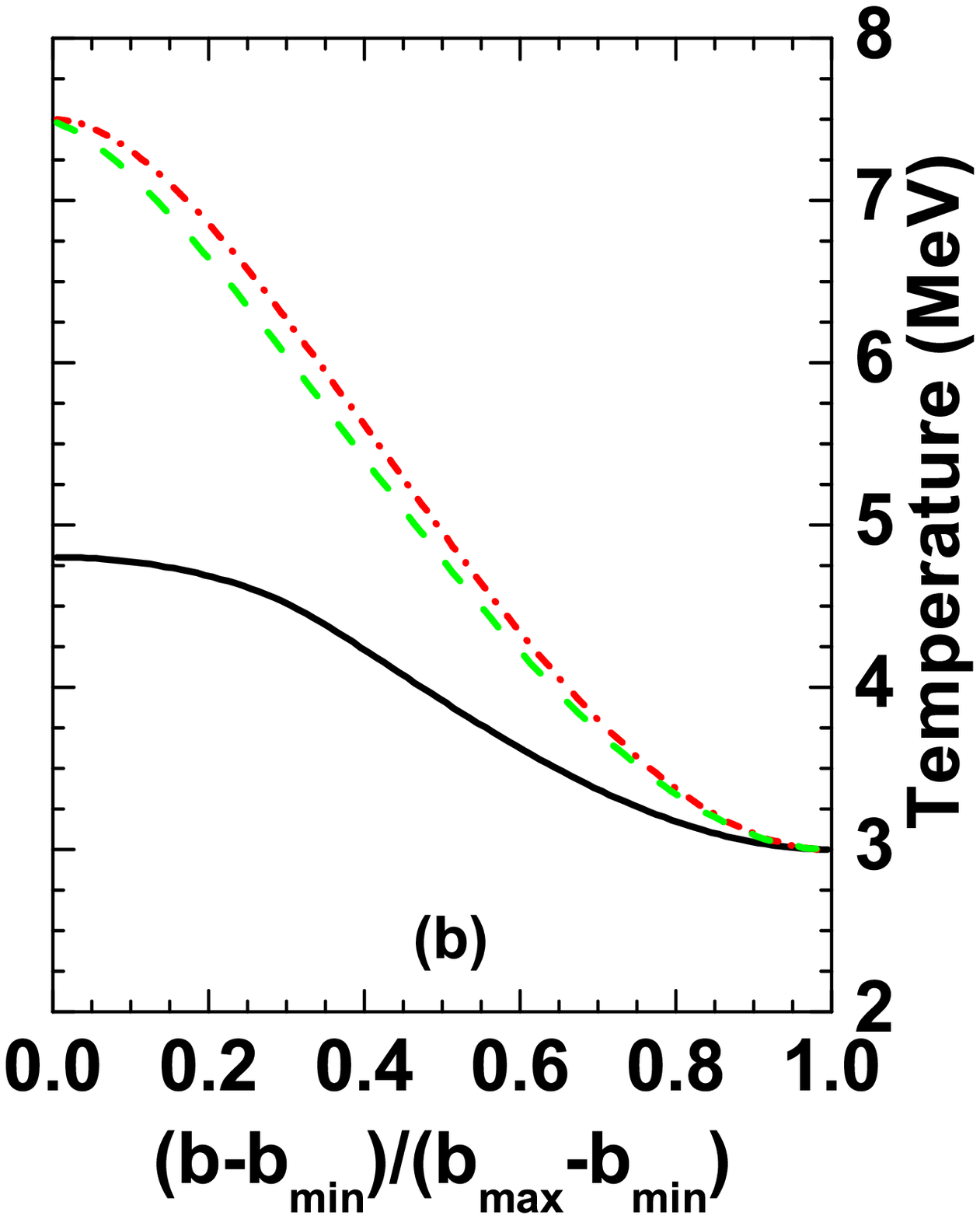}
\caption[$A_s/A_P$ and temperature variation with impact parameter]{Variation of (a) $A_s/A_P$ and (b) temperature with normalized impact parameter $(b-b_{min})/(b_{max}-b_{min})$ for $^{58}$Ni on $^9$Be (solid line), $^{58}$Ni on $^{181}$Ta (dotted line) and $^{124}$Sn on $^{119}$Sn (dashed line) reactions.}
\label{PLF_parameterized_temperature}
\end{center}
\end{figure}
\indent
To get the impact parameter dependent temperature profile i.e. $T=T(b)$ two types of parametrization can be suitable. The simplest case is when the temperature directly depends upon the impact parameter i.e.
\begin{equation}
T(b)=C_0+C_1b+C_2b^2+...
\label{Temperature_Eq1}
\end{equation}
In a more physically modified parametrization the temperature depends on the wound that the projectile suffers in the collision i.e. $1.0-A_s(b)/A_P$, so in this case,
\begin{equation}
T(b)=D_0+D_1(A_s(b)/A_P)+D_2(A_s(b)/A_P)^2+...
\label{Temperature_Eq2}
\end{equation}
\begin{figure}[h!]
\begin{center}
\includegraphics[width=9cm,height=9cm,clip]{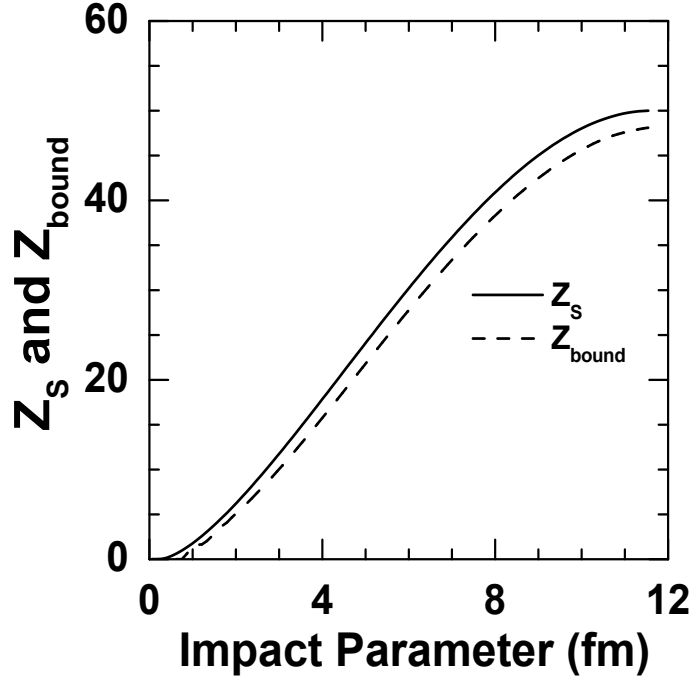}
\caption[$Z_{s}$ and $Z_{bound}$ variation with impact parameter]{Variation of $Z_{s}$ (solid line) and $Z_{bound}$ (dashed line) with impact parameter for $Sn^{107}$ on $Sn^{119}$ reaction.}
\label{Z_s_and_Z_bound_vs_b}
\end{center}
\end{figure}
After calculating different observables of projectile fragmentation by using these temperature profiles and comparing the theoretical results with experimental data, it is observed that linear parameterizations are enough i.e. $C_2$, $C_3$... or $D_2$, $D_3$... are negligible. Since $T(b)=D_0+D_1(A_s(b)/A_P)$ is physically more acceptable than $T(b)=C_0+C_1b$, we finally choose this temperature profile. The values of $D_0$ and $D_1$ are obtained by comparing theoretical model results with experimental data of mass distribution and multiplicity of intermediate mass fragments (fragments having charge between 3 to 20) of different target-projectile combinations. The comparison led to the values $D_0$=7.5 and $D_1$=-4.5 which are used in subsequent calculations \cite{Mallik3,Mallik101,Mallik102}. Further in Chapter 3, the temperature profile is obtained from microscopic static model calculation and Boltzmann--Uehling-Uhlenbeck (BUU) calculation and compared with this parameterized temperature profile. Nice agreement between them proves the acceptability of this simple parameterized formula,
\begin{equation}
T(b)=7.5-4.5(A_s(b)/A_P)
\label{Temperature_Eq3}
\end{equation}
\indent
For three different nuclear reactions $^{58}$Ni on $^9$Be, $^{58}$Ni on $^{181}$Ta and $^{124}$Sn on $^{119}$Sn, the variation of the quantity $A_s/A_P$ obtained after abrasion stage with normalized impact parameter $(b-b_{min})/(b_{max}-b_{min})$ is shown in Fig. 2.2.a where as Fig. 2.2.b represents the freeze-out temperature profile of these three reactions calculated from the formula $T(b)=7.5-4.5(A_s(b)/A_P)$. This parametrization has profound consequences.  This implies that the temperature profile $T(b/b_{max})$ of $^{124}$Sn on $^{119}$Sn is very different from that of $^{58}$Ni on $^9$Be. In the first case $A_s(b)/A_P$ is nearly zero for $b=b_{min}$=0 whereas in the latter case $A_s(b)/A_P$ is $\approx 0.6$ for $b=b_{min}$=0. Even more remarkable feature is that the temperature profile of $^{58}$Ni on $^9$Be is so different from the temperature profile of $^{58}$Ni on $^{181}$Ta. In the latter case $b_{min}=R_{Ta}-R_{Ni}$ and beyond $b_{min}$, $A_s(b)/A_P$ grows from zero to 1 for $b_{max}$. This is very similar to the temperature profile of $^{124}$Sn on $^{119}$Sn.\\
\begin{figure}[t]
\begin{center}
\includegraphics[height=3.1in,width=5.2in,clip]{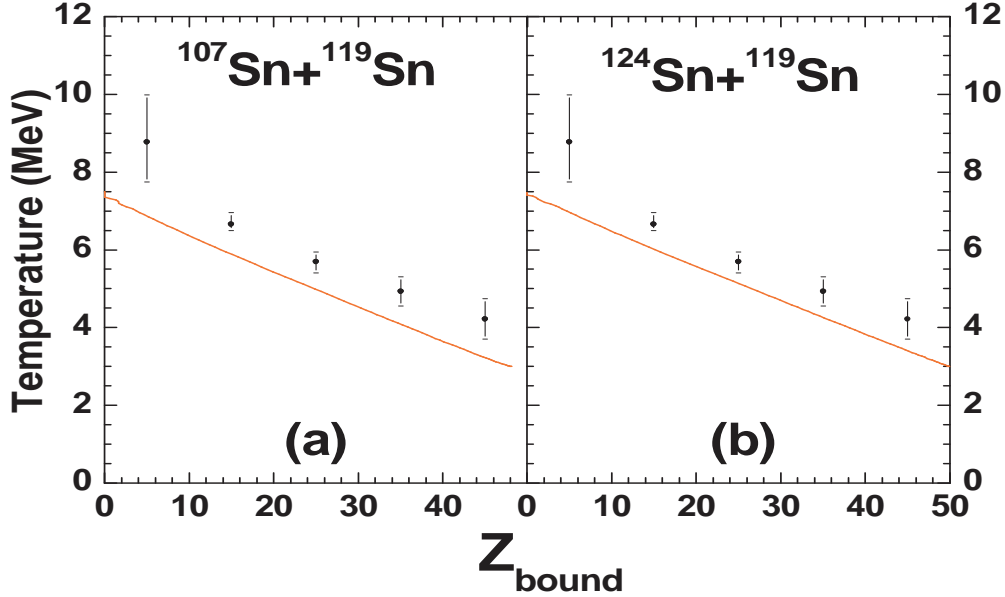}
\caption[Comparison of temperature profiles]{Comparison of theoretically used temperature profile calculated by the formula  $T(b)=7.5-4.5(A_s(b)/A_P)$ (solid lines) with that deduced by Albergo formula from experimental data \cite{Ogul} (circles with error bars) for (a) $^{107}$Sn on $^{119}$Sn and (b) $^{124}$Sn on $^{119}$Sn.}
\label{Chapter2_Fig3}
\end{center}
\end{figure}
\indent
Experimentally neither impact parameter nor mass ($A_{s}$) of the abraded projectile can be measured directly. But in experiments indirect determination of impact parameter is done by measuring $Z_{bound}$ (=$Z_s$ minus charges of all composites with charge $Z=1$). Fig. \ref{Z_s_and_Z_bound_vs_b} shows the variation of $Z_{s}$ and $Z_{bound}$ with impact parameter for $^{107}$Sn on $^{119}$Sn obtained from our projectile fragmentation model.\\
In Fig.\ref{Chapter2_Fig3} the temperatures calculated from the model is plotted with $Z_{bound}$  and compared with experimentally measured temperatures (by Albergo formula \cite{Albergo}) for two different projectile fragmentation reactions $^{107}$Sn on $^{119}$Sn and $^{124}$Sn on $^{119}$Sn. Nice agreement in both cases establishes the validity of the parametrization used.
\section{Results}
The projectile fragmentation model is used to calculate the basic observables of projectile fragmentation like mass distribution, charge distribution, differential charge distribution, isotopic distribution etc. for different nuclear reactions at intermediate energies with different projectile target combinations. The average number of intermediate mass fragments ($M_{IMF}$), the average size of the largest cluster and their variation with bound charge ($Z_{bound}$) are also calculated from this model.
\subsection{Charge Distribution}
The total charge distribution of $^{58}$Ni on $^9$Be reaction calculated from projectile fragmentation model after three different stages are shown in Fig. \ref{Charge_distribution_three_stages}. After abrasion stage only $Z=16$ to $Z=28$ region is populated. PLF having $Z<16$ is not created because projectile is much larger than the target. But due to breaking of the PLF's into different composites in the multifragmentation stage entire charge spectrum is populated. In the evaporation stage, these excited composites only emit light particles and $\gamma$ rays to reach to their stable ground state, therefore evaporation only modifies the cross-section and completes the calculation. Further theoretical results of this chapter are obtained after evaporation stage calculation.\\
\begin{figure}[!h]
\begin{center}
\includegraphics[width=9cm,keepaspectratio=true,clip]{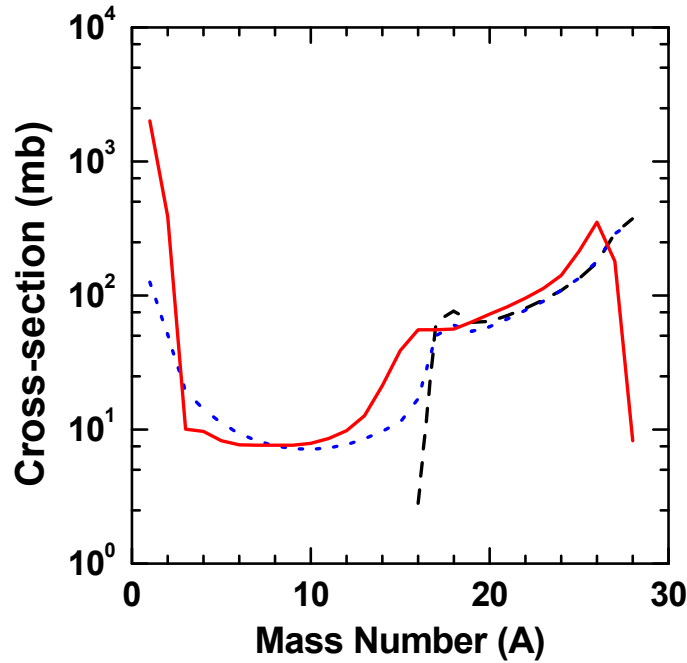}
\caption[Charge distribution after abrasion, multifragmetation and evaporation stage]{Total charge distribution calculated by projectile fragmentation model after abrasion (black dashed), multifragmentation (blue dotted) and evaporation (red solid) stage for $^{58}$Ni on $^{9}$Be reaction.}
\label{Charge_distribution_three_stages}
\end{center}
\end{figure}
\indent
The total charge distributions of different experiments ($^{124}$Xe and $^{136}$Xe on $^{208}$Pb at 1 GeV/nucleon \cite{Henzlova}, $^{124}$Sn on $^{119}$Sn at 600 MeV/nucleon \cite{Ogul,Sfienti} and $^{129}$Xe on $^{29}$Al at 790 MeV/nucleon \cite{Reinhold} are performed at GSI. $^{58}$Ni on $^{9}$Be, $^{64}$Ni on $^{9}$Be, $^{40}$Ca on $^{9}$Be, $^{48}$Ca on $^{9}$Be, $^{58}$Ni on $^{181}$Ta, $^{64}$Ni on $^{181}$Ta, $^{40}$Ca on $^{181}$Ta and $^{48}$Ca on $^{181}$Ta, all reactions at 140 MeV/nucleon done at MSU \cite{Mocko_thesis,Mocko}.) are theoretically calculated from the projectile fragmentation model by using the same temperature profile. This is shown in Fig. \ref{Charge_distribution}. In theoretical calculation cross-section of all fragments ranging from light nucleon to original projectile are calculated separately, but in Fig. \ref{Charge_distribution} the cross-section of the fragments for which experimental data are available are only shown.
\begin{figure}[!h]
\begin{center}
\includegraphics[width=15cm,keepaspectratio=true,clip]{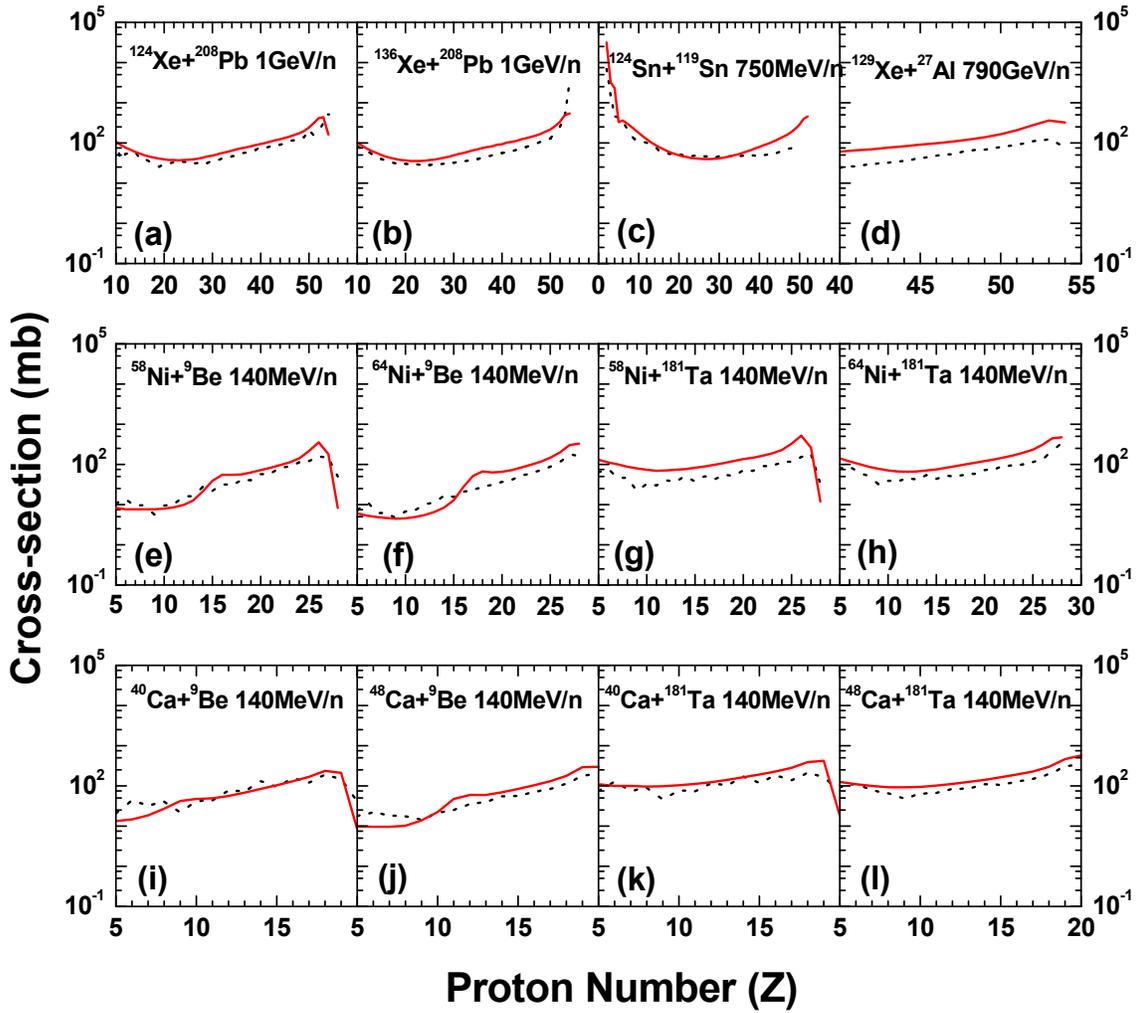}
\caption[Charge distribution in projectile fragmentation]{Theoretical total charge distribution (solid lines) for different projectile fragmentation reactions (see text) compared with experimental data (dashed lines).}
\label{Charge_distribution}
\end{center}
\end{figure}
\subsection{Mass Distribution}
Fig. \ref{Mass_distribution} shows the comparison of theoretical result and experimental data of mass distribution for (a) $^{124}$Xe on $^{208}$Pb, (b) $^{136}$Xe on $^{208}$Pb, (c) $^{58}$Ni on $^{9}$Be (d) $^{64}$Ni on $^{9}$Be (e) $^{58}$Ni on $^{181}$Ta, (f) $^{64}$Ni on $^{181}$Ta (g) $^{40}$Ca on $^{9}$Be (h) $^{48}$Ca on $^{9}$Be (i) $^{40}$Ca on $^{181}$Ta and (j) ($^{48}$Ca on $^{181}$Ta projectile fragmentation reactions. For (a) and (b) the projectile beam energy is 1GeV/nucleon and for (c) to (j) it is 140 MeV/nucleon.\\
\begin{figure}[!h]
\begin{center}
\includegraphics[width=15cm,keepaspectratio=true,clip]{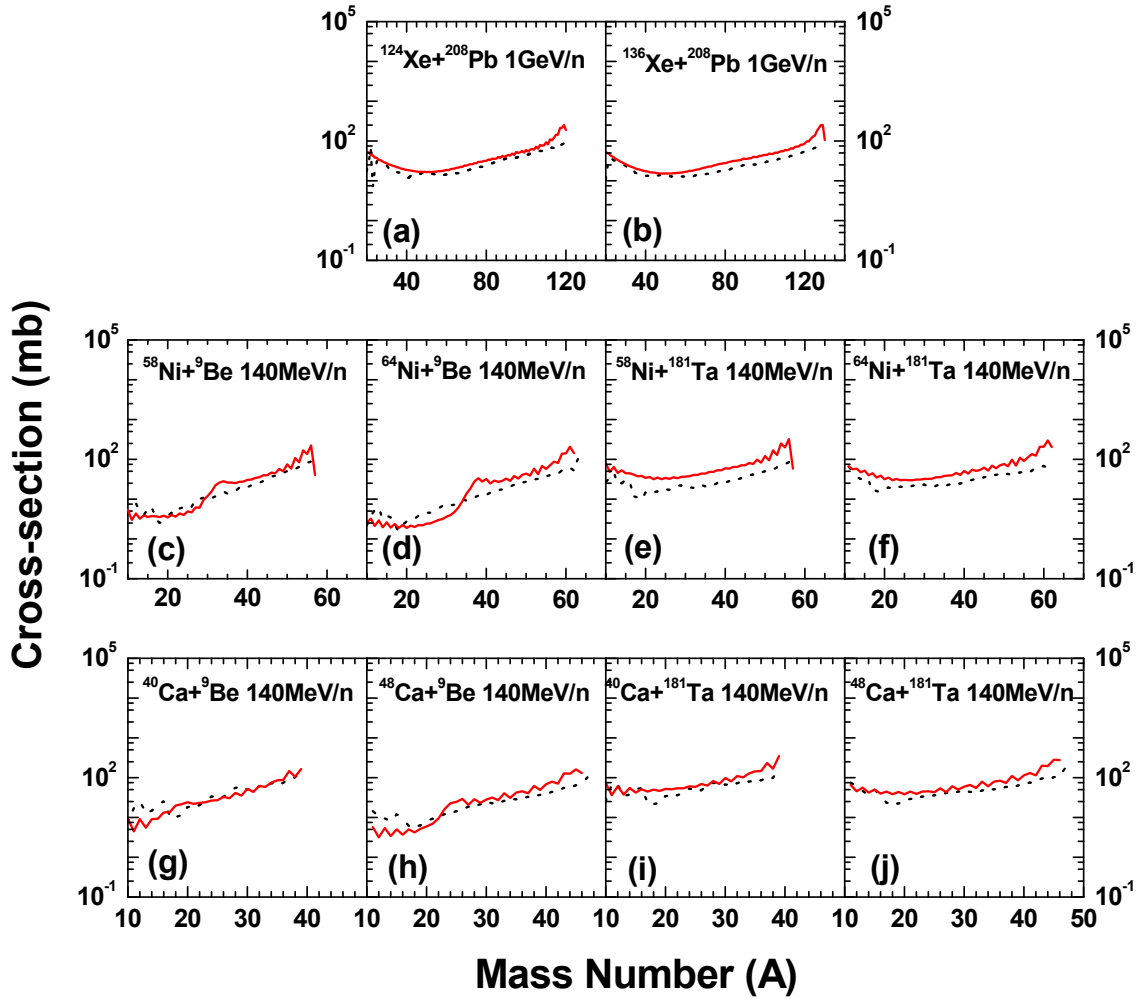}
\caption[Mass distribution in projectile fragmentation]{Theoretical total mass distribution (solid lines) for different projectile fragmentation reactions (see text) compared with experimental data (dashed lines).}
\label{Mass_distribution}
\end{center}
\end{figure}
\indent
It is observed from Fig. \ref{Charge_distribution} and Fig. \ref{Mass_distribution} that, though the projectile beam energies in experiments are widely different, the same temperature profile can explain all the data pretty well. This establishes the concept of limiting fragmentation \cite{LeBrun}.
\subsection{Isotopic Distribution}
Fig \ref{IsotopicNi58Ni64Ni68onBe9} shows the cross-section of different Si and Ca isotopes calculated by statistical projectile fragmentation model from projectile fragmentation reactions of three different projectiles $^{58}$Ni, $^{64}$Ni and $^{68}$Ni on the same target $^{9}$Be. In both cases the production cross-section of very neutron rich isotopes are higher from neutron rich projectiles. Reasonably good agreement has been observed in all cases.\\
\begin{figure}[!h]
\begin{center}
\includegraphics[width=14cm,keepaspectratio=true,clip]{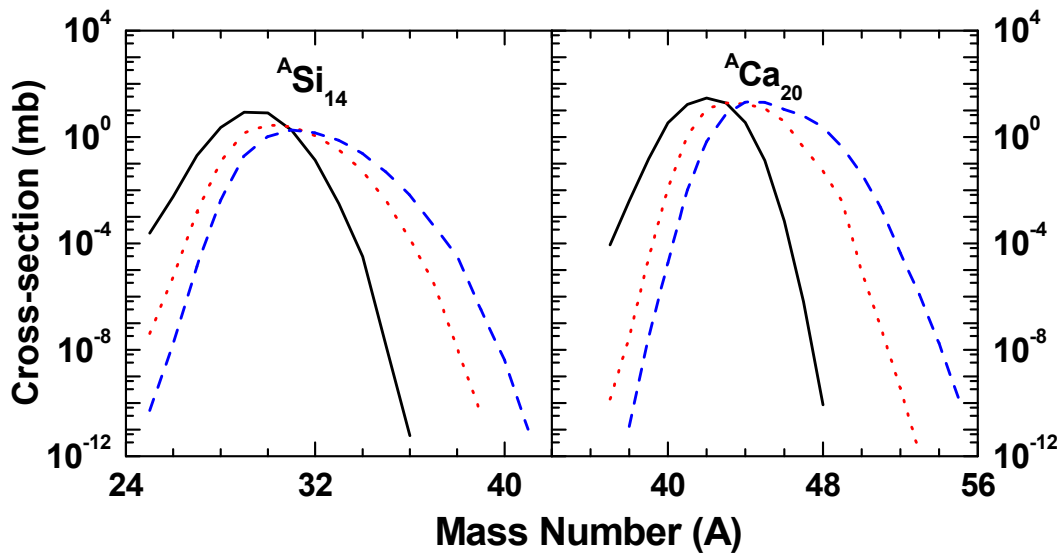}
\caption[Projectile isospin dependence on isotopic distribution]{Theoretically calculated cross-section of $Si$ and $Ca$ isotopes for $^{58}$Ni on $^{9}$Be (black solid lines), $^{64}$Ni on $^{9}$Be (red dotted lines) and $^{68}$Ni on $^{9}$Be (blue dashed lines) projectile fragmentation reactions.}
\label{IsotopicNi58Ni64Ni68onBe9}
\end{center}
\end{figure}
\indent
Fig. \ref{IsotopicNi58Be9}, \ref{IsotopicNi64Be9} and Fig. \ref{IsotopicNi58Ta181} shows the comparison of theoretically obtained isotopic distributions with experimental data for $^{58}$Ni on $^{9}$Be and $^{58}$Ni on $^{181}$Ta reactions at 140 MeV/necleon respectively. Fig. \ref{IsotopicXe124Pb208} shows the same for $^{124}$Xe on $^{208}$Pb reaction at 1 GeV/nucleon.\\
\begin{figure}[!h]
\begin{center}
\includegraphics[width=14.0cm,keepaspectratio=true,clip]{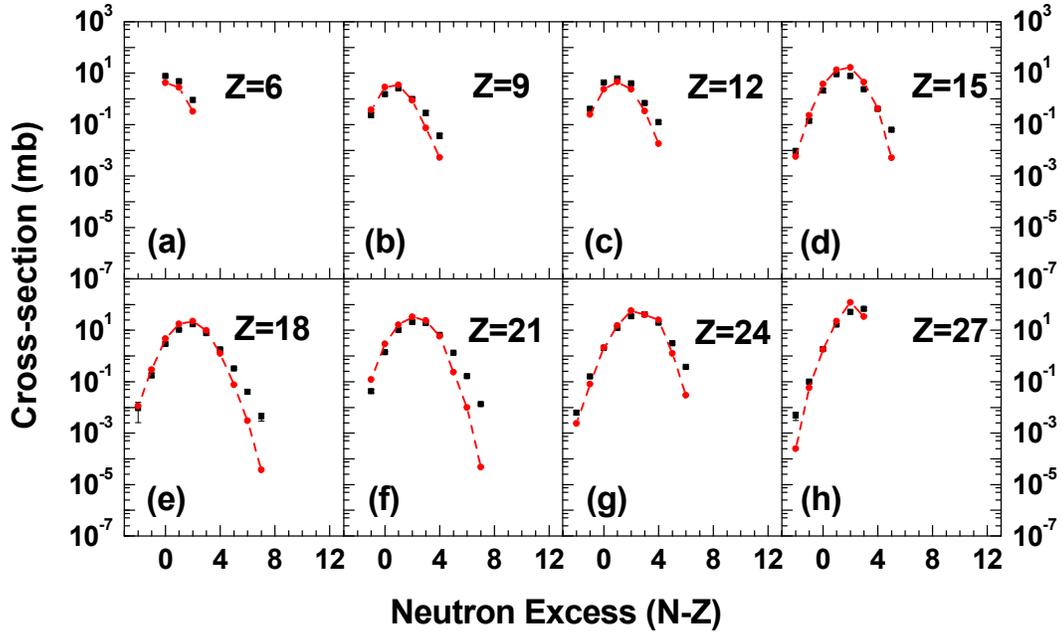}
\caption[Isotopic distribution in projectile fragmentation for $^{58}$Ni on $^{9}$Be reaction]{Theoretical isotopic cross-section distribution (circles joined by dashed lines) for $^{58}$Ni on $^{9}$Be reaction compared with experimental data \cite{Mocko} (squares with error bars).}
\label{IsotopicNi58Be9}
\end{center}
\end{figure}
\begin{figure}[!h]
\begin{center}
\includegraphics[width=14.0cm,keepaspectratio=true,clip]{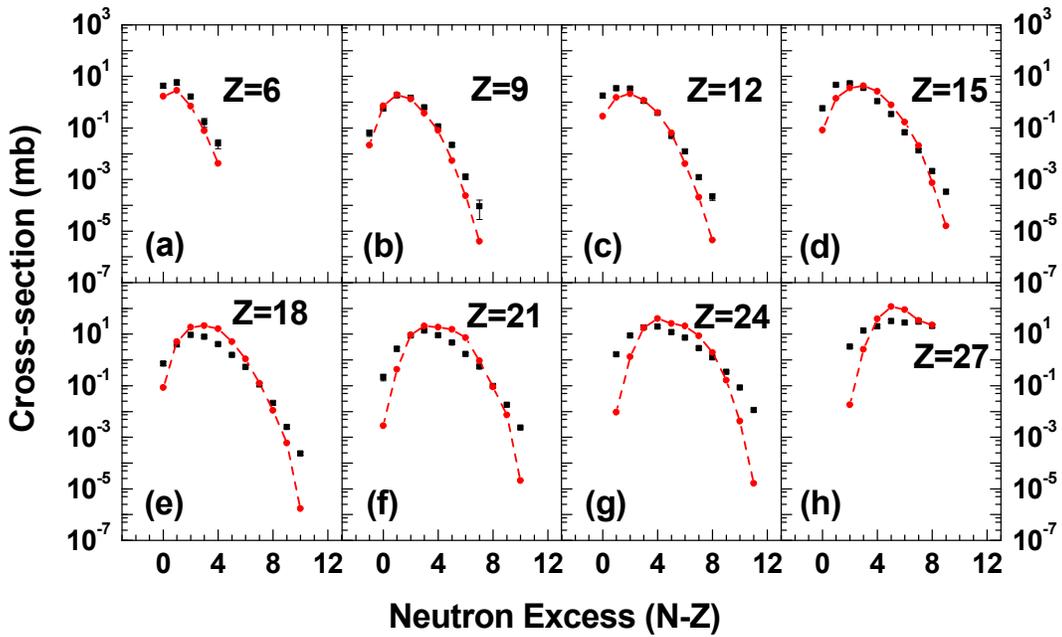}
\caption[Same as Fig. \ref{IsotopicNi58Be9} except that here the projectile is $^{64}$Ni instead of $^{58}$Ni]{Same as Fig. \ref{IsotopicNi58Be9} except that here the projectile is $^{64}$Ni instead of $^{58}$Ni.}
\label{IsotopicNi64Be9}
\end{center}
\end{figure}
\begin{figure}[!h]
\begin{center}
\includegraphics[width=14.0cm,keepaspectratio=true,clip]{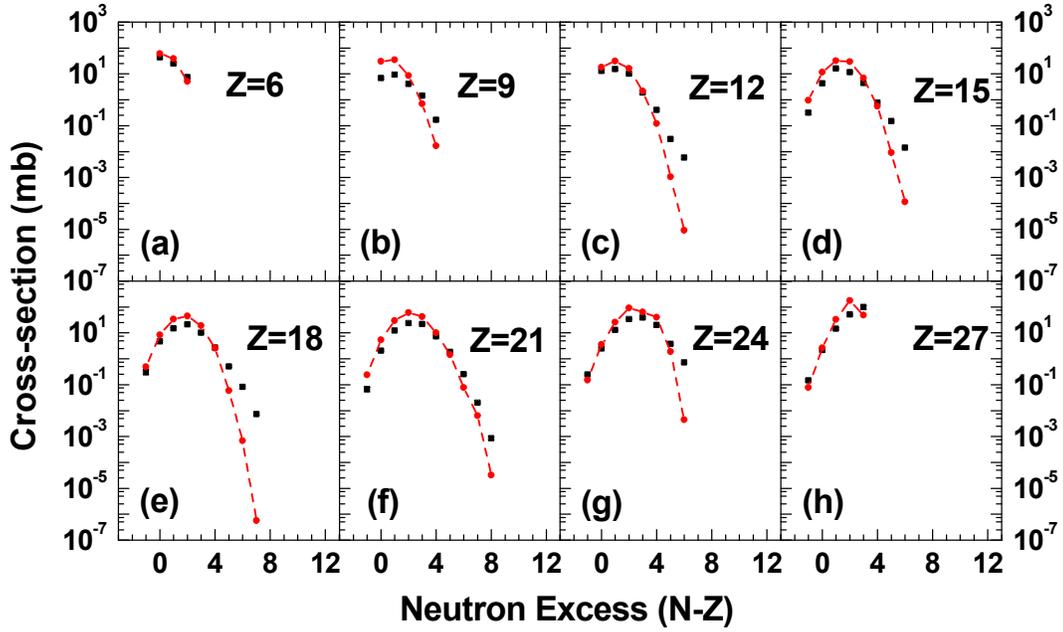}
\caption[Same as Fig. \ref{IsotopicNi58Be9} except that here the target is $^{181}$Ta  instead of $^{9}$Be.]{Same as Fig. \ref{IsotopicNi58Be9} except that here the target is $^{181}$Ta  instead of $^{9}$Be.}
\label{IsotopicNi58Ta181}
\end{center}
\end{figure}
\begin{figure}[!h]
\begin{center}
\includegraphics[width=14.0cm,keepaspectratio=true,clip]{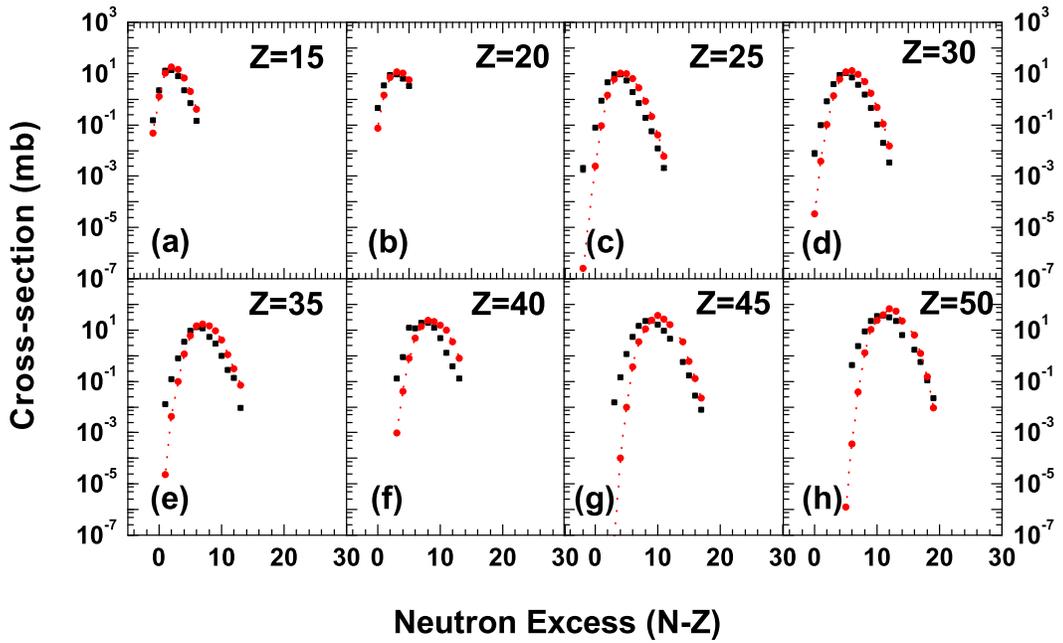}
\caption[Same as Fig. \ref{IsotopicNi58Be9} except that here the reaction is $^{124}$Xe on $^{208}$Pb]{Same as Fig. \ref{IsotopicNi58Be9} except that here the projectile fragmentation reaction is $^{124}$Xe on $^{208}$Pb at 1 GeV/nucleon.}
\label{IsotopicXe124Pb208}
\end{center}
\end{figure}
\subsection{Cross-section and binding energy of neutron rich nuclei}
From the previous section it is clear that the theoretical model reproduces the cross-sections of projectile fragmentation experiments very well. A remarkable feature is the correlation between the measured fragment cross-section ($\sigma$) and the binding energy per nucleon($B/A$). This observation has prompted attempts of parametrization of cross-sections \cite{Mocko2,Mocko3,Chaudhuri_be}.
One very successful parametrization is
\begin{equation}
\sigma=Cexp[\frac{B}{A}\frac{1}{\tau}]
\label{Cross-section_and_binding_energy}
\end{equation}
\begin{figure}[!h]
\begin{center}
\includegraphics[width=9.5cm,keepaspectratio=true,clip]{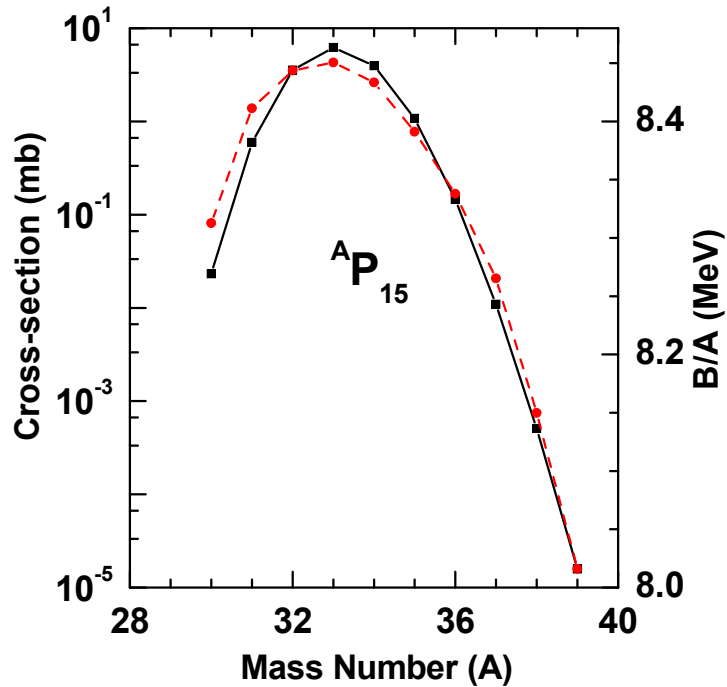}
\label{fig7}
\caption[Variation of fragment cross-section and binding energy with mass number]{Fragment cross-section (circles joined by red dotted line) for  $^{64}Ni$ on $^{9}Be$ reaction and binding energy per nucleon (squares joined by black solid line) plotted as a mass number for $Z=15$ isotopes.}
\label{Cross-section_bindingenergy}
\end{center}
\end{figure}
\indent
Here $\tau$ is a fitting parameter. In this parametrization, the pairing energy contribution in nuclear binding energy has not been considered. The production cross-sections of $Z=15$ isotopes from $^{64}Ni$ on $^{9}Be$ reaction has been calculated from projectile fragmentation model and plotted in log scale in Fig. \ref{Cross-section_bindingenergy} (circles joined by red dotted line). The variation of the theoretical binding energy per nucleon for same isotopes of $Z=15$ in linear scale is also shown in the same figure (squares joined by black solid line). The similar trend of the cross-section curve (in log scale) and binding energy curve (in linear scale) confirms the validity of above parametrization from the projectile fragmentation model. By this method one can interpolate (or extrapolate) the cross-section of an isotope if the binding energy is known. One can also estimate the binding energy of an isotope by measuring its cross-section experimentally.
\subsection{Differential charge distribution}
\begin{figure}[!h]
\begin{center}
\includegraphics[height=3.4in,width=5.5in,clip]{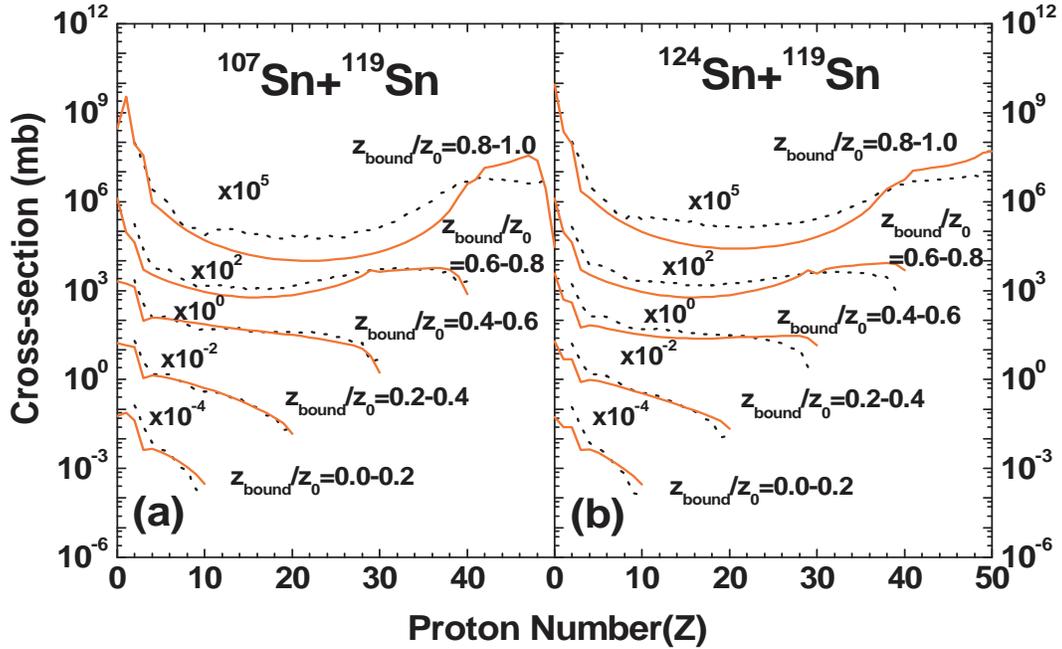}
\caption[Differential charge distribution in projectile fragmentation]{Theoretical differential charge cross-section distribution (solid lines) for (a) $^{107}$Sn on $^{119}$Sn and (b) $^{124}$Sn on $^{119}$Sn reaction compared with the experimental data (dashed lines).}
\label{Differential_charge_distribution}
\end{center}
\end{figure}
The differential charge distributions for different intervals of $Z_{bound}/Z_0$ are calculated by the projectile fragmentation model for $^{119}$Sn and $^{124}$Sn on $^{119}$Sn reactions and compared with experimental data \cite{Ogul}. This is shown in Fig. \ref{Differential_charge_distribution}. For the sake of clarity the distributions are normalized with different multiplicative factors. At peripheral collisions (i.e. $0.8 {\le} Z_{bound}/Z_0 {\le} 1.0$)  due to small temperature of PLF, it breaks into one large fragment and small number of light fragments, hence the charge distribution shows $U$ type nature. But with the decrease of impact parameter the temperature increases, the PLF breaks into larger number of fragments and the charge distributions become steeper. The features of the data are nicely reproduced by the model.
\subsection{$M_{IMF}$ variation with $Z_{bound}$}
The emission of intermediate mass fragments (IMF) is the most important observable of nuclear multifragmentation reactions. In literature, IMF is usually defined as a fragment with charge $3{\le}Z{\le}20$. Sometimes fragments having charge $3{\le}Z{\le}30$ or $3{\le}Z{\le}Z_0/3$ ($Z_0$ is the charge of excited compound nuclear system) are considered as intermediate mass fragments. Generally the upper limit is set not to include fission like fragments and the lower limit is set to exclude the evaporated particles (proton, neutron, alpha etc.). The "rise and fall" nature of intermediate mass fragment multiplicities is also an important signature for nuclear liquid gas phase transition \cite{Dasgupta_Phase_transition,Peaslee,Ogilvie,Tsang}.
\begin{figure}[!h]
\begin{center}
\includegraphics[height=3.4in,width=5.5in,clip]{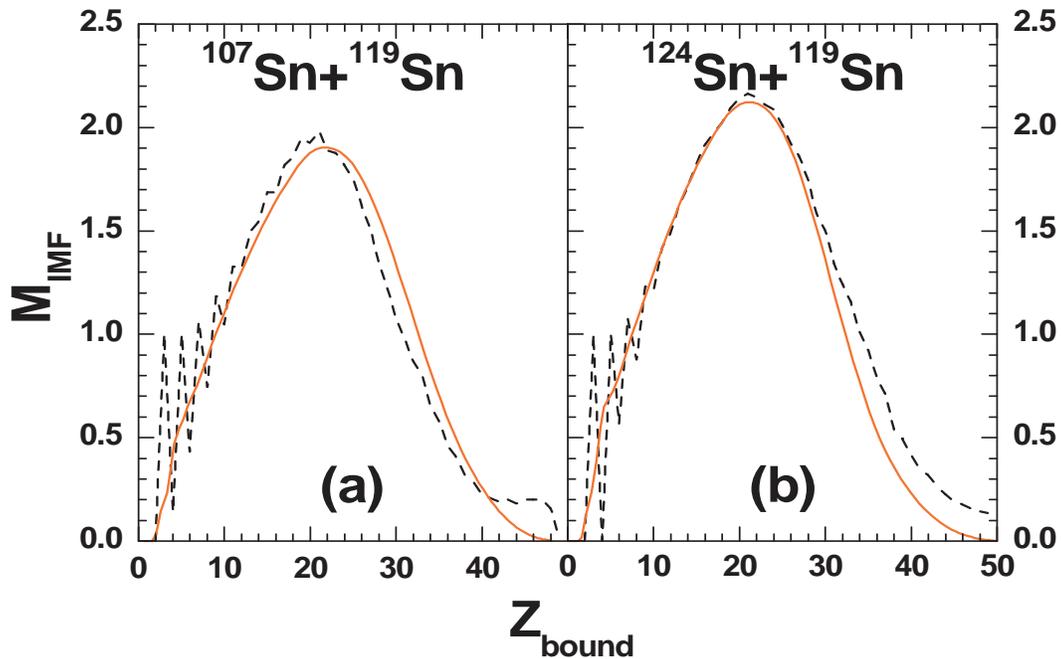}
\caption[$M_{IMF}$ variation with $Z_{bound}$ in projectile fragmentation]{Mean multiplicity of intermediate-mass fragments $M_{IMF}$, as a function of $Z_{bound}$ for (a) $^{107}$Sn on $^{119}$Sn and (b) $^{124}$Sn on $^{119}$Sn reaction obtained from projectile fragmentation model (solid lines). The experimental results are shown by the dashed lines. }
\label{IMF}
\end{center}
\end{figure}
\indent
The variation of the average multiplicity of intermediate mass fragments ($M_{IMF}$) ($3{\le}Z{\le}20$) with $Z_{bound}$ for $^{107}$Sn on $^{119}$Sn and $^{124}$Sn on $^{119}$Sn reactions is shown in Fig. \ref{IMF}. The theoretical calculation reproduces the average trend of the experimental data very well. At small impact parameters, the size of the projectile spectator (also $Z_{bound}$) is small and the temperature of the dissociating system is very high. Therefore the PLF will break into fragments of small charges (mainly $Z=1, 2$). Therefore the IMF production is less. But at mid-central collisions PLF's are larger in size and the temperature is smaller compared  to the  previous case, therefore larger number of IMF's are produced. With further increase of impact parameter, though the PLF size (also $Z_{bound}$) increases,  the temperature is low, hence breaking of dissociating system is inhibited (large fragment remains) and therefore IMF production is less.\\
\indent
The overall feature of the figure is that the general shapes of the theoretical and experimental curves agree. However, there are significant fluctuations in the experimental values of $M_{IMF}$ for low values of $Z_{bound}$ whereas theory completely misses these fluctuations. Experimentally $Z_{bound}$ is obtained event by event and in every event $Z_{bound}$ is an integer. From many events with the same $Z_{bound}$ one can obtain $M_{IMF}$. In the present calculation, although $Z_s$ is an integer but $Z_s-\sum_in_{z=1}(i)$ will usually be non-integer since the $n_{z=1}(i)$'s are generally non-integers. This would be quite wrong if values of $M_{IMF}$ belonging to neighbouring $Z_{bound}$'s differ strongly (as it happens for very small systems) but for large systems the difference would be small and our prescription is adequate for an estimate. In next part the case for $Z_{bound}=3$, $4$ and $5$  will be considered separately.\\
\indent
If $Z_{bound}$=3, it guarantees the formation of a $^A_3$Li nucleus. Thus for $Z_{bound}$=3, $M_{IMF}$ is $1$. If $^A_3$Li decays by a proton emission then $Z_{bound}$ becomes $2$.  Also there is no IMF. If it decays by neutron emission to a particle stable state of a different isotope of Li, then $M_{IMF}$ is $1$. There are several particle stable states of Li so $Z_{bound}=3$, hence $M_{IMF}$=1 is always satisfied.\\
\indent
For $Z_{bound}=5$, there are two possibilities, either one Boron nucleus or a Li nucleus plus a He nucleus. In both the cases $M_{IMF}$=1. If the Boron nucleus sheds a proton, the status drops to $Z_{bound}$=4 and it is equivalent to the $Z_{bound}$=4 case. If the Boron nucleus sheds one or more neutrons to reach a particle stable state it maintains $Z_{bound}=5, M_{IMF}$=1. If Boron decays into a Li and He two things can happen. It can reach a particle stable state of Li i.e. $Z_{bound}$ becomes $5$, $M_{IMF}=1$.  If the Li sheds a proton then it longer have $Z_{bound}$=5. Thus, for $Z_{bound}=5$, $M_{IMF}$ is always $1$.\\
\indent
For, $Z_{bound}$=4, there is a Be nucleus with $N_{IMF}$=1 but it can also decay into two He isotopes which still retains $Z_{bound}$=4 but with $M_{IMF}$=0. Therefore value of $M_{IMF}$ is not same for all $Z_{bound}=4$ cases. To calculate the $M_{IMF}$  more precisely, two modifications are required in the existing model. In projectile fragmentation model except for nuclei up to $^4$He, the liquid-drop model is used for calculating the ground state energy and the Fermi-gas model for determining the excited states. For small PLF's which are inaccurate therefore these are replaced by the experimental values of ground state and excited state energies. Next, the decays of hot composites resulting from CTM are only considered.  By using it, the calculated value of $M_{IMF}$ at $Z_{bound}=4$ is $0.145$ for $^{107}$Sn on $^{119}$Sn reaction and $0.38$ for $^{124}$Sn on $^{119}$Sn reaction. The details of the calculation procedure is described in Ref. \cite{Mallik_unpublished}.
\subsection{Charge of largest cluster and its variation with $Z_{bound}$}
Largest cluster is the most useful order parameter for studying nuclear liquid gas phase transition. The theoretical calculation of largest cluster charge ($Z_{max}$) for the case of fragmentation of a PLF with $N_s$ neutrons and $Z_s$ protons (produced after abrasion stage at impact parameter $b$) at temperature $T$ is discussed in the first part of this subsection and then the variation of $Z_{max}$ with $Z_{bound}$ for projectile fragmentation is shown.\\
\indent
As described in section 2.2.2, there is an enormous number of channels for the case of fragmenting system having $N_s$ neutrons and $Z_s$ protons. Different channels will have different $Z_{max}$. For example, there is a term $\frac{\omega_{1,0}^Z}{Z!}\frac{\omega_{0,1}^N}{N!}$ in the sum of Eq. \ref{Multifragmentation_Eq3}. The probability of occurrence of this channel is $\frac{1}{Q_{N_s,Z_s}}\frac{\omega_{1,0}^Z}{Z!}\frac{\omega_{0,1}^N}{N!}$.The full partition function can be written as $Q_{N_s,Z_s}(\omega_{1,0},\omega_{0,1},\omega_{1,1},....\omega_{i,j}....)$. If $Q_{N_s,Z_s}$ is constructed such that all $\omega$'s except $\omega_{1,0}$ and $\omega_{0,1}$ are zero then this  $Q_{N_s,Z_s}=Q_{N_s,Z_s}(\omega_{1,0},\omega_{0,1},0,0,0....)=\frac{\omega_{0,1}^Z}{Z!}\frac{\omega_{1,0}^N}{N!}$ and this has $Z_{max}=1$. If $Q_{N_s,Z_s}$ is constructed with only three $\omega$'s, then $Q_{N_s,Z_s}=Q_{N_s,Z_s}(\omega_{1,0},\omega_{0,1},\omega_{1,2},0,0,0....)$. This will have $Z_{max}$ sometimes $1$ (as $\frac{\omega_{0,1}^Z}{Z!}\frac{\omega_{1,0}^N}{N!}$ is still there) and sometimes 2 (as, for example, in the term $\frac{\omega_{1,2}^3}{3!}\frac{\omega_{0,1}^{Z-6}}{(Z-6)!}\frac{\omega_{1,0}^{N-3}}{(N)!})$. By using this concept one can write a general formula for determining the largest cluster probability \cite{Chaudhuri_largest_cluster}. To obtain the probability that a given value $Z_{max}$ occurs as the maximum charge, $Q_{N_s,Z_s}(Z_{m})$ can be constructed such that all values of $\omega_{N,Z}$ are set at $0$ when $Z\ge Z_{m}$. So, $Q_{N_s,Z_s}(Z_{max})/Q_{N_s,Z_s}(Z_{0})$ represents the probability that the maximum charge is any value between $1$ and $Z_{m}$. Similarly $Q_{N_s,Z_s}(Z_{m}-1)$ can be constructed, where $\omega_{i,Z}=0$ whenever $Z\ge Z_{m}-1$. The probability that $Z_m$ is $Z_{max}$ is given by
\begin{equation}
p(Z_m)=\frac{Q_{N_s,Z_s}(Z_m)-Q_{N_s,Z_s}(Z_m-1)}{Q_{N_s,Z_s}(Z_S)}
\label{Largest_cluster_probability}
\end{equation}
The average value of largest charge produced from the fragmentation of a PLF ($N_s,Z_s$)
\begin{equation}
Z_{max}=\sum_{Z_m=1}^{Z_m=Z_s}Z_m p(Z_m)
\label{Largest_cluster_charge}
\end{equation}
\begin{figure}[!h]
\begin{center}
\includegraphics[height=3.4in,width=5.5in]{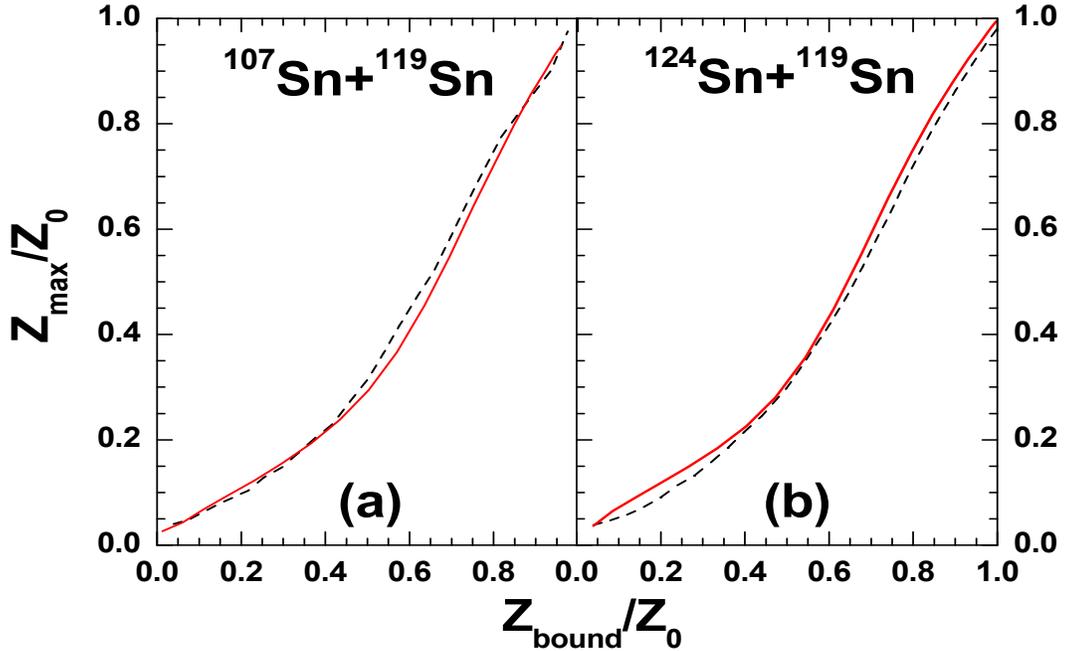}

\caption[Average charge of largest cluster in projectile fragmentation]{$Z_{max}/Z_0$ as a function of $Z_{bound}/Z_0$ for (a) $^{107}$Sn on $^{119}$Sn and (b) $^{124}$Sn on $^{119}$Sn reaction obtained from projectile fragmentation model (solid lines). The experimental results are shown by the dashed lines. }
\label{Largest_cluster}
\end{center}
\end{figure}
To include the effect of secondary decay, the same procedure as described in section 2.2.3 is followed.\\
\indent
Above mentioned method is applied for the fragmentation of each projectile spectator produced at different impact parameters with different temperatures.  Fig. \ref{Largest_cluster} shows the variation of $Z_{max}/Z_0$ with $Z_{bound}/Z_0$ obtained from theoretical projectile fragmentation model for $^{119}$Sn and $^{124}$Sn on $^{119}$Sn reactions at 600 MeV/nucleon and and compared with the experimental data. Nice agreement once again confirms the validity of the model.
\section{Summary}
A projectile fragmentation model has been developed whose origin can be traced back to the Bevalac era. This model consist of three different stages: (i) abrasion, where the PLF mass is calculated by geometrical model (ii) fragmentation of the abraded PLF is studied by the canonical thermodynamical model and (iii) secondary decay of excited fragments is calculated by evaporation model. A very simple impact parameter dependence of input temperature is incorporated in the model which helps to analyze the more peripheral collisions. The model is applied to calculate the charge, isotopic distributions, average number of intermediate mass fragments and the average size of largest cluster at different $Z_{bound}$  of different projectile fragmentation reactions at different energies and comparison with the relevant experimental data has been done.\\
\indent
While the projectile fragmentation model results have reasonable agreement with the various data considered here, it is desirable to push the model for further improvements. The goal will be to find the size and excitation of the initial projectile spectator from microscopic calculations. This is explained in the next chapter.
\vskip3cm
\end{normalsize} 
\chapter{Initial conditions of projectile fragmentation from microscopic calculation}
\begin{normalsize}
\section{Introduction}
In chapter-2 a model for projectile fragmentation has been proposed and different important observables have been studied from this model. The theoretical results obtained from this model have been compared with many experimental data with good success. The initial stage of the model is abrasion, where the PLF mass is determined from geometrical calculation. The PLF will have an excitation energy. It is conjectured that this will depend upon the relative size of the PLF with respect to the projectile, i.e., on $(A_s/A_p)$ where $A_s$ is the size of the PLF and $A_p$ is the size of the whole projectile. Instead of excitation energy the concept of freeze-out temperature $T$ is used and this temperature is not calculated rather it is parameterised with the help of experimental data and used for further multifragmentation stage calculation (by using canonical thermodynamical model). The concept of temperature is quite familiar in heavy ion physics, whether to describe the physics of participants (where the temperature can be very high) or the physics of spectators (where the temperature is expected to be much lower). One standard way of extracting temperatures is the Albergo formula \cite{Albergo}, where temperature is calculated from the measured isotopic yields. Another common technique for obtaining temperature is to measure the kinetic energy spectra of emitted particles. These methods have been widely used in the past (for a review see, for example, \cite{Dasgupta_Phase_transition,Pochodzalla,Agrawal,Trautmann_temperature}). But in both cases, sequential decay from higher energy states \cite{Nayak2}, Fermi motion \cite{Bauer2}, pre-equilibrium emission etc complicate the scenario of temperature measurement and the response of different thermometers is sometimes contradictory \cite{Xi_temperature,Francesca_temperature}.\\
\indent
In this chapter, the extraction of PLF mass and temperature is done first in a simple static model which will be followed by a complete dynamic calculation. In the static model \cite{Mallik7}, a projectile like fragment or projectile spectator (PLF) is assumed to be formed with a deformed shape. The mass and shape of the PLF will be determined with straight-line trajectory of the projectile. One could use the liquid drop has a constant density. In this work a Hamiltonian is used that gives correct nuclear matter binding, compressibility and density distribution in finite nuclei. By knowing PLF mass and excitation, the temperature is calculated by using canonical thermodynamical model. In heavy ion collisions the model produces a dependence of temperature on $(A_s/A_p)$ (i.e. impact parameter) which appears to be correct but the magnitude of the excitation energy falls short.\\
\indent
Though the main reason of PLF excitation is its non-optimum shape but in addition to this, particle migration from participant is also responsible for it. The crooked shape effect is added by the microscopic static model, but the effect of particle diffusion can not be included by any static calculation. To do that, dynamical calculation is needed. Hence the detailed time evolution of projectile and target nucleons have been studied \cite{Mallik9} by a transport model based on Boltzmann-Uehling-Uhlenbeck (BUU) calculation \cite{Dasgupta_BUU1}. At the end of time evolution, at-first one has to identify the PLF and has to separate it from the remaining part. Then by calculating its mass and excitation, the temperature is determined by applying CTM. For different projectile fragmentation reactions and varying projectile energies, the calculated temperature profile agrees quite well with the parameterised temperature profile used for calculations in Chapter-2.\\
\indent
This chapter is structured as follows. In Section 3.2 the development of microscopic static model and its results is explained, while the dynamical model based on BUU calculation is described in Section 3.4. The initial conditions of projectile fragmentation obtained from dynamical model is discussed in Sec. 3.3. Finally a brief summary is presented in Section 3.5.
\section{Microscopic Static Model}
In this model initially the ground state of the projectile nucleus is constructed by Thomas-Fermi (TF) method \cite{Thomas,Fermi,Lee}. The complete details of the procedure of Thomas-Fermi solution is described in Appendix-B. For completeness the prescription is outlined here. The kinetic energy density is given by
\begin{equation}
T(\vec r)=\int d^3p f(\vec{r},\vec{p})p^2/2m
\end{equation}
where $f(\vec{r},\vec{p})$ is the phase space density.\\
For the lowest energy, at each $\vec{r}$, $f(\vec {r},\vec{p})$ to be non-zero from 0 to some maximum $p_F(\vec{r})$.  Therefore one can write,
\begin{equation}
f(r,p)=\frac{4}{h^3}\theta [p_F(r,p)-p]
\end{equation}

The factor 4 is due to spin-isospin degeneracy and using the spherical symmetry of the TF solution the vector sign on $r$ and $p$ can be dropped. This leads to
\begin{equation}
T=\frac{3h^2}{10m}\Big[\frac{3}{16\pi}\Big]^{2/3}\int \rho(r)^{5/3}d^3r
\end{equation}
The Thomas-Fermi equation is given by,
\begin{equation}
\frac{1}{2m}\Big\{\frac{3h^3}{16\pi}\Big\}^{2/3} \rho(r)^{2/3}+ U(\vec {r})-\lambda=0
\label{Laqrange_multiplier}
\end{equation}
Where,  $\lambda$ is the Lagrange multiplier and $U(\vec {r})$ is the interaction potential. Thomas Fermi solutions for three kinds of interactions are given below. The numerical calculation techniques of determining Thomas-Fermi ground states for different interaction potentials are described in Appendix B.
\subsubsection{Skyrme Potential:-}
This is the most commonly used form of nuclear potential which only depend on the local density $\rho({\vec {r}})$. The mathematical form of the Skyrme potential is
\begin{equation}
U(\vec {r})=A^{'} \rho({\vec {r}})+B{'} \rho^\sigma({\vec {r}})
\label{Skyrme_potential}
\end{equation}
Where the first term is attractive and second term is repulsive. The constants $A^{'}$, $B^{'}$ and $\sigma$ should be chosen such that in nuclear matter the minimum energy is obtained at $\rho$=$\rho_0$=0.16 $fm^{-3}$ with energy per nucleon $-16$ MeV and compressibility $201$ MeV \cite{Dasgupta_Phase_transition}. One set of choice is $A^{'}=-2230.0 MeVfm^3$, $B^{'}=2577.85 MeVfm^{7/6}$ and $\sigma=7/6$.\\
\begin{figure}[h!]
\begin{center}
\includegraphics[width=16cm,keepaspectratio=true,clip]{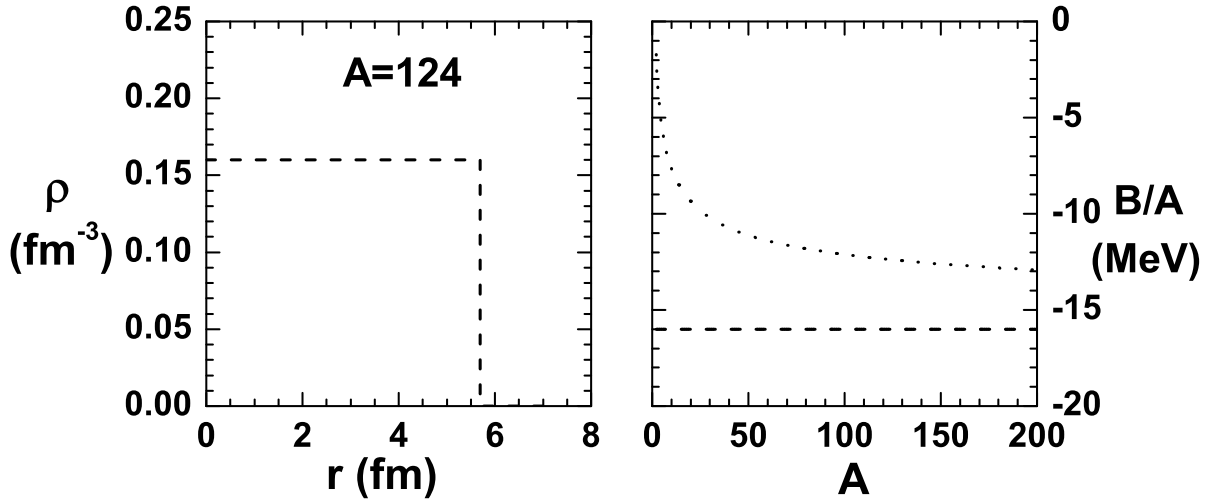}
\caption[Ground state density and energy from TF model for Skyrme interaction]{(a) Density profile for $A=124$ nucleus obtained by solving Thomas-Fermi equation for interaction potential $U(\vec {r})=A^{'} \rho({\vec {r}})+B^{'} \rho^\sigma({\vec {r}})$.
(b) Variation of energy per nucleon with mass number obtained from Thomas-Fermi method (dashed line), compared to a typical liquid drop formula $e=-16+18A^{-1/3}$  (dotted line).}
\label{Skyrme_den}
\end{center}
\end{figure}
If this form of $U(\vec {r})$, is used to solve the Thomas-Fermi euation for finite nuclei, then the ground state energy for finite nuclei of any mass will be also -16 MeV/nucleon and the corresponding density profile indicates sharp surface which are not realistic. Fig. \ref{Skyrme_den}.(a) shows the density profile for $A=124$ nucleus and Fig. \ref{Skyrme_den}.(b) indicates the variation of energy per nucleon with mass calculated for Thomas-Fermi method with $U(\vec {r})$ given in Eq. \ref{Skyrme_potential}. This problem does not arise in quantum mechanical treatment with Skyrme interaction. The theoretical studies of peripheral collisions \cite{Bonasera,Gregoire2,Bonche,Gan} show that the diffuse surfaces of real nuclei play an important role in spectator dynamics. Diffuse surfaces can be generated by two different ways: by adding finite range interactions with zero range potential \cite{Bonche} or by using Gaussian wave-packets to describe individual nucleons \cite{Gregoire2}. In the next two cases, two kinds of finite range terms will be added with the zero range Skyrme interaction.
\subsubsection{Skyrme+Yukawa Potential:-}
To get diffuse nuclear surfaces in semiclassical calculations, in addition to zero range Skyrme interaction, contribution from finite range Yukawa interaction should be included. Therefore the form of the potential will be,
\begin{equation}
U(\vec {r})=A^{''} \rho({\vec {r}})+B^{''} \rho^\sigma({\vec {r}})+\int u_y(\vec{r},\vec{r}')\rho(\vec{r'}) d^3r'
\label{Skyrme+Yukawa_potential}
\end{equation}
where $u_y(\vec{r},\vec{r}')$ is the finite range Yukawa potential
\begin{equation}
u_y(\vec{r},\vec{r}')=V_0\frac{e^{-|\vec{r}-\vec{r'}|/a}}{|\vec{r}-\vec{r'}|/a}
\end{equation}
\begin{figure}[h!]
\begin{center}
\includegraphics[width=16cm,keepaspectratio=true,clip]{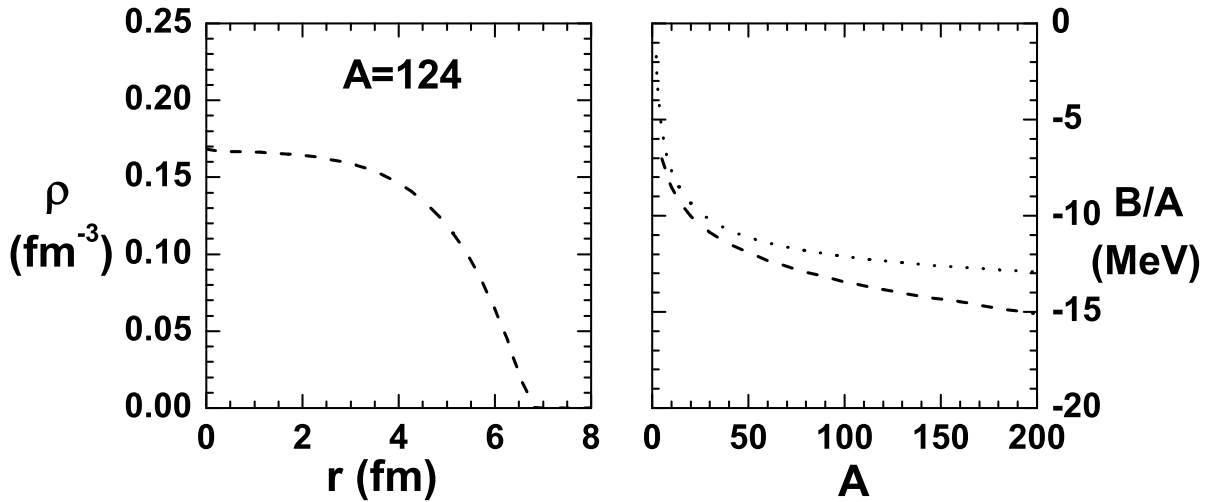}
\caption[Same as Fig. \ref{Skyrme_den} but the interaction is Skyrme+Yukawa type]{Same as Fig. \ref{Skyrme_den}, but instead of Skyrme interaction potential, here Skyrme+Yukawa interaction (Eq. \ref{Skyrme+Yukawa_potential}) is used for Thomas Fermi calculation.}
\label{Skyrme+Yukawa_den}
\end{center}
\end{figure}
The constants are taken as, $A^{''}=-1563.6 MeVfm^3$, $B^{''}=2805.3 MeVfm^{7/6}$, $\sigma=7/6$, $V_0=-668.65 MeV$ and $a=0.45979 fm$. Note that, for infinite nuclear matter, the contribution of Yukawa term to the total energy per nucleon reduces to $2 \pi V_0a^3 \rho_0$ but for finite nuclei the Thomas-Fermi solution produces realistic ground state energies and densities. Fig. \ref{Skyrme+Yukawa_den}.(a) shows the density profile for $A=124$ nucleus and Fig. \ref{Skyrme+Yukawa_den}.(b) indicates the variation of energy per nucleon with mass calculated for Thomas-Fermi method with $U(\vec {r})$ given in eq. \ref{Skyrme+Yukawa_potential}.\\
\indent
The concept of test particles will be introduced here which will be used very frequently in the rest of this thesis. Mathematically it is just a mapping
of a continuous distribution by a set of (usually large) points. The distribution must be positive definite in the domain. The distribution could be in one, two or more dimensions. Consider a distribution $f(x)$ in the domain $x_1$ and $x_2$. In this domain one can choose $N$ discrete points ($N$ large). In the interval $\delta x$ ($\delta x$ not too small) let there be $n$ points $x_i$. Then with suitable choice $f(x)\approx \frac {n}{N\delta x}$ where the points $x_i$ are in the interval $x-\frac{\delta x}{2}$ to $x+\frac{\delta x}{2}$.  Given $f(x)$ the points $x_i$ are usually chosen by Monte-Carlo sampling.\\
\indent
In classical and semi-classical physics there is a phase space density associated with a nucleon. In the ground state of a nucleus the phase space density of a nucleon is confined within a radius $R$ and a Fermi momentum $P_F$. This phase space can be mapped by $N_{test}$ point like objects which have positions $\vec r$ and momenta $\vec p$. These objects are called test particles. Throughout this work each nucleon is represented by 100 test particles ($N_{test}$=100).  For example, the phase space distribution of $^{58}$Ni is described by 5800 test particles.
\subsubsection{Skyrme potential with $\nabla^2$ correction:-}
Though eq. \ref{Skyrme+Yukawa_potential} can produce realistic ground state energies and diffuse nuclear surfaces, but for larger finite nuclei calculation is very time consuming and needs huge computer memory. A $\nabla^2$ correction with the original Skyrme potential can overcome this problem. This form of potential has suggested by Lenk and Pandharipande \cite{Lenk} and is given by
\begin{equation}
U(\vec {r})=A^{'}\rho(\vec{r})+B^{'}\rho^{\sigma}(\vec{r})+\frac{C^{'}}{\rho_0^{2/3}}\nabla_r^2[\frac{\rho(\vec{r})}{\rho_0}]
\label{Lenk_potential}
\end{equation}

\begin{figure}[t]
\begin{center}
\includegraphics[width=16cm,keepaspectratio=true,clip]{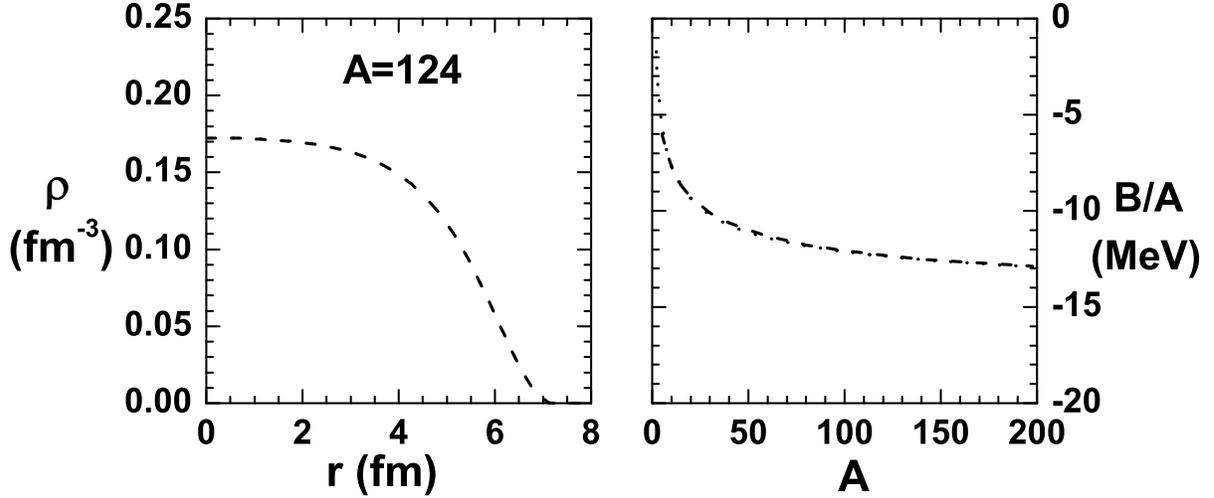}
\caption[Same as Fig. \ref{Skyrme_den} but for Skyrme interaction with $\nabla^2$ correction] {Same as Fig. \ref{Skyrme_den}, but with Skyrme interaction, $\nabla^2$ correction term is added (Eq. \ref{Lenk_potential} in the potential for Thomas Fermi calculation.}
\label{Lenk_den}
\end{center}
\end{figure}
where similar to Eq. \ref{Skyrme_potential}, $A^{'}=-2230.0 MeVfm^3$, $B^{'}=2577.85 MeVfm^{7/6}$ and $\sigma=7/6$ and the constant in correction term $C^{'}=-6.5 MeV$ is very useful for reproducing liquid drop binding energies and realistic densities from Thomas-Fermi calculation (Shown in Fig. \ref{Lenk_den}).\\
\indent
Here, for microscopic static model calculation, Skyrme+Yukawa type of interaction potential (given in Eq. \ref{Skyrme+Yukawa_potential}) is taken. The calculation of Yukawa (and/or Coulomb) potential due to a general mass/charge distribution is very non-trivial and involves iterative procedure. This has been used a great deal in applications involving time-dependent Hartree-Fock theory \cite{Koonin, Press, Varga, Feldmeier}. Eq. \ref{Lenk_potential} is very useful for dynamical model calculations with large nuclei. This is described in Section 3.3. But presently the discussion is restricted for Skyrme+Yukawa type interaction only. The Thomas Fermi phase space distribution will then be modeled by choosing test particles with appropriate positions and momenta using Monte Carlo technique.\\
\indent
After constructing the projectile nucleus microscopically, one can start to prepare the PLF. Assuming the straight-line geometry, one can postulate that due to collision of projectile and target nuclei PLF is separated suddenly from the remaining part and at the time of separation, PLF is created with a crooked shape. Therefore, a PLF can be constructed by removing a set of test particles.  Which test particles will be removed depends upon collision geometry envisaged.  For example, consider central collision of $^{58}$Ni on $^9$Be. Let $z$ to be the beam direction. For impact parameter $b$=0 all test particles in $^{58}$Ni will be removed whose distance from the center of mass of $^{58}$Ni has  $x^2+y^2<{r_9}^2$ where $r_9=2.38$ fm is the radius at half density of $^9$Be. The cases of non-zero impact parameter can be similarly considered.\\
\begin{figure}[t!]
\begin{center}
\includegraphics[width=12cm,keepaspectratio=true,clip]{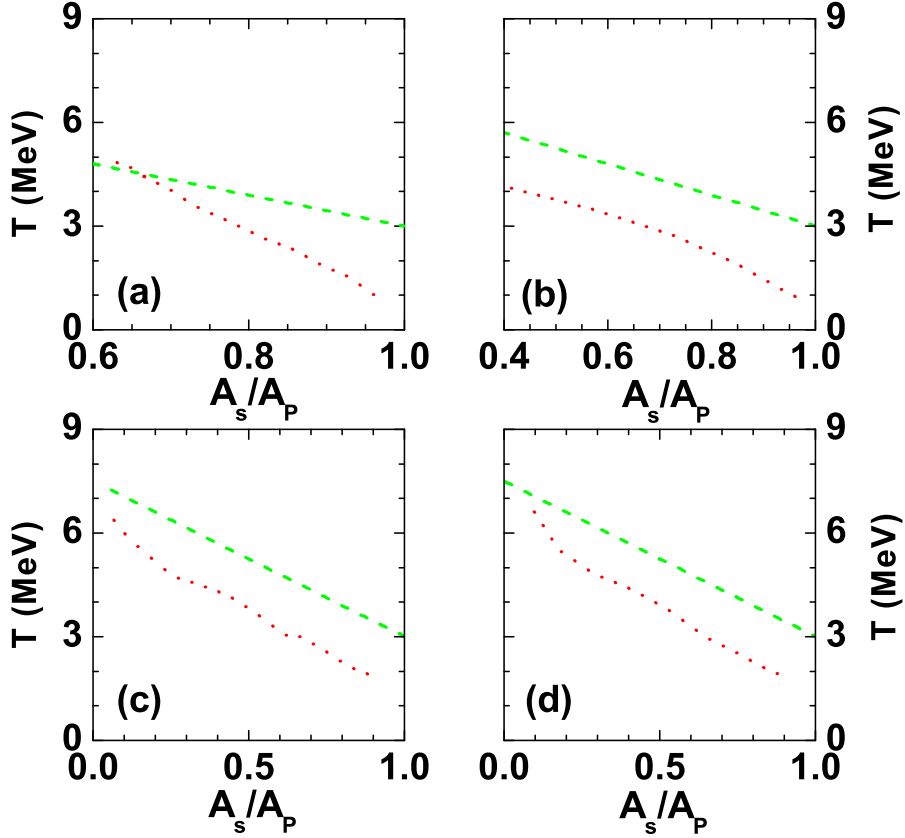}
\caption[Temperature profile obtained from microscopic static model]{Plot of temperature against $A_s/A_P$ for (a)$^{58}$Ni on $^9$Be reaction at 140 MeV/nucleon, (b)$^{58}$Ni on $^{181}$Ta reaction at 140 MeV/nucleon,   (c)$^{126}$Xe on $^{208}$Be reaction at 1 GeV/nucleon and (d) $^{124}$Sn on $^{119}$Sn reaction at 600 MeV/nucleon. The red dotted curves are the microscopic static model calculation result, the green dashed lines are from our previous work described in chapter-2.}
\label{Microscopic_model_temperature}
\end{center}
\end{figure}
The mass number of the PLF is the sum of the number of test particles remaining divided by $N_{test}$.  Similarly the total kinetic energy of the PLF is the sum of kinetic energies of the teat particles divided by $N_{test}$. For evaluating potential energy the Lattice Hamiltonian technique \cite{Lenk} is incorporated in the microscopic static model. This technique is also used for BUU calculation, therefore it is not discussed here, it is discussed in Section 3.3.1. With this method the total energy of PLF can be calculated.\\
\indent
However, to know excitation one needs to calculate the ground state state energy also. This is done by applying TF method for a spherical (ground state) nucleus having mass equal to the PLF mass. Knowing PLF mass and its excitation, the freeze-out temperature is calculated by using CTM. The details are described in Section 3.4.3.  Fig. \ref{Microscopic_model_temperature} shows the temperature profile obtained from microscopic static model calculation and comparison with parameterised temperature profile for (a)$^{58}$Ni on $^9$Be reaction at 140 MeV/nucleon, (b)$^{58}$Ni on $^{181}$Ta reaction at 140 MeV/nucleon, (c)$^{126}$Xe on $^{208}$Be reaction at 1 GeV/nucleon and (d) $^{124}$Sn on $^{119}$Sn reaction at 600 MeV/nucleon. The parameterised temperature profile is essentially the outcome of the experimental result as the temperatures which reproduce experimental results were selected. The present results obtained solely from a Hamiltonian which fits just ground state data does have the correct trend but underestimates the temperature. This is to be expected. Zero coupling have been assumed between participants and spectators. Participants have much more energy density (hence higher temperature).  Energy flows from higher value to lower value and hence temperature of the spectator is expected to rise further. But this can not be studied from this model, to study this one needs to do dynamical calculations. This will be explained in the next sections of this chapter. Also, in this static model, the beam energy does not enter the calculation, the assumption being that it is large enough for straight line trajectories to be valid. It is also assumed that due to this straight-line cuts the division between participant and spectators is very clean. Therefore, it is better to do dynamical calculations where no such assumptions are made.
\section{Dynamical Model}
To calculate the PLF mass and excitation more accurately Boltzmann-Uehling-Uhlenbeck (BUU) model \cite{Dasgupta_BUU1} is used. BUU model has been successfully applied earlier for explaining different experimental observables like transverse momenta, flow, particle production etc \cite{Bauer1,Danielewicz,Dasgupta_BUU2,Dasgupta_BUU3,Wong,Bauer7}. This model is based on the BUU transport equation (also called  Vlasov-Uehling-Uhlenbeck (VUU) equation \cite{Kruse,Uehling}, the Boltzmann-Nordheim equation \cite{Nordeim} or the Landau-Vlasov equation) \cite{Gregoire}). A further description of the BUU model exists in literature but to describe the work at least a shorter version is needed to present here. The precursor of our approach is the cascade model \cite{Cugnon1,Cugnon2}.\\
\indent
In the cascade model each nucleus is considered as a collection of point nucleons distributed within a sphere. No Fermi momenta are assigned. The two nuclei are started towards each other with appropriate beam velocity and impact parameter. The initial position of each nucleon is assigned by Monte-Carlo sampling.\\
\indent
Let $\sigma_{nn}^t(\sqrt{s})$ be the total nucleon-nucleon scattering cross-section at centre of mass energy $\sqrt{s}$. If two nucleons approach each other with a impact parameter less than $b_{max}=\sqrt{\sigma_{nn}^t(\sqrt{s})/\pi}$ they will scatter. After sufficient time the collisions are over, the nucleons are freely streaming and one can consider it as one event. The details of cascade model part of the calculation done here are given in section 3.3.2. The BUU model includes both a mean field and hard collision. Instead of deriving it formally it will be more useful for us to consider it as an extension of the cascade model. Each cascade model run will produce a different result as the positions of the nucleons are generated by Monte-Carlo sampling. To get an average answer many runs are needed. It is advantageous to have $N_{test}$ runs simultaneously. In the cascade model different runs do not communicate with each other. Thus nucleus $1$ hits nucleus $1'$, nucleus $2$ hits nucleus $2'$....nucleus $N_{test}$ hits nucleus $N_{test}$. Now communication between runs is introduced. Those were labelled as nucleons will be called test particles. An isolated nucleus of $A$ nucleons is regarded as a collection of $AN_{test}$ test particles. Because the test particles feel a potential $U\{\rho(r)\}$, the Fermi momentum of the test particles can be assigned. If there are $N'$ test particles in a small volume $(\delta r)^3$, then the density is $\rho(r)=N'/N_{test}(\delta r)^3$. The potential that a test particle feels is dependent upon the density $U=U\{\rho(r)\}$. Thus test particles have an $\vec{r}$ and $\vec{p}$ and are kept bound in an isolated nucleus because of $U\{\rho(r)\}$. Thus in the nuclear reaction, $(A_p+A_t)$ nucleons are represented by $(A_p+A_t)N_{test}$ test particles. As far as collisions go, in usual codes of BUU different runs are still segregated but this is merely a computational trick. The test particles should hit each other with a cross-section $\sigma_{nn}/N_{test}$. By segregating the collisions, $\sigma_{nn}$ can be used which will reduce computation. Thus in BUU the test particles occasionally collide and in between collisions their trajectories are governed by mean field. This mean field propagation is considered in next.

\subsection{Vlasov Propagation}
The propagation of the test particles can be described by Hamilton's canonical equations
\begin{eqnarray}
\frac{d\vec{p}_i}{dt}&=&-\nabla_r U(\rho(\vec{r}_i),t)\nonumber\\
\frac{d\vec{r}_i}{dt}&=&\vec{v}_i\nonumber\\
&& i=1,2,.....,(A_p+A_t)N_{test}
\label{Hamilton_equation}
\end{eqnarray}
where depending upon the original beam velocity $\vec{v}_i$ can be calculated relativistically or non-relativistically. Details of numerical method of computing trajectories of test particles have changed from author to author. One method which was good enough for many purposes is given in \cite{Dasgupta_BUU1}. Higher accuracy in energy and momentum conservation is needed for our work. The Lattice Hamiltonian method is used which was proposed by R. J. Lenk and V. M. Pandharipande \cite{Lenk} and has proven to be phenomenally accurate. According to this method the configuration space is divided into cubic lattices. The lattice points are $l$ fm apart.  Thus the configuration space is discretized into boxes of size $l^3$fm$^3$.  Density at lattice point $r_{\alpha}$ is defined by
\begin{equation}
\rho_L(\vec{r}_\alpha)=\sum_{i=1}^{AN_{test}}S(\vec{r}_{\alpha}-\vec{r}_i)
\label{LH_lattice_density}
\end{equation}
where $\alpha$ stands for values of the three co-ordinates of the lattice point $\alpha=(x_l,y_m,z_n)$ , $\vec{r}_{\alpha}$ is the position of site $\alpha$ and $S(\vec{r})$ is the form factor and given by
\begin{equation}
S(\vec{r})=\frac{1}{N_{test}(nl)^6}g(x)g(y)g(z)
\label{LH_form_factor}
\end{equation}
\begin{equation}
g(q)=(nl-|q|)\Theta (nl-|q|)
\label{LH_g}
\end{equation}
where $l$ is the lattice spacing, $\Theta$ is the Heaviside function and $n$ is an integer which determines the range of $S$. A test particle contributes to the average density $\rho_L$ at exactly $(2n)^3$ lattice sites, and the movement of a particle results in a continuous change in $\rho_L$ at nearby lattice sites. In our calculation we always used $l$=1 fm and $n$=1. Therefore by knowing test particle positions, density of all the lattice points can be calculated from Eq. \ref{LH_lattice_density}, then depending upon the requirement, potential at the lattice points can be calculated from Eq. \ref{Skyrme+Yukawa_potential} or Eq. \ref{Lenk_potential}. Then the positions and momentum are modified using Eq. \ref{Hamilton_equation}.
\subsection{Collision}
To include the effect of collision in the BUU model Monte-Carlo methods are applied which was formerly used in intranuclear cascade model \cite{Cugnon1,Cugnon2}. In cascade calculation, the frequency of collisions is governed by scattering cross-section only, where as in BUU, the Pauli blocking effect is also included. It is assumed that during each time step a test particle collide with another test particle at most once. In each collision, from original frame of reference one goes to the centre of mass frame of the colliding test particles. In the concerned energy regimes the important processes are  elastic collisions and the production and absorptions of $\Delta$ resonances ($\Delta$'s produce pions). Therefore the following channels are considered:
\begin{eqnarray}
n+n\rightarrow n+n &\text{(a)}\nonumber\\
n+n\rightarrow n+\Delta &\text{(b)}\nonumber\\
n+\Delta\rightarrow n+n &\text{(c)}\nonumber\\
n+\Delta\rightarrow n+\Delta &\text{(d)}\nonumber\\
\Delta+\Delta\rightarrow \Delta+\Delta &\text{(e)}\nonumber\\
\label{Reaction_channels}
\end{eqnarray}
Now for two nucleons to collide their distance of closest approach has to be less than the maximum nucleon-nucleon impact parameter $b_{max}$(=$\sqrt{\sigma^t_{max}/\pi})$ where $\sigma^t_{max}$ is the maximum value of total cross-section, which is 55 mb. Since the distance of closest approach is cumbersome to calculate, a pretest is done: two test particles can collide in a given time interval $\delta t$ only if the separation between these test particles at the starting of that time interval is less than $\sqrt{b^2_{max}+c^2\delta t^2}$.\\
The centre-of-mass energy and velocity are,
\begin{equation}
\sqrt{s}=\sqrt{(E_1+E_2)^2-(\vec{p_1}+\vec{p_2})^2}
\end{equation}
\begin{equation}
\vec{\beta}=\frac{\vec{p}_1+\vec{p}_2}{E_1+E_2}
\end{equation}
The momentum of first colliding test particle in the centre-of-mass frame is,
\begin{equation}
\vec{p}=\gamma \bigg{(}\frac{\vec{p}_1\cdot\vec{\beta}}{\beta}-\beta E_1\bigg{)}\frac{\vec{\beta}}{\beta}+\bigg{(}\vec{p}_1-\frac{\vec{p}_1\cdot\vec{\beta}}{\beta}\frac{\vec{\beta}}{\beta}\bigg{)}
\end{equation}
and the momentum of the other colliding particle is $-\vec{p}$. Here $\gamma$ is $1/\sqrt{1-\beta^2}$. The distance between two colliding test particles in the centre-of-mass frame is then:
\begin{equation}
\Delta\vec{r}=(\gamma-1)\bigg{(}(\vec{r_1}-\vec{r_2})\cdot\frac{\vec{\beta}}{\beta}\bigg{)}\frac{\vec{\beta}}{\beta}+(\vec{r_1}-\vec{r_2})
\end{equation}
For two test particles whose separation is less, one has to check if the particles pass the point of closest approach within the time interval. If this distance of closest approach is not greater than $b_{max}$ ie in the time interval $-\delta t/2$ to $\delta t/2$ the two test particles become candidates for collision if:
\begin{equation}
\bigg{|}{\frac{\Delta\vec{r}\cdot\vec{p}}{p}}\bigg{|}\le\bigg{(}\frac{p}{\sqrt{p^2+m^2_1}}-\frac{p}{\sqrt{(p^2+m^2_2}}\bigg{)}\frac{\delta t}{2}
\label{Collision_criteria_1}
\end{equation}
and if:
\begin{equation}
\sqrt{(\Delta \vec{r})^2-(\Delta \vec{r}\cdot\vec{p}/p)^2}\le b_{max}
\label{Collision_criteria_2}
\end{equation}
If collision criteria are satisfied, the collision channel of the colliding test particles is determined by Monte carlo simulation. This is described in Appendix. C.\\
The elastic nucleon-nucleon scattering [channel (a)], cross-section can be taken from fits to experimental data,
\begin{eqnarray}
\sigma^e_{nn \rightarrow nn}(\sqrt{s})&=55     &\text{if $\sqrt{s}\le1.8993$}\nonumber\\
&=\frac{35}{1+100(\sqrt{s}-1.8993)}+20&\text{if $\sqrt{s}>1.8993$}\nonumber\\
\label{Elastic_cross_section}
\end{eqnarray}
(where the energies are expressed in GeV and the cross-section in millibarn, momentum in GeV/c, c=1). For elastic collision, the angle of scattering is determined from differential cross-section given by,
\begin{equation}
\frac{d\sigma}{dt}=ae^{b(\sqrt{s})t(\theta)}
\end{equation}
where $a$ is a proportionality constant, $t(\theta)=-2p^2(1-cos\theta)$, is the negative of the square of the momentum transfer of the colliding test particles, $\theta$ is the polar angle of scattering which can vary from $0$ ($t_{low}=-2p^2$) to $\pi/2$ ($t_{high}=0$) and $b(\sqrt{s})$ is parameterized as
\begin{equation}
b(\sqrt{s})=\frac{6[3.65(\sqrt{s}-1.866)]^6}{1+[3.65(\sqrt{s}-1.866)]^6}
\end{equation}
Therefore for a particular two body elastic collision, the polar angle ($\theta_s$) can be obtained from
\begin{equation}
\int_{t_{low}}^{t(\theta_s)}ae^{bt}dt\Bigg{/}{\int_{t_{low}}^{t_{high}}ae^{bt}dt}=x_1
\label{Elastic_radial_angle}
\end{equation}
The azimuthal angle of elastic scattering is chosen randomly i.e.
\begin{equation}
\phi_s=2\pi x_2
\label{Elastic_azimuthal_angle}
\end{equation}
where $x_1$ and $x_2$ are two random numbers. Thus for each scattering, the linear momentum and energy are conserved but not the angular momentum. The cumulative effect of angular momentum nonconservation has been found to be found small in the intermediate energy domain.\\
The cross-section for channel (b) is:
\begin{eqnarray}
\sigma^i_{nn \rightarrow n\Delta}(\sqrt{s})&=0 &\text{if $\sqrt{s}\le2.015$}\nonumber\\
&=\frac{20(\sqrt{s}-2.015)^2}{0.015+(\sqrt{s}-2.015)^2} &\text{if $\sqrt{s}>2.015$}\nonumber\\
\label{Inelastic_cross_section}
\end{eqnarray}
The mass of the produced $\Delta$ particle can be parameterized as \cite{Dasgupta_BUU4}
\begin{eqnarray}
m_{\Delta}&=1.077+0.75(\sqrt{s}-2.015)&\text{for $2.015<\sqrt{s}\le2.2203$}\nonumber\\
&=1.231&\text{for $\sqrt{s}>2.2203$}\nonumber\\
\label{Delta_mass}
\end{eqnarray}
The cross-section for process (c) can be obtained from (b) by detailed balance:
\begin{equation}
\sigma^i_{n\Delta \rightarrow nn}=\frac{p'^2}{8p^2}\sigma^i_{nn \rightarrow n\Delta}
\end{equation}
where the factor 8 is due to spin-isospin degeneracy and the identical nature of particles in the final states. $p$ and $p'$ represents the momenta of colliding particles before and after the inelastic collision $n+n\rightarrow n+\Delta$ respectively.\\
The inelastic channels are assumed to scatter isotropically and the magnitude of final momentum is fixed by energy conservation. The cross-sections for channels (d) and (e) are taken to be the same as for (a). For successful collisions, after changing the momenta and calculating the final mass (for inelastic collision only) of the particles one has to go back to the original frame of reference.\\
Since the aim of this chapter is to study the excitation and mass of PLF at intermediate energies which will not be affected very much by inelastic collisions (as inelastic collisions become dominant at very high energies and $\Delta$'s mainly produced in the participant region) therefore in most of the calculations the inelastic channels are switched off to avoid unnecessary computations. However an estimation of pion production is done for $^{124}$Sn on $^{119}$Sn reaction at 600 MeV/nucleon. This is described in Sec 3.4.4 where all of the above mentioned collision channels are included in the calculation.
\subsubsection{Pauli Blocking}
Consider two test particles come within the distance of closest approach and due to collision they change from ($\vec{r}$,$\vec{p}$), ($\vec{r_2}$,$\vec{p_2}$) to ($\vec{r}$,$\vec{p}'$), ($\vec{r_2}$,$\vec{p_2}'$). Since the colliding particles are Fermions, therefore by calculating Uehling-Uhlenbeck term one has to check whether the final states are allowed or not i.e. the collision will actually take place or it will be Pauli blocked.\\
\indent
To obtain Uehling-Uhlenbeck term for intermediate energy heavy ion reactions, the phase space densities about the final states ($\vec{r}$,$\vec{p_1}'$) and ($\vec{r_2}$,$\vec{p_2}'$) are required. Therefore a radius $r_p$ around $\vec{r}$ in configuration space and radius $p_p$ around $\vec{p}'$ in momentum space is selected, such that $N_p$ test particles inside this phase space volume imply complete filling at normal nuclear matter density \cite{Dasgupta_BUU1,Aichelin2} i.e.
\begin{equation}
\frac{4}{h^3}\int_0^{r_p}\int_0^{p_p} d^3rd^3p=\frac{N_p}{N_{test}}
\label{Pauli_blocking1}
\end{equation}
$N_p$ should be small so that one is examining the phase space densities near the collision points. But it can not be taken to be so small that fluctuations inherent in Monte-Carlo become severe. For $N_{test}=100$ past work indicate that $N_p=8$ is a good choice. From Eq. \ref{Pauli_blocking1} it is clear that, specifying $N_p$ does not determine both $r_p$ and $p_p$, one has to add the extra condition. In this thesis, $r_p/p_p=R/P_F$ is used where $R$ is the hard sphere radius of the static nucleus and $P_F$ is the Fermi momentum at normal nuclear matter density. Now the blocking factor at ($\vec{r}$,$\vec{p}'$) can be defined as $f_1=N_1/N_p$, where $N_1$ is the number of test particles (excluding the colliding test particle at ($\vec{r}$,$\vec{p}'$)) within a radius $r_p$ around $\vec{r}$ in configuration space and a radius $p_p$ around $\vec{p}'$ in momentum space. So the collision probability factor is $=1-f_1$.  Similarly for the second particle, blocking factor is $f_2=N_2/N_p$ and collision probability factor is $=1-f_2$. Therefore the probability of scattering is taken to be $(1-f_1)(1-f_2)$ and this is calculated by Monte-Carlo method by the usual way.\\
What is described in section 3.3 is a numerical method for solving the BUU equation \cite{Huang_stat_mech}. This equation is,
\begin{eqnarray}
\frac{\partial f}{\partial t}+\vec{v}\cdot \overrightarrow{\nabla}_r f-\overrightarrow{\nabla}_r U\cdot \overrightarrow{\nabla}_p f &=&\frac{1}{(2\pi)^6}\int d^3\vec{p}_2d^3\vec{p'}_{2}d\Omega\frac{d \sigma}{d \Omega} \vec{v}_{12}\nonumber\\
&& \times \big\{f'f'_2(1-f)(1-f_2)-ff_2(1-f')(1-f'_2)\big\}\nonumber\\
&& \times (2\pi)^3 \delta^3 (\vec{p}+\vec{p}_2-\vec{p'}-\vec{p'}_2)
\label{BUU_equation}
\end{eqnarray}
The right hand side is the collision integral including Pauli blocking. The left hand side gives the Vlasov propagation (Eq. \ref{Hamilton_equation}) and the equivalence is shown in more explicit detail in ref. \cite{Dasgupta_BUU1}.
\section{Initial Conditions from Dynamical Model}
In the recently developed projectile fragmentation model (explained in details in Chapter 2) the PLF size was calculated by assuming straightline geometry and instead of excitation energy of the PLF, the concept of freeze-out temperature for multifragmentation stage is used. The freeze-out temperature was not calculated, it was fitted from experimental data. It was assumed that the temperature will depend upon the relative size of the PLF with respect to the projectile, i.e., on $(A_s(b)/A_p)$ where $A_s$ is the size of the PLF at impact parameter $b$ and $A_p$ is the size of the whole projectile and the parametrization was expressed by Eq. \ref{Temperature_Eq3}.\\
\indent
The objective of this section is to calculate the PLF size and its excitation over the entire impact parameter range directly from a transport model based on Boltzmann-Uehling-Uhlenbeck (BUU) equation. At each impact parameter by knowing PLF size and excitation, Canonical Thermodynamical Model (CTM) is used to deduce the freeze-out temperature. Then calculated temperature profile is compared with the earlier parameterized temperature profile.
\subsection{Identification of PLF}
\begin{figure}[b!]
\begin{center}
\includegraphics[width=12cm,keepaspectratio=true,clip]{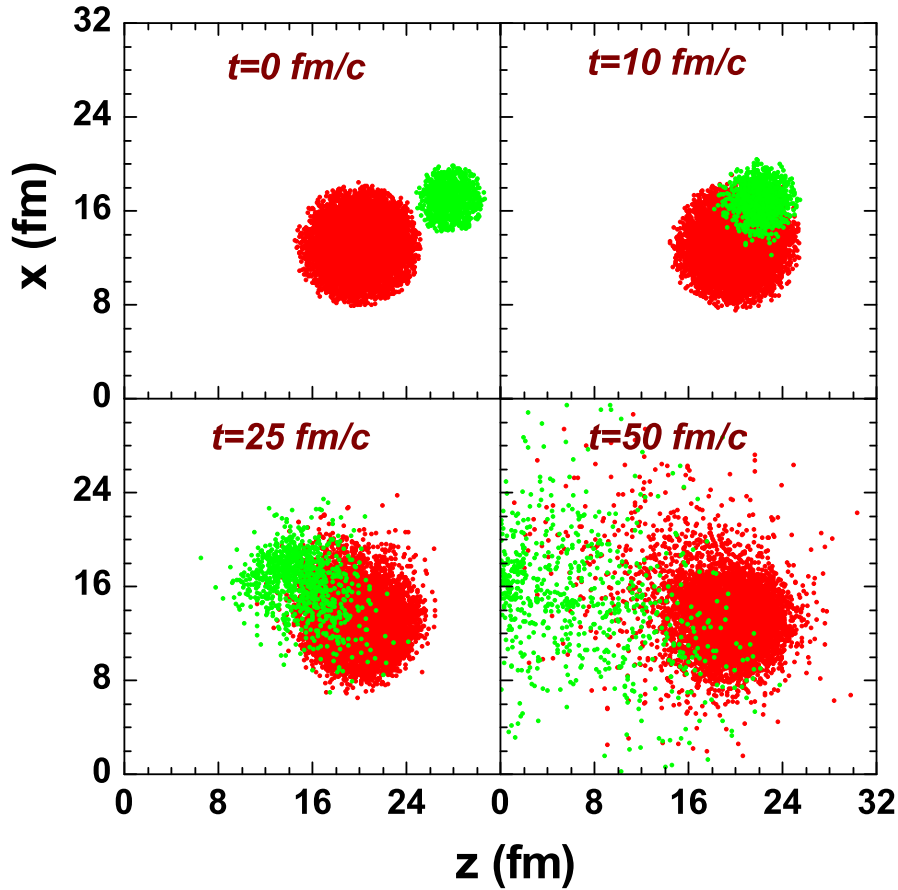}
\caption[Time evolution study for the initial stage of projectile fragmentation]{Time evolution of $^{58}$Ni (red) and $^{9}$Be (green) test particles for $140$ MeV/nucleon at an impact parameter $b=4$ fm.}
\label{PF_test_particles}
\end{center}
\end{figure}
The transport model calculation is started by choosing an impact parameter and boost the test particles of one nucleus with appropriate velocities in its Thomas-Fermi ground state towards the test particles of the other nucleus, also in its ground state. Firstly, it can be chosen to study $^{58}$Ni on $^9$Be reaction with beam energy 140MeV/nucleon which was experimentally investigated at Michigan State University (MSU) and also studied theoretically from the projectile fragmentation model (described in chapter 2). The transport calculations are done in a 25$\times 25\times31fm^3$ box and for calculating the mean field the form given in \ref{Skyrme+Yukawa_potential} is used. The configuration space is divided into $1fm^3$ boxes. It is useful to work in the projectile frame. Initially the projectile and target test particles are centered at (13$fm$, 13$fm$, 20$fm$) and (13+$b fm$,13$fm$, 27$fm$) respectively and the target test particles are moving with the beam velocity in the negative $z$ direction. In this energy domain Vlasov propagation is treated non-relativistically. We exemplify our method with collision at impact parameter b=4 fm. Fig. \ref{PF_test_particles} shows the test particles at t=0 fm/c (when the nuclei are separate), t=10 fm/c, t=25 fm/c and t=50 fm/c (Be has traversed the original Ni nucleus). The calculation was started with the center of Ni at 20 fm; at the end a large blob remains centered at 20 fm.  Clearly this is the PLF.  However for a quantitative estimate of the mass of the PLF and its energy requires further analysis.  This type of analysis was done for each pair of ions and at each impact parameter and details vary from case to case. This is exemplified for b=4 fm only.\\
\indent
For the analysis, it is convenient to introduce $z$ component of density which is defined as
\begin{equation}
\rho_z(z)=\sum_{lm}\rho_L(\vec{r}_\alpha)=\sum_{lm}\sum_{i=1}^{AN_{test}}S(\vec{r}_{\alpha}-\vec{r}_i)
\label{PF_density_eqn}
\end{equation}
where $\rho_L(\vec{r}_\alpha)$ is the density at a lattice point $\vec{r}_{\alpha}$ (described in section 3.3.1). In Fig. \ref{PF_density}, $\rho_z(z)$ is plotted as a function of $z$ at $t$=0 (when the the nuclei start to approach each other) and at t=50 fm/c (when $^9$Be has traversed $^{58}$Ni).
\begin{figure}[t!]
\begin{center}
\includegraphics[width=9cm,height=9cm,clip]{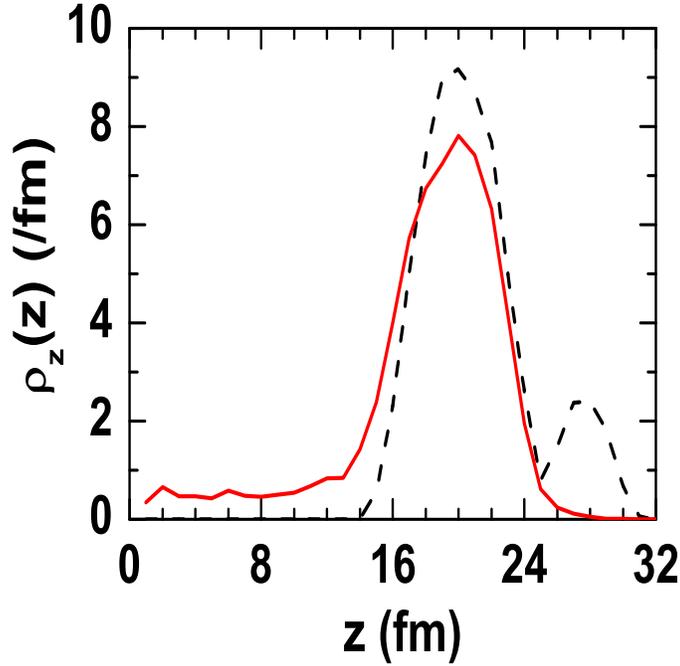}
\caption[$\rho_z(z)$ variation with z from dynamical model calculation]{$\rho_z(z)$ variation with z at t=0 fm/c (black dashed line) and 50 fm/c (red solid line) for $140$ MeV/nucleon $^{58}$Ni on $^{9}$Be reaction studied at an impact parameter $b=4$ fm.}
\label{PF_density}
\end{center}
\end{figure}
Fig. \ref{PF_kinetic_enery_momenta} adds more details to the situation at 50 fm/c where kinetic energy per nucleon ($\mu$) and z component of momentum per nucleon ($\nu$) is plotted as a function of z. The kinetic energy density is defined as
\begin{equation}
T_L(\vec{r}_\alpha)=\sum_{i=1}^{AN_{test}}T_iS(\vec{r}_{\alpha}-\vec{r}_i)
\label{PF_kinetic_energy_density_eqn}
\end{equation}
where $T_i$ is the kinetic energy of the i-th test particle. Therefore the kinetic energy per nucleon:
\begin{equation}
\mu(z)=\frac{T_z(z)}{\rho_z(z)}=\frac{\sum_{lm}(T_z)_L(\vec{r}_\alpha)}{\sum_{lm}\rho_L(\vec{r}_\alpha)}
\label{PF_kinetic_energy_per_nucleon}
\end{equation}
Similarly one can introduce a density for the z-th component of momentum (actually $p_zc$ is used rather than $p_z$)
\begin{equation}
(p_zc)_L(\vec{r}_\alpha)=\sum_{i=1}^{AN_{test}}(p_zc)_iS(\vec{r}_{\alpha}-\vec{r}_i)
\label{PF_momentum_density_eqn}
\end{equation}
Therefore, $p_z(z)c$ per nucleon can be expressed as:
\begin{equation}
\nu(z)=\frac{p_z(z)c}{\rho_z(z)}=\frac{\sum_{lm}(p_zc)_L(\vec{r}_\alpha)}{\sum_{lm}\rho_L(\vec{r}_\alpha)}
\label{PF_momentum_per_nucleon}
\end{equation}
At far right, $\mu$ and $\nu$ are very small which indicates PLF regions (since the calculation is done in the projectile frame therefore the PLF have very low z-component of momentum and kinetic energy). Progressively towards left one has the participant zone characterised by a higher $\mu$ and lower value of $\nu$. Closer to the left edge one has target spectators.\\
\begin{figure}[t!]
\begin{center}
\includegraphics[width=14cm,keepaspectratio=true,clip]{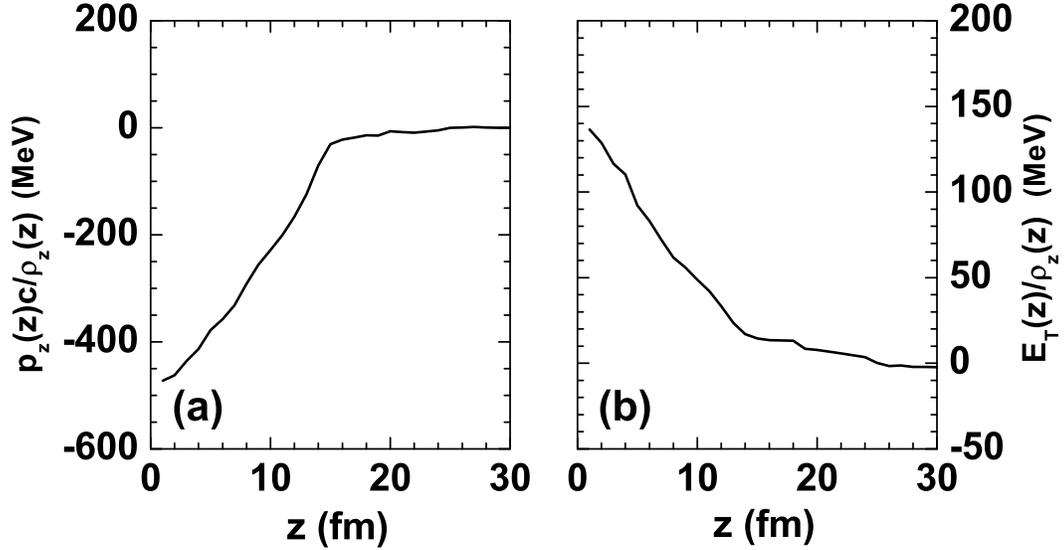}
\caption[Momentum per nucleon and total energy per nucleon variation in position space]{Variation of (a) Momentum per nucleon $\nu(z)$ and (b) total energy per nucleon $\mu(z)$ for $140$ MeV/n $^{58}$Ni on $^{9}$Be reaction at an impact parameter $b=4$ fm studied at $t=50$ fm/c.}
\label{PF_kinetic_enery_momenta}
\end{center}
\end{figure}
\subsection{Mass and Excitation of PLF}
In order to specify the mass number and energy per nucleon of the PLF one needs to specify which test particles belong to the PLF and which to the rest (participant and target spectators).  The configuration box stretches from $z$=0 to $z$=31 $fm$. If all test particles in this range are included, one gets the full system with the total particle number 67(58+9) and the total energy of beam plus projectile in the projectile frame. Let us consider constructing a wall at z=0 and pulling the wall to the right. If the wall is pulled, the test particles positioned on the left of the wall are left out.  With the test particles to the right of the wall one can compute the number of nucleons and the total energy per nucleon $E_{wr}$.  The number of particles goes down and initially the energy per nucleon $E_w$ will go down also as initially the target spectators and then the participants are being left out. At some point one enters the PLF and if pulls a bit further, part of the PLF is cut off and a non-optimum shape is formed. So the energy per nucleon $E_{wr}$ will rise.  The situation is shown in Fig. \ref{PF_energy}.  The point which produces this minimum is a reference point.  The test particles to the right are taken to belong to PLF; those, to the left are taken to represent the participants and target spectators. Not surprisingly, this point is in the neighborhood where both $\mu$ and $\nu$ flatten out. The energy per nucleon $E_{wr}$ at the reference point is PLF excited state energy.\\
\begin{figure}[h!]
\begin{center}
\includegraphics[width=10cm,keepaspectratio=true,clip]{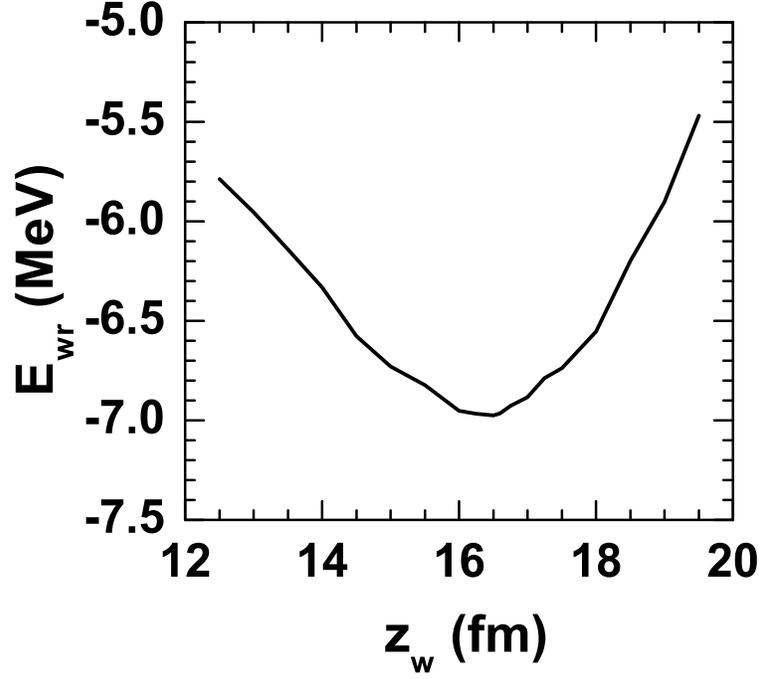}
\caption[Determination of Mass and excitation in projectile fragmentation]{Energy per nucleon ($E_w$) of the test particles remains right side of the separation (z) for $140$ MeV/nucleon $^{58}$Ni on $^{9}$Be reaction at an impact parameter $b=4$ fm studied at $t=50$ fm/c.}
\label{PF_energy}
\end{center}
\end{figure}
\indent
Now, to calculate the excitation energy of the system over the entire range of impact parameter the ground state energy of the PLF is required. Hence at each impact parameter, the Thomas Fermi method is used again for a spherical (ground state) nucleus having mass equal to the PLF mass and its total energy is calculated. Subtracting ground state energy from excited state energy excitation is obtained. For $^{58}$Ni on $^{9}$Be reaction at 140 MeV/nucleon, the variation of PLF mass ($A_s$) and excitation per nucleon ($E^*$) with impact parameter, obtained from this calculation is shown in Fig. \ref{PLF_mass_and_excitation}. As expected, with the increase of impact parameter, the total amount of mass which are driven out from the original projectile decreases, hence PLF mass increases. Also with the decrease of centrality, the deformation of the PLF decreases, therefore PLF excitation decreases.\\
\begin{figure}[h!]
\begin{center}
\includegraphics[width=14cm,keepaspectratio=true,clip]{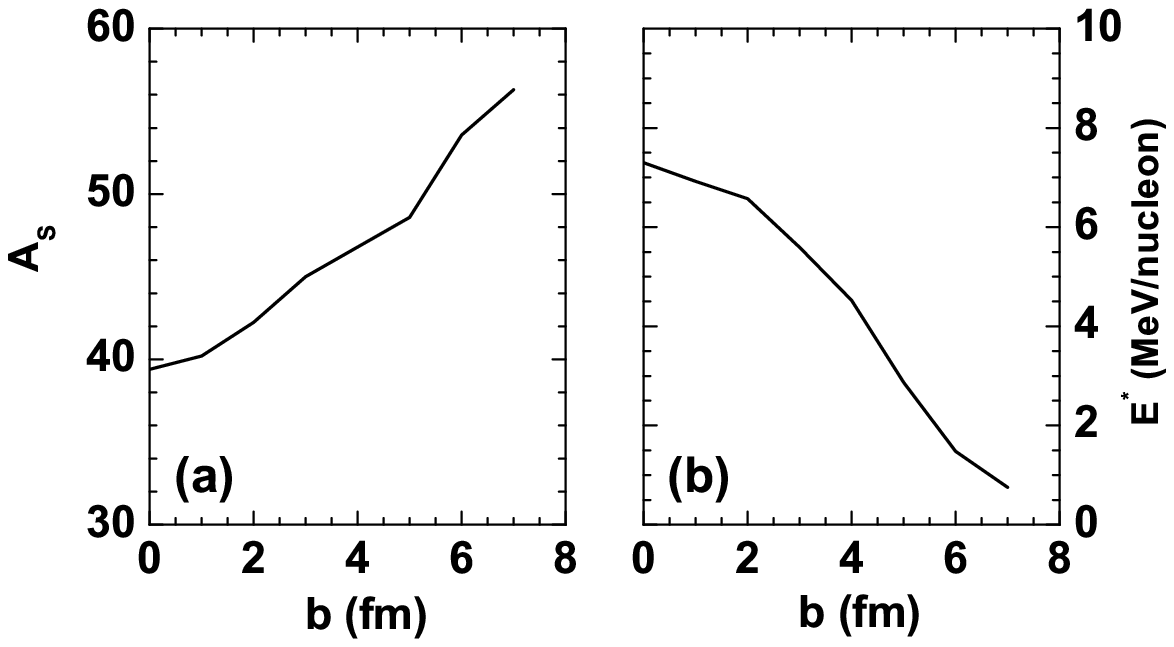}
\caption[PLF mass and excitation per nucleon from dynamical model]{Variation of (a) PLF mass ($A_s$) and (b) it's excitation per nucleon ($E^*$) with impact parameter obtained from BUU calculation for $^{58}$Ni+$^{9}$Be reaction at 140 MeV/nucleon.}
\label{PLF_mass_and_excitation}
\end{center}
\end{figure}

\subsection{Temperature of PLF}
\begin{figure}[t!]
\begin{center}
\includegraphics[width=14cm,keepaspectratio=true,clip]{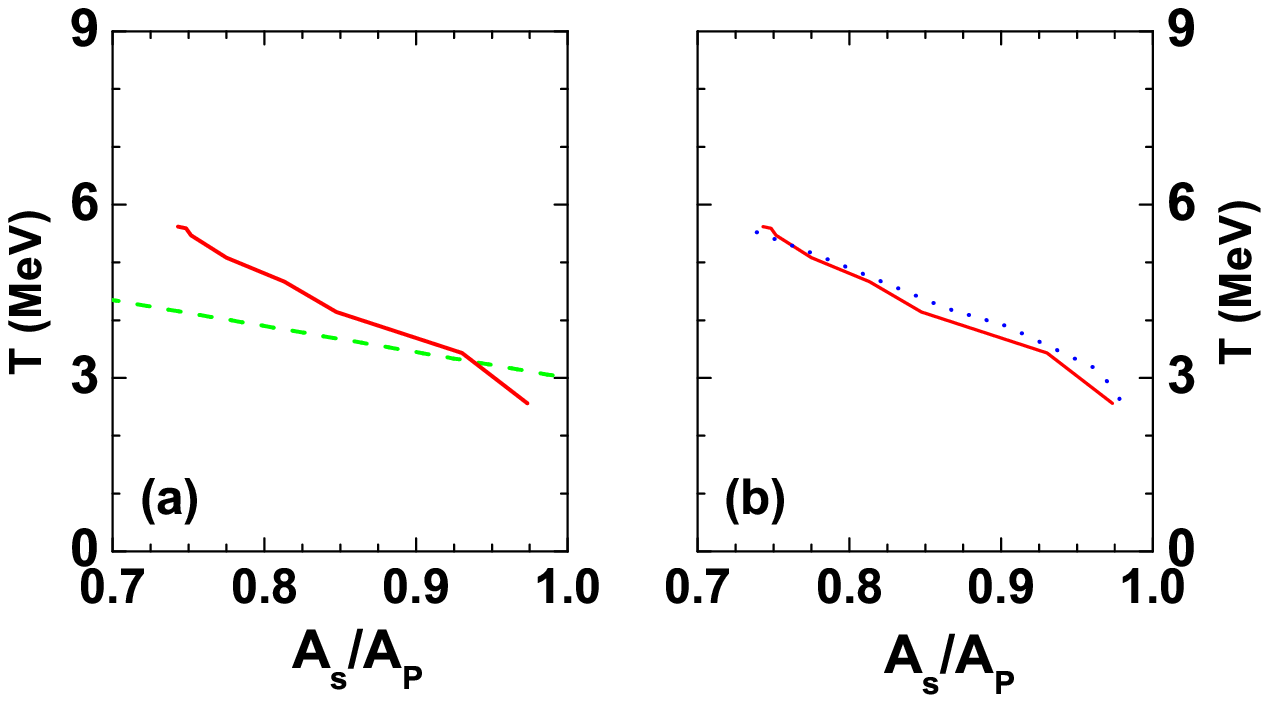}
\caption[Temperature profile for  $^{58}$Ni+$^{9}$Be reaction from dynamical model calculation]{(a) Temperature profile for  $^{58}$Ni+$^{9}$Be reaction at 140 MeV/nucleon obtained from BUU model calculation (red solid line)  compared with that calculated from general formula [Eq. \ref{Temperature_Eq3}] (green dashed line).
(b) Comparison of temperature profile obtained from BUU calculation for $^{58}$Ni+$^{9}$Be reaction at 140 MeV/nucleon (red solid line) and at 400 MeV/nucleon (blue dotted line).}
\label{PLF_Ni+Be_Temperature}
\end{center}
\end{figure}
The main motivation of this chapter is to determine the freeze-out temperature corresponding to this excitation. The canonical thermodynamic model (CTM) \cite{Das} described in chapter 2 can be used to calculate average excitation per nucleon for a given temperature, mass number and charge at a given impact parameter. For CTM calculation, the neutron to proton ratio of compound nuclear system is considered as same as that of original projectile. The fragmentation of the compound nuclear system $N_s,Z_s$ ($N_s+Z_s=A_s$, which is obtained from BUU calculation) is repeated at different temperatures and finally the appropriate temperature is determined which can reproduce the earlier calculated excitation $E^{*}(N_0,Z_0)$ i.e.
\begin{equation}
E^{*}_{N_s,Z_s}(b)-B_{N_s,Z_s}=\sum_{I,J}\langle n_{I,J}(b)\rangle[\frac{3T(b)}{2}+\frac{AT^2(b)}{\epsilon_0}-B_{I,J}]
\end{equation}
this relation is satisfied. Here  $A=I+J$ and $\langle n_{I,J}(b)\rangle$ is the multiplicity of the fragments having $I$ neutrons and $J$ protons produced by the multifragmentation from the system $N_s,Z_s$ at impact parameter $b$ and temperature $T$.\\
\indent
The variation of deduced freeze-out temperature with $A_s/A_0$ is shown by the red solid line of Fig. \ref{PLF_Ni+Be_Temperature}(a). Now the next aim is to compare this with universal temperature profile (Eq. \ref{Temperature_Eq3}) which was parameterized from experimental data and used for calculations of Chapter 2. It is displayed by the green dashed line of Fig. \ref{PLF_Ni+Be_Temperature}(a). The entire calculation is repeated at beam energy 400 MeV/nucleon. Comparison of the temperature profiles obtained from BUU calculation at 140 MeV/nucleon and 400 MeV/nucleon is shown in Fig. \ref{PLF_Ni+Be_Temperature}(b). Any experiment  at 400 MeV/nucleon is not known, this is done merely to check if in BUU, PLF physics is sensitive to beam energy.  Geometrical model assumes it is not. Similar temperature profile for PLF at different beam energies also ensures projectile fragmentation at 140 MeV/nucleon is in the limiting fragmentation region. Fig. \ref{PLF_Ca+Be_&_Ni_Ta_Temperature}(a) displays the temperature profile obtained from BUU calculation and its comparison with universal temperature profile for another projectile fragmentation reaction $^{40}$Ca on $^{9}$Be at 140 MeV/nucleon.\\
\begin{figure}[t!]
\begin{center}
\includegraphics[width=14cm,keepaspectratio=true,clip]{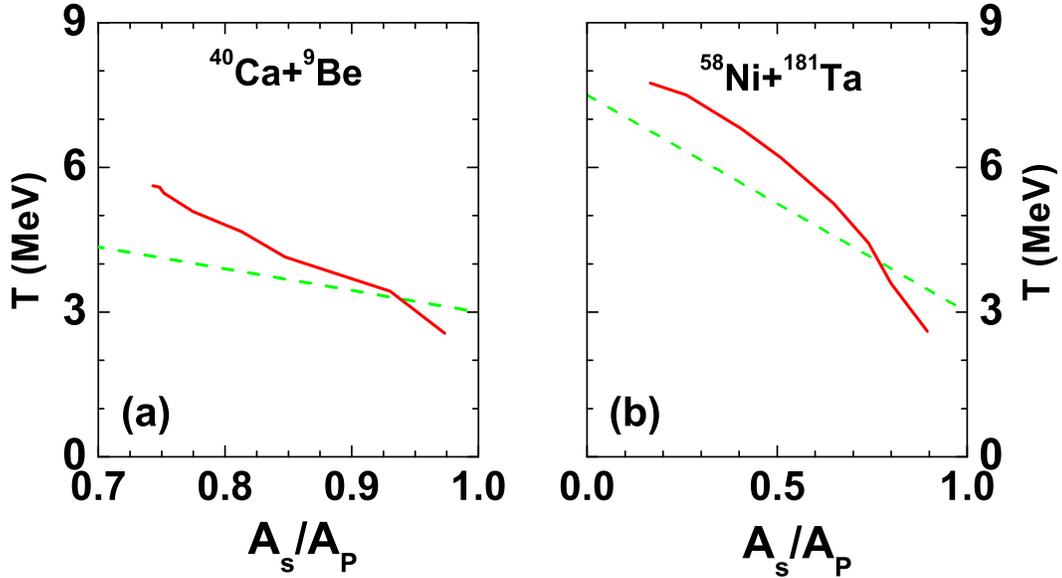}
\caption[Temperature profile for (a) $^{40}$Ca+$^{9}$Be and (b)$^{58}$Ni+$^{181}$Ta reactions]{(a) Temperature profile for (a) $^{40}$Ca+$^{9}$Be and (b)$^{58}$Ni+$^{181}$Ta reaction both at energy 140 MeV/nucleon. Red solid lines are for BUU model calculation results and green dashed lines are parametrization given in eq. \ref{Temperature_Eq3}.}
\label{PLF_Ca+Be_&_Ni_Ta_Temperature}
\end{center}
\end{figure}
After getting satisfactory results for $^{58}$Ni on $^{9}$Be reaction, the BUU calculation is tried to extend for determining the initial conditions of projectile fragmentation when the projectile and target nuclei are much heavier (for example $^{124}$Sn on $^{119}$Sn). Vlasov propagation with Skyrme plus Yukawa interaction for large ion collisions is not practical.  Given nuclear densities  on lattice points, one is required to generate the potential which arises from the Yukawa interaction.  Standard methods require iterative procedures involving matrices.  In the case of $^{58}$Ni on $^{9}$Be, in the early times of the collision, the matrices are of the order of 1000 by 1000: as the system expands the matrices grow in size reaching about 7000 by 7000 at $t=50$ fm/c. If large systems are wanted to be done by the same method, very large computing efforts are required. To overcome this problem the Skyrme interaction potential can be used with $\nabla^2$ term (Eq. \ref{Lenk_potential}) which will produce realistic ground state energies and diffuse nuclear surfaces with much less computational effort.\\
\begin{figure}[b!]
\begin{center}
\includegraphics[width=14cm,keepaspectratio=true,clip]{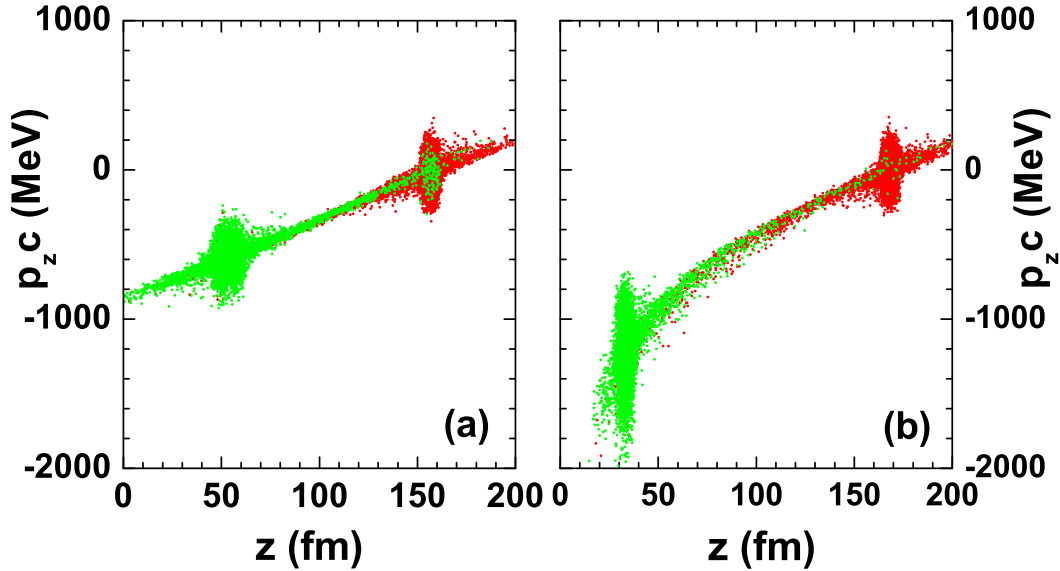}
\caption[$p_zc$ vs $z$ variation of the test particles]{$p_zc$ vs $z$ variation of projectile (red) and target (green) test particles at t=200 fm/c for $^{124}$Sn on $^{119}$Sn reaction studied at an impact parameter $b=8$ fm with energy (a)$200$ MeV/nucleon (non-relativistic kinematics) and (b) $600$ MeV/nucleon (relativistic kinematics).}
\label{Sn_Sn_z_pz_distribution}
\end{center}
\end{figure}
\indent
Two cases of large colliding systems: $^{124}$Sn on $^{119}$Sn and $^{58}$Ni on $^{181}$Ta are simulated. For these cases the calculations are done in a 200$\times 200\times200fm^3$ box and initially the projectile and target test particles are centered at (100$fm$, 100$fm$, 170$fm$) and (100+$b fm$,100$fm$, 190$fm$) respectively. Fig. \ref{Sn_Sn_z_pz_distribution}.(a) shows scatter of test particles in the $z,p_zc$ plane for $^{124}$Sn on $^{119}$Sn at time t=200 fm/c for beam energy 200 MeV/nucleon and impact parameter 8 fm/c.  The plot, as before, is in the projectile frame and identifies projectile like spectator, participant zone and target like spectator. Similar to the previous calculation, here the Vlasov propagation is non-relativistic but collisions are treated relativistically. Experimental data for $^{124}$Sn on $^{119}$Sn at 600 MeV/n are available \cite{Ogul} which was reproduced from our projectile fragmentation model in Chapter 2. For 600 MeV/nucleon beam energy, relativistic kinematics is used for propagation of test particles. This means the following. In the rest frame of each nucleus, the Fermi momenta of test particles is calculated in the standard fashion except that once they are generated they are treated like relativistic momenta. Relativistic kinetic energy per nucleon in the rest frame of the nucleus, on the average, becomes only slightly different from  the non-relativistic value (about 0.3 MeV per nucleon). As before, the calculations are performed in the rest frame of the projectile and the transformation of momenta of test particles of the target to the projectile frame is relativistic. In between collisions, the test particles move with $\dot{\vec{r}}=(\vec{p}c/e_{rel})c$ instead of $\vec{p}/m$. Similarly the change of momentum in test particles induced by the mean field is considered to be the change in relativistic momentum. However these changes made little difference since in the projectile frame the PLF test particles move slowly. The scatter of test particles in the $z,p_zc$ plane for $^{124}$Sn on $^{119}$Sn reaction at time t=200 fm/c for 600 MeV/nucleon and impact parameter 8 fm/c is shown in Fig. \ref{Sn_Sn_z_pz_distribution}.(b). The temperature profile obtained from BUU calculation for $^{124}$Sn on $^{119}$Sn at 600 MeV/nucleon and its comparison with parameterised temperature profile is shown in Fig. \ref {PLF_Sn_Sn_Temperature}.(a). Fig.\ref {PLF_Sn_Sn_Temperature}.(b) shows the comparison of temperature profiles calculated from BUU model for 200 MeV/n (non-relativistic kinematics) and 600 MeV/n (relativistic kinematics). Identical temperature profiles for $^{124}$Sn on $^{119}$Sn projectile fragmentation at different beam energies also ensures the validity of limiting fragmentation. Fig. \ref{PLF_Ca+Be_&_Ni_Ta_Temperature}(b) displays the temperature profile obtained from BUU calculation and its comparison with universal temperature profile for $^{58}$Ni on $^{181}$Ta at 140 MeV/nucleon.
\begin{figure}[h!]
\begin{center}
\includegraphics[width=14cm,keepaspectratio=true,clip]{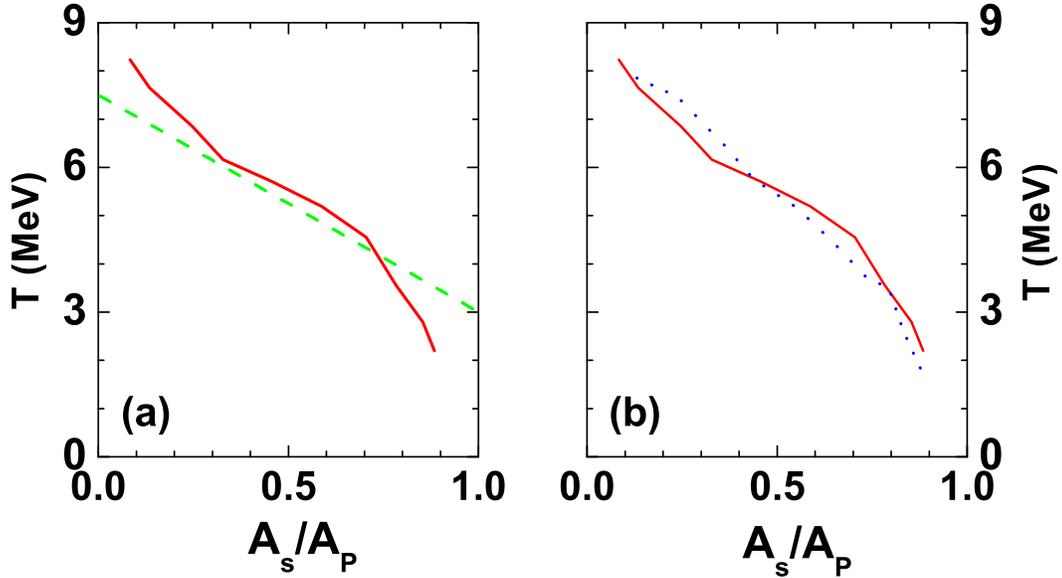}
\caption[Temperature profile for $^{124}$Sn+$^{119}$Sn reaction]{(a) Temperature profile for $^{124}$Sn+$^{119}$Sn reaction at 600 MeV/nucleon obtained from BUU model calculation (red solid line) compared with that calculated from general formula [eq. \ref{Temperature_Eq3}] (green dashed line).
(b) Comparison of temperature profile obtained from BUU calculation for $^{124}$Sn+$^{119}$Sn reaction for $600$ MeV/nucleon by using relativistic kinematics (red solid line) and $200$ MeV/n by using non-relativistic kinematics (green dashed line).}
\label{PLF_Sn_Sn_Temperature}
\end{center}
\end{figure}
\subsection{Effect of Inelastic Channels}
In section 3.3.2, it is assumed that depending upon the kinematics the particles can collide both elastically and inelastically and possible channels of collision are given in \ref{Reaction_channels}. The $\Delta$ particles (which produce $\pi$'s) are mainly produced in the participant zone i.e. the region of violent collisions and the aim is to find the properties of PLF. Therefore to avoid unnecessary extra computations the inelastic channels have been switched off. For collision at 140 MeV/n, $\Delta$ production should be minuscule but at 600 MeV/nucleon the occurrence of inelastic channels may not be negligible. The produced pions can cause secondary reactions and change properties of PLF to a certain extent.\\
\indent
To get an estimate of pion production, for $^{124}$Sn on $^{119}$Sn at 600 MeV/n, the time evolution is studied by adding the inelastic channels for a range of impact parameters $3$ fm to $10$ fm. One can assume that number of pions is equal to the number of $\Delta$'s when almost all collisions are over. For brevity the result is quoted for $b=6$ fm. The total number of $n_{\Delta}$'s is about 5 in each event. This number is the sum of $n_{\Delta}$'s in the participant region and $n_{\Delta}$'s in the spectator regions. However $n_{\Delta}$ in the PLF region is small about $0.1$. Hence if it is assumed that pion emerges from the decay of $\Delta$'s in the participant region stays in the participant region and those decaying from $\Delta$'s in the PLF stay in the PLF, this would mean that there would be only one pion in the PLF in one out of ten events. Therefore the chance of secondary reactions is very low.\\
\indent
Another model would be that when two nucleons collide, occasionally they produce a pion with a given rapidity. Pions with rapidities close to that of the PLF then thermalise in the PLF. The effect on the temperature of the PLF in this model would be very hard to compute. Similar model has been employed to calculate hypernucleus production where a $\Lambda$ particle is produced in the participant but with a rapidity close to that of the PLF. However in all the known applications the PLF is assigned a guessed temperature and the modification of temperature due to $\Lambda$ absorption is not considered \cite{Dasgupta_hypernuclei2,Botvina_hypernuclei,Botvina_hypernuclei2}.
\section{Summary}
This chapter focusses on the study of initial conditions of projectile fragmentation i.e. the PLF mass and it's excitation from microscopic models. Initially a microscopic static model is developed and the PLF is constructed by removing the particles from the given projectile depending upon the target radius and impact parameter. The PLF mass and it's excitation is obtained microscopically from it's crooked shape. Then in order to go beyond the static model, transport model simulation based on Boltzmann-Uehling-Uhlenbeck (BUU) equation is performed. In transport model, two nuclei in their Thomas-Fermi ground state are boosted towards each other with appropriate velocities at a given impact parameter. The position and momenta of test particles are updated at each instant of time by considering Vlasov propagation as well as collision. For potential energy calculation of $^{58}$Ni+$^{9}$Be reaction at $140$ MeV/nucleon and $400$ MeV/nucleon and $^{40}$Ca+$^{9}$Be at $140$ MeV/nucleon zero range Skyrme interaction and finite range Yukawa interaction is considered. To avoid huge computation for large ion collisions (e.g. $^{58}$Ni+$^{181}$Ta reaction at $140$ MeV/nucleon and $^{124}$Sn on $^{119}$Sn reaction at $600$ MeV/nucleon and $200$ MeV/nucleon) Lenk Pandharipande mean field is used. The time evolution calculation is stopped when the PLF is completely separated from the remaining part. By identifying the PLF region and by knowing the number of test particles in the PLF region the PLF mass is obtained and from the position and momentum of each test particle excited state energy is calculated. Subtracting ground state energy from excited state energy the excitation is obtained. Then Canonical Thermodynamical Model (CTM) is used to deduce the freeze-out temperature from the calculated excitation. This procedure is repeated for the entire impact parameter range. It is observed that the PLF masses at different impact parameters calculated from transport model are comparable to that obtained from geometric calculation. Nice agreement between the deduced temperature profile and earlier used parameterized temperature profile is obtained for different projectile fragmentation reactions at different energies.\\
\indent
In addition to projectile fragmentation, central collision multifragmentation reactions around Fermi energy domain are also very useful for producing exotic nuclei and for studying nuclear liquid gas phase transition. Therefore it will be interesting to study central collision multifragmentation reactions by combining statistical and dynamical model. This will be discussed in the next chapter.
\vskip3cm
\end{normalsize} 
\chapter{Hybrid model for multifragmentation around Fermi energy domain}
\begin{normalsize}
\section{Introduction}
In addition to projectile fragmentation, central collision multifragmentation reactions around fermi energy domain are also extensively used for producing neutron rich isotopes. As described in chapter 1, the dynamical models of nuclear multifragmentation are based on more microscopic calculations where the time evolution of projectile and target nucleons are studied but the problems of the dynamical model calculation is that (i) clusterization technique in dynamical models is ambiguous (will be described in chapter 7) (ii)the calculation is very time consuming. In statistical models the clusterization technique is nicely incorporated but the disadvantage of statistical model is that the calculation starts by assuming some initial conditions (like temperature, excitation energy, freeze-out volume, fragmenting source size etc.). Similar to the case of projectile fragmentation, as described in chapter 2, these conditions are either parameterized or obtained from some experimental observables.\\
\indent
In this chapter a hybrid (dynamical+statistical) model \cite{Mallik11} is developed for explaining multifragmentation reaction around the Fermi energy domain. Only central collisions are treated by this model. Initially the excitation of the colliding system is calculated by using dynamical Boltzmann-Uehling-Uhlenbeck (BUU) approach \cite{Dasgupta_BUU1} with proper consideration of pre-equilibrium emission. Then the fragmentation of this excited system is calculated by Canonical Thermodynamical model (CTM) \cite{Das}. The decay of excited fragments, which are produced in multifragmentation stage is calculated by an evaporation model \cite{Mallik1} based on Weisskopf theory \cite{Weisskopf}. This hybrid model also estimates the freeze-out temperature of the heavy ion collisions in the Fermi energy domain. The idea of setting the initial conditions for a statistical model from a dynamical model is of course not new; see for example Barz et al. \cite{Barz}. In Statistical Model of Multifragmentation (SMM), the initial conditions are fixed by some measured data. In the hybrid model, which is described in this chapter, the initial conditions for the thermodynamical model are set up almost entirely by the transport model calculation.\\
\indent
This chapter is structured as follows. In section 5.2 the technique for determining the excitation energy from BUU model is explained, then computation with the statistical model and the method of extraction of temperature is briefly described in section 5.3. Theoretical results obtained from hybrid model for $^{129}$Xe+$^{119}$Sn reaction at different projectile energies and its comparison with experimental data are presented in section 5.4. Finally this work is summarised in section 5.5.
\section{Excitation Energy from BUU model}
The excitation energy ($E^*$) of the fragmenting system can be calculated from the projectile beam energy ($E_{beam}$) by direct kinematics assuming that the projectile and the target fuse together completely. In that case the energy is determined from the centre of mass energy and the Q-value of the reaction. But this value is too high as a measure of the excitation energy of the system which multifragments. That is primarily because the pre-equilibrium particles which carry a significant part of the energy are unaccounted for in the multifragmenting system.\\
\begin{figure}[t!]
\begin{center}
\includegraphics[width=12cm,keepaspectratio=true,clip]{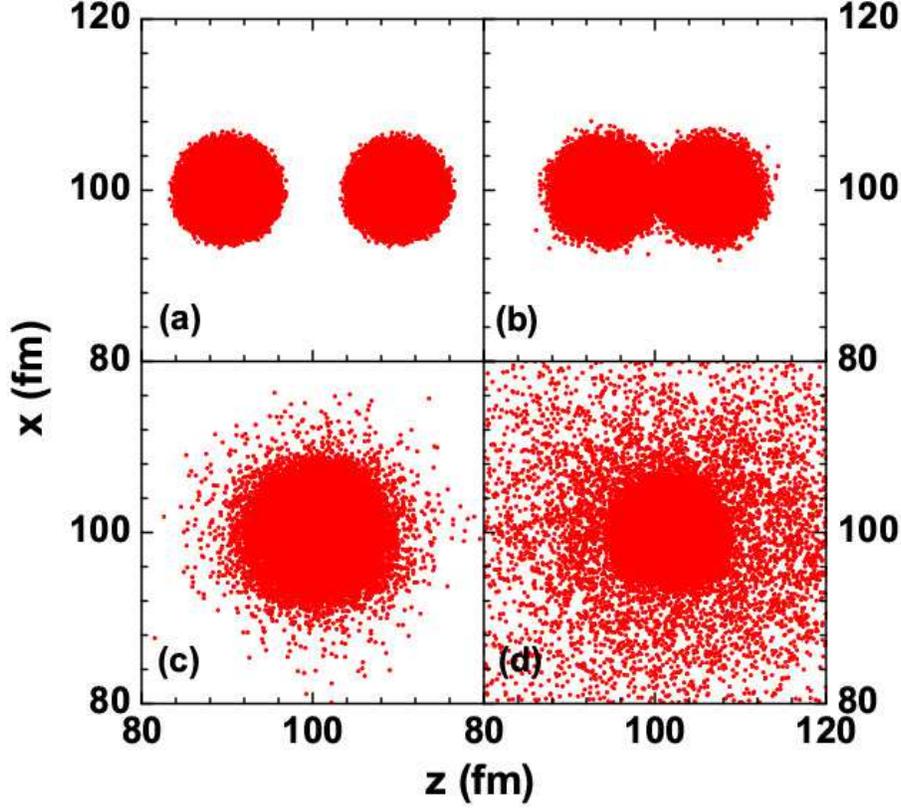}
\caption[Time evolution study for the initial stage of central collision]{Evolution of test particles at (a) 0 fm/c, (b) 25 fm/c (c) 75 fm/c and (d) 200 fm/c in center of mass frame for 45 MeV/nucleon $^{129}$Xe on $^{119}$Sn reaction.}
\label{Central_collision_time_evolution}
\end{center}
\end{figure}
\indent
To get a better measure of excitation of the fragmenting system one need to do a BUU calculation where the pre-equilibrium particles can be identified and thus eliminated in order to calculate the excitation energy per nucleon. The dynamical model calculation is started when two nuclei in their respective ground states approach each other with specified velocities. The mean field potential is taken from Ref. \cite{Lenk}. This is described in details in Eq. \ref{Lenk_potential}. Initially the Thomas-Fermi solutions for ground states \cite{Lee} are constructed. The Thomas-Fermi phase space distribution will then be modeled by choosing test particles with appropriate positions and momenta using Monte Carlo simulations (described in Appendix B). Each nucleon is represented by 100 test particles ($N_{test}=100$).\\
\begin{figure}[t!]
\begin{center}
\includegraphics[width=14cm,keepaspectratio=true,clip]{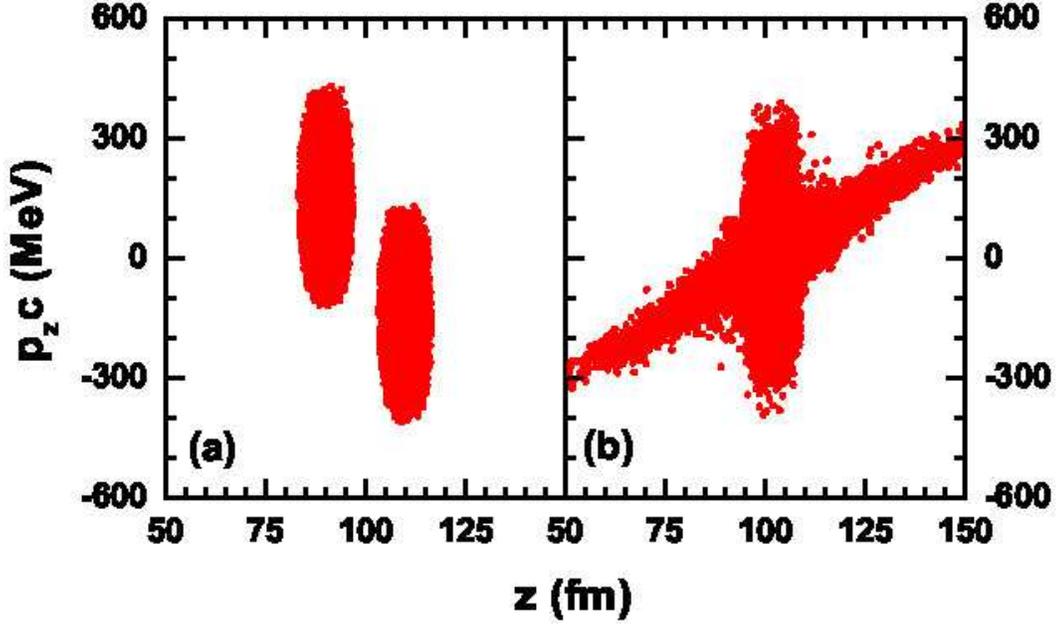}
\caption[$p_zc$ vs $z$ variation for central collison]{$p_zc$ vs $z$ variation of test particles at (a) $t=0$ fm/c and (b) 200 fm/c for $^{129}$Xe on $^{119}$Sn reaction at 45 MeV/nucleon.}
\label{Central_collision_z_pz_distribution}
\end{center}
\end{figure}
\indent
BUU transport model calculation method is exemplified with central collision reactions $^{129}$Xe+$^{119}$Sn at projectile beam energy $45$ MeV/nucleon. A three dimensional box of volume $200\times 200\times 200 fm^3$ is constructed in the configuration space and entire space is divided into $1 fm^3$ boxes. Initially the center of $^{129}$Xe and $^{119}$Sn nuclei are kept at ($100$fm, $100$fm, $90$fm) and ($100$fm, $100$fm, $110$fm) respectively in the centre of mass frame and they are boosted towards each other along $z$ direction. The test particles move in a mean-field $U(\rho(\vec{r}))$ and will occasionally suffer two-body collisions when two of them pass close to each other and the collision is not blocked by Pauli principle. The mean-field propagation is done by using the lattice Hamiltonian method (see section 3.3.1) which conserves energy and momentum very accurately \cite{Lenk}. Two body collisions are calculated as in section 3.3.2, except that the pion channels are closed, as there will not be any pion production in this energy region.\\
\indent
Fig. \ref{Central_collision_time_evolution} shows the test particles at $t=0$ fm/c (when the nuclei are separate), $25$ fm/c (the nuclei start to overlap), 75 fm/c (the time when violent collisions occur) and $200$ fm/c (almost all collisions are completed). From the figure it is clear that for $t=200$ fm/c some test particles are far distant from the central dense region. These fit into the category of pre-equilibrium emission. Fig. \ref{Central_collision_z_pz_distribution} shows scatter of test particles in $z-p_zc$ plane of the centre of mass frame at $t=0$ and $200$ fm/c. The particles produced in pre-equilibrium emission are also clearly visible at $t=200$ fm/c in Fig. \ref{Central_collision_z_pz_distribution}. In different multifragmentation experiments, it is observed that after pre-equilibrium emission accounts for $20\%$ to $25\%$ of the total mass, hence $75\%$ to $80\%$ creates the fragmenting system \cite{Xu,Frankland,Verde}. The test particles which create $80\%$ of the total mass (i.e. $A_0=198$) are chosen from the most central dense region. Knowing the momenta of the selected test particles, the kinetic energy is calculated and from the positions of these selected test particles the potential energy is calculated by using Eq. \ref{Lenk_potential}. By adding kinetic and potential energy the energy of the fragmenting system is obtained.  Fig. \ref{Central_collision_excitated_state_energy_vs_time} shows the variation of excited state energy of the central dense region (i.e. $80\%$ of the total  test particles) with time. Here total energy is always constant but as time progresses, pre-equilibrium particles having high kinetic energy, are escaping from the central dense region, therefore the energy of the central dense region is decreasing. It is clear that after $t=100$ fm/c, the energy becomes independent of time. Hence, one can stop BUU calculation at any time after t=100 fm/c and consider the corresponding energy as excited state energy. To get the excitation, the ground state energy of the fragmenting system is needed. For this Thomas Fermi method is again applied for a spherical nucleus of mass $A=198$ ($80\%$ of $^{129}$Xe+$^{119}$Sn mass). Subtracting ground state energy from the calculated energy above, the excitation energy is obtained.
\begin{figure}[t!]
\begin{center}
\includegraphics[width=9cm,keepaspectratio=true,clip]{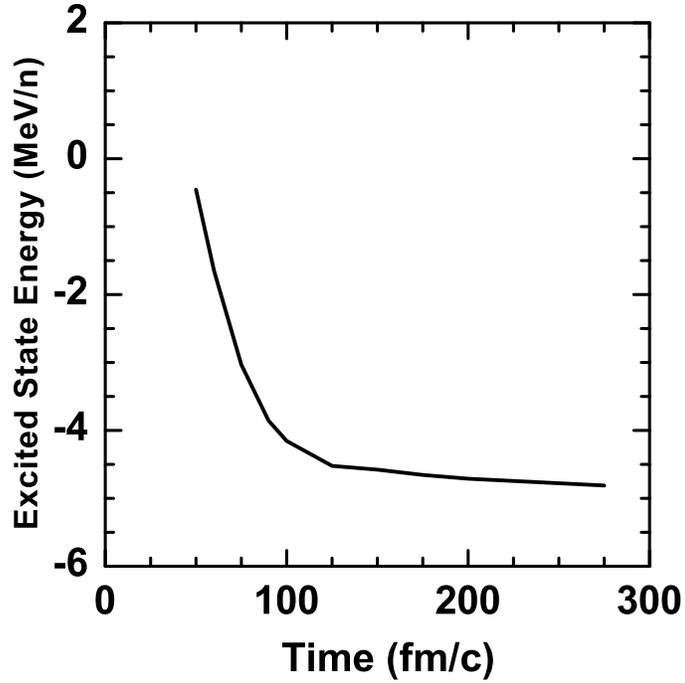}
\caption[Variation of energy of the central dense region with time]{Variation of energy of the central dense region (containing $80\%$ of total test particles) with time obtained from dynamical BUU calculation for $^{129}$Xe on $^{119}$Sn reaction at 45 MeV/nucleon.}
\label{Central_collision_excitated_state_energy_vs_time}
\end{center}
\end{figure}
\section{Computations with the statistical model: Extraction of temperature}
\begin{figure}[b!]
\begin{center}
\includegraphics[width=15cm,keepaspectratio=true,clip]{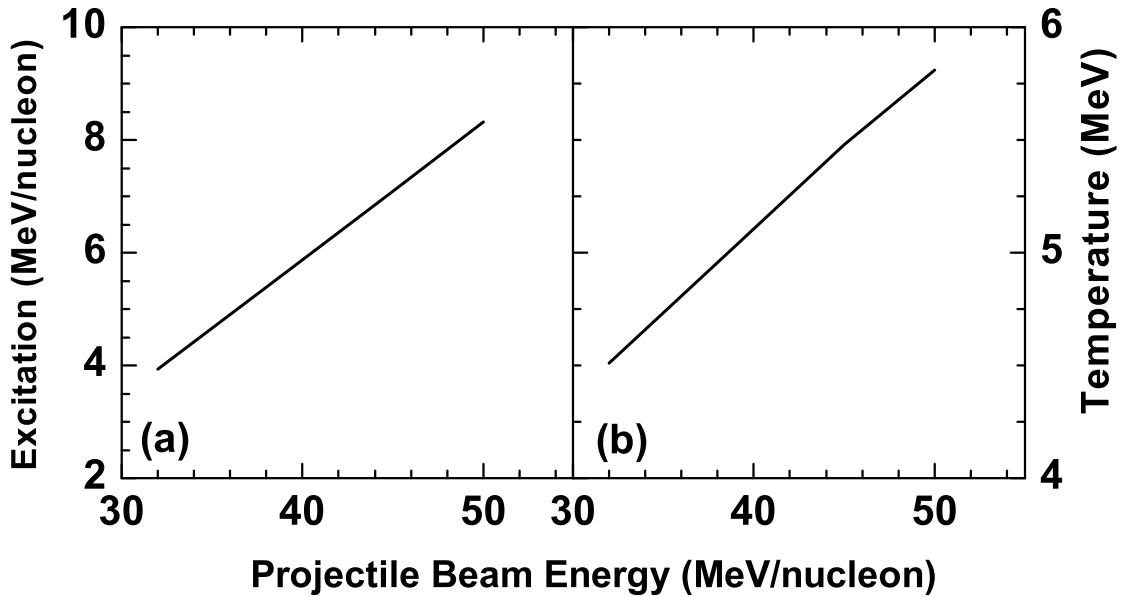}
\caption[Excitation and temperature dependence on projectile energy]{Left Panel indicates the variation of excitation energy per nucleon with projectile beam energy per nucleon obtained from dynamical BUU model. The Canonical Thermodynamical Model (CTM) can calculate average excitation energy per nucleon for a given freeze-out temperature, mass number and charge. Therefore to know the required freeze-out temperature corresponding to each excitation (obtained from BUU calculation) CTM is used. The variation of freeze-out temperature with projectile beam energy is shown in the right panel.}
\label{Central_collision_exitation_and_temperature}
\end{center}
\end{figure}
The process of extraction of excitation energy of the fragmenting system from the BUU model for central collision reactions is described in the previous section. The next task will be to obtain the freeze-out temperature. The canonical thermodynamic model (CTM) \cite{Das} can be used to calculate average the excitation per nucleon for a given temperature, charge and mass. This is described in section 3.4.3 for projectile fragmentation. The two main differences between the CTM calculation for projectile fragmentation and for central collision are as follows \\
\indent
(i) in a projectile fragmentation reaction, CTM calculation was done for different abraded PLFs which were produced at different impact parameters. But in a central collision reaction, we will use CTM for the the fragmentation of a single source, which is produced after pre-equilibrium emission.\\
\indent
(ii) in projectile fragmentation, there is no violent collision in PLF, therefore the expansion of the excited PLF is lesser. But in central collision, due to collision of the projectile and target nuclei the produced excited system is initially compressed and then expanded to a larger freeze-out volume. Therefore the freeze-out volume for projectile fragmentation is taken to be $3$ times the normal nuclear volume but for central collision multifragmentation reaction the freeze-out volume to normal volume ratio is chossen to be $6$. These two ratios are obtained from experimental measurements and theoretical data fitting.\\
Therefore, the freeze-out temperature for central collision reaction can be obtained by using CTM at a temperature $T$ which satisfies the condition,
\begin{equation}
E^{*}(N_0,Z_0)=\sum_{I,J}\langle n_{I,J}\rangle\bigg{[}\frac{3T}{2}+\frac{AT^2}{\epsilon_0}-B(I,J)\bigg{]}
\end{equation}
where $E^{*}$ is the excited state energy obtained from BUU calculation and $\langle n_{I,J}\rangle$ is the multiplicity of the fragments having $I$ neutrons and $J$ protons produced by the multifragmentation from the system $N_0,Z_0$ at temperature $T$, $B$'s denote the binding energies. The composites that follow from CTM will further decay by evaporation. The details of how this can be done is given in section 2.2.3.\\
\indent
The complete hybrid model calculation is done for the same $^{129}$Xe+$^{119}$Sn pair for projectile beam energies $32$, $39$, $45$ and $50$ MeV/nucleon. The variation of the calculated excitation energy with projectile beam energy is shown in the left diagram of Fig. \ref{Central_collision_exitation_and_temperature}. The freeze-out temperature is found out from this excitation energy. Thus the freeze-out temperature for a given beam energy is obtained. This is plotted on the right side of Fig. \ref{Central_collision_exitation_and_temperature}.\\
\section{Results}
Important properties of nuclear multifragmentation like charge distribution, isotopic distribution, largest cluster probability distribution and average size of largest cluster are studied theoretically by using the hybrid model. To check the accuracy of the hybrid model, some results are also compared with the available experimental data.
\subsection{Isotopic Distribution}
The average multiplicities ($\langle n_{N,Z} \rangle$) of different $C$, $Si$ and $Ni$ isotopes are calculated by the hybrid model for $^{129}$Xe+$^{119}$Sn central collision multifragmentation reaction at projectile beam energies $32$, $39$, $45$ and $50$ MeV/nucleon. This is shown in Fig. \ref{Central_collision_isotopic_distribution}.
\begin{figure}[h!]
\begin{center}
\includegraphics[width=\textwidth,keepaspectratio=true,clip]{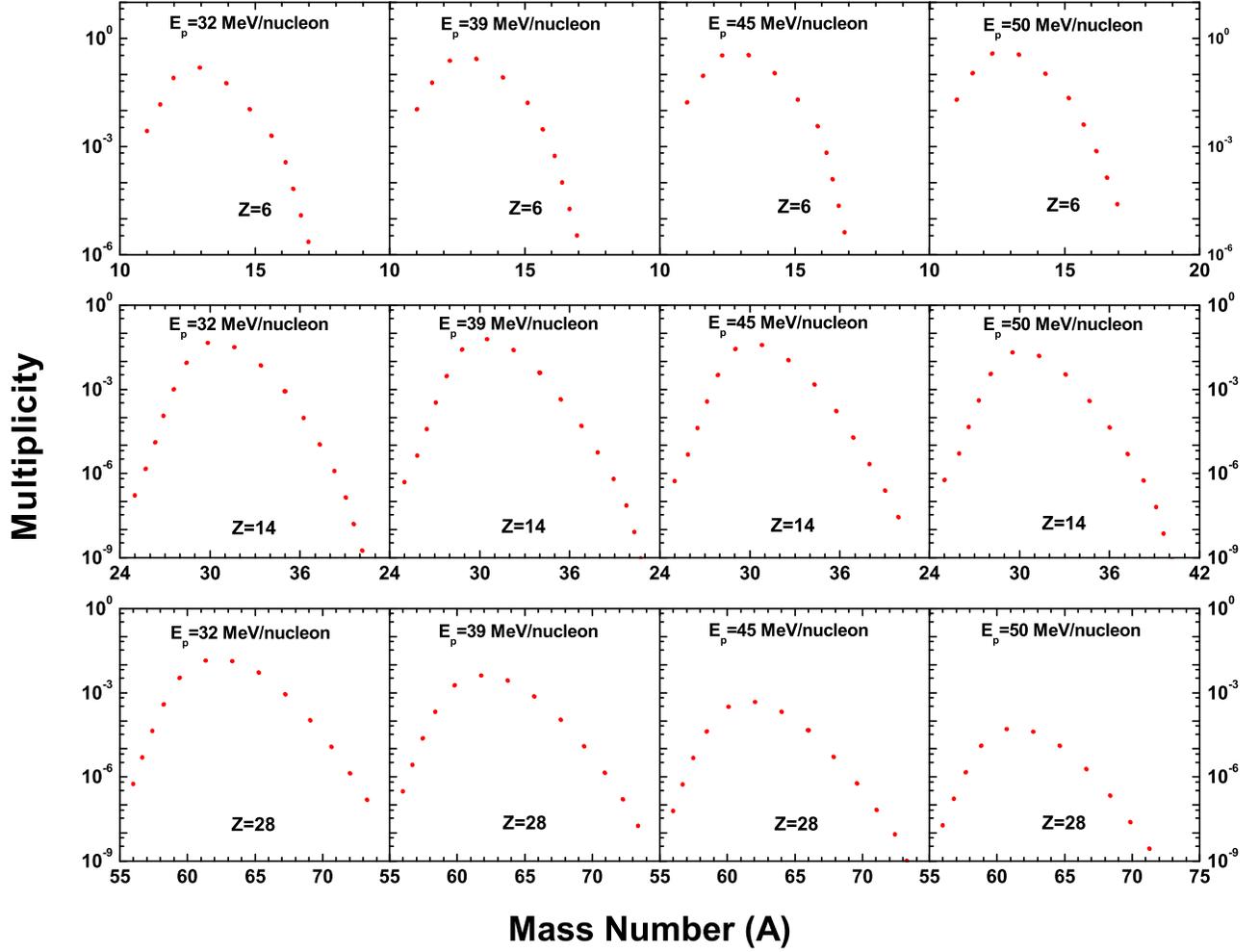}
\caption[Isotopic distribution in central collision reaction]{Theoretical isotopic distributions of fragments having $Z=6$ (upper panel), $14$ (middle panel) and $28$ (lower panel) produced from $^{129}$Xe on $^{119}$Sn reaction at $32$, $39$, $45$ and $50$ MeV/nucleon.}
\label{Central_collision_isotopic_distribution}
\end{center}
\end{figure}
\subsection{Charge Distribution}
The total charge multiplicities $\langle n_Z \rangle=\sum_N \langle n_{N,Z} \rangle$ are computed from the hybrid model for $^{129}$Xe+$^{119}$Sn reaction at different projectile beam energies mentioned above and compared with the experimental data. This is shown in Fig. \ref{Central_collision_charge_distribution}. The experiments \cite{Hudan,Bonnet_NPA} are done by the INDRA collaboration in GANIL. From Fig. \ref{Central_collision_charge_distribution}, one can conclude that, with the increase of energy (i.e. increase of temperature), fragmentation is more, therefore multiplicities of higher fragments gradually decrease.
\begin{figure}[h!]
\begin{center}
\includegraphics[width=14cm,keepaspectratio=true,clip]{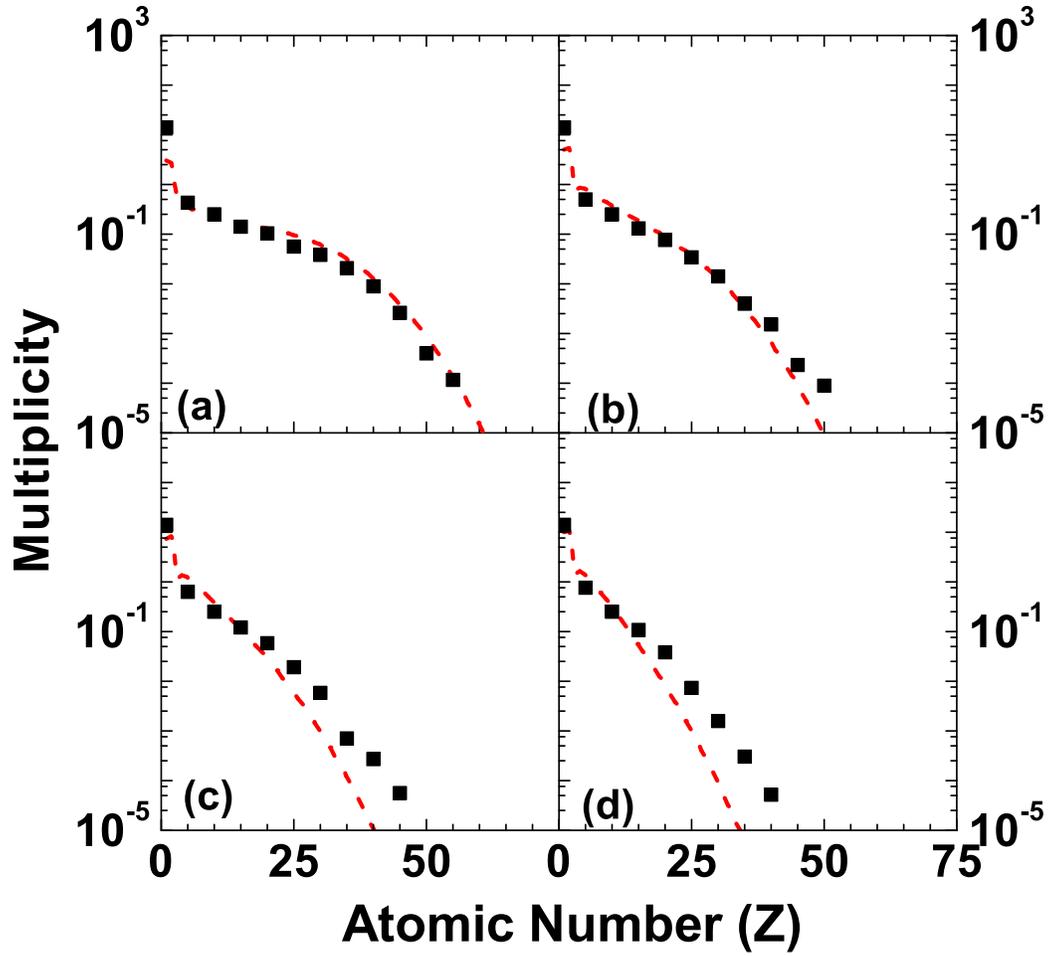}
\caption[Charge distribution in central collision reaction]{Theoretical charge distribution (red dotted lines) for $^{129}$Xe on $^{119}$Sn reaction at (a) 32 MeV/nucleon (b) 39 MeV/nucleon
(c) 45 MeV/nucleon and (d) 50 MeV/nucleon. The experimental data are shown by black squares.}
\label{Central_collision_charge_distribution}
\end{center}
\end{figure}
\subsection{Largest Cluster Probability Distribution}
To study the largest cluster probability distribution for central collisions, the same procedure is followed as described in section 2.4.7. Fig. \ref{Central_collision_largest_cluster_probability_distribution} represents the largest cluster probability distribution at different energies. Since with the increase of energy breaking increases the peak of the largest cluster probability distribution shifts towards the lower atomic number side and the width of the distribution gradually decreases. Reasonable agreement with data is observed.
\begin{figure}[h!]
\begin{center}
\includegraphics[width=14cm,keepaspectratio=true,clip]{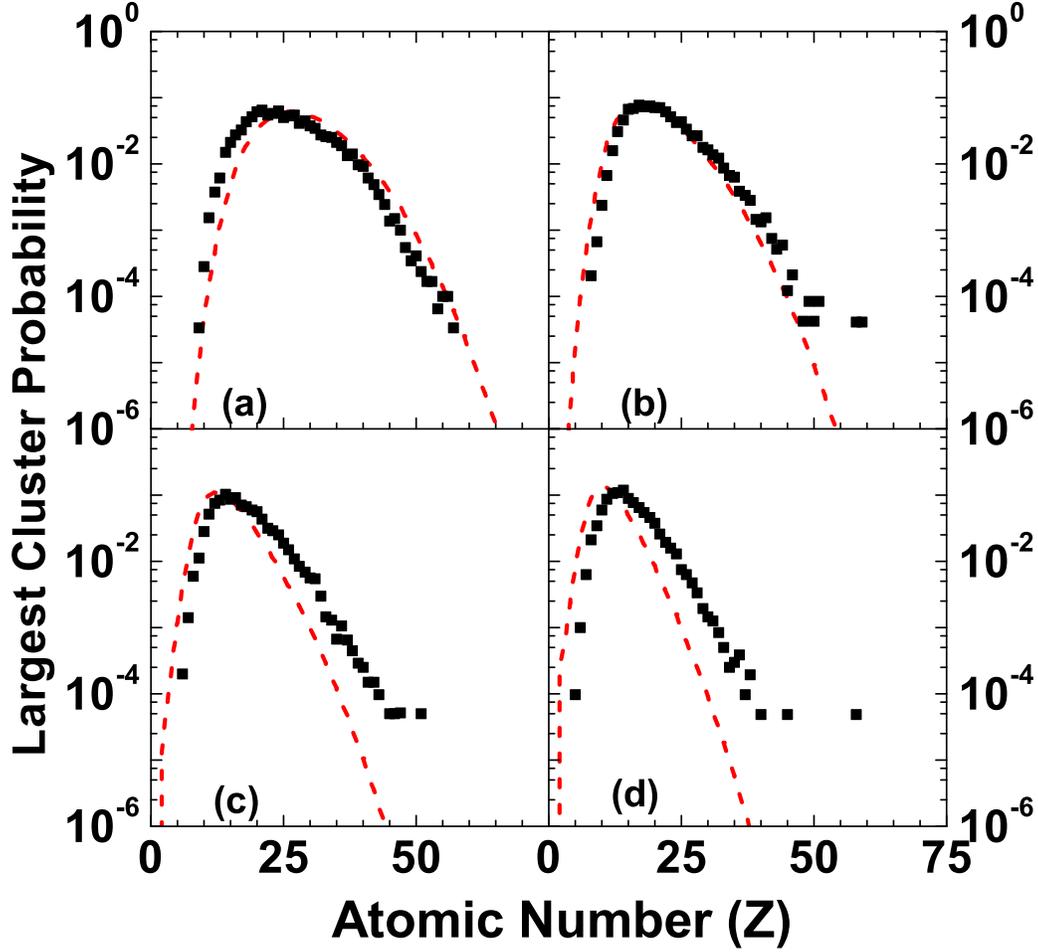}
\caption[Largest cluster probability in central collision reaction]{Theoretical largest cluster probability distribution (red dotted lines) for $^{129}$Xe on $^{119}$Sn reaction at (a) 32 MeV/nucleon (b) 39 MeV/nucleon (c) 45 MeV/nucleon and (d) 50 MeV/nucleon. The experimental data are shown by black squares.}
\label{Central_collision_largest_cluster_probability_distribution}
\end{center}
\end{figure}
\subsection{Average charge of Largest Cluster}
The variation of average charge of largest cluster $\langle Z_{Largest}\rangle$ with projectile beam energy is shown in Fig. \ref{Central_collision_largest_cluster_size_vs_energy}. For fragmenting system, the value of $80\%$ of total mass is adopted from experimental observations. But the present calculations (see Fig. \ref{Central_collision_largest_cluster_size_vs_energy}) show that this was a reasonable choice. A plot of $\langle Z_{Largest}\rangle$ is shown which agrees fairly well with data. Now $\langle Z_{Largest}\rangle$ depends upon the size of the fragmenting system as well as the temperature of the fragmenting system. The larger the fragmenting system, the larger is the $\langle Z_{Largest}\rangle$. The higher the temperature, the smaller is $\langle Z_{Largest}\rangle$. Now the temperature also depends upon what percentage of nucleons are left out as pre-equilibrium particles. The choosen value of $80\%$ gives a combination of temperature and fragmenting mass that seems to be just about right. One could do a detailed best "fit" but this was not attempted.\\
\begin{figure}[!h]
\begin{center}
\includegraphics[width=10cm,keepaspectratio=true,clip]{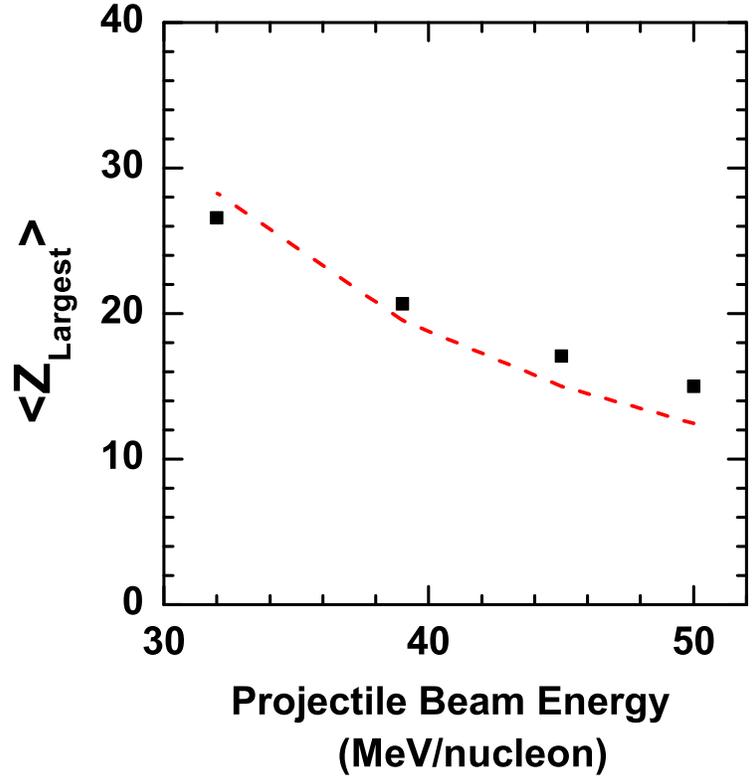}
\caption[Largest cluster charge dependence on projectile energy]{Variation of average charge of largest cluster with projectile beam energy obtained from hybrid model calculation (red dotted lines) for $^{129}$Xe on $^{119}$Sn reaction. The experimental data are shown by black squares.}
\label{Central_collision_largest_cluster_size_vs_energy}
\end{center}
\end{figure}
\indent
Radial flow has not been considered in the analysis so far. One reason is that the collision energy being only about 50 MeV/nucleon, the initial compression is small so any radial flow must also be small. The best signature for radial flow will be in the velocity distribution but in this work only multiplicity distribution and largest cluster probability distribution have been calculated. Neither CTM nor SMM can incorporate radial flow easily but in Lattice gas model, where flow is easily incorporated. it was found that even for significant radial flow, multiplicity distributions are hardly affected \cite{Das_LCG}.
\section{Summary}
In this work a hybrid (dynamical+statistical) model is developed for explaining nuclear multifragmentation reactions near the Fermi energy regime. BUU transport model is applied for obtaining the excitation energy per nucleon of the multifragmenting system produced in the central collision reactions. The canonical thermodynamic model is then used to determine the temperature which would lead to this excitation energy. With this temperature the canonical thermodynamic model calculation is performed for studying various important obserbables of nuclear multifragmentation like multiplicities of different composites, probability distribution of the largest cluster etc. for central collisions of $^{129}$Xe on $^{119}$Sn at beam energies of (a) 32 MeV/nucleon, (b) 39 MeV/nucleon, (c) 45 MeV/nucleon and (d) 50 MeV/nucleon. Theoretical results are compared with the available experimental data. The agreement with data is pleasing.\\
\indent
For explaining the multifragmentation stage in central collision reaction as well as in projectile fragmentation, canonical model is very successful. In addition to the canonical model, calculations based on grand canonical ensembles are also commonly used for describing multifragmentation phenomena. Hence, it will be very useful to study the conditions of convergence of these two statistical ensembles for the fragmentation of finite nuclei. This will be discussed in the next chapter.
\vskip3cm
\end{normalsize} 
\chapter{Ensemble equivalence in nuclear multifragmentation}
\begin{normalsize}
\section{Introduction}
Statistical models have been extensively used to study the physics of intermediate energy heavy ion collisions to study the physics of intermediate energy heavy ion collisions, since they are most successful for explaining the clusterization technique in multifragmentation and different observables have been routinely compared to experimental data.
The fragmentation of the nucleus into available channels (depends on phase space) can be analyzed with different statistical ensembles (microcanonical, canonical and grand canonical).
From a technical point of view, it is always much simpler to calculate an observable in the grandcanonical ensemble than in the canonical one. On the other side, in realistic modelling of nuclear fragmentation, the appropriate ensemble is rather the canonical \cite{Das} or the microcanonical \cite{Pratt} one. Indeed nuclear systems that can be formed in the laboratory are isolated systems which are not coupled to an external energy or particle bath. The excited nuclear systems which can be described via statistical models typically constitute only a subsystem of the total interacting system, meaning that conservation laws on particle number and energy are not strict. However, energy and particle numbers can in principle be measured, and statistical ensembles with a fixed number of particles and energy can be obtained by an appropriate sorting of experimental data. Therefore the partitioning into available channels can be solved in the canonical model \cite{Das} where the number of particles in the nuclear system is finite (as it would be in experiments). Even when the number of particles is fixed one can replace a canonical model by a grand canonical model where the particle number fluctuates but the average number is constrained to the given value \cite{Chaudhuri2,Reif}.\\
\indent
It is well known that the underlying physical assumption behind canonical and grand canonical models is fundamentally different, and in principle they agree only in the thermodynamical limit \cite{Reif}, that is, when the number of particles become infinite. For example, for one kind of particle (nucleon) and for arbitrarily large nuclear system (therefore approximates the thermodynamical limit) \cite{Chaudhuri3}, it was observed that results agree with each other under certain conditions. This equivalence is generally known not to be valid for nuclear systems of finite size.\\
\indent
The aim of this chapter is to show that the results from the canonical and the grand canonical models can agree even for finite nuclei under certain conditions and hence the multiplicity of the fragments leading to charge and mass distributions from the canonical and grand canonical distributions under varying conditions will be investigated. This led us to identify the conditions \cite{Mallik4} under which results from both the models converge.
\section{Theoretical formalism}
In canonical model, the average multiplicity of the fragments with $N$ neutrons and $Z$ protons produced from the fragmentation of a nucleus of $N_{0}$ neutrons and $Z_{0}$ protons is given by
\begin{equation}
\langle n_{N,Z}\rangle_{c}=\omega_{N,Z}\frac{Q_{N_{0}-N,Z_{0}-Z}}{Q_{N_{0},Z_{0}}}
\label{Canonical_multiplicity}
\end{equation}
The derivation of the Eq. \ref{Canonical_multiplicity} is already described in details in section 2.2.2.\\
\indent
Similar to the canonical model, in grand canonical model one assumes that the system with $A_{0}$ nucleons and $Z_{0}$
protons at temperature $T$, has expanded to a volume higher than the normal nuclear volume and fragments are formed which remain in a thermodynamical (statistical) equilibrium at this freeze-out condition. In the grand canonical model \cite{Chaudhuri2}, if the neutron chemical potential is $\mu_{n}$ and the proton chemical potential is $\mu_{p}$, then statistical equilibrium implies \cite{Reif} that the chemical potential of a composite with $N$ neutrons and $Z$ protons is $\mu_{n}N+\mu_{p}Z$. The average number of composites with $N$ neutrons and $Z$ protons is given by \cite{Chaudhuri2}
\begin{equation}
\langle n_{N,Z}\rangle_{gc}=e^{\beta\mu_{n}N+\beta\mu_{p}Z}\omega_{N,Z}
\label{grandcanonical_multiplicity}
\end{equation}
Here $\omega_{N,Z}$ is the partition function of the composite with $N$ neutrons and $Z$ protons and it is calculated in the similar way of canonical model as described in section 2.2.2. The chemical potentials $\mu_{n}$ and $\mu_{p}$ are determined by solving neutron and proton conservation equations
\begin{eqnarray}
N_{0}=\sum Ne^{\beta\mu_{n}N+\beta\mu_{p}Z}\omega_{N,Z} \nonumber \\
Z_{0}=\sum Ze^{\beta\mu_{n}N+\beta\mu_{p}Z}\omega_{N,Z}
\label{grandcanonical_particle_conservation}
\end{eqnarray}
This amounts to solving for an infinite system but we emphasize that this infinite system can break-up into only certain kinds of species as are included in the above two equations. One can look upon the sum on $N$ and $Z$ as a sum over $A$ and a sum over $Z$. In principle $A$ goes from 1 to $\infty$ and for a given $A$, $Z$ can go from 0 to $A$. Here for a given $A$ we restrict $Z$ by the same drip lines used for the canonical model.
\section{Conditions of convergence of statistical ensembles}
This section focuses on how the equivalence (or non-equivalence) can be varied by changing the temperature or freeze-out volume of the fragmenting nucleus or by varying the source size or isospin asymmetry of the source.

\begin{figure}[h!]
\begin{center}
\includegraphics[width=13.0cm,keepaspectratio=true,clip]{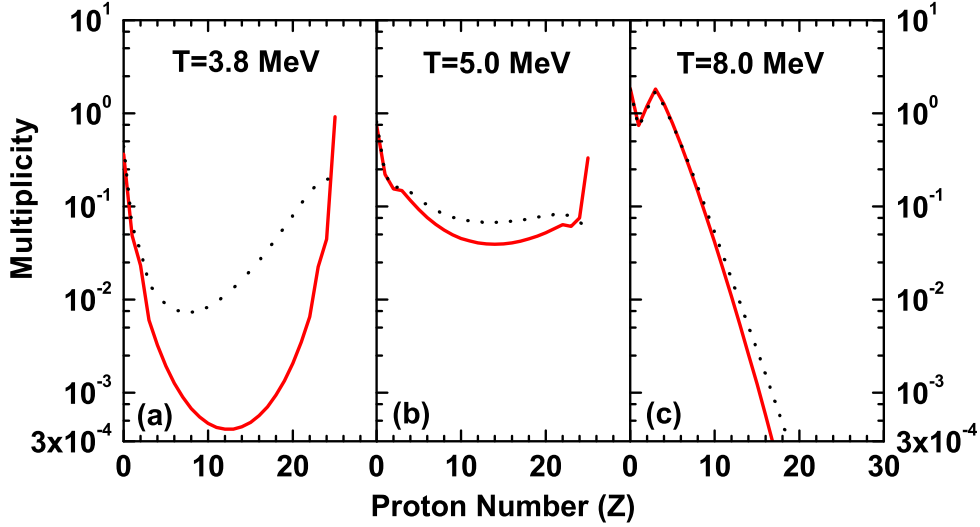}
\caption[Temperature ($T$) dependence of ensemble equivalence]{Total charge distribution of $A_0=60$, $Z_0=25$ system from canonical (red solid lines) and grand canonical model (black dotted lines) at same freeze-out volume $V_f=3V_0$ but three different temperatures (a) $3.8$ MeV , (b) $5$ MeV and (c) $8$ MeV.}
\label{Ensemble_equivalence_temperature_dependence}
\end{center}
\end{figure}
\subsection{Effect of temperature}
The total charge distribution $\langle n_{Z}\rangle=\sum_{N}\langle n_{N,Z}\rangle$ obtained from both the ensembles at three different temperatures $3.8$ MeV, $5$ MeV and $8$ MeV from disassembly of a particular source $(Z_{0}=25,A_{0}=60)$ at a fixed freeze-out volume $3V_{0}$ is compared in Fig. \ref{Ensemble_equivalence_temperature_dependence}. The difference in result is maximum at the lowest temperature $3.8$ MeV where fragmentation is less and the disassembly of the nucleus results in more of 'liquid-like' fragments or higher mass fragments. As one increases the temperature, fragmentation increases, the number of such higher mass fragments decrease (at the expense of the lower mass ones) and the results from the canonical and grand canonical ensembles begin to converge. This is easily seen at the two higher temperatures. At $8$ MeV the results from both the ensembles are very close to each other since fragmentation is maximum at this temperature, the nucleons and the lower mass fragments dominating the distribution.
\subsection{Effect of freeze-out Volume}
The effect of increasing the freeze-out volume (decreasing the density) is equivalent to that of increasing the temperature and this is seen in Fig. \ref{Ensemble_equivalence_volume_dependence}.\\
\indent
Here the calculation is repeated for the same source $(Z_{0}=25,A_{0}=60)$ at a fixed temperature $T=5$ MeV but for three different freeze-out volumes $V_f=3V_0$, $4V_0$ and $5V_0$. It is seen that results from both the ensembles agree with each other as one increases the freeze-out volume when the nucleus fragments more into smaller pieces.
\begin{figure}[t!]
\begin{center}
\includegraphics[width=13.0cm,keepaspectratio=true,clip]{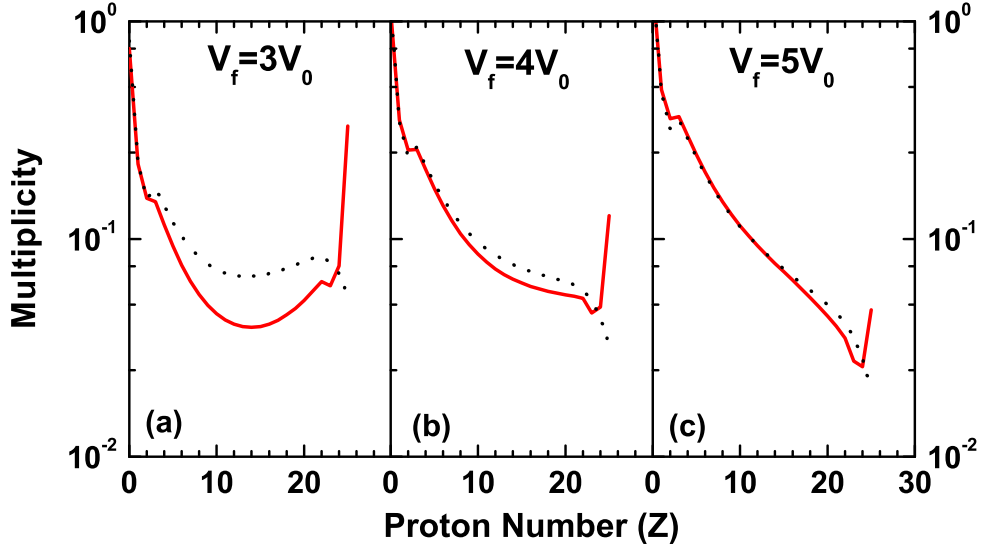}
\caption[Freeze-out volume $V_f$ dependence of ensemble equivalence]{Total charge distribution of $A_0=60$, $Z_0=25$ system at $T=5.0$ MeV by using canonical (red solid lines) and grand canonical model (black dotted lines) for three different freeze-out volumes (a) $V_f=3V_0$, (b) $V_f=4V_0$ and (c) $V_f=5V_0$.}
\label{Ensemble_equivalence_volume_dependence}
\end{center}
\end{figure}
\subsection{Effect of isospin asymmetry}
\begin{figure}[b!]
\begin{center}
\includegraphics[width=13.0cm,keepaspectratio=true,clip]{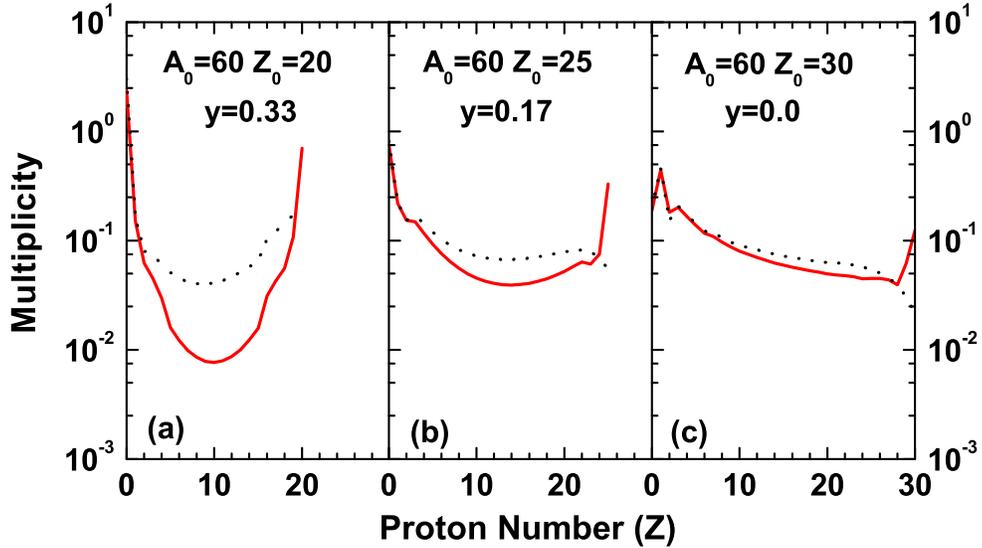}
\caption[Isospin asymmetry $y$ dependence of ensemble equivalence]{Total charge distribution at $T=5.0$ MeV and $V_f=3V_0$ from canonical (red solid lines) and grand canonical model (black dotted lines) of the sources having same $A_0=60$ but different isospin asymmetry (a) $y=0.33$, (b) $0.17$ and (c) $0$.}
\label{Ensemble_equivalence_asymmetry_dependence}
\end{center}
\end{figure}
Convergence of canonical and grand canonical model result is also seen if we vary the source asymmetry $y=(N_{0}-Z_{0})/(N_{0}+Z_{0})$ keeping the temperature fixed  $5$ MeV, freeze-out volume at $3V_{0}$ and source size at $A_0=60$. Fig. \ref{Ensemble_equivalence_asymmetry_dependence} shows the charge distribution for three nuclei having $y$ = 0.33, 0.17 and 0 respectively. We observe that the difference in results between both the ensembles is maximum when the asymmetry is more (Fig. \ref{Ensemble_equivalence_asymmetry_dependence}.(a)) and the difference is least for the symmetric nucleus (Fig. \ref{Ensemble_equivalence_asymmetry_dependence}.(c)).\\
\begin{figure}[b!]
\begin{center}
\includegraphics[width=13.0cm,keepaspectratio=true,clip]{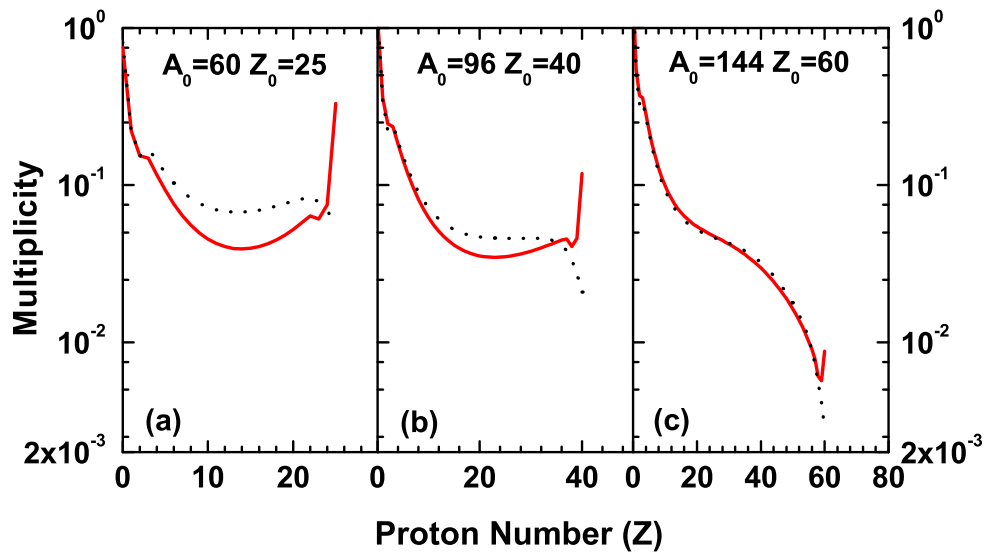}
\caption[Source size ($A_0$)dependence of ensemble equivalence]{Total charge distribution at $T=5.0$ MeV and $V_f=3V_0$ by using canonical (red solid lines) and grand canonical model (black dotted lines) for three different source sizes $A_0=$ (a)$60$, (b)$96$ and (c)$144$ each having same isospin asymmetry  $y=0.17$.}
\end{center}
\label{Ensemble_equivalence_size_dependence}
\end{figure}
\subsection{Effect of fragmenting source size}
This convergence effect is also seen if one keep both temperature, freeze-out volume and the asymmetry parameter fixed but increase the source size (mass) as shown in Fig. \ref{Ensemble_equivalence_size_dependence}. The difference in result between both the ensembles is maximum when the source size is minimum as expected and the results become close to each other for a large nucleus. One can say that the nucleus fragments more and more as one increases the source size (keeping other parameters fixed) and the effect is similar to that of increasing the temperature keeping the source size fixed.
\section{Reasons for convergence of statistical ensembles}
In order to investigate the reasons of converge of statistical ensembles for finite nuclei, the ratio (normalized) of higher mass fragments formed to that of the total number of fragments (total multiplicity) is calculated. The fragment whose size is more than 0.8 times $A_{0}$ (more than $80\%$ of the source in size) is considered as higher mass fragment, i.e. the ratio ($\eta$) is defined as
\begin{equation}
\eta=\frac{\sum_{A>0.8A_{0}}^{A_{0}}\langle n_{N,Z}\rangle}{\sum_{A=1}^{A_{0}}\langle n_{N,Z}\rangle}
\label{Higher_mass_fragments_equation}
\end{equation}
\begin{figure}[h!]
\begin{center}
\includegraphics[width=12cm,keepaspectratio=true,clip]{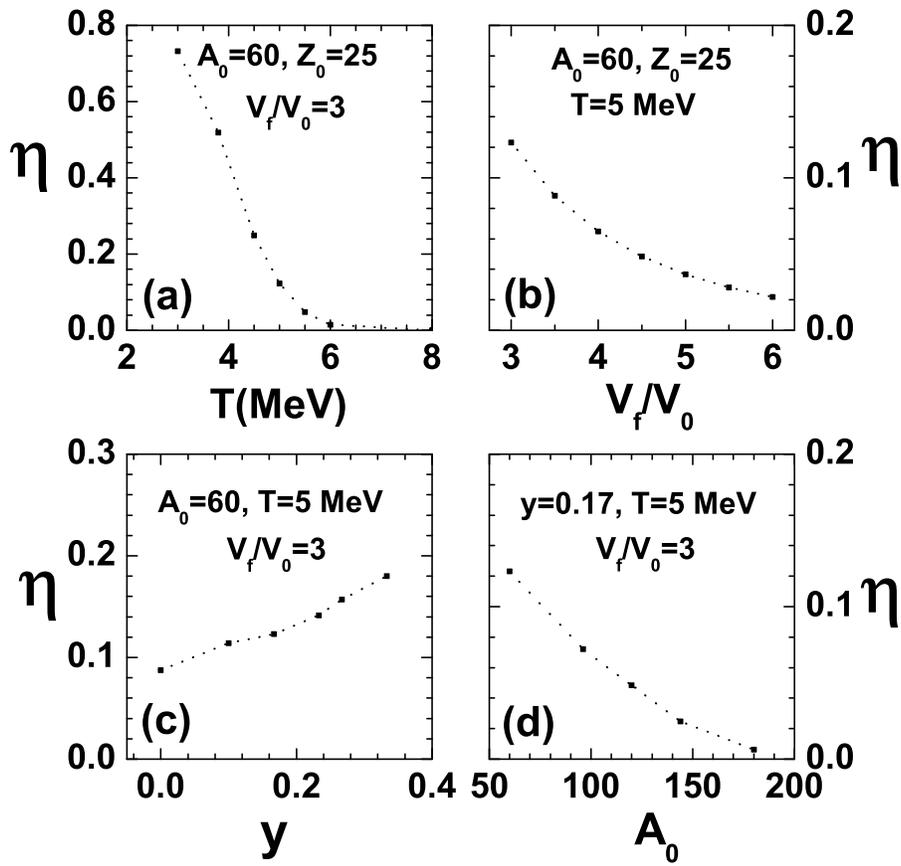}
\caption[Variation of $\eta$ with $T$, $V_f$, $y$ and $A_0$] {Variation of $\eta$ with (a) temperature, (b) freeze-out volume, (c) isospin asymmetry and (d) source size from grand canonical model.}
\label{Ensemble_equivalence_higher_fragments_grand_canonical}
\end{center}
\end{figure}
This criteria of choosing the higher mass fragments is not very rigid and can be relaxed. It has been checked that even if the ratio $0.75$ or $0.85$ instead of $0.8$ the trend of the results remain same. This calculation has been done in both canonical and grand canonical models and the results are similar. The results from the grand canonical model is shown in Fig. \ref{Ensemble_equivalence_higher_fragments_grand_canonical}.  In Fig \ref{Ensemble_equivalence_higher_fragments_grand_canonical}.a the variation of this ratio is displayed as a function of temperature (keeping source size, freeze-out volume and asymmetry fixed) and it is seen that the ratio decreases with  increase in $T$. This shows that for a source with lower values of $T$, the fraction of higher mass fragments formed as a result of fragmentation is more as compared to those with higher $T$ values. It is being emphasize that the difference in the charge distributions from the canonical and grand canonical ensembles is mainly caused by the presence of the higher mass fragments in the distribution. The lesser is the fraction of the higher mass fragments, the deviation in results between both the ensembles will be less and this is exactly what we saw in Fig. \ref{Ensemble_equivalence_temperature_dependence}. Similar effect is seen when one plots this ratio (Fig. \ref{Ensemble_equivalence_higher_fragments_grand_canonical}.b) as a function of $V_{f}/V_{0}$ keeping other parameters fixed. It is clearly seen that with increase in the freeze-out volume, the fraction of higher mass fragments decrease and this causes the results between both the ensembles to be very close when $V_{f}$ is maximum as shown in Fig. \ref{Ensemble_equivalence_volume_dependence}.\\
\indent
$\eta$ is being also plotted as a function of the asymmetry parameter $y$ of the source, the source size ($A_{0}=60$), temperature (5 MeV) and freeze-out volume ($3V_{0}$) being kept fixed and it is seen that the ratio increases with $y$ (Fig \ref{Ensemble_equivalence_higher_fragments_grand_canonical}(c)). So here it is being observed that the less is the asymmetry of the source, less is the number of large fragments and hence fragmentation of the nucleus is more. In this scenario, when the nucleus is more symmetric the results from the two ensembles agree to a very good extent than when the nucleus is less symmetric as seen in Fig \ref{Ensemble_equivalence_asymmetry_dependence}. The same effect is seen (Fig \ref{Ensemble_equivalence_higher_fragments_grand_canonical}(d)) if one increases the source size keeping the other parameters fixed and it is being asserted that the effect of increasing the source size is similar to that of increasing the temperature or freeze-out volume or decreasing the asymmetry of the source as far as convergence between both the ensembles is considered.\\
\indent
The differences in results between the canonical and the grand canonical ensemble is mainly because of the presence of the higher mass fragments in the fragmentation of a nucleus. If the conditions are such that the fragmentation is more and there are only lower mass clusters, then the particle number fluctuation in the grand canonical model is very less \cite{Das}. In canonical model, particle number is strictly conserved and there is no such fluctuation. Hence the results from both the ensembles agree to a much better extent. The same condition is also valid for convergence between micro canonical and canonical ensembles where energy plays the role of the extensive variable instead of the total number of particles. The more the nucleus disintegrates, the less will be the fluctuation in energy and better will be the convergence between the micro canonical and the canonical ensembles.
\section{Transformation between statistical ensembles: Case of One component system}
In the previous sections of this chapter it is shown that, even for finite nuclei ensembles may be close to equivalence under certain conditions. In this section, an approximate expression will be developed allowing to transform the results for observables from one ensemble to the other. But for present study, the ensemble transformation will be restricted to the one component case only i.e. neutrons are protons are being considered in the same footing (so there is no asymmetry term and 50\% of the total particles are considered as charged i.e. each cluster of size $A$ has an effective charge $Z_{eff}=A/2$). For two components, this transformation procedure will be more complicated. That is reserved for the future. Similar to, Eq. \ref{Canonical_multiplicity} and \ref{grandcanonical_multiplicity}, in one component canonical and grand canonical model, the average multiplicity of a cluster of size $A$ produced from the fragmentation of a nucleus of $A_0$ can be written as:

\begin{equation}
\langle n_A\rangle_{c}=\omega_A \frac{Q_{A_0-A}}{Q_{A_0}}
\label{One_component_canonical_multiplicity}
\end{equation}

\begin{equation}
\langle n_A\rangle_{gc}=\omega_A \exp \alpha_A
\label{One_component_grand_canonical_multiplicity}
\end{equation}

In section 5.3 and 5.4, for the grand canonical model calculation of the real nuclei, the size of the largest cluster was restricted as same as the fragmenting source size (i.e. $N_{max}=N_0$ and $Z_{max}=Z_0$). But ideally in grand canonical ensemble the cluster size may vary from $0$ to $\infty$. For systems with finite volume, the upper limit of largest cluster size can not be $\infty$, it will be the maximum number of particles which can be spaced within this volume. For example, if one considers that fragmentation occurs in a freeze-out volume $V_f=\Xi V_0$ ($V_0$ is the volume of a nucleus of mass number $A_0$ at the normal nuclear density), then the upper limit of largest cluster size will be $(\Xi-1)A_0$. For further calculations of this chapter, this will be used. The effect of changing of the upper limit of largest cluster size on few observables in multifragmentation  (for one component case) can be found in Ref \cite{Chaudhuri3,Das3}.\\
\indent
There are two ways of computing the grand canonical average number of particles $\langle A_0 \rangle$ at fugacity $\alpha=\beta\mu$,
where $\beta$ is the inverse temperature and $\mu$ the chemical potential. The first way needs the calculation of the grand canonical partition sum $Q_\alpha$
\begin{equation}
\langle A_0 \rangle_\alpha=\frac{\partial ln Q_\alpha}{\partial \alpha}
\end{equation}
While the second way uses the definition of the particle number distribution in the grand canonical ensemble
\begin{equation}
\langle A_0 \rangle_\alpha=\sum_{A_0=0}^{\infty} A _0P_\alpha (A_0)
\end{equation}
This distribution is given by
\begin{equation}
P_\alpha (A_0)=Q_{\alpha}^{-1} Q_{A_0} \exp \alpha A_0
\label{proba}
\end{equation}
and implies the knowledge of the canonical partition sum \cite{Pathria}.
It is to be noted that the knowledge of $Q_{\alpha}$ is not really necessary in this last equation because it can be deduced from the condition of normalization of probabilities.
The same kind of relations is established for the particle variance:
\begin{equation}
\sigma^2_\alpha=\frac{\partial^2 ln Q_\alpha}{\partial \alpha^2}=
\sum_{A_0=0}^{\infty} \left (A_0-\langle A_0\rangle_\alpha \right )^2  P_\alpha (A_0) \label{eq2}
\end{equation}

This analytical connection between canonical and grand canonical suggests that one should be able to extract grand canonical results from canonical ones and viceversa \cite{Mallik6}, provided the probability distribution
is completely described by a limited number of moments. This is particularly true if this distribution is a gaussian (defined only by mean value and variance) as it is shown below.\\
\indent
A given inverse temperature $\beta$ and a given volume $V$ is considered which is supposed to be fixed and is omitted from all the notations. We will concentrate on  a generic observable of interest $R$  which can be computed either in the canonical ($R_c$) or in the grand canonical ($R_{gc}$) ensemble is considered.
Starting from the exact relation connecting canonical and grand canonical:

\begin{equation}
R_{gc}(\alpha)=\sum_{A_0} P_\alpha (A_0) R_c(A_0)
\label{eq3}
\end{equation}
By doing a Taylor development of $R_c(A_0)$ around $A_0=\langle A_0 \rangle_\alpha$ truncated at second order:
\begin{eqnarray}
R_{c}(A_0) &\approx& R_{c}(A_0=\langle A_0 \rangle_\alpha)+(A_0-\langle A_0 \rangle_\alpha)  \frac{\partial R_{c}}{\partial A_0}\bigg{|}_{A_0=\langle A_0 \rangle_\alpha}  \nonumber \\
&+&
\frac 12  (A_0-\langle A_0 \rangle_\alpha)^2  \frac{\partial^2 R_{c}}{\partial A^2_0}\bigg{|}_{A_0=\langle A_0 \rangle_\alpha}
\label{eq4}
\end{eqnarray}

Substituiting this in Eq. \ref{eq3}:

\begin{equation}
R_{gc}(\alpha) \approx R_{c}(\langle A_0 \rangle_\alpha) +\frac 12  \sigma^2_{gc}(\alpha) \frac{\partial^2 R_{c}}{\partial \langle A_0 \rangle^2_\alpha}\bigg{|}_{A_0=\langle A_0 \rangle_\alpha}
\label{eq5}
\end{equation}

where $\sigma^2_{GC}(\alpha)$ is given by Eq. \ref{eq2}.\\
\indent
This result indicates that the difference between  the two predictions does not only increase with increasing particle number fluctuation (which is linked to the system size and temperature, and independent of the observable $R$), \{but also with increasing convexity of the observable\} \cite{maras}. An expression similar to Eq. \ref{eq5}, but which would express the micro canonical, or at least the canonical result as a function of the grand canonical one, would be most welcome. The extension of the formalism to the implementation of energy conservation is reserved for a future work, and focus here is on the mass conservation constraint. Now, similar to $R$, another observable is defined $S(A_0)= \frac{\partial^2 R(A_0)}{\partial A^2_0}$. The Taylor expansion Eq. \ref{eq4} gives:

\begin{eqnarray}
S(A_0) &\approx& S(\langle A_0 \rangle_\alpha) +\  (A_0-\langle A_0 \rangle_\alpha)  \frac{\partial S}{\partial A_0}\bigg{|}_{A_0=\langle A_0 \rangle_\alpha} \nonumber \\
&+&
\frac 12  (A_0-\langle A_0 \rangle_\alpha)^2  \frac{\partial^2 S}{\partial A^2_0}\bigg{|}_{A_0=\langle A_0 \rangle_\alpha}
\label{eq6}
\end{eqnarray}

and the grand canonical estimation of $S$ is given by

\begin{equation}
S_{gc} \approx S(\langle A_0 \rangle_\alpha) +\frac 12  \sigma^2_{gc}(\alpha) \frac{\partial^2 S}{\partial A^2_0}\bigg{|}_{A_0=\langle A_0 \rangle_\alpha}
\label{eq7}
\end{equation}

Eq. \ref{eq7} is substituted in Eq. \ref{eq5} and the limit of small particle number fluctuations,
$ \sigma^2_{gc}/\langle A_0 \rangle^2_\alpha < 1$ is considered. This limit is not realized in phase transitions, but otherwise it should be correct.
Within this limit terms of the order of $\sigma^4_{gc}$ can be neglected and hence:

\begin{equation}
R_{c}(\langle A_0 \rangle_\alpha) \approx R_{gc}(\alpha)   -\frac 12  \sigma^2_{gc}(\alpha) \frac{\partial^2 R_{gc}}{\partial \langle A_0 \rangle_\alpha^2}\bigg{|}_{\alpha=\alpha(A_{gc})}
\label{eq8}
\end{equation}

This is the desired expression, since the r.h.s. of this formula can be entirely calculated in the grandcanonical ensemble.\\
\begin{table}[b!]
\begin{center}
\begin{tabular}{|c|c|c|c|c|}
\hline
& \multicolumn{2}{|c|}{Canonical result} & \multicolumn{2}{c|}{Grand canonical result} \\
\cline{2-5}
Observable & From Canonical & From eq.(\ref{eq8}) & From Grand & From eq.(\ref{eq5}) \\
 & model &  &Canonical model &\\
\hline
$\langle n \rangle_{total}$ & 18.034 & 18.028 & 17.798 & 17.809\\
\hline
$\langle n \rangle_{A=1}$ & 1.0778 & 1.0774 & 1.0740 & 1.0745\\
\hline
$\langle n \rangle_{A=50}$ & 0.0200 & 0.0201 & 0.0223 & 0.0222\\
\hline
$A_{max}$ & 39.896 & 39.920 & 38.773 & 38.844\\
\hline
\end{tabular}
\end{center}
\caption[Application of Eq. \ref{eq5} and \ref{eq5} in $\langle n \rangle_{total}$, $\langle n \rangle_{A=1}$, $\langle n \rangle_{A=50}$ and   $A_{max}$]{The total average multiplicity, multiplicity of monomers, clusters of $A=50$ particles, and average size of the largest clusters for a system of $200$ nucleons, a freeze-out volume $V_f=6V_0$ and a temperature $T=5$ MeV, as calculated in the different ensembles are compared. The approximation Eq. \ref{eq8} of the canonical result from the grand canonical ensemble, and the approximation  Eq. \ref{eq5}  of the grand canonical result from the canonical ensemble are also given. }
\label{table1}
\end{table}
\indent
Since the fragmentation model is exactly solvable both in the canonical and in the grand canonical ensemble, it constitutes an ideal playing ground to test the quality of the approximate transformations Eqs. \ref{eq8} and \ref{eq5} in different thermodynamic situations. If these transformations can be validated in well defined thermodynamic regions and/or for well defined observables of interest, the natural continuation of this work will be to exploit such transformations to account for situations where no analytical solution exists. In particular, applying the constraint of energy or angular momentum conservation requires numerically heavy Monte-Carlo techniques with all the associated convergence problems, while an approximate implementation of these conservation laws through appropriate Lagrange multipliers (the analogous of the grand canonical ensemble) can be easily implemented. Table \ref{table1} shows the performance of Eq. \ref{eq8} and \ref{eq5} for a representative system with a total number of particles $A_0=200$  a fixed temperature $T=5$ MeV, and a fix ed freeze-out volume $V_f=6V_0$. As already described in previous chapters, these values are typical for applications to experimental multifragmentation data.\\
\begin{table}[b!]
\begin{center}
\begin{tabular}{|c|c|c|c|c|c|}
\hline
& & \multicolumn{2}{|c|}{Canonical result} & \multicolumn{2}{c|}{Grand canonical result} \\
\cline{3-6}
Observable & T & From Canonical & From eq.(\ref{eq8}) & From Grand & From eq.(\ref{eq5}) \\
 & (MeV) & model &  &Canonical model &\\
\hline
& 3 & 3.344 & 3.242 & 3.213 & 2.881 \\
\cline{2-6}
$\langle n \rangle_{total}$ & 5 & 18.034 & 18.028 & 17.798 & 17.809\\
\cline{2-6}
& 7 & 38.648 & 38.647 & 38.456 & 38.457\\
\cline{2-6}
& 10 & 55.174 & 55.176 & 55.035 & 55.035\\
\hline
& 3 & 39.896 & 39.920 & 38.773 & 38.844\\
\cline{2-6}
$A_{max}$ & 5 & 18.034 & 18.028 & 17.798 & 17.809\\
\cline{2-6}
& 7 & 16.625 & 16.624 & 16.541 & 16.542\\
\cline{2-6}
& 10 & 10.352 & 10.352 & 10.322 & 10.322\\
\hline
\end{tabular}
\end{center}
\caption[Application of Eq. \ref{eq5} and \ref{eq5} in $\langle n \rangle_{total}$ and $A_{max}$ calculation at different $T$]{Total average multiplicity  and average size of the largest clusters for a system of $200$ nucleons and a freeze-out volume $V_f=6V_0$ at different temperatures. The approximation Eq. \ref{eq8} of the canonical result from the grand canonical ensemble, and the approximation  Eq. \ref{eq5}  of the grand canonical result from the canonical ensemble are also given.}
\label{table2}
\end{table}
\indent
It can be observed that the predictions of the two ensembles are very close for the different observables like average multiplicities at $A=1$ and $50$, total multiplicity, average size of largest cluster etc. The expressions of average multiplicities at different $A$ for canonical and grandcanonical model is already described in Eq. \ref{One_component_canonical_multiplicity} and \ref{One_component_grand_canonical_multiplicity} respectively.
where as the total multiplicity can be obtained by summing up all the multiplicities of the different sizes.
The average size of the largest cluster can be computed  in the two ensembles as

\begin{equation}
A_{max}^{gc}=  \sum_{A=1}^\infty A \left ( 1 - e^{-\langle n\rangle^{A}_{gc}}\right )\prod_{A'>A}
 e^{-\langle n\rangle^{A'}_{gc}}
\end{equation}
and
\begin{equation}
A_{max}^{c}=  \sum_{A_m=1}^{A_0} A \frac{Q_{A_0}(A_m)-Q_{A_0}(A_m-1)}{Q_{A_0}(A_0)}.
\end{equation}
\indent
In this last expression, $Q_{A_0}(A_m)$ is the canonical partition sum of $A_0$ particles where all $\omega_k$ with $k>A_m$ have been set to zero (already described in 2.4.7 for two component system). The residual differences can be very well accounted for the transformation relations among the ensembles. The good performance of Eqs. \ref{eq8} and \ref{eq5} can be understood from the inspection of Fig. \ref{Grand_canonical_probability_distribution}. This figure shows the behavior as a function of the particle number of the canonical multiplicity and size of the largest cluster, as well as the grand canonical particle number distribution. It is seen that at $T=5$ MeV the grand canonical distribution, though large and non-gaussian as it is expected in the multifragmentation regime, is still a normal distribution and the canonical observables variation is approximately linear in the $N$ interval where the distribution is not negligible. The  performance of the transformation formulas  is worse for $A_{max}$ (0.6 \%) than for $\langle n_{total} \rangle$ (0.3\%), but this can be understood from the fact that the difference between the two ensembles is more important for this highly exclusive observable. Conversely, at $T=3$ MeV the grand canonical distribution strongly deviates from a gaussian, and presents several peaks.\\
\begin{figure} [h!]
\begin{center}
\includegraphics[width=12cm,height=12cm,clip]{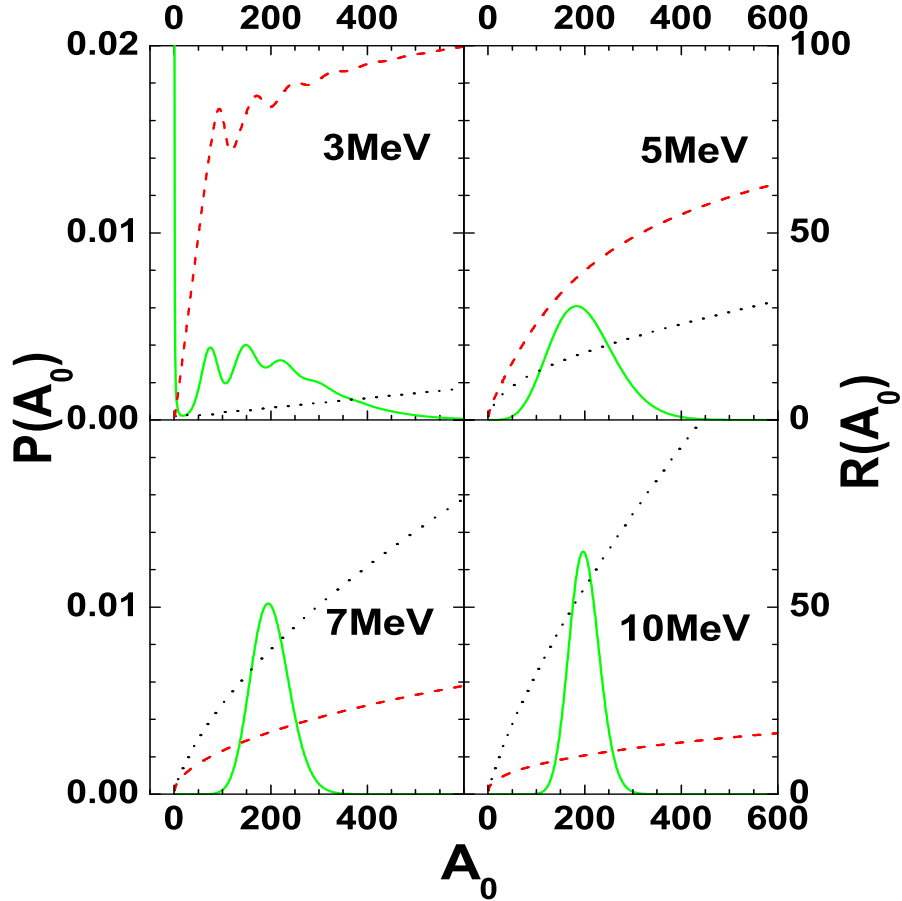}
\caption[Canonical and grand canonical model predictions for different observables]{Canonical and grand canonical model predictions in the freeze-out volume $V_f=6V_0$  at different temperatures, $T=$ 3, 5, 7 and 10 $MeV$. Full lines: grand canonical particle number distributions $\{P(A_0)\}$. Dashed lines: average size of the largest cluster in the canonical ensemble as a function of the total particle number. Dotted lines: average total cluster multiplicity in the canonical ensemble as a function of the total particle number.}
\end{center}
\label{Grand_canonical_probability_distribution}
\end{figure}
\begin{table}[b!]
\begin{center}
\begin{tabular}{|c|c|c|c|c|c|}
\hline
& & \multicolumn{2}{|c|}{Canonical result} & \multicolumn{2}{c|}{Grand canonical result} \\
\cline{3-6}
Observable & Particle & From Canonical & From eq.(\ref{eq8}) & From Grand & From eq.(\ref{eq5}) \\
 & number & model &  &Canonical model &\\
\hline
& 50 & 4.709 & 4.6794 & 4.450 & 4.538\\
\cline{2-6}
$\langle n \rangle_{total}$ & 100 & 9.142 & 9.129 & 8.899 & 8.924\\
\cline{2-6}
& 200 & 18.034 & 18.028 & 17.798 & 17.809\\
\cline{2-6}
& 400 & 46.354 & 46.335 & 45.710 & 45.712\\
\hline
& 50 & 25.446 & 25.257 & 23.363 & 23.127\\
\cline{2-6}
$A_{max}$ & 100 & 32.957 & 32.764 & 31.221 & 31.047\\
\cline{2-6}
& 200 & 16.625 & 16.624 & 16.541 & 16.542\\
\cline{2-6}
& 400 & 35.829 & 35.826 & 35.596 & 35.601\\
\hline
\end{tabular}
\end{center}
\caption[Application of Eq. \ref{eq5} and \ref{eq5} in $\langle n \rangle_{total}$ and $A_{max}$ calculation at different $A_0$]{Total average multiplicity  and average size of the largest clusters for a system of  volume $V=6V_0(200)$ at  a temperature $T=4$ MeV for different particle numbers. The approximation Eq.(\ref{eq8}) of the canonical result from the grand canonical ensemble, and the approximation  Eq.(\ref{eq5})  of the grand canonical result from the canonical ensemble are also given.}
\label{table3}
\end{table}
\indent
Indeed at low temperature, the equilibrium partitions are dominated by the most bound clusters which lie between $A=75$ and $A=125$ according to the employed liquid drop mass formula. Integer numbers of the most bound clusters therefore maximize the particle number distribution. This effect, combined with the decrease at high $A_0$ due to the chemical potential constraint, which imposes the average $\langle A_0 \rangle=200$ particle number, and the excluded volume effect, which suppresses the high multiplicity events, leads to the multi-modal shape of the distribution function. As a consequence, it is expected that Eq. (\ref{eq8}) would give a poor approximation of the canonical thermodynamics at $T=3$ MeV. This is confirmed by table \ref{table2}, which displays the grand canonical approximation of the canonical ensemble for the chosen observables as a function of the temperature. One can see that the approximation is extremely precise at high temperature, where the distributions are gaussian and the observables vary linearly with the particle number, while larger deviations (3\% for both $\langle n \rangle_{total}$ and  $A_{max}$) are observed at $T=3$ MeV.  Similar observations can be drawn from the inspection of table \ref{table3}, which displays the performance of the grand canonical estimation Eq.(\ref{eq8}) at  fixed temperature $T=4$ MeV as a function of the particle number. The particle number fluctuation increases with decreasing average particle number in the grand canonical ensemble. As a consequence, the grand canonical approximation worsens with decreasing $A_0$, while it is almost perfect for $A_0=400$.\\
\begin{figure} [h!]
\begin{center}
\includegraphics[width=12cm,height=12cm,clip]{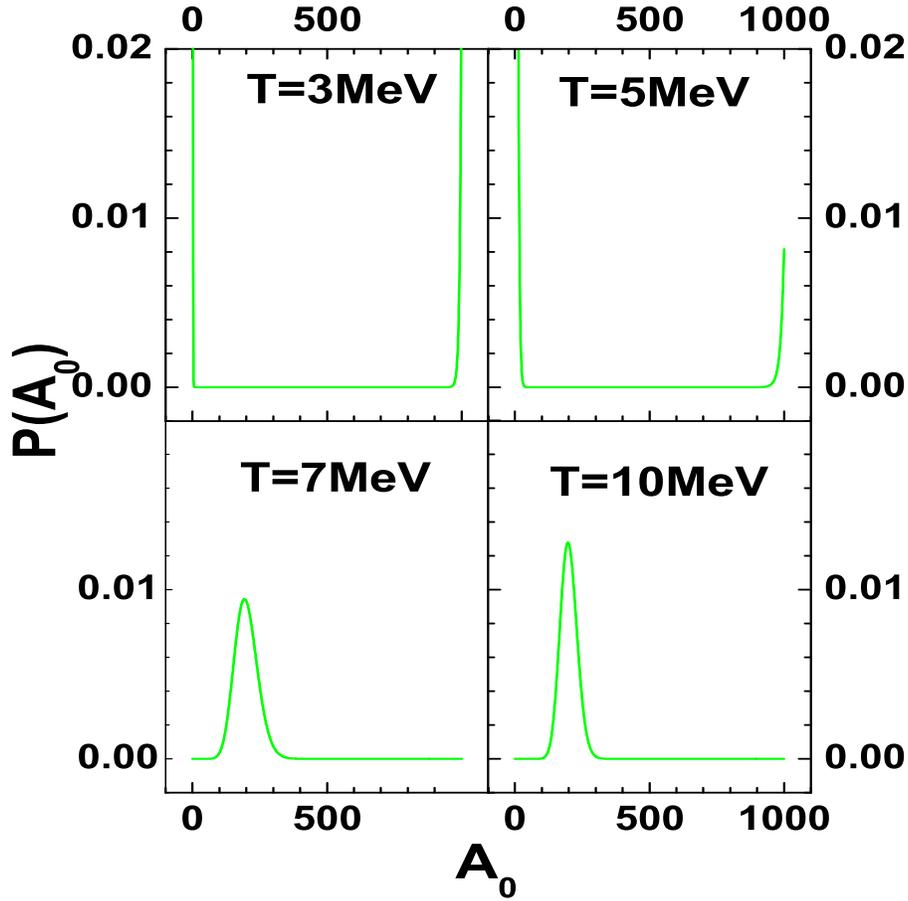}
\caption[Grand Canonical particle number distributions when Coulomb is switched off]{Grand Canonical particle number distributions (Coulomb switched off) at different temperatures at $\langle A_{0}\rangle=200$ .}
\end{center}
\label{Grand_canonical_probability_distribution_without_Coulomb}
\end{figure}
\indent
Globally speaking, these results show that the equivalence among the different statistical descriptions is approximately verified, and transformation equations are remarkably good in correcting the residual small differences. This is surprising in such small systems, especially considering that the thermodynamic limit of nuclear matter presents a first order phase transition. To show this, the uncharged case is considered where the Coulomb energy is artificially switched off (as the Coulomb interaction prevents from obtaining a thermodynamic limit for the liquid fraction at finite density, and thus quenches the phase transition). Some selected results are shown in table \ref{table4}. One can see that Eq.(\ref{eq8}) badly fails in this case up to a temperature of around $T=5$ MeV. This temperature domain comprises the phase transition regime, as it can be seen in Fig. \ref{Grand_canonical_probability_distribution_without_Coulomb}. This figure displays the grand canonical particle number distribution at different temperatures. The distribution is two-peaked, and the high mass peak corresponds to the maximum cluster size $A_{max}=1000$ which is allowed in the calculation in order to avoid divergences of the partition sum. This peak physically corresponds to the nuclear liquid fraction, while the peak at $A=1$ corresponds to the nuclear gas fraction.\\
\begin{table}[h!]
\begin{center}
\begin{tabular}{|c|c|c|c|c|c|}
\hline
& & \multicolumn{2}{|c|}{Canonical result} & \multicolumn{2}{c|}{Grand canonical result} \\
\cline{3-6}
Observable & T & From Canonical & From eq.(\ref{eq8}) & From Grand & From eq.(\ref{eq5}) \\
 & (MeV) & model &  &Canonical model &\\
\hline
& 3 & 1.0718 & 0.211 & 0.2654 & 0.996\\
\cline{2-6}
$\langle n \rangle_{total}$ & 5 & 3.748 & 0.543 & 2.539 & 8.146 \\
\cline{2-6}
& 7 & 36.103 & 36.127 & 35.831 & 35.832\\
\cline{2-6}
& 10 & 54.325 & 54.324 & 54.178 & 54.179\\
\hline
& 3 & 199.91 & 181.107 & 181.123 & 199.786\\
\cline{2-6}
$A_{max}$ & 5 & 191.97 & 183.138 & 178.85 & 168.329\\
\cline{2-6}
& 7 & 20.031 & 20.043 & 20.067 & 20.062\\
\cline{2-6}
& 10 & 10.737 & 10.737 & 10.707 & 10.707\\
\hline
\end{tabular}
\end{center}
\caption[Same as table \ref{table2} but without Coulomb interaction]{ Same as table \ref{table2} except that here the Coulomb force is artificially switched off.}
\label{table4}
\end{table}
\indent
It is interesting to remark that a similar method has proved to give excellent results in the case of free ideal gases \cite{kosov} and this formula has been also introduced and used to study statistical ensemble effects in one-dimensional metallic alloys \cite{maras}.
\section{Summary}
This chapter analyzes the charge distributions of fragments formed in nuclear multifragmentation in both canonical and grand canonical versions of the multifragmentation model. Both models are typically used to study experimental data from heavy-ion collisions at intermediate energies. Results from both models are shown to be  in agreement for finite nuclei provided the nucleus fragments predominantly into nucleons and low mass clusters. It has been observed that this condition is achieved under certain conditions of temperature, freeze-out volume, source size and source asymmetry. The main message to be conveyed from this chapter is that while canonical and grand canonical models have very different underlying physical assumption, the results from both the models can be in agreement with each other provided the contribution of higher mass fragments in nuclear disassembly is insignificant. This condition can be achieved either by increasing the temperature or freeze-out volume of the fragmenting nucleus or by increasing the source size, or by decreasing the asymmetry of the source. In fact when all these four conditions are satisfied then one obtains the best convergence between the two models. On the other hand, when the temperature and freeze-out volume are low, nucleus is small and more asymmetric then fragmentation of the nucleus is least; in these cases higher mass fragments dominate the distribution and the results from both the ensembles will be very different. It is also to be noted that the convergence between the micro canonical and the canonical ensemble will also be achieved under the similar conditions as those between the canonical and the grand canonical ensembles. For the one component system, analytical expressions have been derived which allow to extract canonical results from a grand canonical calculation and vice versa. The validity of this analytical expression is checked for different observables of multifragmentation. These analytic expressions of ensemble transformation do not work only when the system experiences a first order phase transition where the ensembles are irreducibly inequivalent.\\
\indent
The models discussed so far, will be applied in the next two chapters for throwing light into two important aspects of nuclear physics: study of symmetry energy and nuclear phase transition from nuclear multifragmentation.
\vskip3cm
\end{normalsize} 
\chapter{Symmetry Energy from nuclear multifragmentation}
\begin{normalsize}
\section{Introduction}
In the nuclear multifragmentation reactions, the neutron-proton composition of the break-up fragments is dictated by the symmetry term of the equation of state and hence the study of the multifragmentation process allows one to obtain information about the symmetry term. Different formulas have been proposed in the literature which connect the measurable fragment isotopic and isobaric observables of multifragmentation reactions to the symmetry energy of excited nuclei and these have been applied to the analysis of heavy-ion collision data. Most useful prescriptions of determining symmetry energy are (i)isoscaling source formula, (ii) isoscaling fragment formula, (iii) fluctuation formula and (iv) isobaric yield ratio formula. These formulas have all been deduced using the grand canonical version of the nuclear multifragmentation model assuming an equilibrium scenario for the break-up stage of the disintegrating system. They have been used to analyze experimental data from different projectile fragmentation as well as central collision reactions and the extracted values for the symmetry energy coefficient  $C_{sym}$ ranges between 15 and 30 MeV \cite{Souliotis, Shetty2, Fevre, Henzlova1, Shetty3,Buyukcizmeci_symmetry_energy,Iglio}. But as described in the previous chapter, the model based on canonical ensemble is better suited compared to grand canonical model for describing intermediate energy nuclear reactions where baryon and charge numbers are conserved. Therefore, it will be very interesting to study how different prescriptions of symmetry energy measurement are affected by the effect of particle fluctuation \cite{Mallik8}. In addition to that, in this chapter a comparative analysis of the predictive power of the different existing formulas of symmetry energy measurement both at the primary stage and also after evaporation for projectile fragmentation reactions as well as their relative agreement with experimental data will be done\cite{Mallik5}.\\
\indent
The organization of this chapter is as follows. In Section 7.2 we give a brief introduction to the different prescriptions for determining symmetry energy where as section 7.3 describes the ensemble dependence of the symmetry energy measuring formulae. The symmetry energy measurement in projectile fragmentation is explained in section 7.4. Finally we summarise the result in Section 7.5.
\section{Prescriptions for determining symmetry energy}
There exist different formulas in the literature using which the symmetry energy coefficient has been extracted from the fragment yields obtained from intermediate energy heavy ion reactions. A short review of these existing formulae is made in the next subsection.
\subsection{Isoscaling}
The phenomena of Isoscaling is most widely used to study the nuclear EOS and to extract the symmetry energy coefficient from multifragmentation reactions \cite{Tsang1,Tsang2,Botvina_isoscaling,Chaudhuri_isoscaling,Colonna,Colonna2,Ogul_symmetry_energy}. It has been observed both theoretically and experimentally that the ratio of yields from two different reactions(having different isospin asymmetry), exhibit an exponential relationship as a function of the neutron($N$) and proton($Z$) number and this is termed as 'isoscaling'. Two fragmentation reactions "$1$" and "$2$" at a given energy are being considered whose fragmenting systems have different mass $A_{01}$ and $A_{02}$ ($A_{02}>A_{01}$) but same charge $Z_1=Z_2=Z_0$.\\
Considering grand canonical ensemble, the yield of fragments having $N$ neutrons and $Z$ protons produced in the reaction "$1$" at temperature $T$ can be written as,
\begin{equation}
\langle {n_1}_{N,Z}\rangle_{gc}=\frac{V}{h^3}(2\pi mT)^{3/2}A^{3/2}\times\exp[-\frac {{F(N,Z)-{\mu_{n_1}}N-{\mu_{p_1}}Z}}{T}]
\label{yield_1st_reaction}
\end{equation}
where $F(N,Z)$ is the Helmholtz Free energy of the fragment, which is given by Eq.\ref{Multifragmentation_free_energy} and $\mu_{n_1}$ and $\mu_{p_1}$ are neutron and proton chemical potentials which depend on the isospin asymmetry of the fragmenting reaction (reaction $1$). For, reaction $2$ if one assumes the break up takes place at the same temperature and density, then the yield of the same type of fragment $(N,Z)$ is given by,
\begin{equation}
\langle {n_2}_{N,Z}\rangle_{gc}=\frac{V}{h^3}(2\pi mT)^{3/2}A^{3/2}\times\exp[-\frac {{F(N,Z)-{\mu_{n_2}}N-{\mu_{p_2}}Z}}{T}]
\label{yield_2nd_reaction}
\end{equation}
where $\mu_{n_1}$ and $\mu_{p_1}$ are neutron and proton chemical potentials of the 2nd fragmenting reaction.
Therefore the ratio of yields of the same type of fragment $(N,Z)$ originating from two different sources is given by,
\begin{eqnarray}
R_{21}&=& \langle {n_2}_{N,Z}\rangle_{gc}/\langle {n_{1}}_{N,Z}\rangle_{gc}\nonumber\\
&=& C\exp(\frac{\mu_{n_2}-\mu_{n_1}}{T}N+\frac{\mu_{p_2}-\mu_{p_1}}{T}Z)\nonumber\\
&=& C\exp(\alpha N+\beta Z)
\label{isoscaling}
\end{eqnarray}
\begin{figure}[b!]
\begin{center}
\includegraphics[width=13cm,keepaspectratio=true,clip]{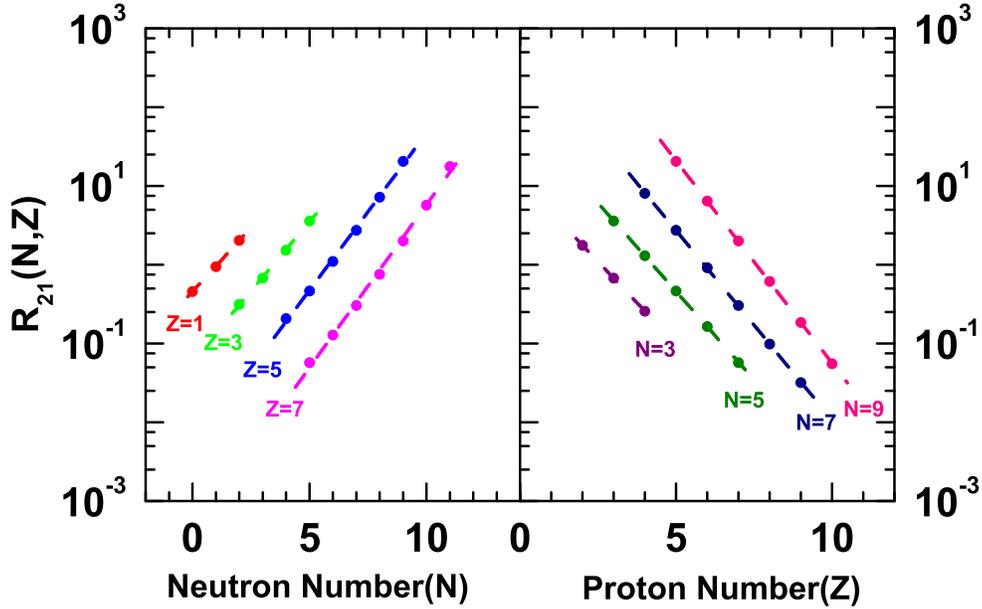}
\caption[Isoscaling]{Ratios($R_{21}$) of multiplicities of fragments $(N,Z)$ where mass and charge of the fragmenting system for reaction 1 are $A_0=58$ and $Z_0=28$ respectively and these for reaction 2 is are $A_0=64$ and $Z_0=28$. The left panel shows the ratios as function of neutron number $N$ for fixed $Z$ values, while the right panel displays the ratios as function of proton number $Z$ for fixed neutron numbers ($N$). The lines drawn through the points(circles) are best fits of the calculated ratios. Theoretical calculations are done by using Canonical thermodynamical model. The temperature used for both the reactions is 5.0 MeV and freeze-out volume is $V_f=3V_0$.}
\label{Isoscaling_diagram}
\end{center}
\end{figure}
Where $\alpha$ and $\beta$ are the isoscaling parameters and $C$ is a normalization factor. In theoretical models as well as in experiments, $R_{21}$ is calculated from the ratio of $\langle {n_2}_{N,Z}\rangle$ and $\langle {n_1}_{N,Z}\rangle$ and then isoscaling parameters $\alpha$ and $\beta$ are obtained by linear fitting of $lnR_{21}$ with $N$ (at constant $Z$) and $Z$ (at constant $N$) respectively. Fig. \ref{Isoscaling_diagram} shows, the isoscaling results obtained from Canonical thermodynamical model calculation at temperature $T=5$ MeV and freeze-out volume $V_f=3V_0$ for two fragmenting sources having same proton numbers $Z_0=28$ but different mass number $A_{01}=58$ and $A_{02}=64$. The ratio $R_{21}$ is plotted as function of the neutron number for $Z=1$, $3$, $5$ and $7$  in the left panel whereas the right panel displays the ratio as function of the proton number $Z$ for $N=3$, $5$, $7$ and $9$. It is seen that the primary fragments exhibit very well the linear isoscaling behaviour for the lighter fragments  over a wide range of isotopes and isotones. The lines in the figures are the best fits of the calculated $R_{21}$ ratios(open triangles) to Eq. \ref{isoscaling}. They are essentially linear and parallel on the semi log plot. For a given set of fragmenting sources, the dependence of isoscaling parameter on temperature and freeze-out volume is already described in Ref. \cite{Chaudhuri_isoscaling} and \cite{Dasgupta_isoscaling} respectively. That will not be repeated here.\\
\indent
Based on isoscaling, two different formulae are developed for extracting the symmetry energy:(i) Isoscaling source formula and (ii) Isoscaling fragment formula.\\
\subsubsection{Isoscaling Source Formula}
The formula that connects the symmetry energy coefficient with the isospin asymmetry of the source was first proposed in the framework of the Expanding evaporating source (EES) model \cite{Tsang1}. This is based on the grand canonical ensemble, and assuming thermodynamic equilibrium at the time of fragmentation of two systems having same charge $Z_0$ but different masses $A_{01}$, $A_{02}$ ($A_{01}<A_{02}$) at the same temperature T, $C_{sym}$ is given by
\begin{equation}
C_{sym}(Z)=\frac{\alpha(Z) T}{4 \left[ \left( \frac {Z_0}{A_{01}}\right)^2-
\left( \frac {Z_0}{A_{02}}\right)^2\right] }.
\label{Isoscaling_source_formula}
\end{equation}
and
\begin{equation}
C_{sym}(N)=\frac{\beta(N) T}{4 \left[ \left( \frac {N_{01}}{A_{01}}\right)^2-
\left( \frac {N_{01}}{A_{02}}\right)^2\right] }.
\label{Isoscaling_source_formula2}
\end{equation}
where $\alpha(Z)$ ($\beta(N)$) is the isoscaling parameter of fragments having $Z$ protons (N neutrons). The suffix 0 indicates the fragmenting source. In Eq. \ref{Isoscaling_source_formula2} $N_{01}$, $N_{02}$ are the neutron numbers of the fragmenting sources 1 and 2 respectively. This formula is referred to as the isoscaling (source) formula and these has been extensively used on experimental data \cite{Souliotis, Shetty2, Fevre, Henzlova1, Shetty3}.\\

\subsubsection{Isoscaling Fragment Formula}
The isoscaling (fragment) formula which is also derived \cite{Ono_isoscaling,Ono_isoscaling2} from the grand canonical ensemble assumes equilibrium at the breakup stage of two fragmenting sources in identical thermodynamical states that differ in their isospin content (different masses $A_{01}$, $A_{02}$ but same charge $Z_0$). It is given by
\begin{equation}
C_{sym}(Z)=\frac{\alpha(Z) T}{4 \left[ \left( \frac {Z}{<A_1(Z)>}\right)^2-
\left( \frac {Z}{<A_2(Z)>}\right)^2\right] }
\label{Isoscaling_fragment_formula}
\end{equation}
In this expression it is assumed that the isotopic distributions are essentially Gaussian and that the free energies contain only bulk terms. Here, $\alpha(Z)$ is the isoscaling parameter of fragments having $Z$ protons, $\langle A_{i}(Z)\rangle$ is the average mass number of a fragment of charge $Z$ produced by source $i$=1 (less neutron rich), 2 (more neutron rich) i.e.
\begin{equation}
\langle A_{i}(Z)\rangle=\frac{\sum_{N}(N+Z)\langle {n_i}_{N,Z}\rangle}{\sum_{N}\langle {n_i}_{N,Z}\rangle}
\label{Average_A_in_fragment_formula}
\end{equation}
where $\langle {n_i}_{N,Z}\rangle$ is the average multiplicity of fragments having $N$ neutrons and $Z$ protons produced from the $i^{th}$ reaction.\\
Eq. \ref{Isoscaling_fragment_formula} is  similar in structure to the isoscaling (source) formula, connects the symmetry energy coefficient to the isotopic composition of fragments instead of the isotopic composition of sources as in Eq. \ref{Isoscaling_source_formula}.
Similarly, for $\beta(N)$, one can get,
\begin{equation}
C_{sym}(N)=\frac{\beta(N) T}{4 \left[ \left( \frac {N}{<A_1(N)>}\right)^2-
\left( \frac {N}{<A_2(N)>}\right)^2\right] }
\label{Isoscaling_fragment_formula2}
\end{equation}
Here, $\langle A_{i}(N)\rangle$ can be obtained from
\begin{equation}
\langle A_{i}(N)\rangle=\frac{\sum_{Z}(N+Z)\langle {n_i}_{N,Z}\rangle}{\sum_{Z}\langle {n_i}_{N,Z}\rangle}
\label{Average_A_in_fragment_formula2}
\end{equation}
\subsection{Fluctuation formula}
In the last section, it was discussed that in order to get information about the symmetry energy from isoscaling, two different nuclear reactions are required. An alternate expression (fluctuation formula) has been derived in Ref. \cite{Raduta2, Chaudhuri_symmetry_energy}, which can connect the symmetry energy of a cluster of size $A$ to the width of its isobaric distribution of a single nuclear reaction. From eq. \ref{yield_1st_reaction}, the yield of fragments having isospin asymmetry $I=(N-Z)$ can be written as
\begin{equation}
\langle {n}_{I,A}\rangle_{gc} \approx exp^{-\frac{-C_{sym}(I-I_0)^2}{AT}}
\label{Fluctuation_formula_eq_1}
\end{equation}
where $I_0$ is the most probable value of isospin asymmetry for a given value of cluster size ($A$). Eq. \ref{Fluctuation_formula_eq_1} is comparable to a Gaussian distribution of mean value $x_0$ and standard deviation $\sigma$
\begin{equation}
f(x) \propto exp^{-\frac{-(x-x_0)^2}{2\sigma^2}}
\label{Fluctuation_formula_eq_2}
\end{equation}
Comparing above two equation one can write
\begin{equation}
\frac{C_{sym}}{T}=\frac{A}{2\sigma^2}
\label{Fluctuation_formula_eq_3}
\end{equation}
For a given $A$, if the yields of fragments for three different values of isospin asymmetry $I_1$, $I_2$ and $I_3$ are $\langle {n}_{I_1}\rangle_{gc}$, $\langle {n}_{I_2}\rangle_{gc}$, $\langle {n}_{I_3}\rangle_{gc}$  respectively, one can extract $\sigma$ from the relation
\begin{equation}
\frac{1}{2\sigma^2}=\frac{(I_1-I_2)(\ln \langle {n}_{I_1,A}\rangle_{gc}-\ln \langle {n}_{I_3,A}\rangle_{gc})-(I_1-I_3)(\ln \langle {n}_{I_1,A}\rangle_{gc}-\ln \langle {n}_{I_2,A}\rangle_{gc})}{(I_1-I_2)(I_1-I_3)(I_2-I_3)}
\label{Fluctuation_formula_eq_4}
\end{equation}
Hence, substituting eq. \ref{Fluctuation_formula_eq_4} in eq. \ref{Fluctuation_formula_eq_3}, the symmetry energy co-efficient to temperature ratio can be expressed as
\begin{equation}
\frac{C_{sym}}{T}=A\bigg[\frac{(I_1-I_2)(\ln \langle {n}_{I_1,A}\rangle_{gc}-\ln \langle {n}_{I_3,A}\rangle_{gc})-(I_1-I_3)(\ln \langle {n}_{I_1,A}\rangle_{gc}-\ln \langle {n}_{I_2,A}\rangle_{gc})}{(I_1-I_2)(I_1-I_3)(I_2-I_3)}\bigg]
\label{Fluctuation_formula}
\end{equation}
Generally out of three $I$'s, the maximum yield position of Gaussian distribution is taken as one value ($I_1$) and other two ($I_2$ and $I_3$) are selected on the two sides of maximum yield position.
\subsection{Isobaric yield ratio formula}
Isobaric yield ratio  \cite{Huang,Ma1,Ma2,Souza,Ma3} is also a well known method for extracting the liquid drop model parameters (symmetry energy coefficient for example) from multifragmentation reactions.The ratio of yields of two different types of fragments having same mass number $A$ but different isospin asymmetry $I=N-Z$ and $I^{'}=N^{'}-Z^{'}$ originating from single nuclear reaction is given by,
\begin{equation}
R[I^{'},I,A]=\langle n_{I,A}\rangle_{gc}/\langle n_{I^{'},A}\rangle_{gc}
\label{Isobaric_yield_ratio_formula_eq_1}
\end{equation}
For $I^{'}=I+2$, by using eq. \ref{yield_1st_reaction} one can get
\begin{equation}
R[I+2,I,A]=\frac{1}{T} \exp\bigg\{ \frac{a^{*}_c(A-I-2)}{A^{1/3}}-\frac{4C_{sym}(I+1)}{A}+(\mu_{n}-\mu_{p}) \bigg\}
\label{Isobaric_yield_ratio_formula_eq_2}
\end{equation}
Similar expression was obtained in Ref. \cite{Huang} from the modified Fisher model \cite{Minich,Fisher} by neglecting the mixing entropy terms.\\
With the choice of $I=-1$ the ratio will be
\begin{equation}
ln R[1,-1,A]=\frac{\mu_{n}-\mu_{p}}{T}+\frac{a^{*}_c}{T}A^{2/3}
\label{Isobaric_yield_ratio_formula_eq_3}
\end{equation}
For $I=1$, eq. \ref{Isobaric_yield_ratio_formula_eq_2} will be
\begin{equation}
\frac{C_{sym}}{T}=-\frac{A}{8} \bigg( \ln R[3,1,A]-\ln R[1,-1,A] \bigg)-\frac{2a^{*}_c}{A^{1/3}T}
\label{Isobaric_yield_ratio_formula_eq_4}
\end{equation}
Since $2a^{*}_c/A^{1/3}T$ is much smaller, therefore by neglecting it one can calculate the symmetry energy coefficient to temperature ratio by using the relation
\begin{equation}
\frac{C_{sym}}{T} \approx -\frac{A}{8} \bigg( \ln R[3,1,A]-\ln R[1,-1,A] \bigg)
\label{Isobaric_yield_ratio_formula}
\end{equation}
This is known as isobaric yield ratio formula.\\
\indent
It is interesting to mention that isobaric yield ratio formula can be obtained directly from fluctuation formula for $I_1=1$, $I_2=3$ and $I_3=-1$ i.e if the the yield of the isobaric distribution in fluctuation method is maximum at $I_1=1$, then the fluctuation method and isobaric yield ratio method results are completely identical.
\section{Effect of particle fluctuations on symmetry energy measurement formulae}
\begin{figure}[b!]
\begin{center}
\includegraphics[width=10.0cm,keepaspectratio=true,clip]{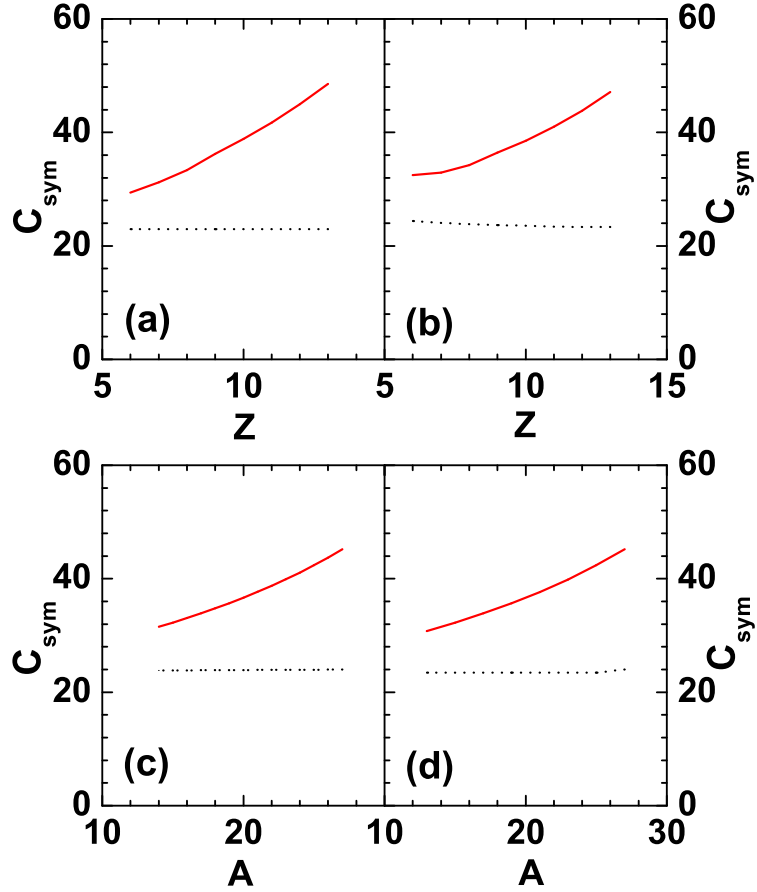}
\caption[$C_{sym}/T$ of excited fragments obtained from different prescriptions]{Symmetry energy coefficient obtained from canonical (red solid lines) and grand canonical model (black dotted lines) at $T=$5 MeV and input $C_{sym}=$23.5 MeV. (a) and (b) represents the variation of extracted $C_{sym}$ with proton number $Z$ by using Eq. \ref{Isoscaling_source_formula}  and Eq. \ref{Isoscaling_fragment_formula}
 respectively from two sources having same $Z_0=24$ but different $A_0=66$ and $54$ where as (c) and (d) indicates the variation of extracted $C_{sym}$ with
mass number $A$ calculated by using Eqs. \ref{Fluctuation_formula} and \ref{Isobaric_yield_ratio_formula} for $Z_0=24$ and $A_0=54$. }
\label{Primary_symmetry_energy_from_four_methods}
\end{center}
\end{figure}
The isotopic and isobaric yield distributions are related to the nuclear symmetry energy coefficient through different prescriptions, which have already been described briefly in the previous section. These relations are derived using the yields of the fragments  obtained in the grand canonical ensemble. We first compare the symmetry energy coefficient obtained using the different prescriptions in both the canonical and the grand canonical ensemble \cite{Mallik5}. In this calculation we have used the statistical models (canonical or grand canonical) in order to obtain the yields of the composites formed after multifragmentation (second stage) of the hot single source. The sources used for the first two methods [Fig \ref{Primary_symmetry_energy_from_four_methods}. (a) and \ref{Primary_symmetry_energy_from_four_methods}. (b)] are $A_{01}=55$ and $A_{02}=60$ and $Z_0=25$, while for the later two methods [Fig \ref{Primary_symmetry_energy_from_four_methods}. (c) and \ref{Primary_symmetry_energy_from_four_methods}. (d)] in which a single source is required, the source $A_0 =55$ and $Z_0=25$ is used. The temperature used for the calculation is $5$ MeV. These are the results for the primary fragments and no secondary decay is used for the de-excitation of the excited primary fragments. There are some differences in the results from the canonical and the grand canonical ensemble, and these differences are almost same in all the methods used. In the canonical ensemble, the extracted symmetry energy coefficient changes with the fragment mass or charge and this variation is more or less same for all the four methods used for extraction of this coefficient. The value of extracted symmetry energy coefficient varies between 25 and 50 MeV while the input symmetry energy coefficient used is fixed at 23.5 MeV.\\
\indent
In contrast, in the results from the grand canonical ensemble, $C_{sym}$ is independent of the fragment mass or charge. The value of $C_{sym}$ extracted
from the grand canonical ensemble lies between $23$ and $24$ MeV and hence matches almost exactly with the input value used for the calculation.
This difference in results between the two ensembles is mainly because these formulas are all derived using the prescription of the grand canonical ensemble and hence when this ensemble is used to extract the value, the results agree with the input value as expected. The results from the canonical ensemble deviate from that of the grand canonical ensemble in general for finite nuclei. Hence extraction of $C_{sym}$ leads to values that can differ widely from the input value used.\\
\indent
Now, it is important to study the conditions under which the convergence of canonical and grand canonical model calculations can be achieved in case of isoscaling and isobaric yield ratio and how they coincide with the formulae (Eq. \ref{isoscaling} and \ref{Isobaric_yield_ratio_formula_eq_2})derived in the last section (using the grand canonical ensemble) \cite{Mallik8}.
\subsection{Source size dependence of isoscaling and isobaric yield ratio}
The neutron and proton chemical potentials are the key features in the Eq. \ref{isoscaling} and \ref{Isobaric_yield_ratio_formula_eq_2}. Hence in order to study the convergence we have studied the dependence of neutron and proton chemical potentials on both source size and source asymmetry and the results are displayed in Fig. \ref{Chemical_potential_source_&_asymmetry_dependence}. In grand canonical model the $\mu_p$ and $\mu_n$ are determined from proton and neutron conservation conditions (described in the chapter 6). In canonical ensemble though chemical potential does not come into picture directly, but we can define them as $\mu=-\frac{\partial F_t}{\partial N}$. So in our case,
\begin{equation}
\mu_p= F_t(N_0,Z_0-1)-F_t(N_0,Z_0)
\end{equation}
\begin{equation}
\mu_n= F_t(N_0-1,Z_0)-F_t(N_0,Z_0)
\end{equation}
where $F_t(N_0,Z_0)=-TlnQ_{N_0,Z_0}$ is the total free energy and $Q_{N_0,Z_0}$ is the total partition function of the fragmenting system containing $N_0$ neutrons and $Z_0$ protons (described in chapter 2).\\
\indent
In Fig. \ref{Chemical_potential_source_&_asymmetry_dependence}(a), the variation of both $\mu_n$ and $ \mu_p$ with source proton number $Z_0$  is shown for two values of  the asymmetry parameter $y=(N_0-Z_0)/(N_0+Z_0)=$ 0.11 and 0.27. The source size $A_0$ is varied from 44 to 264 ($Z_0$ from 16 to 96) and it is seen that both the neutron and the proton chemical potential remains almost constant as one increases the source size irrespective of the value of isospin asymmetry ($y$). In the same figure results are shown from both the canonical and the grand canonical models and it is seen that the chemical potentials calculated from both the ensembles are almost equal except for the very small sources where they are slightly different. In Fig. \ref{Chemical_potential_source_&_asymmetry_dependence}(b), the variation of chemical potentials with the asymmetry parameter $y$ is shown where $y$ is varied from 0 to 0.33. The source size $A_0$ is kept fixed at 60 whereas $Z_0$ varies  from 20 ($y$=0.33) to 30 ($y$ =0). The change of $\mu_n$ and $\mu_p$ with $y$ is almost linear for both the ensembles and also the results from the canonical and the grand canonical ensembles are more or less  the same except for higher $y$ values where they are slightly different.
\begin{figure}[h!]
\begin{center}
\includegraphics[width=11.0cm,keepaspectratio=true,clip]{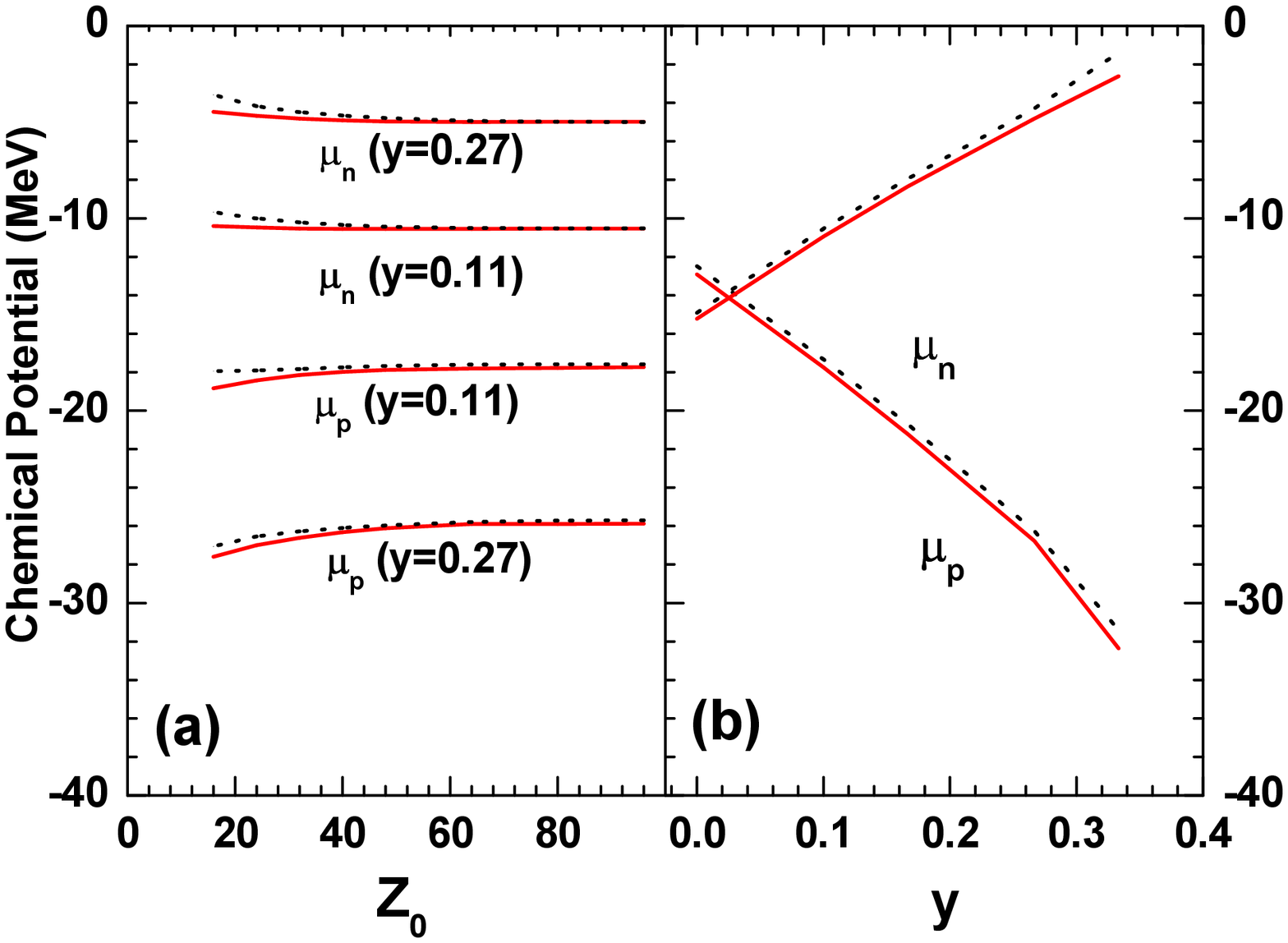}
\caption[$\mu_p$ and $\mu_n$ variation with source size and source asymmetry]{Variation of proton chemical potential ($\mu_p$) and neutron chemical potential ($\mu_n$) with (a) source size (at constant source asymmetry $y=0.27$ and $0.11$) (b) source asymmetry (for fixed source size $A_0=60$) from canonical (red solid lines) and grand canonical models (black dotted lines).}
\label{Chemical_potential_source_&_asymmetry_dependence}
\end{center}
\end{figure}
\indent
Since it is seen from Fig. \ref{Chemical_potential_source_&_asymmetry_dependence}(a) and \ref{Chemical_potential_source_&_asymmetry_dependence}(b) that the chemical potentials are almost same for both models for the entire range of source size and source asymmetry, hence for all other results to be presented for this section, the chemical potentials obtained from grand canonical model are being used.\\
\begin{figure}[b!]
\begin{center}
\includegraphics[width=9.0cm,keepaspectratio=true,clip]{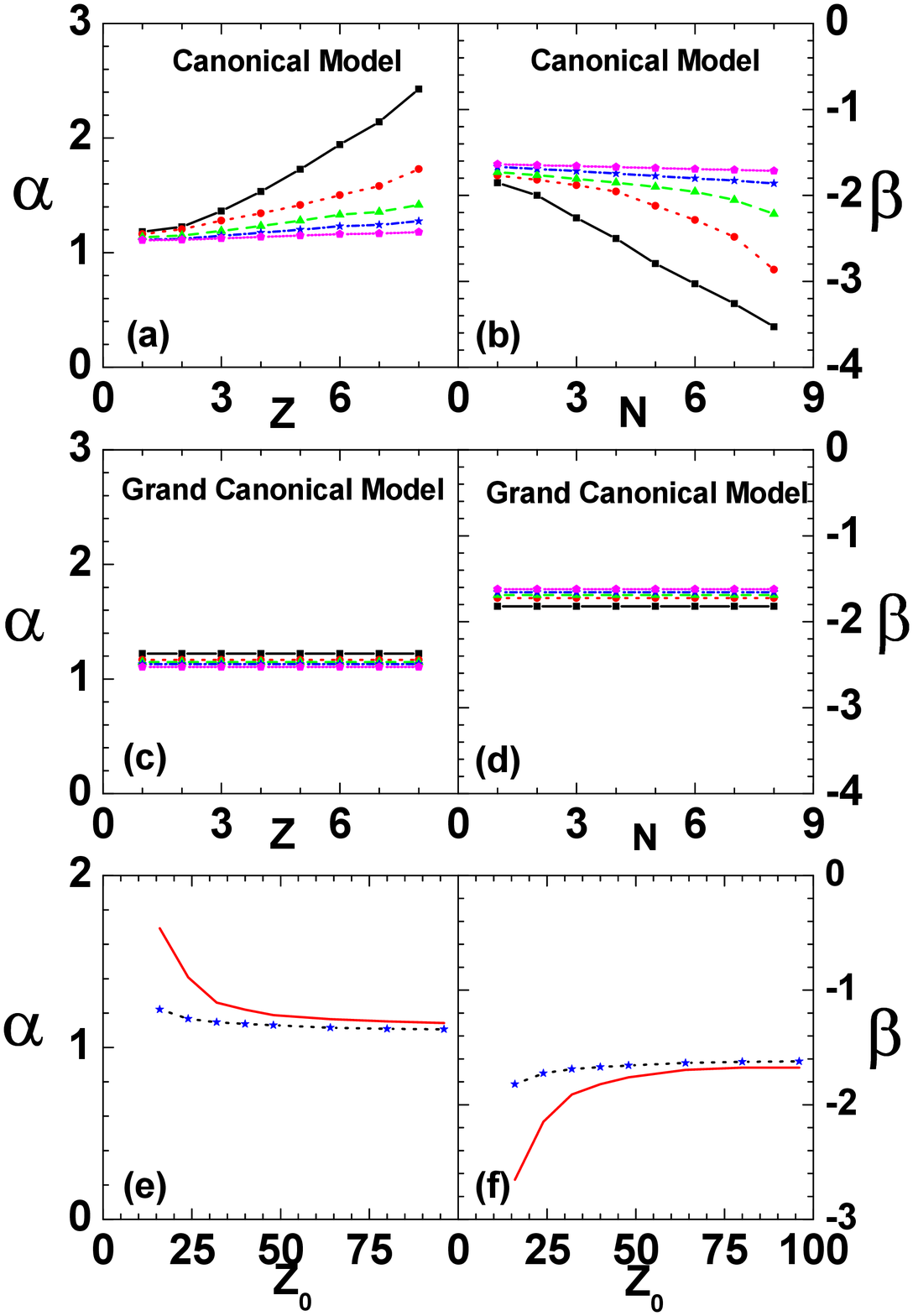}
\caption[Source dependence of isoscaling parameters]{Variation of isoscaling parameter $\alpha$ ($\beta$) with fragment proton number $Z$ ($N$) for each pair of sources having same isospin asymmetry
$0.27$ and $0.11$ but different charges $Z_0=16$ (black squares joined by solid lines), 24 (red circles joined by dotted lines), 32 (green circles joined by dashed lines) 48 (blue stars joined by dash dotted lines) and 96 (magenta pentagons joined by short dotted lines) from canonical [2(a), 2(b)] and grand canonical [2(c), 2(d)] model. 2(e) and 2(f) shows the variation of the isoscaling parameters $\alpha$ and $\beta$ respectively with source charge ($Z_0$) obtained from canonical model (red solid lines), grand canonical model (black dotted lines) and that calculated from the formulae $\alpha=(\mu_{n_2}-\mu_{n_1})/T$ and $\beta=(\mu_{p_2}-\mu_{p_1})/T$ (blue stars).}
\label{Isoscaling_source_depedence}
\end{center}
\end{figure}
\indent
The isoscaling parameters $\alpha$ and $\beta$ (Eq. \ref{isoscaling}) depend only on the difference in chemical potentials of  more neutronrich and less neutron rich fragmenting systems and on the temperature at freeze-out. For isoscaling studies, a pair of sources with same $Z_0$ value is required. We have kept the asymmetry value of the more neutron-rich source to be 0.27 and that of the less neutron-rich one to be 0.11. The study was done for different source sizes ranging from $Z_0$ =16 to $Z_0$ =96, the y values being the same for each pair. Since for a particular pair of reaction the difference in chemical potentials are the same, therefore it is expected that the isoscaling parameters  $\alpha$ ($\beta$) would remain constant throughout the entire $Z$ ($N$)  regime of the fragments. It can also be concluded from Fig. \ref{Chemical_potential_source_&_asymmetry_dependence}(a) that  at constant temperature and same $y$ value of the fragmenting sources, $\mu_p$ and $\mu_n$ are almost independent of the source size. On the contrary, from theoretical calculation by the canonical model it is observed  from Fig. \ref{Isoscaling_source_depedence}(a) and \ref{Isoscaling_source_depedence}(b) that $\alpha$ increases with the increase of  fragment proton number $Z$ and $\beta$ decreases with increase of fragment neutron number $N$ and the change is more for smaller fragmenting sources. On the other hand, $\alpha$ and $\beta$ values calculated from the grand canonical model (Fig. \ref{Isoscaling_source_depedence}(c) and \ref{Isoscaling_source_depedence}(d)) are independent of the fragment size. The dependence on source size is also very small as compared to the canonical results (Fig. \ref{Isoscaling_source_depedence}(a) and \ref{Isoscaling_source_depedence}(b)).Therefore if we take the average of $\alpha$ (or $\beta$) in the range  of $Z=1$ to $8$ (or $N=1$ to $8$) for each source  and compare those with that calculated from the formula $\alpha=(\mu_{n_2}-\mu_{n_1})/T$ or $\beta=(\mu_{p_2}-\mu_{p_1})/T$, then it is observed that the average values obtained from canonical and grand canonical model are different for the smaller fragmenting sources and the difference decreases substantially as one increases the source size as seen in Fig. \ref{Isoscaling_source_depedence}(e) and \ref{Isoscaling_source_depedence}(f)). The values of the isoscaling parameters  $\alpha$ and $\beta$  calculated from the slopes of the ratio $R_{21}$ of the grand canonical model coincides exactly with those calculated from the formula. This is seen from the dotted lines and the stars in Fig. \ref{Isoscaling_source_depedence}(e) \& \ref{Isoscaling_source_depedence}(f). This is what is expected since the formulae connecting the isoscaling parameters with the difference in chemical potentials is deduced from the grand canonical ensemble and hence results from the later exactly coincide with those calculated from the formulae.\\
\begin{figure}[b!]
\begin{center}
\includegraphics[width=10.0cm,keepaspectratio=true,clip]{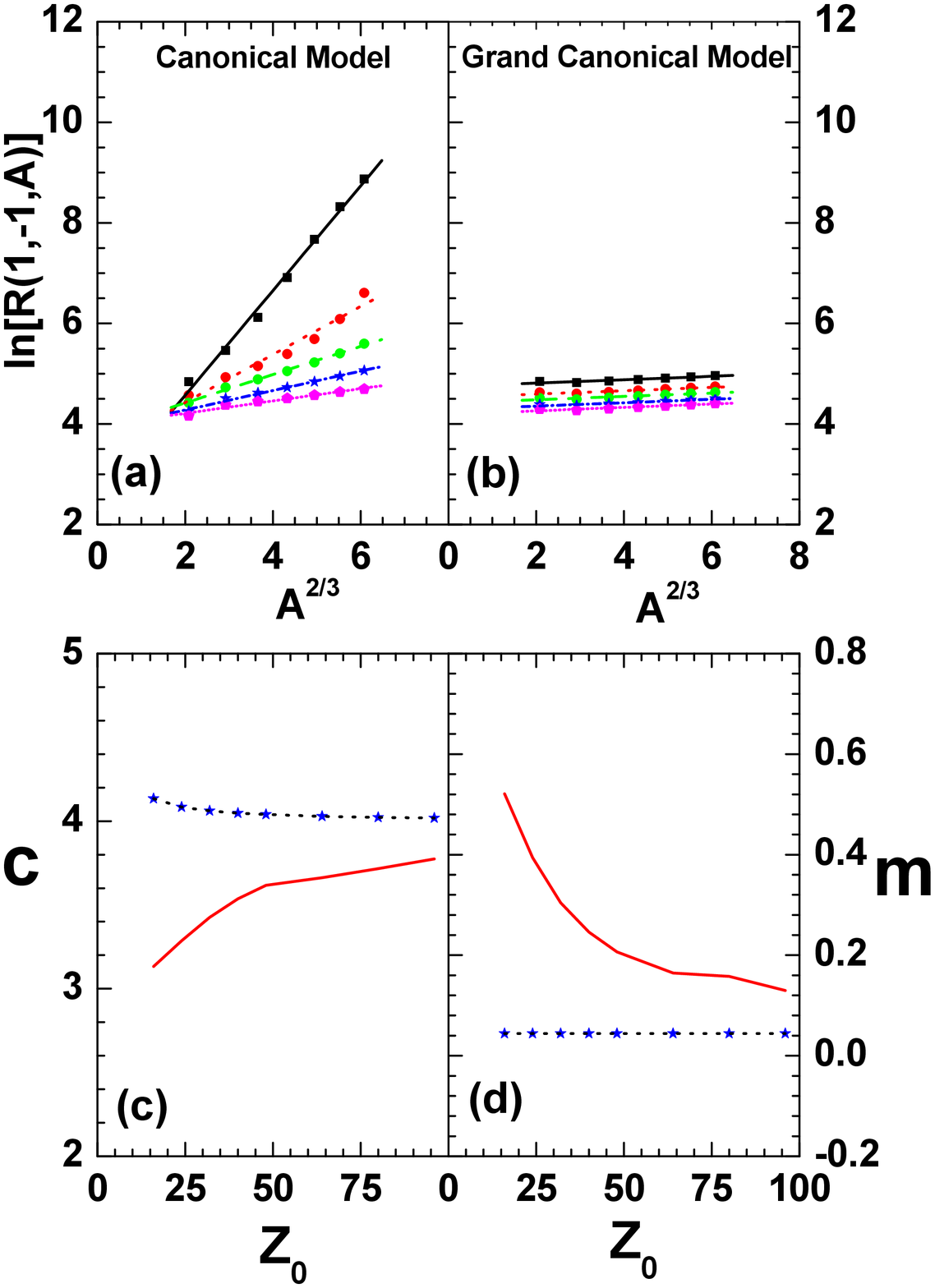}
\caption[Source dependence of isobaric yield ratio parameters]{Variation of isobaric yield ratio $ln R[1,-1,A]$ with $A^{2/3}$ for sources having same isospin asymmetry 0.27 but different charge $Z_0=16$ (black squares), 24 (red circles), 32 (green circles) 48 (blue stars) and 96 (magenta pentagons) from canonical (a) and grand canonical (b) model. Here the lines connecting the points represent the linear fittings. (c) and (d) shows variation of the isobaric yield ratio parameters $c$ and $m$ respectively with source charge ($Z_0$) obtained from canonical model (red solid lines), grand canonical model (black dotted lines) and that calculated from the formulae $c=\Delta\mu/T$ and $m=a^*_c/T$ (blue stars).}
\label{Isobaric_yield_ratio_source_depedence}
\end{center}
\end{figure}
\indent
For the case of isobaric yield ratio Eq. \ref{Isobaric_yield_ratio_formula} is used to extract symmetry energy from fragment yields and in Eq. \ref{Isobaric_yield_ratio_formula_eq_3}, $lnR[1,-1,A]$ calculated for odd  $A$  nuclei (since $N - Z = 1 or -1$), varies linearly with $A^{2/3}$ by
an equation like $y=mx+c$ with $m=a^{*}_c/T$ and $c=\Delta\mu/T$ (where $a^{*}_c=a_{c}\{1-(V_{0}/V_{f})^{1/3}\}$ and $\Delta\mu= \mu_n-\mu_p$).Therefore, in Fig. \ref{Isobaric_yield_ratio_source_depedence}(a) and \ref{Isobaric_yield_ratio_source_depedence}(b) the variation of the isobaric yield ratio $lnR[1,-1,A]$ with the fragment sizes is displayed where each line represents  a particular source having different size but same isospin asymmetry 0.27. The difference $\mu_p - \mu_n$  is almost independent of the source size (Fig. \ref{Chemical_potential_source_&_asymmetry_dependence}(a)); hence c's should be almost equal and m's are exactly equal for all the fragmenting sources. Therefore the plot of the ratio $lnR[1,-1,A]$  originating from all the fragmenting sources (as used in Fig. \ref{Chemical_potential_source_&_asymmetry_dependence}(a)) should almost coincide. But from the canonical model, the variation of $lnR[1,-1,A]$ with $A^{2/3}$ is different for different sources as shown in Fig. \ref{Isobaric_yield_ratio_source_depedence}(a). The slopes of the lines from different sources vary with the source size although the slope should be exactly equal according to the formula. This deviation arises because from Eq. \ref{Isobaric_yield_ratio_formula_eq_3} the slope $m=a^*_c/T$ is derived from the grand canonical model and  the same may not hold true for the canonical results. The results from the grand canonical model are shown in Fig. \ref{Isobaric_yield_ratio_source_depedence}(b). Here it is seen that the slopes are exactly equal irrespective of the source size. The calculated values of the slope $m$ and  the y-intercept $c$ obtained from linear fitting of the lines from canonical models (Fig. \ref{Isobaric_yield_ratio_source_depedence}(a)) and those calculated from formula $c=\Delta\mu/T$ and $m=a^*_c/T$ are not same for smaller fragmenting sources, but are close for  the larger sources. This is shown in Fig. \ref{Isobaric_yield_ratio_source_depedence}(c)  and  \ref{Isobaric_yield_ratio_source_depedence}(d). The reason for this deviation is that the formulae are derived using the grand canonical ensemble and hence they are in general not true for the canonical model results. For larger sources the fragmentation is more, therefore the particle number fluctuation in grand canonical model is very less \cite{Das}. In canonical model, particle number is strictly conserved and there is no such fluctuation. Hence the isoscaling parameters and isobaric yield ratios obtained from the canonical and grand canonical model become closer compared to that from the smaller fragmenting sources. The values of $c$ and $m$ obtained by fitting the lines from the grand canonical model (Fig. \ref{Isobaric_yield_ratio_source_depedence}(b)) coincide exactly with the values given by the formula. This is shown by the dashed line and the symbols in Fig. \ref{Isobaric_yield_ratio_source_depedence}(c) and Fig. \ref{Isobaric_yield_ratio_source_depedence}(d).
\begin{figure}[h!]
\begin{center}
\includegraphics[width=10.0cm,keepaspectratio=true,clip]{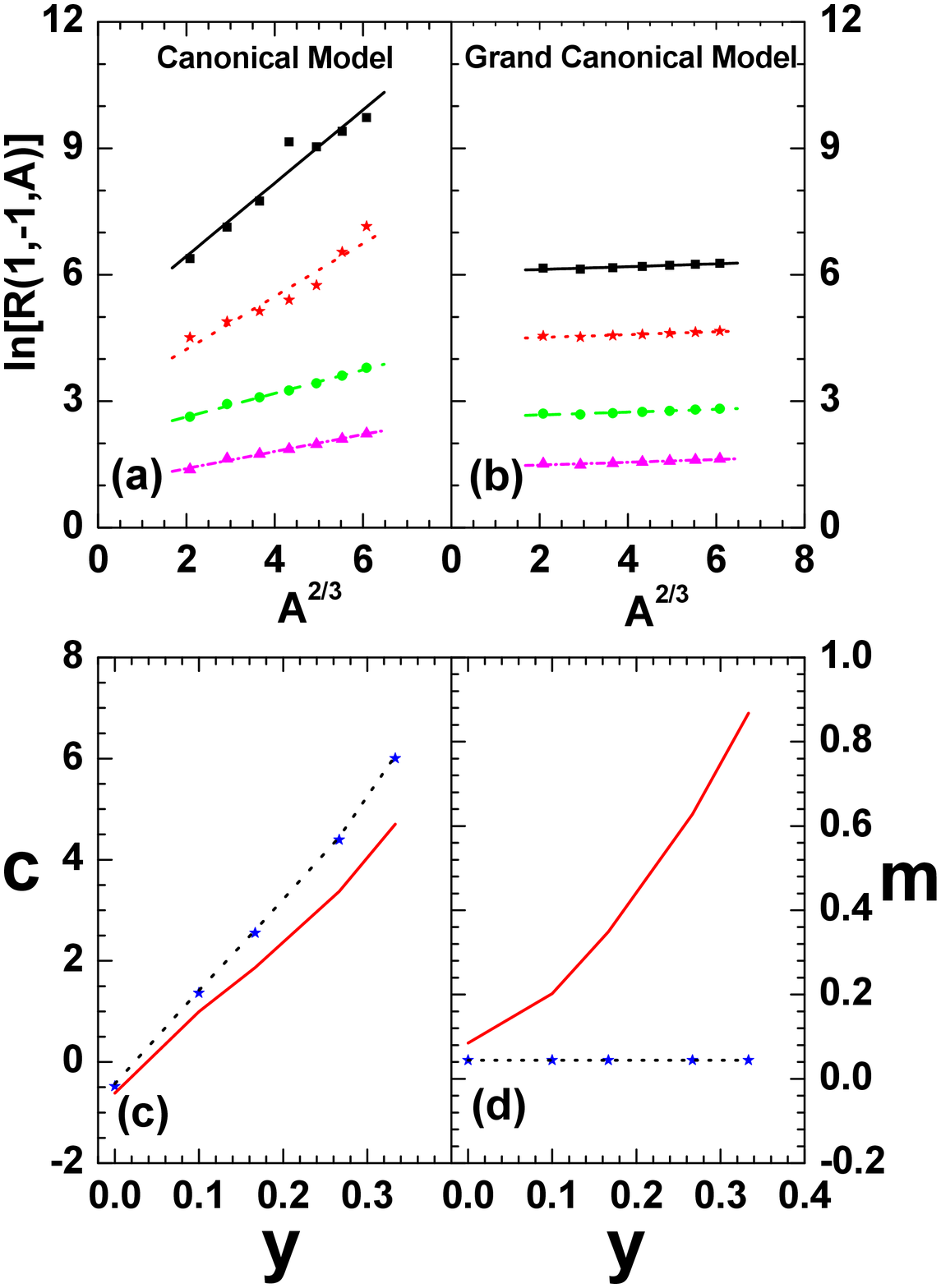}
\caption[Isospin dependence of isobaric yield ratio parameters]{Variation of isobaric yield ratio $ln R[1,-1,A]$ with $A^{2/3}$ for sources having same mass $A_0=60$ but different isospin asymmetry 0.1 (black squares), 0.17 (red stars), 0.27 (green circles) and 0.33 (magenta triangles) from canonical (a) and grand canonical (b) model. Here the lines connecting the points represent the linear fittings. (c) and (d) shows the variation of isobaric yield ratio parameters $c$ and $m$ respectively with source isospin asymmetry ($y$) obtained from canonical model (red solid lines), grand canonical model (black dotted lines) and that calculated from the formulae $c=\Delta\mu/T$ and $m=a^*_c/T$ (blue stars).}
\label{Isobaric_yield_ratio_isospin_asymmetry_depedence}
\end{center}
\end{figure}
\subsection{Isospin asymmetry dependence of isobaric yield ratio}
In Fig. \ref{Isobaric_yield_ratio_isospin_asymmetry_depedence} we show the effect of variation of the source asymmetry $y$ on the isobaric yield ratio parameters. In Fig. \ref{Isobaric_yield_ratio_isospin_asymmetry_depedence}(a) we plot the ratio $lnR[1,-1,A]$ with $A^{2/3}$ where $A$ is the mass number of the fragment. The different lines on the plot corresponds to different sources with $y$ values ranging from 0.33 to 0. The slopes of these lines according to Eq. \ref{Isobaric_yield_ratio_formula_eq_3} is equal to $m=a^*_c/T$ and hence should not depend on its $y$ value. The value of the y-intercept $c$ of these lines will be different since it is equal to $(\mu_n-\mu_p)/T$ which depend on $y$ as seen from Fig \ref{Chemical_potential_source_&_asymmetry_dependence}(b). It is seen from Fig. \ref{Isobaric_yield_ratio_isospin_asymmetry_depedence}(a) that the slopes are different for different sources from the canonical model calculation, the deviation being more for the source which is more asymmetric. For the grand canonical model(Fig. \ref{Isobaric_yield_ratio_isospin_asymmetry_depedence}(b)), the slopes are exactly equal for each source as expected from the formulae. The values of the parameters $c$ and $m$ are plotted in Fig. \ref{Isobaric_yield_ratio_isospin_asymmetry_depedence}(c) and \ref{Isobaric_yield_ratio_isospin_asymmetry_depedence}(d) respectively. It is seen that results from the canonical model differs from that of the formulae for higher values of $y$ and they become close as $y$ value approaches 0 or in other words the source becomes symmetric. The results from the grand canonical ensemble coincide exactly with that from the formulae. It has been already studied that results from the canonical and grand canonical models converge more as the fragmenting system becomes more symmetric as the particle fluctuation in grand canonical model becomes less in such cases. Similar effect is obtained earlier for charge distribution described in chapter 6.

\section{Symmetry energy from projectile fragmentation model}
The symmetry energy in projectile fragmentation reactions are extracted \cite{Mallik5} by using the prescriptions mentioned above. Since for finite nuclei, the canonical ensemble is physically more acceptable than the grand canonical ensemble, in the projectile fragmentation model (as described chapter 2) the canonical ensemble is used for the fragmentation of the excited PLF. In this case, the reduced symmetry energy coefficient is used. The reduced symmetry energy coefficient is nothing but the ratio of the symmetry energy coefficient to temperature i.e, $C_{sym}/T$. Recent studies on this parameter using different methods can be found in ref. \cite{Huang,Ma2,Chen,Marini}. This is mainly because it is difficult to estimate temperature both from experimental yields and from the model results. Hence it is better to express the results in the form of $C_{sym}/T$ instead of using only $C_{sym}$ since unambiguous extraction of temperature is very difficult.  In the isoscaling (source) formula [Eq. (\ref{Isoscaling_source_formula})], symmetry energy is related to the isoscaling parameter $\alpha$  and the $Z/A$ value of the two sources. The other three formula depend solely on the properties of the fragments. In the isoscaling (fragment) formula [Eq. (\ref{Isoscaling_fragment_formula})], $C_{sym}/T$ depends on the isoscaling parameter and on the $Z/\langle A \rangle$ values of the fragments. In the fluctuation formula [Eq. (\ref{Fluctuation_formula})], the coefficient depends on the width of the isobaric distribution of the fragments and their mass  while in the isobaric yield ratio [Eq. (\ref{Isobaric_yield_ratio_formula})], it depends on the ratio of the fragment yields and their mass. In the break-up stage of the multifragmentation reaction, the yields of the primary fragments can be used to deduce the values of $C_{sym}/T$ from all the four formulas.  The projectile fragmentation reactions involved are $^{58}$Ni on $^{9}$Be and $^{64}$Ni on $^{9}$Be at 140 MeV/n \cite{Mocko_thesis,Mocko}. From Eqs. (\ref{Isoscaling_source_formula}) and (\ref{Isoscaling_fragment_formula}), the variation of the reduced symmetry energy coefficient $C_{sym}/T$ with fragment charge $Z$ is shown in Fig. \ref{Symmetry_energy_four_formula_primary_projectile_fragmentation}(a). In  the same figure the results from Eq. (\ref{Fluctuation_formula}) and (\ref{Isobaric_yield_ratio_formula}) are shown as functions of the fragment mass $A$. The first two methods [Eq. (\ref{Isoscaling_source_formula}) and (\ref{Isoscaling_fragment_formula})] depend on the isoscaling parameter which is calculated using two  projectiles $Ni^{58}$ and $Ni^{64}$ having same value of Z. The fluctuation method [Eq. (\ref{Fluctuation_formula})] and the isobaric yield ratio method [Eq. (\ref{Isobaric_yield_ratio_formula})] depend on one source for the calculation of the reduced symmetry energy coefficient and the projectile $Ni^{58}$ has been used. It is seen that the results from the primary fragments are close to each other for all four formula and $C_{sym}/T$ increases with the fragment mass or charge. Since these values are for the break-up stage before the final de-excitation, no comparison has been made to the experimental data.\\
\begin{figure}[h!]
\begin{center}
\includegraphics[width=12.0cm,keepaspectratio=true,clip]{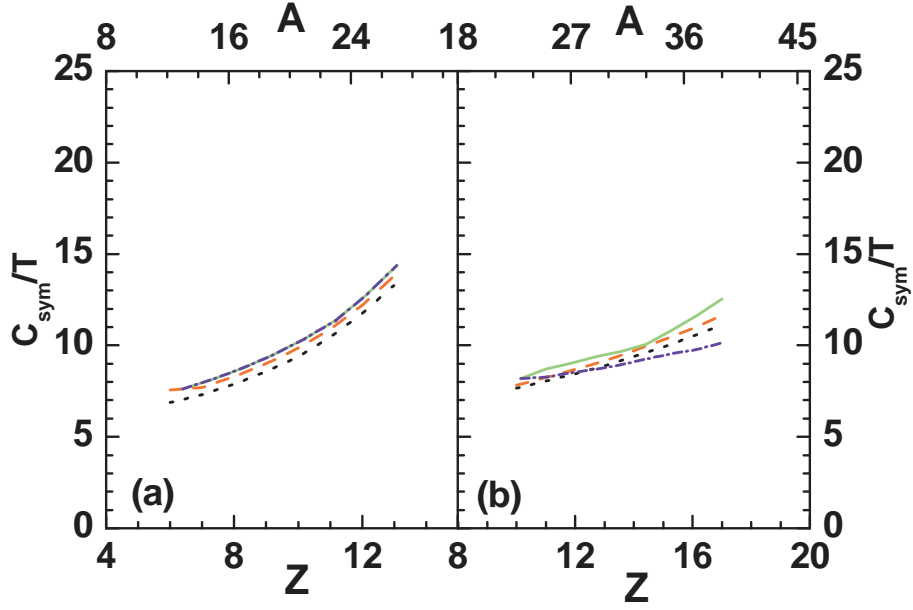}
\caption[$C_{sym}/T$ of excited fragments in projectile fragmentation]{Comparison of $C_{sym}/T$ calculated by using (i) Eq. \ref{Isoscaling_source_formula} (black dotted lines) (ii) Eq. \ref{Isoscaling_fragment_formula} (red dashed lines) (iii) Eq. \ref{Fluctuation_formula} (green solid lines) and (iv) Eq. \ref{Isobaric_yield_ratio_formula} (blue dash dotted lines) for primary fragments in projectile fragmentation reactions $Ni$ on $Be$ (left panel) and $Xe$ on $Pb$ (right panel). For (i) and (ii), $C_{sym}/T$ is calculated from two sets of reactions $^{58}$Ni on $^{9}$Be and $^{64}$Ni on $^{9}$Be (left panel) and $^{124}$Xe on $^{208}$Pb and $^{136}$Xe on $^{208}$Pb (right panel) and plotted against $Z$ where as for (iii) and (iv), $C_{sym}/T$ is calculated for only neutron less reactions and plotted against $A$. }
\label{Symmetry_energy_four_formula_primary_projectile_fragmentation}
\end{center}
\end{figure}
\indent
Same calculation  is repeated for different projectiles ($^{124}$Xe and $^{136}$Xe) [Fig. \ref{Symmetry_energy_four_formula_primary_projectile_fragmentation}(b)]. The projectile fragmentation reactions involved were $^{124}$Xe on $^{208}$Pb and $^{136}$Xe on $^{208}$Pb at 1 GeV/n \cite{Henzlova}. The beam energy is 1 GeV/n which is much higher than the previous value of 140 MeV/n \cite{Mocko_thesis,Mocko}. The trend of the results remains almost the same irrespective of the beam energy. The results from all four formulas are close to each other for the primary fragments (at the break-up stage) for a wide range of beam energies from 140 MeV/n to 1 GeV/n.
\indent
In the next stage the results for the secondary fragments i.e, after the evaporation of the excited primary fragments are investigated. In each figure, the results from the primary fragments are also shown for the sake of convenience in comparison. While comparing the isoscaling results before and after evaporation, from Ref. \cite{Mallik1} it is evident that the isoscaling is valid for a limited range of isotopes for the secondary fragments as compared to the primary ones and due to the effect of secondary decay the magnitude of the isoscaling parameters change. The result for the isoscaling (source) formula [Eq. (\ref{Isoscaling_source_formula})] is shown in \ref{Symmetry_energy_Ni_on_Be_reaction} (a). Here $C_{sym}/T$ decreases after evaporation because the isoscaling parameter $\alpha$ decreases after evaporation for the temperature range used here while the denominator of the right-hand side of Eq. (\ref{Isoscaling_source_formula}), which depends only on the source sizes remain unchanged. In both results, from the  model and from  the experimental data, the reduced symmetry energy coefficient does not depend very much on the fragment size. The numbers obtained from the experimental data are less than those obtained from our model for the projectile fragmentation. The results for the isoscaling (fragment) formula [Eq. (\ref{Isoscaling_fragment_formula})] is plotted in Fig. \ref{Symmetry_energy_Ni_on_Be_reaction} (b). In this case, $C_{sym}/T$ increases after secondary decay and the results more or less agree with those extracted from the experimental data.
\begin{figure}[h!]
\begin{center}
\includegraphics[width=12.0cm,keepaspectratio=true,clip]{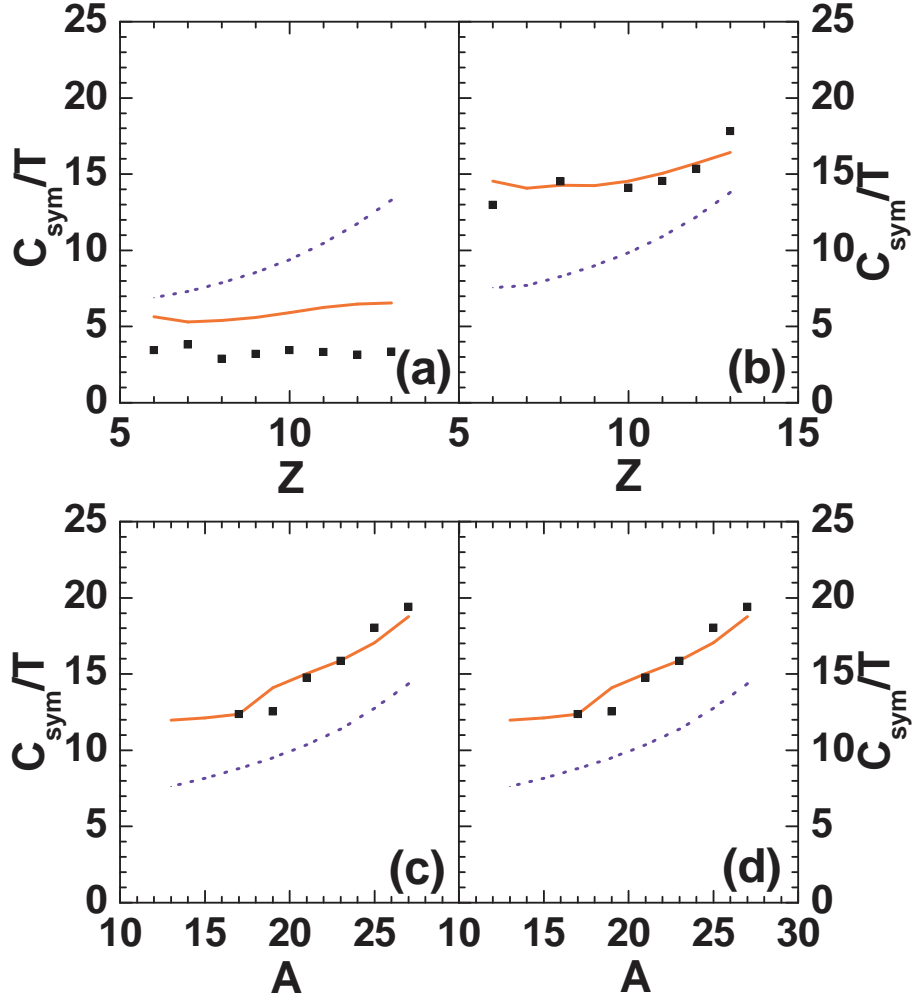}
\caption[$C_{sym}/T$ calculated from $^{58}$Ni and $^{64}$Ni on $^{9}$Be reaction at 140 MeV/nucleon]{ Variation of $C_{sym}/T$ with atomic number $Z$ calculated by using (a) Eq. \ref{Isoscaling_source_formula} and (b) Eq. \ref{Isoscaling_fragment_formula} for $^{58}$Ni on $^{9}$Be and $^{64}$Ni on $^{9}$Be reactions. Panels (c) and (d) depict the variation of $C_{sym}/T$ with mass number $A$ for $^{58}$Ni on $^{9}$Be reaction calculated from Eqs. \ref{Fluctuation_formula} and \ref{Isobaric_yield_ratio_formula} respectively. Experimental data (black squares) compared with theoretical results: primary  fragments (blue dotted lines) and secondary fragments (red solid lines). }
\label{Symmetry_energy_Ni_on_Be_reaction}
\end{center}
\end{figure}
\begin{figure}[b!]
\begin{center}
\includegraphics[width=12.0cm,keepaspectratio=true,clip]{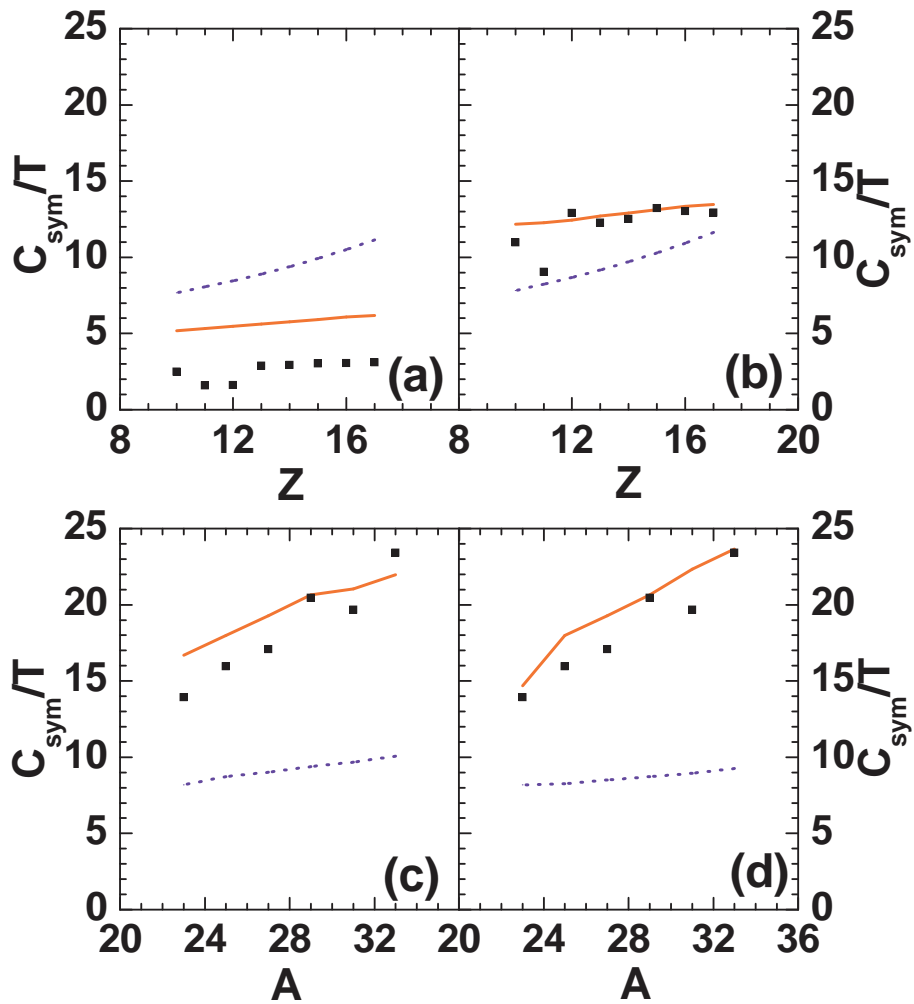}
\caption[$C_{sym}/T$ calculated from $^{124}$Xe and $^{136}$Xe on $^{208}$Pb reaction at 1 GeV/nucleon]{Same as Fig. \ref{Symmetry_energy_Ni_on_Be_reaction} except that here the projectile fragmentation reactions involved are $^{124}$Xe and $^{136}$Xe on $^{208}$Pb instead of $^{58}$Ni and $^{64}$Ni on $^{9}$Be. }
\label{Symmetry_energy_Xe_on_Pb_reaction}
\end{center}
\end{figure}
\indent
The isoscaling (fragment) formula depends on the $Z/\langle A \rangle$ values of the fragments [see Eq. (\ref{Isoscaling_fragment_formula})]. For the less neutron rich source, this quantity does not change very much after evaporation but for the more neutron-rich nuclei, this quantity increases after evaporation since $Z/\langle A \rangle$ decreases as the  peak of the isotopic distribution shifts to the left (lower values of A) after evaporation. Hence the denominator of the right-hand side of Eq. (\ref{Isoscaling_fragment_formula}) decreases. The isoscaling parameter $\alpha$ in the numerator also decreases after evaporation but the denominator decreases much more and hence $C_{sym}/T$  increases after secondary decay as is seen from the Fig. \ref{Symmetry_energy_Ni_on_Be_reaction}(b). This is in contrast to the results from the isoscaling (source) formula where the denominator is independent of the property of the fragments. In Fig. \ref{Symmetry_energy_Ni_on_Be_reaction} (c) we have plotted the results from the fluctuation formula [Eq. (\ref{Fluctuation_formula})]. Here also the trend of the results is same as in Fig. \ref{Symmetry_energy_Ni_on_Be_reaction} (b).  $C_{sym}/T$
increases after secondary decay, and this is due to the fact that $\sigma^2$ which is a measure of the width of the isobaric distribution, decreases after secondary decay and one can see from  Eq. (\ref{Fluctuation_formula}) that if this decreases then  $C_{sym}/T$ will increase and this is exactly what happens as it is observed from Fig. \ref{Symmetry_energy_Ni_on_Be_reaction} (c). The experimental values obtained in this case also are quite close to those obtained from the model. The results from the isobaric yield ratio method is plotted in Fig. \ref{Symmetry_energy_Ni_on_Be_reaction} (d). In this case also the result is similar to that of the previous two cases and the reduced symmetry energy coefficient increases after evaporation.  The numbers extracted from  the experimental data are close to those from the theoretical calculation. A similar trend of results from the isobaric yield ratio method in Antisymmetrized molecular dynamics (AMD) model+GEMINI calculations is reported in \cite{Huang}.  In this case as seen from Eq. (\ref{Isobaric_yield_ratio_formula}), $C_{sym}/T$ depends on the ratios of yields of different isobars. The reduced symmetry energy coefficient increases after evaporation because the width of the isobaric distribution decreases due to secondary decay, which results in decrease of the yield $Y(-1,A)$ and  $Y(3,A)$ in Eq. (\ref{Isobaric_yield_ratio_formula}) while the yield $Y(1,A)$ remains almost unchanged. In the results from Eq. (\ref{Isoscaling_fragment_formula}), (\ref{Fluctuation_formula}) and (\ref{Isobaric_yield_ratio_formula}), $C_{sym}/T$ increases after evaporation, as compared to the results obtained from the primary fragments.\\
\indent
This calculation is repeated for another projectile fragmentation reaction (Xe on Pb) at 1 GeV/n \cite{Henzlova} and this is shown in Fig. \ref{Symmetry_energy_Xe_on_Pb_reaction}. The  observations and results are similar to those of the previous reaction in spite of the vast difference in the projectile energy and widely different target-projectile combination.\\
\indent
All the four formulas are derived from the yields of the grand canonical ensemble and they hold good for the breakup stage of the reaction. Hence any attempt to deduce the value of the symmetry energy coefficient from the yields obtained after evaporation might lead to the wrong conclusion. Neither the experimental yields (which are the values after evaporation from the excited fragments) nor the yields of the secondary fragments from the model should be used to deduce the values of the symmetry energy coefficient. It might be possible to deduce the value of $C_{sym}$ from the break-up stage of the reaction, i.e., from the hot primary fragments, but since it is difficult to access this stage from an experimental point of view, no attempts are made to do such calculations.
\section{Summary}
In this chapter, the symmetry energy to temperature ratio is determined from different measurable fragment isotopic and isobaric observables of nuclear multifragmentation. The values of $C_{sym}$ obtained from the primary fragments for the canonical and the grand canonical ensembles for a single source at a fixed temperature are compared by using the four different prescriptions. At break up stage, for the canonical model calculation, $C_{sym}$ varies with the fragment size and differs from the input $C_{sym}$ value which is equal to 23.5 MeV. In the grand canonical model, this value is independent of the fragment size and is almost equal to the input value used. This is because  all the four formulas are deduced using the grand canonical ensemble at the break-up stage of multifragmentation. Hence it is possible to get back the value of the input symmetry energy coefficient using only the grand canonical model for calculating the yields of primary fragments at  the break-up stage. For better understanding, the source dependence of isoscaling parameters and source and isospin dependence of isobaric yield ratio parameters are studied from both canonical and grand canonical models. The results of canonical and grand canonical ensembles differ in general for finite nuclei and are found to converge only when fragmentation of the nucleus is more and it happens for the sources having larger mass or less asymmetry.\\
\indent
The value of $C_{sym}/T$ is also extracted  from the yields of projectile fragmentation reactions using the model for projectile fragmentation. The results from the primary fragments are close to each other for all the four formulas used. In this model, where the canonical ensemble is used to calculate the yields of the hot primary fragments, the values of $C_{sym}/T$ obtained from the secondary fragments after evaporation are close to those obtained from the experimental yields but they differ from those  obtained from the primary fragments and from the input value of $C_{sym}$ used in the model.  The message  which is conveyed is that in order to deduce the value of the symmetry energy coefficient using the existing prescriptions, it is advisable to use the grand canonical model to obtain the yield of the fragments and one should use the yields at the break-up stage (primary fragments) of the reaction. The experimental yields (which are from the "cold" fragments) should not be used to deduce the value of the symmetry energy coefficient since the formulas used for the deduction are all valid at the equilibrium stage or the breakup stage of the reaction and secondary decay disturbs the equilibrium scenario of the breakup stage. Attempts to deduce the value of $C_{sym}$ from the secondary fragments or from the experimental yields might lead to values that are very different from the actual value.
\vskip3cm
\end{normalsize} 
\chapter{Phase transition in nuclear multifragmentation from dynamical model}
\begin{normalsize}
\section{Introduction}
The subject of phase transition in nuclear matter has already been introduced in chapter 1. Over the last thirty years a great deal of effort has been made in order to develop the theoretical models for phase transition for nuclear matter. Simultaneously experimental efforts were made to identify the signals for phase transition from data of intermediate energy heavy ion collisions. It is highly non-trivial to extract signals of phase transition from data. As described in chapter 1, phase transition occurs in very large systems but because of Coulomb interaction very large nuclei are not available in the laboratory. The primary interest in nuclear physics has been to learn about characteristics of phase transition in systems interacting with nuclear forces only. This prompted the study of models where the only interactions are of nuclear force type, suitably simplified but retaining the most important features.\\
\indent
There has been an enormous amount of statistical model studies on liquid-gas phase transition in heavy ion collisions at intermediate energy. The standard methods of statistical model studies on liquid-gas phase transition assume that because of two body collisions nucleons equilibrate in a given volume and then dissociate into composites of different sizes ‘$a$’ according to availability of phase space. This chapter focuses on whether results of transport model calculation at intermediate energy can point to signatures of phase transition.\\
\indent
This chapter is structured as follows. Some important signals of nuclear liquid gas phase transition obtained from statistical model (CTM) studies is described in section 7.2. BUU transport model with fluctuation is introduced in section 7.3. Section 7.4 contains the modifications in the fluctuation included BUU model which is essential for studying liquid gas phase transition. Comparisons of some important results obtained from the existing model and modified model is presented in section 7.5. Signatures of liquid gas phase transition obtained from the transport model calculation and their comparison with statistical model calculations are displayed in section 7.6. Pauli blocking in the BUU model of fluctuation is discussed in section 7.7. Finally the summary is presented in section 7.8.
\section{Liquid gas phase transition from statistical model}
Canonical thermodynamical model (CTM) has been extremely used for studying nuclear liquid gas phase transition \cite{Das,Das_CTM_specific_heat}. The model is already described in some detail in Chapter-2. The transport equations that are to be solved in this chapter have a finite nucleus hitting another finite nucleus and CTM deals with systems which have a finite number of particles. Hence trying to connect transport model with CTM is natural. As the primary interest is phase transition in nuclear matter due to the nuclear force alone, most theoretical models have considered symmetric nuclear matter where the Coulomb force is switched off \cite{Dasgupta_Mekjian,Bugaev}. Throughout this chapter the same practice will be followed.\\
\begin{figure} [ht]
\begin{center}
\includegraphics[width=\textwidth,keepaspectratio=true,clip]{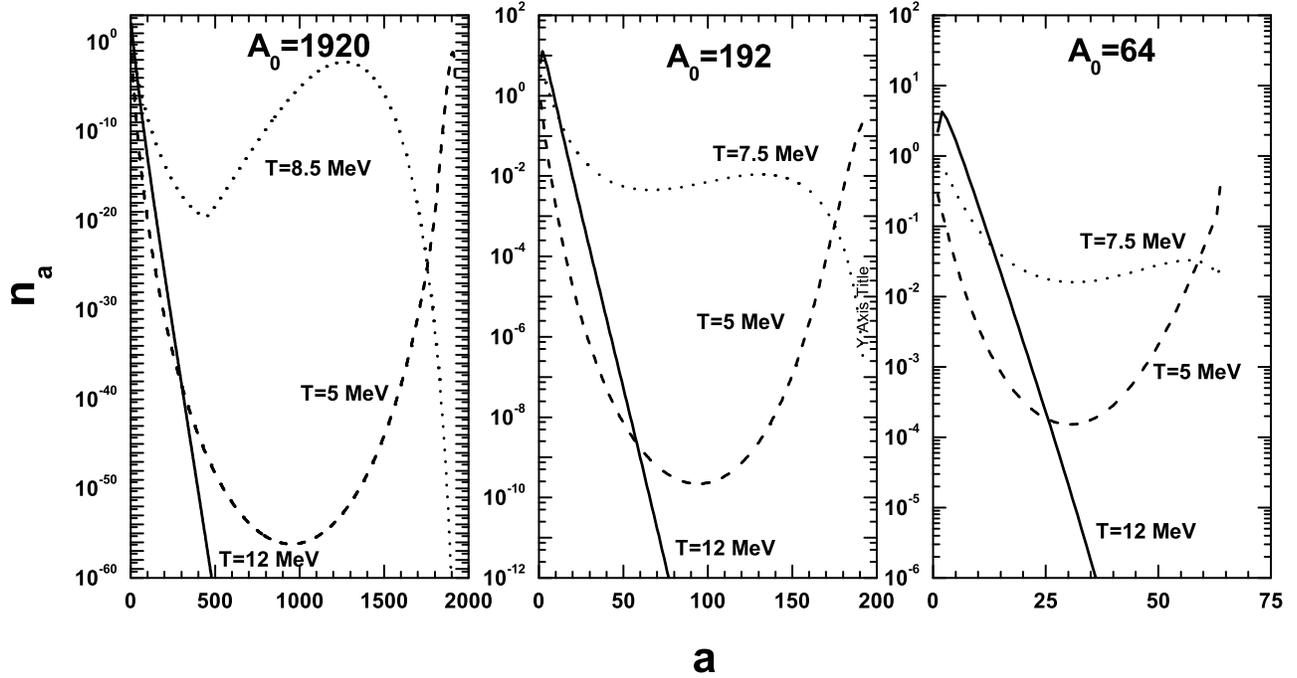}
\caption[Mass distribution calculated by 1 component CTM]{Mass distribution calculated by 1 component canonical thermodynamical model for three different fragmenting sources of masses $1920$ (left panel), $192$ (middle panel) and $64$ (right panel).}
\label{Mass_distribution_statistical_3sources}
\end{center}
\end{figure}
\indent
In CTM, the number of fragments $n_a$ for a given composite of size $a$ is dictated by availability of phase space only (equilibrium statistical mechanics). The expression for $n_a$ is already described in \ref{One_component_canonical_multiplicity} for one kind of particles (nucleons). The shapes of $n_a$ vs a for different $A_0$ and different temperatures are shown in Fig. \ref{Mass_distribution_statistical_3sources}. Note that for each $A_0$ at low temperatures $n_a$ vs $a$ is of $U$ shape; $n_a$ drops with $a$ and reaches a minimum, then rises to a maximum again, finally dropping off to zero. As the temperature rises, the height of the second maximum drops finally disappearing at some temperature. We demonstrate in the next figure that the temperature at which this happens will become the temperature at which first order phase transition happens. Thus the slope of $n_a$ vs $a$, called the multiplicity distribution already forecasts a possibility of a first first order phase transition.\\
\begin{figure} [!h]
\begin{center}
\includegraphics[width=2.4in,height=2.4in,clip]{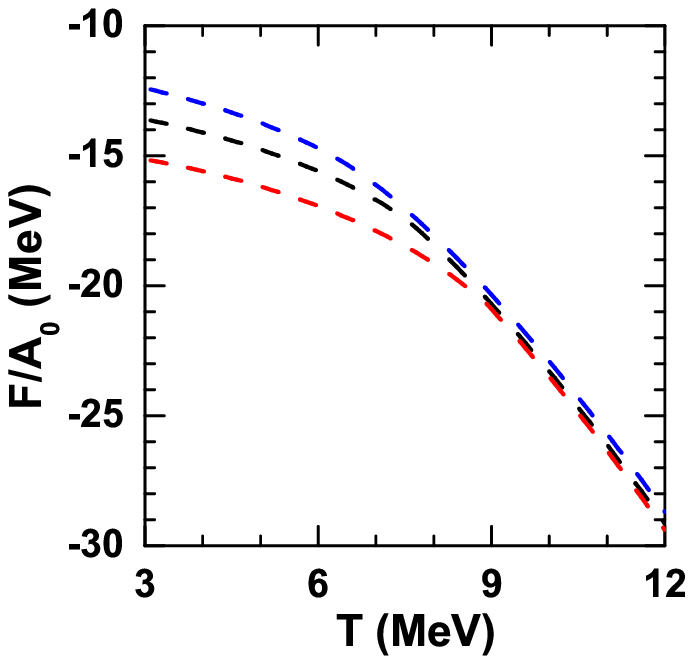}
\includegraphics[width=2.4in,height=2.4in,clip]{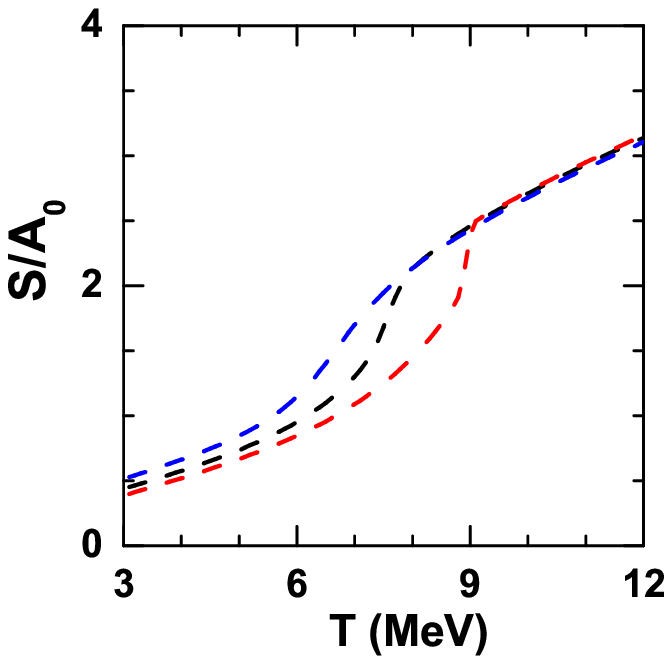}\\
\includegraphics[width=2.4in,height=2.4in,clip]{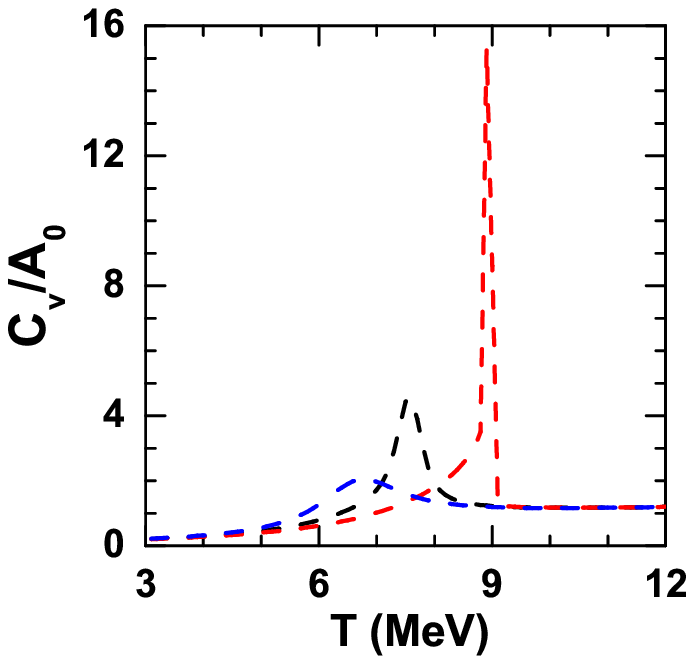}
\includegraphics[width=2.4in,height=2.4in,clip]{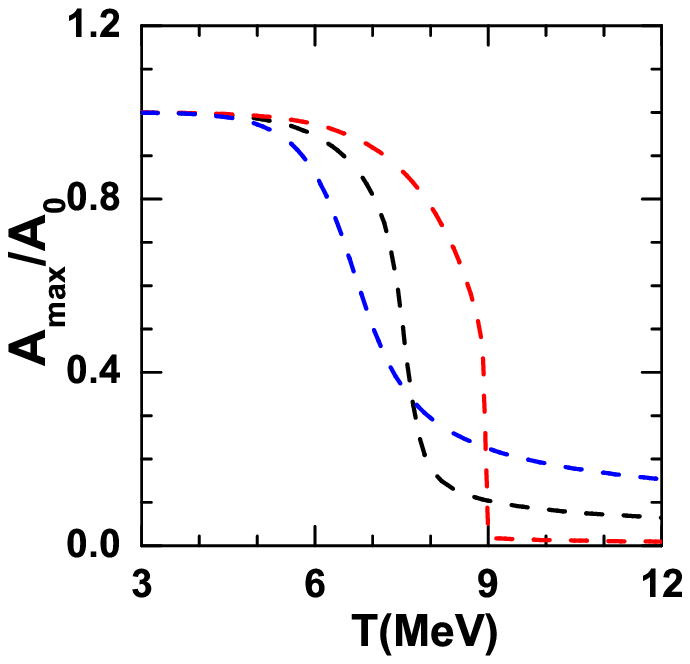}
\caption[Phase transition from statistical model]{Variation of Helmholtz's free energy per nucleon (upper left panel), entropy per nucleon (upper right panel), specific heat per nucleon (bottom left panel) and average size of largest cluster (bottom right panel) with temperature obtained from 1-component canonical thermodynamical model for three different fragmenting sources of mass $A_0=64$ (blue lines), $192$ (black lines) and $1920$ (red lines).}
\label{Phase_transition_statistical}
\end{center}
\end{figure}
\indent
This postulate is substantiated by four graphs in Fig. \ref{Phase_transition_statistical}. The first is free energy $F$ against $T$ for the three value of $A_0$. $F$ is continuous against $T$, but a break in the derivative (entropy) $(\frac{\partial F}{\partial T})_V$ appears to develop at a particular temperature. The break in the derivative is almost not there for low mass $A_0=64$, appears to be developing for $A_0=192$ and is more prominent for $A_0=1920$. A break in the first order derivative implies a first order phase transition. This will be reflected in sudden increase in entropy. As the system size increases the entropy is seen to change abruptly. In each case, the specific heat at constant volume per particle $\frac{C_v}{A_0}$ shows peak (for an infinite system this peak would go to $\infty$). As expected, the temperature where the specific heat is maximum also coincides with the temperature at which the maximum in the high side of mass number ($a$) of the mass distribution just disappears. Now in experiments (as well as in dynamical model calculation) it is difficult to measure free energy, entropy or specific heat. An useful order parameter for studying nuclear liquid gas phase transition is average size of largest cluster $A_{max}$.  Calculating the size of the largest cluster it is found that $A_{max}/A_0$ approaches to $1$ in one phase and approaches a small number at another phase. Similar to entropy, the change of $A_{max}/A_0$ is smooth for $A_0=64$, appears to be developing for $A_0=192$ and more sudden for $A_0=1920$.  This is shown in  Fig. \ref{Phase_transition_statistical} (d). Therefore to study phase transition we can not take very small fragmenting source size like $A_0=64$. On the other hand, in real nuclei, due to presence of Coulomb force a nucleus of very high mass like $A_0=1920$ can not be formed. Hence for dynamical model study for phase transition the calculation will be restricted for $A_0=192$ only. This can be produced by collisions of a projectile of mass $A_p=120$ and target of mass $A_t=120$ ($20\%$ pre-equilibrium emission).\\
\indent
The multiplicity distribution of Fig. \ref{Mass_distribution_statistical_3sources} will not come out of the BUU calculation described in chapter 3. The standard BUU describes properties of the average of all events. It was used very successfully to fit data on flow, transverse momenta etc. To get multiplicity distribution we need an event by event description, not just the average over all events. The first seminal work on this appeared in a paper by Bauer. et. al \cite{Bauer}. Implementation of that model will be described in the next section.
\section{BUU with fluctuation}
To get an event by event description Bauer et. al. \cite{Bauer} proposed the following model. Due to central collision between projectile nucleus of mass $A_p$ and target nucleus of mass $A_t$, for each event two body collisions are checked between $(A_p+A_t)N_{test}$ test particles. Test particle cross-sections are reduced to $\sigma_{nn}/N_{test}$; the collisions are further reduced by a factor $N_{test}$ but if a collision happens between two test particles $i$ and $j$ then not only these two change momenta but in addition $N_{test}-1$ test particles closest to $i$ in phase space suffer the same momentum change as $i$; also $N_{test}-1$ test particles closest to $j$ in phase space are given the same momentum change as $j$. Physically this corresponds to nucleons colliding. At the end of all collisions one event has been completed. In between collisions test particles move in their own mean field (Vlasov propagation). For the second event new Monte-Carlo sampling of $A_p$ on $A_t$ will be started at time zero, similarly for event 3, event 4 etc. To select $2(N_{test}-1)$ closest test particles of the original colliding test particles and to calculate their momentum change following prescription is used.\\
\indent
Let the test particles that move with $i$ are denoted by $i_s$, with $s=0$ to $s=N_{test}-1$. To identify the closest test particles we need to define the distance in phase-space:
\begin{equation}
d_{0s}^2=\frac{(\vec{r_{io}}-\vec{r_{is}})^2}{R^2}
+\frac{(\vec{p_{io}}-\vec{p_{is}})^2}{p_F^2}
\label{Phase_space_distance}
\end{equation}
Here $R$ is the radius of the static nucleus and $p_F$ is the Fermi momentum. The test particles that move with $j$ will be labelled $j_s$. The test particles $j_s$ are then chosen from the rest of the test particles. Now, if original $\Delta \vec{p}$ is added to $N_{test}$ test particles closest to $i$ and $-\Delta \vec{p}$ is added to $N_{test}$ test particles closest to $j$, then this will conserve the total momentum but not the total energy. To overcome this problem, let the average momentum of $N_{test}$ test particles closest to $i$ be defined as $<\vec{p_i}>=\frac{\sum \vec{p_{is}}}N_{test}$, similarly $<\vec{p_j}>$.  One then considers a collision between $<\vec{p_i}>$ and $<\vec{p_j}>$ and obtain a new $\Delta \vec{p}$ for $<\vec{p_i}>$ and -$\Delta \vec{p}$ for $<\vec {p_j}>$.  This $\Delta \vec{p}$ is added to all $\vec {p_{is}}$ and $-\Delta {\vec{p}}$ to all $\vec{p_j}$, which conserves total momentum and total energy simultaneously.\\
\indent
The original BUU prescription described in chapter 3 is simpler, but to get event by event information, when the fluctuation is added, the calculation of the collision part becomes very time consuming and in this case within each time step two body collision is need to be checked between $(A_p+A_t)N_{test}$ test particles. As discussed in section 4.1, to study nuclear liquid gas phase transition one needs to simulate collisions between fairly large nuclei. Therefore, it is very difficult to handle this operation with the existing model. Hence one has to modify the transport model so that it can be used for fairly large nuclei.
\section{Modification in existing BUU model of fluctuation}
The modified method \cite{Mallik10} lies midway between the original BUU calculation \cite{Dasgupta_BUU1} and the model of Bauer et. al. \cite{Bauer,Gallego_thesis}. In the modified method, $N_{test}$ Monte-Carlo simulations of $A_p$ nucleons with positions and momenta and $N_{test}$ simulations of $A_t$ nucleons with positions and momenta is to be done as before . As in cascade calculation for nucleon-nucleon collisions 1 on 1'(event1), 2 on 2'(event2) etc are considered with cross-section $\sigma_{nn}$. For event 1, within each time step, $nn$ collisions only between 1 and 1' (i.e. between first $(A_p+A_t)$ test particles) will be considered. The collision is checked for Pauli blocking. If a collision between $i$ and $j$ in event 1 is allowed, ref. \cite{Bauer} is to be followed and $N_{test}-1$ test particles closest to $i$ are to be picked and the same momentum change $\Delta \vec{p}$ of them as ascribed to $i$ is to be given. Similarly $N_{test}-1$ test particles closest to $j$ are to be selected and these  are to be ascribed the momentum change $-\Delta \vec{p}$, the same as suffered by $j$. As a function of time this is continued till event 1 is over. For Vlasov propagation all test particles are utilised. For event 2 one has to return to time $t$=0, the original situation (or a new Monte-carlo sampling for the  original nuclei), follow the above procedure but consider $nn$ collisions only between 2 and 2' \{i.e. between $(A_p+A_t)+1$ to $2(A_p+A_t)$ test particles\}. This can be repeated for as many events as one needs to build up enough statistics.\\
\begin{figure} [!h]
\begin{center}
\includegraphics[width=1.9in,height=1.9in,clip]{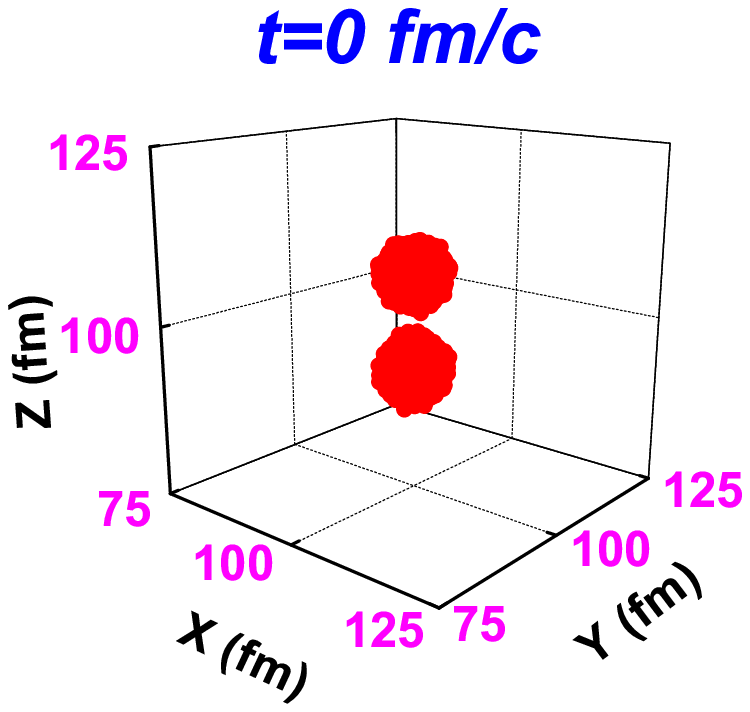}
\includegraphics[width=1.9in,height=1.9in,clip]{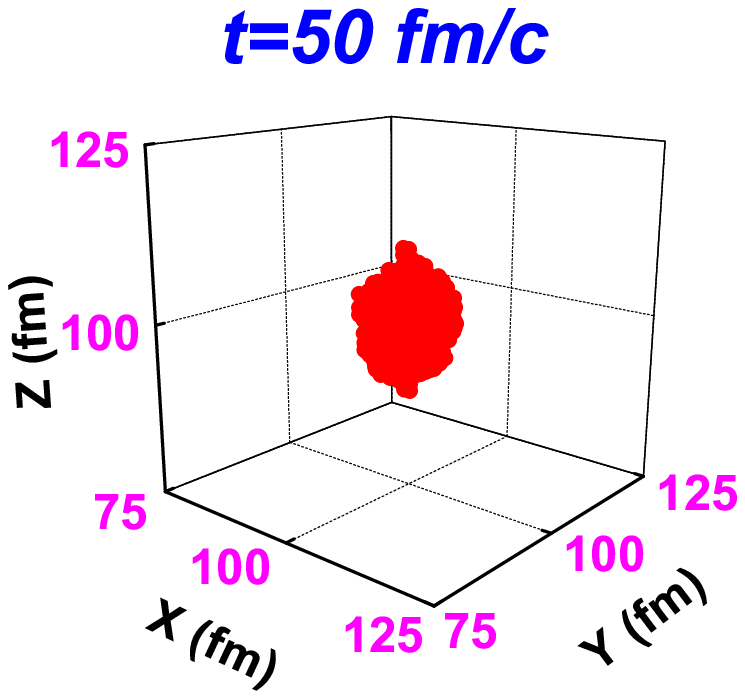}
\includegraphics[width=1.9in,height=1.9in,clip]{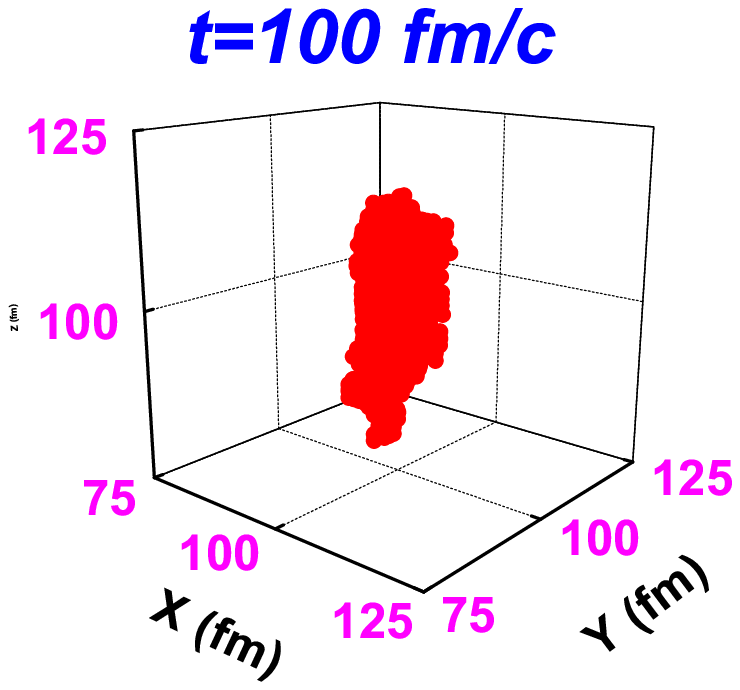}\\

\includegraphics[width=1.9in,height=1.9in,clip]{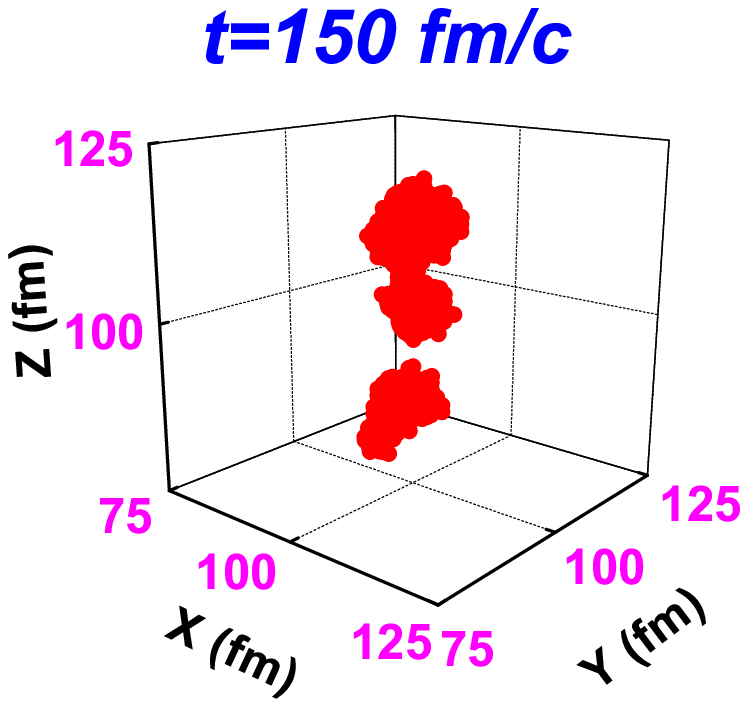}
\includegraphics[width=1.9in,height=1.9in,clip]{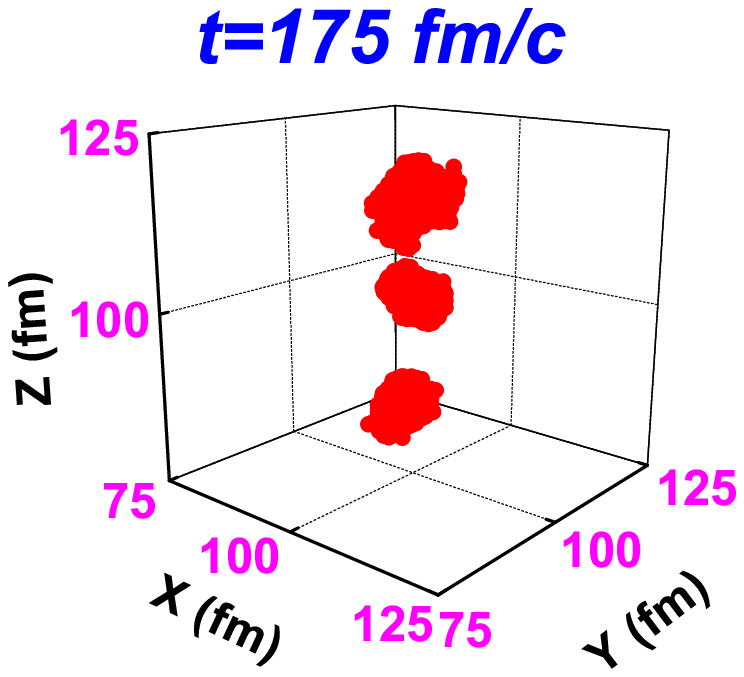}
\includegraphics[width=1.9in,height=1.9in,clip]{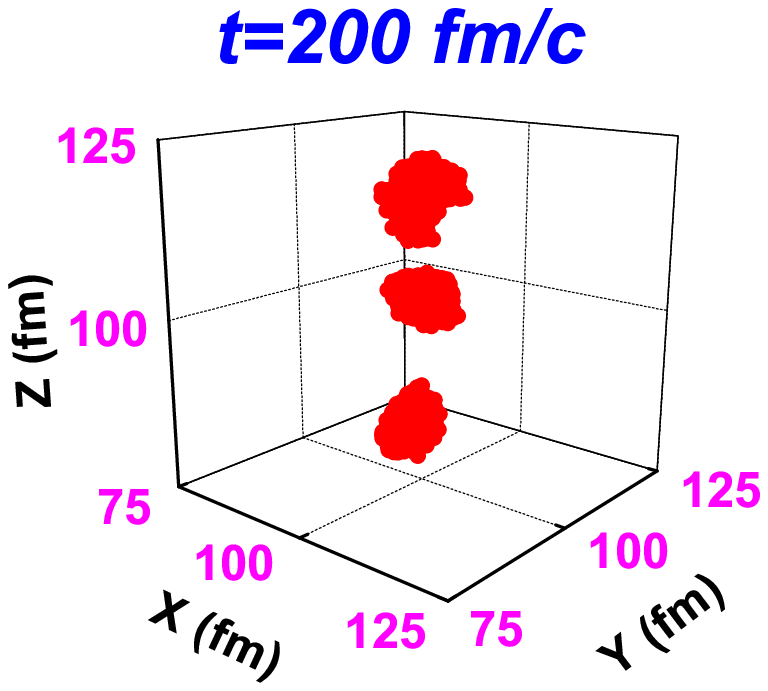}
\caption[Time evolution and clusterization in BUU model]{Time evolution and clusterization of the test particles for the first event of $A_p=40$ on $A_t=40$ reaction calculated by using fluctuation included modified BUU model at beam energy 50 MeV/nucleon.}
\label{Dynamical_clusterization}
\end{center}
\end{figure}
\indent
The advantage of this method over the existing method is that here, for one event, $nn$ collisions need to be considered between $(A_p+A_t)$ test particles whereas in existing method, collisions need to be checked between $(A_p+A_t)N_{test}$ test particles. Hence, in the modified calculation, total number of combinations for two-body collision is reduced by a factor of 1/$N_{test}^2$. Since typically $N_{test}$ is of the order of 100 this is a huge saving in computation and has allowed us to treat mass as large as $120$ on $120$ over a substantial energy range.\\
\indent
One bonus of this prescription is that one sees some common ground between the BUU approach and the ``quantum molecular dynamics'' approach.  In the latter nucleons are represented by Gaussians in phase space; the centroids have a $\vec{r}$ and a $\vec{p}$ which are originally generated by Monte-Carlo. These collide.  This corresponds to ``nucleons'' colliding in our prescription. As the centroids move after collision, they drag the Gaussians along.  The Gaussian wave packets in position and momentum space provide the mean-field and Pauli blocking.  The Gaussians do not change their shapes or widths.  These are very strong restrictions and can lead to very different mean field propagation. The Vlasov propagation has much more flexibility  and originates from more fundamental theory.
\begin{figure}[!h]
\begin{center}
\includegraphics[width=\columnwidth,keepaspectratio=true,clip]{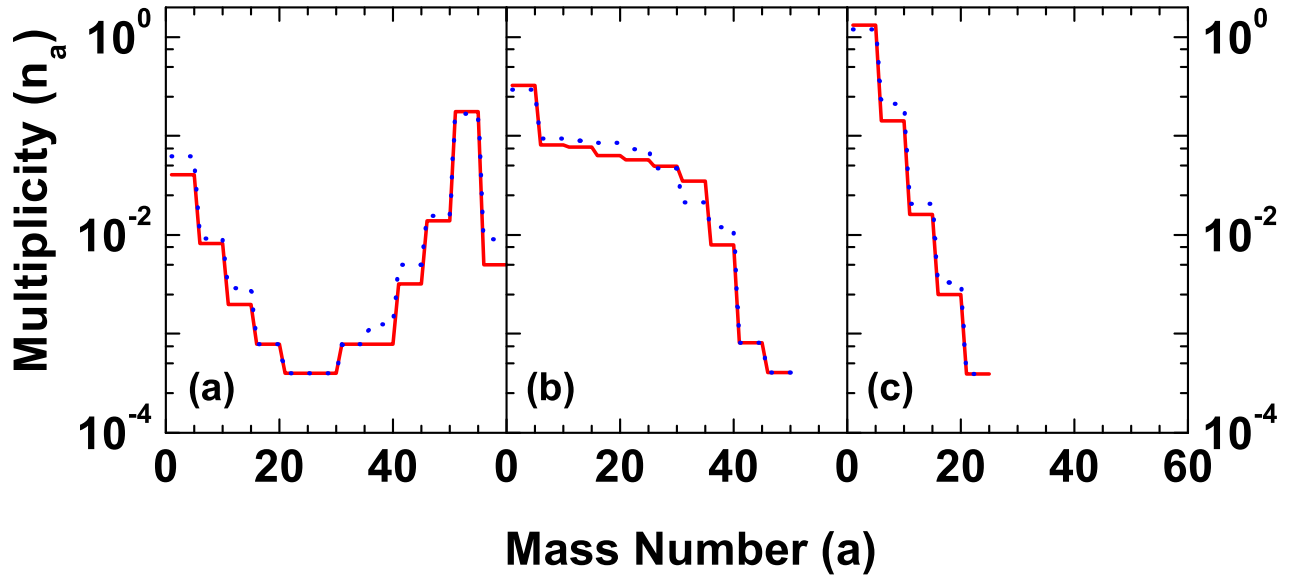}
\caption[Mass distribution from two methods]{Comparison of mass distribution  calculated according to the existing (blue dotted lines) and the modified (red solid lines) BUU prescription. The average value of 5 mass units are shown. The cases are for central collision of mass 40 on mass 40 for different beam energies (a) 25, (b) 50 and (c) 100 MeV/nucleon.}
\label{Mass_distribution_two_methods_BUU}
\end{center}
\end{figure}
\section{Comparison of two prescriptions}
Before applying the modified method in phase transition study, at first one has to check whether modified prescription results are comparable with the existing BUU prescription or not. For this, simulations have to be done by using the both methods for a central collision reaction of $A_p=40$ on $A_t=40$. In both case, calculations are done in a 200$\times 200\times 200 fm^3$ box. The configuration space is divided into $1fm^3$ boxes. For Vlasov propagation, mean field is calculated by using Eq. \ref{Lenk_potential}. The time evolution is studied up to $t=200 fm/c$. At the end of time evolution, the position and momenta of test particles are obtained. To construct clusters from it, one has to consider contiguous boxes with test particles that propagate together for a long time. These test particles are considered to be part of the same cluster. The contiguous boxes have at least one common surface and the nuclear density exceeds a minimum value ($d_{min}$). Different $d_{min}$ values as 0.002, 0.005, 0.01, 0.015 and 0.02 fm$^{-3}$ are tried to check the sensitivity of this parameter. It is observed that the fragment multiplicity distribution does not change very much with $d_{min}$, therefore $d_{min}=$0.01 fm$^{-3}$ will be used for further calculations. Fig. \ref{Dynamical_clusterization} shows the time evolution of the test particles and clusterization at different times for the first event of $A_p=40$ on $A_t=40$ reaction calculated by using fluctuation included modified BUU model at beam energy of 50 MeV/nucleon.\\
\begin{figure} [!h]
\begin{center}
\includegraphics[width=11cm,height=9cm,clip]{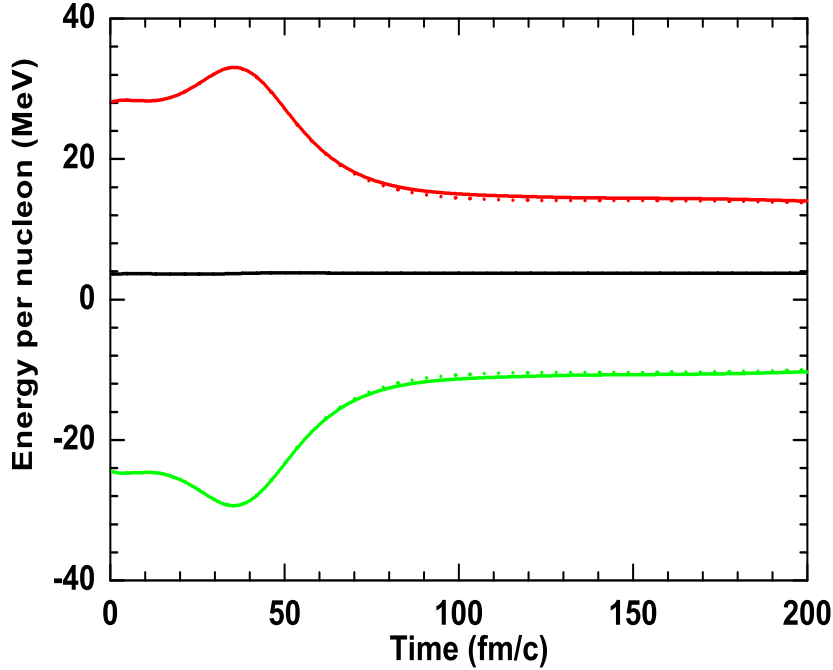}
\caption[Energy distribution from two methods]{Variation of kinetic (red), potential (green) and total (black) energy per nucleon with time calculated in the centre of mass frame according to the existing (dotted lines) and the modified (solid lines) BUU prescription for $A_p=40$ on $A_t=40$ reaction at beam energy 50 MeV/nucleon.}
\label{Energy_vs_time_two_methods}
\end{center}
\end{figure}
\indent
Fig. \ref{Mass_distribution_two_methods_BUU} shows the comparison of mass distribution obtained from existing and modified BUU prescription for three different beam energies 25, 50 and 100 MeV/nucleon. In both the cases, 500 events are simulated at each energy. It shows that the results from both the calculations are identical. The mass distributions are studied at the end of calculation i.e. at $t=200 fm/c$. In Fig. \ref{Energy_vs_time_two_methods} the variation of kinetic energy, potential energy and total energy per nucleon (energies are averaged over 500 events) with time obtained from two methods is shown, which shows not only at $t=200 fm/c$, but in the entire time region both methods give identical result.\\
The results obtained from two methods are similar, because,\\
(a) The number of collisions in an event are statistically the same.\\
(b) In the original formulation the objects that collided were picked from a fine grain sampling of phase-space density. In the modified method these are picked from a coarse grain sampling of the same phase-space density. But many events are needed so statistically it should not matter.\\
(c) Characteristics of scattering are the same.\\
(d) The same Vlasov propagation is used.\\
\indent
The modified method is regarded as a very convenient short cut to numerical modeling of the existing method.  The theoretical formulation of section 4.2 is more appealing and more democratic but numerically our method gives indistinguishable results.\\
Therefore for studying phase transition from dynamical BUU model, the reactions between large nuclei will be simulated by using the modified method.
\section{Liquid gas phase transition from dynamical model}
\begin{figure}[b!]
\begin{center}
\includegraphics[width=14cm,keepaspectratio=true,clip]{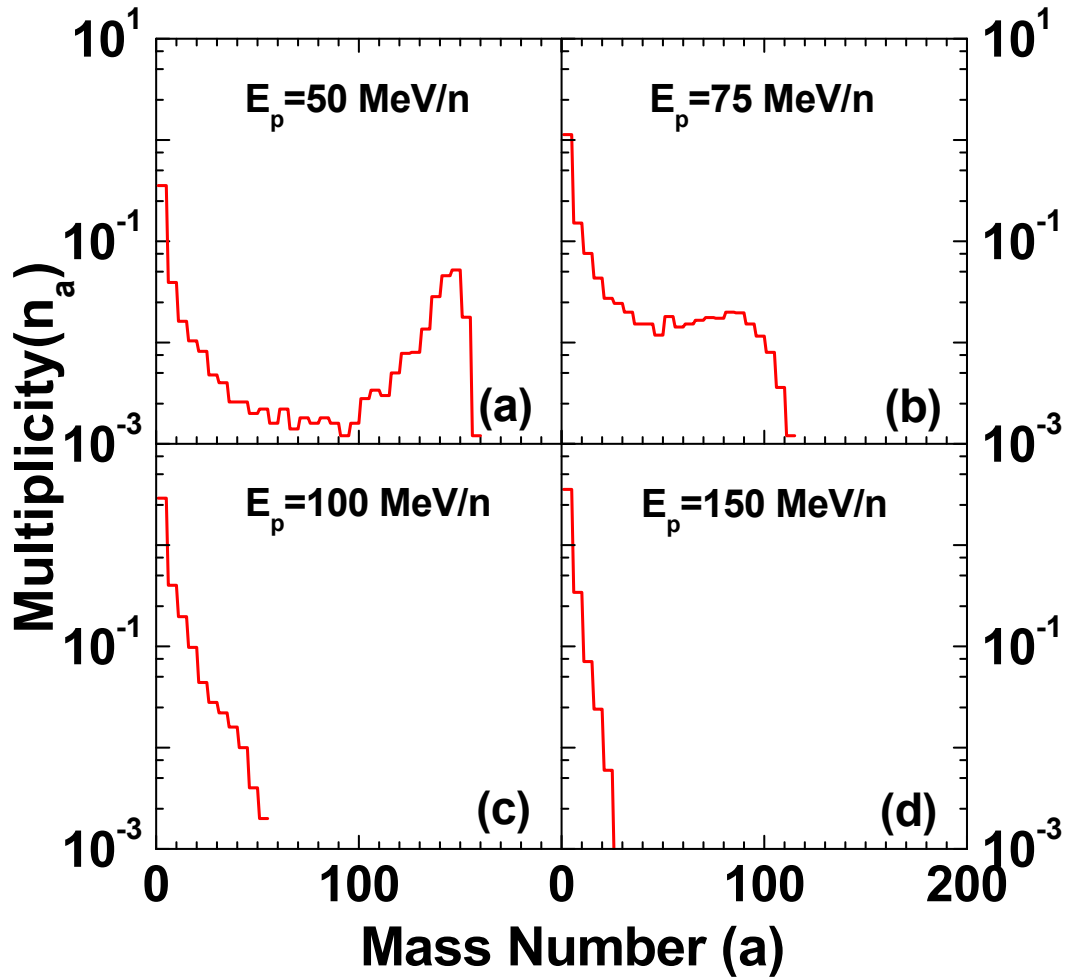}
\caption[Mass distribution from dynamical model]{Mass distribution for $A_p=120$ on $A_t=120$ reaction at beam energies (a)50 MeV/nucleon, (b)75 MeV/nucleon (c)100 MeV/nucleon and (d)150 MeV/nucleon.  The average value of 5 mass units are shown. At each energy 1000 events are chosen.}
\label{Mass_distribution_BUU}
\end{center}
\end{figure}
To study liquid gas phase transition from dynamical model, initially the mass distribution for $A_p=120$ on $A_t=120$ reaction are calculated at four different beam energies. This is shown in Fig. \ref{Mass_distribution_BUU}. For each energy 1000 events are taken. The results of averages for groups of five consecutive mass numbers are shown. At $E_p=50$ MeV/nucleon it is "$U$" shaped i.e. the system is in liquid phase. At $E_p=75$ MeV/nucleon, the second maxima is still visible but in addition to that sufficient amount of intermediate mass fragments are also produced i.e. it is in liquid-gas co-existence phase. The second maximum is gone when beam energy rises to 100 MeV/nucleon. The disappearance of the second maxima indicates that the system is in gas phase. The remarkable feature is that the evolution of this shape was predicted earlier by the canonical thermodynamic model (CTM) \cite{Dasgupta_Phase_transition,Das}. In order to proceed further it is expedient to establish a connection with canonical thermodynamic model (CTM) \cite{Das} which has provided in the past evidences of first order phase transition in intermediate energy heavy ion collisions (be it with the assumption that equilibrium is established). For transport models the natural variable is the beam energy.  For CTM the natural variable is the temperature $T$.  For illustration, in Fig. \ref{Mass_distribution_Statistical} the multiplicity distributions for a system of 192 particles in CTM at temperatures of 6.5 MeV, 7.5 MeV, 10 MeV and 14 MeV are shown.  The calculations with BUU and CTM are so different that the similarity in the evolution of the shape in multiplicity distribution is very striking.\\
\begin{figure}[h!]
\begin{center}
\includegraphics[width=14cm,keepaspectratio=true,clip]{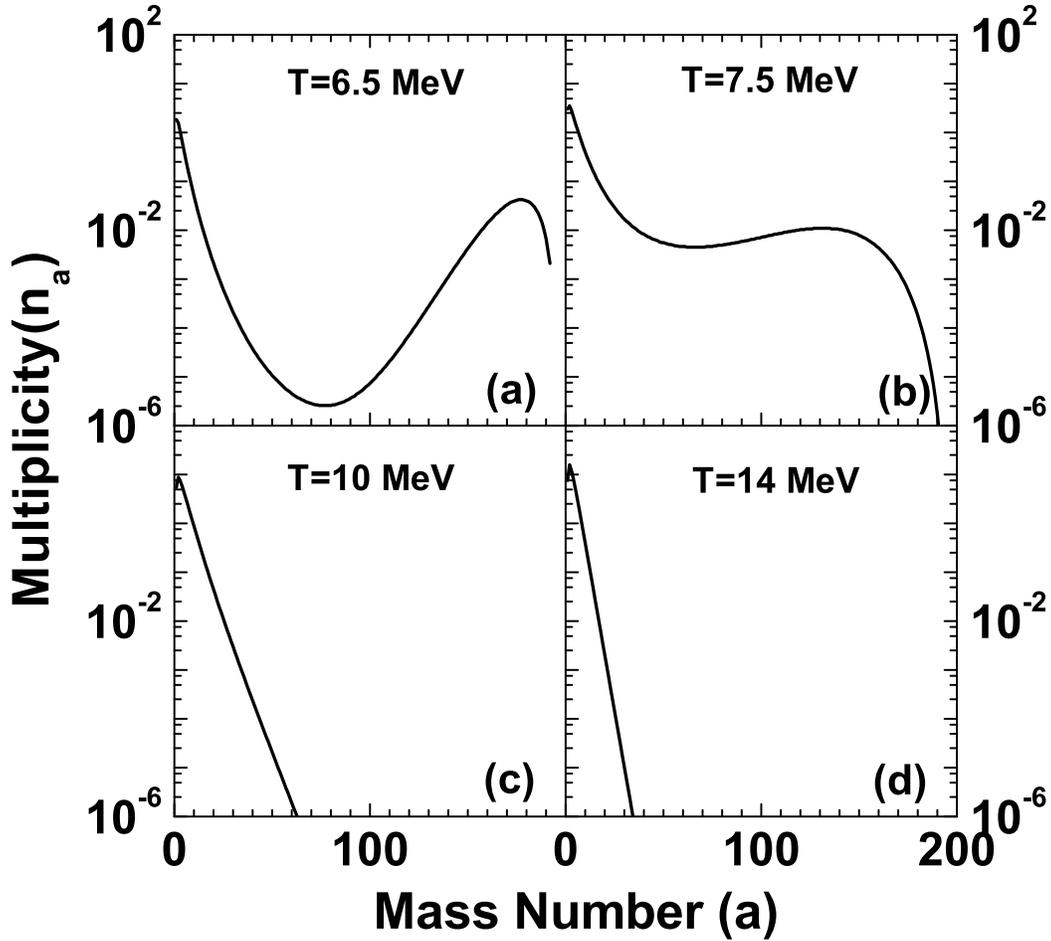}
\caption[Mass distribution from statistical model]{Mass distribution from Canonical Thermodynamical Model (CTM) calculation for fragmentation of a system of mass $A_0=192$ at temperature (a)6.5 MeV, (b)7.5 MeV (c)10 MeV and (d)14 MeV.}
\label{Mass_distribution_Statistical}
\end{center}
\end{figure}
\indent
To proceed further with the correspondence between the two models, one needs to establish a connection between $E_p$ of BUU and temperature $T$ of CTM. Temperature $T$ of CTM will give an average excitation energy $E^*$ of the multifragmenting system in its centre of mass \cite{Das}. Now the excitation energy in CTM and the beam energy in the lab are related but some physics input is needed to go from one to the other. The details of excitation calculation for central collision multifragmentation reaction is already described in chapter-4.\\
\indent
Fig. \ref{Phase_transition_dynamical}  gives some CTM results and also makes a comparison of one CTM result with transport model result. The top left diagram is $E^*$ vs.$T$ in CTM for 192 particles ($A_0$=192=80$\%$ of 240). This approximates usual $E^*$ vs $T$ for first order phase transition. There is a boiling point temperature $T$ which remains constant as energy increases. Since the fragmenting system is very finite, the slope $dE^*/dT$ is not infinite but high.  The lower left diagram is again drawn in CTM.  Here $A_{max}$ is the average value of the largest cluster.  A high value of $A_{max}/A_0$ means liquid phase and low values means gas phase. The criteria of deciding which composites belong to the gas phase and which to the liquid phase are discussed in detail in the two previous papers \cite{Chaudhuri2,Chaudhuri1}.\\
\begin{figure}[b!]
\begin{center}
\includegraphics[width=14cm,keepaspectratio=true,clip]{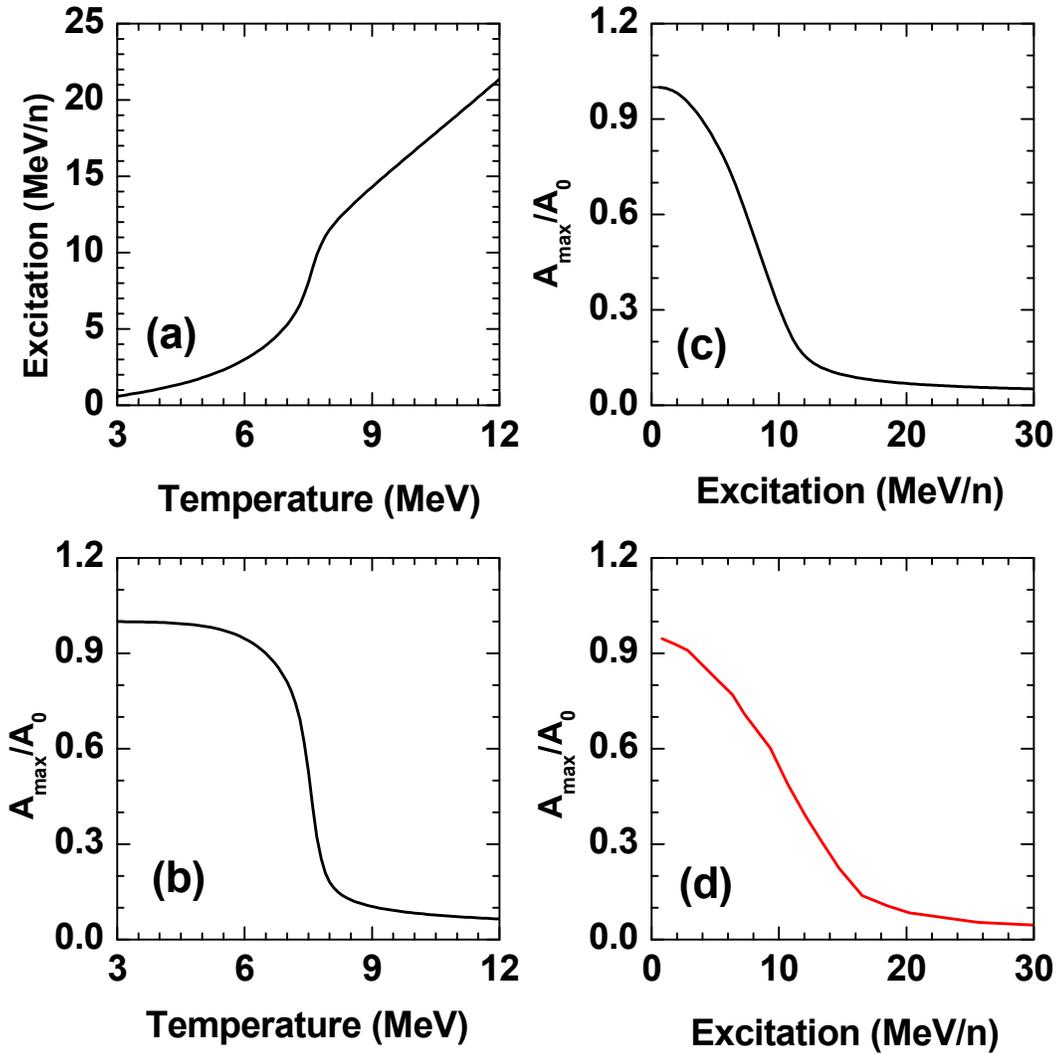}
\caption[Signals of Phase transition from dynamical model]{Top left curve (a) is a Canonical Thermodynamical Model (CTM) calculation for $E^*$ vs. T for $A_0=192$. Between 6 MeV and 7.5 MeV temperatures, $E^*$ rises quickly. The $dE^*/dT$ slope increases sharply with mass size $A_0$ and is indicative of first order phase transition. Bottom left curve (b) is also a CTM curve showing that the size of largest cluster drops sharply between 6 MeV and 7.5 MeV. Again this is first order liquid gas phase transition. Top right (c) is also with CTM but $A_{max}/A_0$ is plotted against excitation energy per nucleon instead of temperature. The change of liquid to gas is necessarily slower, the range of energy for the change is dictated by latent heat. Bottom right (d) is the calculation from BUU model.}
\label{Phase_transition_dynamical}
\end{center}
\end{figure}
\indent
In the bottom left diagram, one sees more dramatically that in a short temperature interval liquid has transformed into gas. The only input in the transport model is the beam energy. The common dynamical variable in both transport model and CTM is $E^*$. Of course $E^*$ in CTM is an average excitation but in BUU, $E^*$ is exact, it is microcanonical. In the top right corner of Fig. \ref{Phase_transition_dynamical} is the plot of $A_{max}/A_0$ as a function of $E^*$ in CTM.  The transformation from liquid to gas is more gradual, essentially spanning the energy range across which, liquid transforms totally into gas.  Even for a large system, where the transformation of liquid to gas as a function of temperature is very abrupt, the transformation as a function of energy per particle will be quite smooth. The bottom right in Fig. \ref{Phase_transition_dynamical} is from the transport model calculation. The similarity with the CTM graph is close enough that one can conclude the transport model calculation gives evidence of liquid-gas phase transition. CTM calculation predicts that nuclear liquid gas transition is first order. In dynamical model, till now the temperature can not be calculated directly. Hence one can not determine the free energy, therefore the transport model can not predict the order of phase transition directly. But as the behaviour of mass distribution and largest cluster variation
with excitation are similar in statistical and dynamical models, one can conclude that dynamical model also suggests that nuclear liquid gas phase phase transition is of first order.
\section{Discussion of Pauli blocking in the fluctuation model}
The one important feature of the model that raised concerns and led to a lot of work to propose alternative methods for calculations \cite{Rizzo,Napolitani} may be discussed now. This is related to dangers of crossing fermionic occupation limits in the model here (as in the model of ref. \cite{Bauer}.). As mentioned in section 3.2, if Pauli blocking allows two test particles $i$ and $j$ to collide then to represent the scattering of two actual nucleons, not only $i$ and $j$ but also $N_{test}-1$ test particles closest to $i$ and $N_{test}-1$ closest to $j$ will suffer momentum change $\Delta {\vec{p}}$ and $-\Delta {\vec{p}}$ respectively. Since the $2(N_{test}-1)$ test particles are moved without verifying Pauli blocking there may be cases where one exceeds the occupation limits for fermions.\\
\begin{figure}[h!]
\begin{center}
\includegraphics[width=9.5cm,keepaspectratio=true,clip]{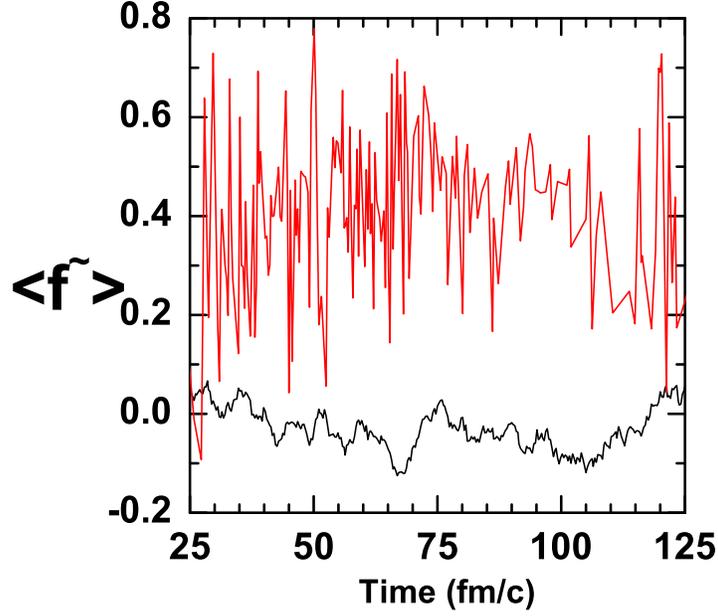}
\caption[Checking of Pauli blocking in the fluctuation model]{Variation of average availability factor (see text) with time (red line) for $A_p=120$ on $A_t=120$ reaction at beam energy 100 MeV/nucleon. The lower curve (black dotted line)is the average availability factor $\langle \tilde{f} \rangle$ at the phase space points of arbitrarily chosen 120 test particles in an isolated mass 120 nucleus as they move in time.}
\label{Pauli_blocking}
\end{center}
\end{figure}
\indent
Initially the two ions have very compact occupation at two different corners of phase space. Collisions make a far wider region of phase space available
to nucleons so this problem may not be severe. An accurate estimation of exceeding the fermionic limit of occupation at various parts of phase space
is very hard to compute in the present problem but some measures are relevant. For 120 on 120 at 100 MeV/nucleon beam energy (50 MeV/nucleon beam energy is studied also) one event is followed as a function of time. In any allowed collision, for each of $2N_{test}$ particles, the availability factor $\tilde{f}=1-f=1.0-N/8$ is calculated separately, where $N$ is the number of test particles (excluding that particle) within the cell in six dimensional phase space (see section 3.3.3). $\tilde{f}$=0 represents the limit of fermionic occupation. If $\tilde{f}$ is negative one has crossed the quantum limit and are in the classical regime. Any positive number between 0 and 1 will accommodate additional fermion.  For each collision there are 200 $\tilde{f}$'s to be calculated so for each collision an average $\tilde{f}$ can be obtained and that is plotted in Fig. \ref{Pauli_blocking}. The results are shown for $t$=25 to $125 fm/c$ when most of the action takes place. For reference average $\tilde{f}$ is also plotted for randomly chosen 120 test particles in a static mass 120 nucleus as they move around in time. This number should ideally be 0 and not fluctuate. The fluctuations from the value 0 reflects uncertainties, probably due to fluctuation in initial Monte-Carlo simulations. This degree of uncertainty must be also present in the values of $\tilde{f}$ we have plotted for collisions.  In spite of these uncertainties the predominantly positive values of $\tilde{f}$ as displayed  in Fig. \ref{Pauli_blocking} lead to believe that the general trends found in the original calculation will hold.\\
\indent
If in a collision every test particle moved to location where $\tilde{f}$ is positive, all the occupations will stay within fermionioc limits. In case there is a test particle which does not satisfy this one can try to improve the situation by discarding that test particle and choosing the next available test particle to be part of the cloud. Complications arise because when some of the previously chosen test particles are discarded for new ones the average momentum of the clouds will change, new $\Delta\vec{p}$ will have to be used so the final resting spots obeying energy and momentum conservation will change too.  An iterative procedure needs to be formulated but convergence may be slow.\\
\indent
Alternative methods have been proposed.  The two papers which give procedural details of moving two clouds of test particles from initial positions to final positions with a stricter adherence to fermionic limits are Refs. \cite{Rizzo,Napolitani}.  Multiplicity distributions are not given so the present calculation results can not be compared. Even if the multiplicity distributions turn out to be similar, higher order correlations can be very different.  The present work extended the first proposed model of fluctuations in BUU to a larger system at many energies and a very interesting lesson has been learned.  The gross features of multiplicity distribution do resemble strongly the results from equilibrium statistical models which have proven very successful in explaining experimental data.

\section{Summary}
An enormous amount of experimental and theoretical work exists on phase coexistence or liquid-gas phase transition in heavy ion collisions at intermediate energy. The standard methods of theoretical studies on liquid-gas phase transition at intermediate energy collisions assume that because of two body collisions nucleons equilibrate in a given volume and then dissociate into composites of different sizes according to the availability of phase space. This chapter focuses on whether the results of the transport model calculations (BUU) at intermediate energy can reveal signatures of phase transition.\\
\indent
In order to study that, one need to simulate collisions between fairly large nuclei. Therefore a simplified yet accurate method of BUU transport model has been developed which allows calculation of fluctuations in systems much larger than what was considered feasible in a well-known and already existing model. BUU transport model simulation has been performed for central collisions of mass 120 on mass 120 at laboratory beam energy in the range 20 MeV/nucleon to 200 MeV/nucleon. The calculations produce clusters. The distribution of clusters is remarkably similar to that obtained in equilibrium statistical model and provides evidence of phase transition.
\vskip3cm
\end{normalsize} 
\chapter{Summary, discussions and future outlook}
\begin{normalsize}
\section{Summary and discussions}
The study of nuclear multifragmentation is important for understanding the reaction mechanism in heavy-ion collisions at intermediate and high energies. In the present thesis, different aspects of statistical and dynamical models of nuclear multifragmentation has been investigated in detail with the objective to concentrate about (i) production of exotic nuclei which are normally not available in the laboratory, (ii) nuclear liquid-gas phase transition and (iii) nuclear symmetry energy from heavy ion collisions at intermediate energies. A general overview of nuclear multifragmentation was presented in chapter 1.\\
\indent
In chapter 2, a model for projectile fragmentation was developed which is grounded on the traditional concepts of heavy-ion reaction (abrasion) as well as the model of multifragmentation (Canonical thermodynamical model) and secondary decay. This model is in general applicable and implementable in the limiting fragmentation region. An impact parameter dependent temperature profile was introduced in the model for projectile fragmentation which could successfully explain experimental data of different target projectile combinations of widely varying projectile energy. The observables which were calculated and were compared to experimental data includes charge distribution, isotopic distribution, intermediate mass fragment multiplicity and the average size of the largest cluster.\\
\indent
In order to extract the initial conditions (size and excitation) of projectile spectator from more basic approach, at first a microscopic static model was developed  in chapter 3. In this model, the PLF was assumed to form with a deformed shape. The mass and shape of the PLF were determined assuming straightline trajectory of the projectile. In heavy ion collisions, this model can be used to extract an impact parameter dependent excitation energy. Then in order to include the dynamical effects, the microscopic static model was expanded to a transport model based on Boltzmann-Uehling-Uhlenbeck equation. In the transport model, two nuclei in their Thomas-Fermi ground state were boosted towards each other and the time evolution of the test particles was studied. At the end of time evolution, PLF mass and excitation were calculated and then the Canonical thermodynamical model was used to deduce the freeze-out temperature. It was observed that the PLF mass at different impact parameters calculated from the microscopic static model and the BUU model were comparable to that obtained from geometric abrasion calculation. It was quite gratifying that detailed BUU calculations was borne out two striking features  of temperature profile in the PLF.  These are : (a) temperatures are of the order of 5 to 6 MeV and (b) there is a very definitive dependence on the intensive quantity $A_s/A_P$, temperature falling as this increases (similar to the parameterised temperature profile used in chapter 2).\\
\indent
In chapter 4, a hybrid model was developed for studying nuclear multifragmentation around the Fermi energy domain. In this model, the excitation of the fragmenting system was calculated by using the dynamical BUU approach with proper consideration of pre-equilibrium emission. Then the fragmentation of this excited system was studied by the Canonical thermodynamical model and finally the decay of the excited fragments, which were produced in multifragmentation stage, was calculated by the evaporation model. This model was used to calculate various observables such as cross-sections of different composites, probability distribution of the largest cluster etc for central collisions of $^{129}$Xe on $^{119}$Sn at different beam energies around the Fermi energy domain and these results were compared with experimental data. There is no adjustable parameter in the model, and the calculated values of the different observables are pleasingly close to the experimental data. This model is very useful for estimating the freeze-out temperature of the heavy ion collisions in the Fermi energy domain.\\
\indent
Statistical models based on canonical and grand canonical ensembles have been extensively used to study multifragmentation. The basic assumption of canonical and grandcanonical ensembles are fundamentally different, and in principle they agree in the thermodynamical limit when the number of particles become infinite. In chapter 5, the conditions under which the observables obtained from the canonical and the grand canonical ensembles converge, were explored for the fragmentation of finite nuclei i.e. much away from the thermodynamical limit. The results from the two ensembles can be made to converge either by increasing the temperature or freeze-out volume or by increasing the source size or by decreasing the asymmetry of the source. For one component systems, an analytical expression was developed which can extract the canonical results from grand canonical calculation and vice versa. The validity of this analytical formula was checked successfully by using different observables of multifragmentation like total multiplicity, average size of largest cluster etc. This transformation method is not applicable when the system experiences a first order phase transition where the grand canonical particle number distribution is highly non-gaussian.\\
\indent
In chapter 6, the ratio of symmetry energy coefficient to temperature $C_{sym}/T$ was extracted from different prescriptions using the isotopic as well as the isobaric yield distributions obtained from different projectile fragmentation reactions. It was found that the values extracted from our theoretical calculation agree with those extracted from the experimental data but they differ very much from the input value of the symmetry energy used. The best possible way to deduce the value of the symmetry energy coefficient is to use the fragment yield at the breakup stage of the reaction. During the break up stage, source dependence of isoscaling parameters and source and isospin dependence of isobaric yield ratio parameters were also examined in the framework of the canonical and the grand canonical models. It was found that the results obtained from the two ensembles are in general different but when the nucleus fragments more, results from both the ensembles converge. Generally it is better to use the grand canonical model for the fragmentation analysis. This is because the formulas that are used for the deduction of the symmetry energy coefficient have all been derived in the framework of the grand canonical ensemble which is valid only at the break-up (equilibrium) condition. The yield of "cold" fragments either from the theoretical models or from experiments when used for extraction of the symmetry energy coefficient using these prescriptions might lead to the wrong conclusion.\\
\indent
The main focus of Chapter 7 was on whether the results of transport model calculations (BUU) at intermediate energy can reveal signatures of nuclear liquid gas phase transition. Phase transition signals were extracted from multiplicity distributions of fragments resulting from heavy ion collisions. For studying the multiplicity distribution (i.e. to get event by event description in BUU), a computationally feasible method was developed which can simulate collisions between fairly large nuclei. The distribution of clusters obtained from this modified model was found to be remarkably similar to that obtained from the equilibrium statistical model and provides evidence of first-order phase transition.\\
\\
\\
\\
\\
\\
\\
\\
\\
\\
\\
\\
\\
\\
\\
\\
\\
\\
\\
\\
\\
\\
\\
\\
\section{Future outlook}
Our study of the statistical and dynamical models of nuclear multifragmentation reactions at intermediate energies opens up the following directions
of research which can be attempted in future. In the present thesis, the initial conditions of projectile fragmentation at higher energies (140 MeV/nucleon to 1 GeV/nucleon) and central collision multifragmentation reactions in the Fermi energy domain have been obtained from the BUU transport model calculation. In BUU model calculation, the neutrons are protons are considered in the same footing (so there is no asymmetry term). In the Vlasov part, the potential depends upon nucleon density ($\rho$) only.  In a more realistic version, the potential would depend upon $\rho_n+\rho_p$ and $\rho_n-\rho_p$.  One would have to keep a track of isotopic spin of the test particles.  With distinct $\rho_n$ and $\rho_p$, computation would increase very significantly but this is within the realm of feasibility.\\
\indent
The similarity with of the transport model calculation results with that of the statistical model results (CTM) is close enough to let us conclude that the former gives evidence of liquid-gas phase transition (described in chapter 7). To find closer correspondence between transport model calculations and equilibrium statistical model, it will be best if we can deduce at least an approximate value of temperature for each beam energy.  For an interacting system this is very non-trivial. Formulae like $\frac{E^*}{A}=\frac{3T}{2}$ are obviously inappropriate. One might try to exploit the thermodynamic identity $T=(\frac{\partial E}{\partial S})_V$. This requires obtaining a value of the entropy for an interacting system. This can be a important direction of our work in future.\\
\indent
The bimodal behaviour of the order parameter is an important signature of first order phase transition. As described in chapter 2,4 and 8, the largest cluster is an important order parameter for studying nuclear liquid gas phase transition. It has been recently proposed \cite{Chomaz_bimodality,Gulminelli_bimodality,Gulminelli_bimodality2} that a bimodal behavior of the largest cluster distribution in nuclear multifragmentation is also a measurable signature of first order phase transition. Different experimental observations are reported \cite{Lopez_bimodality,Bonnet_bimodality,Pichon_bimodality,dago} regarding this and a debate on its ultimate physical interpretation was raised. To address these issues some theoretical simulations are done mainly by using statistical models \cite{Chaudhuri3,Chaudhuri_bimodality2} and quantum molecular dynamics approach \cite{Fevre_bimodality}. It will be very interesting to study bimodality from BUU transport model calculation.\\
\indent
In chapter 5, analytical expressions have been derived for one component model, in order to extract the canonical results from grand canonical calculation and vice versa. Since, the actual nuclei are formed of neutrons and protons, it will be important to develop similar analytical expressions for two-component systems. In addition to that, one can generalize this method in order to get the microcanonical result out of a canonical calculation and vice versa. The equations for extracting different observables like isoscaling, isobaric yield ratio etc are derived from the grand canonical framework but microcanonical or canonical models are better suited for explaining intermediate energy heavy ion reactions. Therefore these ensemble transformation relations will be very useful for extracting these observables more accurately.\\
\indent
The study of the physics of hypernuclei physics opens the unique opportunity to throw light on the hyperon-nucleon and the hyperon-hyperon interaction \cite{Bielich}. The stability of hypernuclei beyond the neutron and proton driplines (normal nuclear chart) \cite{Khan} is a subject which is important in recent theoretical and experimental activities. The knowledge of the structure of normal nuclei as well as the extension of the nuclear chart into the strangeness sector gets valuable input from the results of hypernuclei study \cite{Samanta}. In nuclear reactions at high energies, hyperons are produced in the participant zone and the produced hyperons have an extended rapidity distribution and some of these are absorbed in the much colder spectator part and hypernuclei are formed \cite{Dasgupta_hypernuclei2,Botvina_hypernuclei}. The fragmentation of excited hypernuclear system formed in heavy ion collisions has been already described by the canonical thermodynamical model extended to three component systems (proton, neutron and hyperon) \cite{Dasgupta_hyperon1}. Production cross-section of normal as well as strange nuclei have been calculated and some preliminary signatures of liquid-gas phase transition have been obtained for hypernuclei (this work is not included in the thesis, it can be found in the recent paper \cite{Mallik12}). But the limitation of this model is that the calculation is done at the equilibrium condition and hence features like how many hyperons get absorbed in the PLF or how it changes the excitation of the PLF these can not be calculated by the existing model. Therefore it will be interesting to determine from dynamical BUU calculation with proper inclusion of hyperon production channels and interactions. Then this hybrid hypernucler fragmentation model can be used in future for studying different theromodynamical and dynamical variables extracted from the fragmentation of hypernuclei.\\
\vskip3cm
\end{normalsize} 
\addcontentsline{toc}{chapter}{Appendix A: Participant volume calculation from abrasion model}
\chapter*{Appendix A: Participant volume calculation from abrasion model}
\begin{normalsize}
\begin{figure}[b!]
\begin{center}
\includegraphics[width=10cm,keepaspectratio=true,clip]{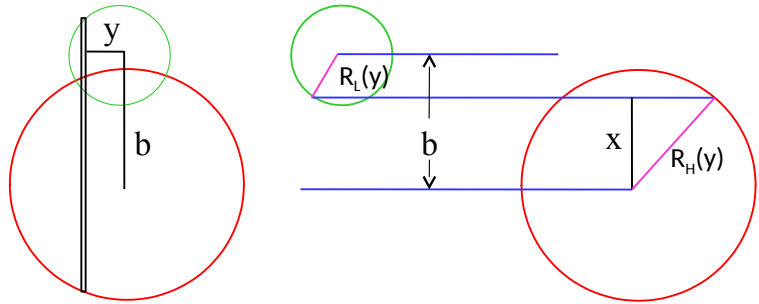}
\caption{Geometrical analysis of abrasion stage.}
\label{Collision_geometrical}
\end{center}
\end{figure}
The prescription given in Ref. \cite{Dasgupta1} is followed for calculating the volume of the participant region. The geometrical picture of the collision between two unequal nuclei is shown in Fig. \ref{Collision_geometrical}. The heavier nucleus (of radius $R_H$) collides with the lighter nucleus (of radius $R_L$) when the impact parameter $b$ is between $0$ and $R_H+R_L$. The entire collision can be visualized as a result of the collisions between thin discs (formed in the $x-z$ plane) of thickness $\delta y$ at a distance $y$ from the centre ( shown in Fig. \ref{Collision_geometrical}.(a)). The quantity $y$ can be varied from $0$ to $y_{max}$. The value $y_{max}$ depends upon the impact parameter ($b$), if $b< \sqrt{R_H^2-R_L^2}$ then $y_{max}=R_L$ else it can be determined from the expression,
\begin{equation}
\sqrt{R^2_H-y^2_{max}}+\sqrt{R^2_L-y^2_{max}}=b
\end{equation}
which gives,
\begin{equation}
y_{max}=\sqrt{R^2_L-\frac{R^2_L-R^2_H+b^2}{4b^2}}
\end{equation}
For $b \ge \sqrt{R_H^2-R_L^2}$, due to the collision of two thin discs, the surface area cut from the disc of heavier nuclei is $2\int_{x_{low}}^{x_{high}} zdx$ where $x_{low}=b-R_L(y)$ and $x_{high}=R_H$. By considering the contribution from all discs the volume of the heavy nucleus that goes into the participant region can be written as
\begin{equation}
V_H(b)=4\int_0^{y_{max}}\int_{x_{low}}^{x_{high}} \sqrt{R^2_H(y)-x^2}dx dy
\end{equation}
Similarly the volume of the light nucleus that goes into the participant region,
\begin{equation}
V_L(b)=4\int_0^{y_{max}}\int_{x_{low}}^{x_{high}} \sqrt{R^2_L(y)-(b-x)^2}dx dy
\end{equation}
For $b<\sqrt{R_H^2-R_L^2}$, the lighter nucleus cuts a portion of the heavier one to form the participant region. In this case there is only one spectator part which comes from the heavier nucleus. The volume of the collidng part of the heavier nucleus can be expressed as,
\begin{equation}
V_H(b)=4\int_0^{y_{max}}\int_{x_{low}}^{x_{high}} \sqrt{R^2_H(y)-x^2}dx dy
\end{equation}
where $x_{low}=b-R_L(y)$ and $x_{high}=b+R_L(y)$
\vskip3cm
\end{normalsize} 
\addcontentsline{toc}{chapter}{Appendix B: Thomas Fermi Method}
\chapter*{Appendix B: Thomas Fermi Method}
\begin{normalsize}
In 1920's Llewellyn Thomas and Enrico Fermi introduced an approximate semiclassical method for explaining the electron density and ground state energy of atoms \cite{Thomas,Fermi}. In this method the atom is described as uniformly distributed electron around the nucleus in a six dimensional phase space and the ground state energy of the atom is expressed as a function of local density. Later on, the Thomas-Fermi method is successfully extended for determining the ground state properties of nuclear matter as well as finite nuclei. The aim of this section is to calculate the ground state energy and corresponding density profile from Thomas-Fermi method for finite nuclei with diffuse surface. Consider a nucleus of $A_0$ nucleons. The total energy (non-relativistic) can be written as,
\begin{equation}
E=\int f(\vec {r},\vec{p})\frac{p^2}{2m}d\vec {r}d\vec {p}+\int V(\vec {r}) d\vec {r}
\label{Total_Energy}
\end{equation}
where $V(\vec {r})$ is the potential energy density at position $\vec {r}$ and $f(\vec {r},\vec{p})$ is the phase space density near $\vec {r},\vec {p}$. The goal of the Thomas-Fermi method is to minimise this energy subject $\rho({\vec {r}})=\int f(\vec {r},\vec{p})d\vec {p}$ under the condition of total particle number conservation i.e.
\begin{equation}
\int \rho({\vec {r}})d\vec {r}=A_0
\label{Total_Mass}
\end{equation}
Using Lagrange multiplier ($\lambda$), one can then do an unconstrained minimization of the quantity
\begin{equation}
E^{'}=\int f(\vec {r},\vec{p})\frac{p^2}{2m}d\vec {r}d\vec {p}+\int V(\vec {r}) d\vec {r}
+\lambda\Big(A_0-\int \rho({\vec {r}})d\vec {r}\Big)
\label{Laqrange_multiplier}
\end{equation}
For lowest energy state, at each $\vec{r}$, $f(\vec {r},\vec{p})$ is to be non-zero from 0 to some maximum $p_F(\vec{r})$. Thus,
\begin{equation}
f(r,p)=\frac{4}{h^3}\theta [p_F(r,p)-p]
\end{equation}

The factor 4 is due to spin-isospin degeneracy and assuming spherical symmetry, one can drop the vector sign on $r$ and $p$. This leads to
\begin{equation}
E^{'}=\frac{3}{10m}\Big\{\frac{3h^3}{16\pi}\Big\}^{2/3}\int \rho(r)^{5/3}d^3r+\int V(\vec {r}) d\vec {r}
+\lambda\Big(A_0-\int \rho({\vec {r}})d\vec {r}\Big)
\label{Laqrange_multiplier2}
\end{equation}
Under the variation $\rho({\vec {r}})$ to $\rho({\vec {r}})+\delta \rho({\vec {r}})$, the change in $E^{'}$ becomes,
\begin{equation}
\delta E^{'}=\int d\vec {r} \bigg[\frac{1}{2m}\Big\{\frac{3h^3}{16\pi}\Big\}^{2/3} \rho(r)^{2/3}+ U(\vec {r})
-\lambda \bigg]\delta \rho({\vec {r}})
\label{Laqrange_multiplier3}
\end{equation}
Where the potential $u(\vec {r})$ is the functional derivative of the potential energy density with respect to $\rho({\vec {r}})$. Since $\delta \rho({\vec {r}})$ is arbitrary, Eq. \ref{Laqrange_multiplier3} will be only zero if at each $\vec {r}$, the following condition holds good.
\begin{equation}
\frac{1}{2m}\Big\{\frac{3h^3}{16\pi}\Big\}^{2/3} \rho(r)^{2/3}+ U(\vec {r})-\lambda=0
\label{Laqrange_multiplier4}
\end{equation}
This is known as Thomas-Fermi equation,                                                                                                                                                            and by solving this equation ground state density profile and ground state energy can be obtained. To solve the eq. \ref{Laqrange_multiplier4}, one has to start from a guess density. For example, initially we can start with Myers density profile \cite{Myers}, which is given by,
\begin{equation}
\rho_{guess}(r)=\rho_M[1-[1+\frac{R}{a}]\exp(-R/a)\frac{sinh(r/a)}{r/a}],  r<R
\end{equation}
\begin{equation}
\rho_{guess}(r)=\rho_M[(R/a)cosh(R/a)-sinh(R/a)]\frac{e^{-r/a}}{r/a},  r>R
\end{equation}
where $\rho_M=1.18A^{1/3}$ fm and $a=1/\sqrt{2}$ fm. $\rho_M$ and $a$ determines the equivalent sharp radius and width of the surface respectively.\\
Therefore, for Skyrme+Yukawa Potential (Eq. \ref{Skyrme+Yukawa_potential}) the Thomas-Fermi equation can be written as,
\begin{equation}
\frac{1}{2m}\Big\{\frac{3h^3}{16\pi}\Big\}^{2/3} \rho(r)^{2/3}+A\rho(\vec{r})+B\rho^{\sigma}(\vec{r})=\lambda-\int u_y(\vec{r},\vec{r}')\rho(\vec{r'}) d^3r'
\label{Thomas_Fermi_Skyrme+Yukawa}
\end{equation}
To solve this equation numerically, $r$ and $\rho_{guess}(r)$ can be discretized into $\{r_i\}$,$\{\rho_{i_{guess}}\}$.  Eq. \ref{Laqrange_multiplier4} is solved by using Newton-Raphson method at each $i$ for a particular guess value of $\lambda=\lambda_1$. After iterations, the densities $\{\rho_{1_i}\}$ are obtained which satisfies the condition $\int \rho_1(r)d^3r=A_1$ (say). Similarly for $\lambda_2=\lambda_1+\delta$ ($\delta$ is a small number) and $\{\rho_{i_{guess}}\}$ one can repeat the earlier steps and another density profile $\rho_2(r)$ can be obtained which satisfies $\int \rho_2(r)d^3r=A_2$ (say). Then a better guess of $\lambda$ can be obtained from the equation $\lambda_0=\lambda_1+\{(A_0-A_1)/(A_2-A_1)\}(\lambda_2-\lambda_1)$. Again by using Newton-Raphson method with inputs as $\lambda_0$ and $\rho_{guess}(r)$, $\rho_0(r)$ is determined. If, $\int \rho_0(r)d^3r$ will be $A_0$, then the calculation can be stopped, otherwise, all the steps (by changing $\rho_{guess}(r)=\rho_0(r)$ and $\lambda_1=\lambda_0$) are repeated until the required number of particles $A_0$ is obtained. This method has been used in the past to construct Thomas-Fermi solutions relevant heavy ion collisions \cite{Gallelgo_PRC}.\\
For Skyrme potential with $\nabla^2$ correction (Eq. \ref{Lenk_potential}) the Thomas-Fermi equation will be
\begin{equation}
\frac{1}{2m}\Big\{\frac{3h^3}{16\pi}\Big\}^{2/3} \rho(r)^{2/3}+A^{\prime}\rho(\vec{r})+B^{\prime}\rho^{\sigma}(\vec{r})+\frac{C}{\rho_0^{2/3}}\nabla_r^2[\frac{\rho(\vec{r})}{\rho_0}]-\lambda=0
\label{Laqrange_multiplier_Lenk}
\end{equation}
Where $A^{\prime}=\frac{A}{\rho_0}$ and $B^{\prime}=\frac{B}{\rho_0^{\sigma}}$.By substituting $y(r)=r\rho(r)$, Eq. \ref{Laqrange_multiplier_Lenk} becomes,
\begin{equation}
\frac{1}{r}\frac{d^2y}{dr^2}+\frac{1}{2mC}\big(\frac{3h^3}{16\pi}\big)^{2/3}+\big(\frac{y}{r}\big)^{2/3}+\frac{A^{\prime}}{C}\frac{y}{r}+
\frac{B^{\prime}}{C}\big(\frac{y}{r}\big)^{\sigma}=\frac{\lambda}{C}
\label{Thomas-Fermi_Lenk}
\end{equation}
So, $y(r)$ vanishes both at $r=0$ and $r=\infty$ i.e. Eq. \ref{Thomas-Fermi_Lenk} becomes a boundary value problem. Therefore similar to the earlier case, in a finite mesh, $r$ and $y$ can be discretized into $\{r_i\}$, $\{y_i\}$, $i=$0,1,2,....$M$ ($r_M=Mh$) and the boundary conditions are implemented as $y_0=0$ and $y_M=0$. $\frac{d^2y}{dr^2}$ can be expressed in terms of finite differences,
\begin{equation}
\frac{d^2y(r_i)}{dr^2}=\frac{1}{h^2}(y_{i+1}+y_{i-1}-2y_i)
\label{Second_derivative}
\end{equation}
Substituting Eq. \ref{Second_derivative} into Eq. \ref{Thomas-Fermi_Lenk}, leads to a system of nonlinear algebraic equations for $\{y_i\}$, which can be solved by multidimensional Newton's method for $\lambda_1$. Then by applying the same procedure as mentioned in the earlier case one can determine the ground state density.\\
Once the density profile is obtained from Thomas-Fermi method, the next aim is to calculate the initial position and momenta of the test particles. To do that Monte-Carlo method is used. The radial distance of the particle from centre can be assigned by choosing,
\begin{equation}
\frac{\int^r_0 \rho(r)4\pi r^2dr}{\int^R_0 \rho(r)4\pi r^2dr}=x_1
\label{radial_distance}
\end{equation}
where $R$ is the radial distance at which $\rho(r)$ becomes $0$ and $x_1$ is random number. Similarly $\theta$ and $\phi$ can be determined by,
\begin{equation}
\frac{\int^{\theta}_0 sin\theta d\theta}{\int^{\pi}_0 sin\theta d\theta}=x_2
\label{radial_theta}
\end{equation}
\begin{equation}
\frac{\int^{\phi}_0 d\phi}{\int^{\phi}_0 d\phi}=x_3
\label{radial_phi}
\end{equation}
Hence $\theta=cos^{-1}(1-2x_2)$ and $\phi=2\pi x_3$. Here $x_2$ and $x_3$ are also two random numbers.
By knowing $\rho(r)$, one can determine Fermi momentum at $r$ from the relation,
\begin{equation}
p_F(r)=h\bigg[\frac{3\rho(r)}{16\pi}\bigg]^{1/3}
\label{Fermi_momentum}
\end{equation}
Hence, the momenta of a test particle at position $r$ can be calculated by,
\begin{equation}
\frac{\int^p_0 4\pi p^2dp}{\int^{p_F(r)}_0 4\pi p^2dp}=x_4
\label{radial_distance_momentum}
\end{equation}
i.e. $p=p_Fx^{1/3}_4$. Polar and azimuthal angle in momentum space can be obtained from relations $\theta_p=cos^{-1}(1-2x_5)$ and $\phi_p=2\pi x_6$ respectively, where $x_4$,$x_5$ and $x_6$ are random numbers.\\
Finally in the cartesian co-ordinate system the position and momenta of the test particles are
\begin{eqnarray}
r_x=r sin\theta cos\phi \nonumber\\
r_y=r sin\theta sin\phi \nonumber\\
r_z=r cos\theta \nonumber\\
p_x=p sin\theta_p cos\phi_p \nonumber\\
p_y=p sin\theta_p sin\phi_p \nonumber\\
p_z=p cos\theta_p \nonumber\\
\label{position_momentum}
\end{eqnarray}
By knowing the positions and momenta of the test particles, one can determine the potential and kinetic energies respectively.
\vskip3cm
\end{normalsize} 
\addcontentsline{toc}{chapter}{Appendix C: Algorithm for the Collision Part of BUU model}
\chapter*{Appendix C: Algorithm for the Collision Part of BUU model}
\begin{normalsize}
The prescription given in Ref. \cite{Dasgupta_BUU1} is followed for selecting the collision channels . Once the collision criteria is satisfied one has to decide through which channel it will proceed. To do that following sequence of steps are followed.\\
\indent
1. A random number ($h$) is generated between 0 and 1 and the elastic scattering cross-section is computed from Eq. \ref{Elastic_cross_section}. If
$h \le \frac{\sigma^e_{nn \rightarrow nn}(\sqrt{s})}{55}$ the collision is considered as elastic (i.e. $n+n\rightarrow n+n$, or $n+\Delta\rightarrow n+n$ or $\Delta+\Delta\rightarrow \Delta+\Delta$)  and the radial and azimuthal angle of scattering are calculated from Eq. \ref{Elastic_radial_angle} and \ref{Elastic_azimuthal_angle} respectively.\\
Otherwise ($h>\frac{\sigma^e_{nn \rightarrow nn}(\sqrt{s})}{55}$), one has to proceed for the next step which checks the possibility of inelastic collision.\\
\indent
2. For examining inelastic collision following possibilities have to be considered,\\
(i) If $\sqrt {s} \le 2.015$ GeV, both test particles represent nucleons and there is not enough energy to generate a $\Delta$.\\
(ii)If the mass of both particles is greater than 938 MeV (i.e. both test particles represents $\Delta$), no inelastic scattering can occur.\\
(iii) If one of the colliding test particles represents a nucleon and another represents a $\Delta$ particle, one has to check for $\Delta$ absorption.\\ If $h \le \frac{\sigma^e_{nn \rightarrow nn}(\sqrt{s})+\sigma^i_{n \Delta \rightarrow nn}(\sqrt{s})}{55}$, the inelastic collision ($n+\Delta\rightarrow n+n$) is successful and the magnitudes of final momenta are calculated by energy conservation condition and the directions are assumed to be isotropic.\\
If $h>\frac{\sigma^e_{nn \rightarrow nn}(\sqrt{s})+\sigma^i_{n \Delta \rightarrow nn}(\sqrt{s})}{55}$, the scattering is not allowed and the momenta of the colliding particles will not change.\\
(iv) If both of the colliding test particles represent nucleons, $\Delta$ production may occur. To study it, the random number is compared with
$\frac{\sigma^e_{nn \rightarrow nn}(\sqrt{s})+\sigma^i_{nn \rightarrow n \Delta}(\sqrt{s})}{55}$. When $h \le \frac{\sigma^e_{nn \rightarrow nn}(\sqrt{s})+\sigma^i_{nn \rightarrow n \Delta}(\sqrt{s})}{55}$, $\Delta$ production is allowed and the mass of the $\Delta$ particle is calculated from Eq. \ref{Delta_mass}. By knowing the masses, the magnitudes of momenta of the final particles are determined from energy conservation relation. The radial and azimuthal angles are determined by assuming that scattering is isotropic.\\
(v) For $h>\frac{\sigma^e_{nn \rightarrow nn}(\sqrt{s})+\sigma^i_{nn \rightarrow n \Delta}(\sqrt{s})}{55}$, the collision is forbidden and the properties of the colliding particles remain unchanged.\\
\vskip3cm
\end{normalsize} 

\end{document}